\documentclass[preprint]{ptephy}
\preprintnumber{} 
\usepackage{graphicx}
\usepackage{placeins}
\usepackage{caption}
\usepackage{lineno}
\usepackage{color}
\usepackage{multirow}
\usepackage{url}
\RequirePackage{ifthen}

\graphicspath{{./figure/}}

\usepackage{subcaption} 
\usepackage{xspace}   

\newcommand{\degree}{^\circ}

\newcommand{\neut}[1]{\ensuremath{\nu_{#1}}}
\newcommand{\numu}{\neut{\mu}}
\newcommand{\nue}{\neut{e}}
\newcommand{\nutau}{\neut{\tau}}
\newcommand{\goesto}{\ensuremath{\rightarrow}}

\newcommand{\dms}[1][NULL]{\ifthenelse{\equal{#1}{NULL}}{
	\ensuremath{\Delta m^2}}{\ensuremath{\Delta m^2_{#1}}}}
\newcommand{\sstt}[1][NULL]{\ifthenelse{\equal{#1}{NULL}}{
	\ensuremath{\sin^2 2\theta}}{\ensuremath{\sin^2 2\theta_{#1}}}}

\newcommand{\nova}{NO$\nu$A\xspace}
\newcommand{\nn}{$\nu$:$\bar{\nu}$\xspace}

\newcommand{\sqtt}{$\sin^2(\theta_{23})=0.5$\xspace}

\newcommand{\qtt}{$\sin^2(\theta_{23})$\xspace}

\newcommand{\dc}{$\Delta \chi^2$\xspace}

\newcommand{\dcp}{$\delta_{CP}$\xspace}

\newcommand{\sdnz}{$\sin\,\delta_{CP} \neq 0$\xspace}

\newcommand{\os}{$1\sigma$\xspace}

\begin{document}

\title{Neutrino Oscillation Physics Potential of the T2K Experiment}

\newcommand{\noaffiliation}{}
\newcommand{\INSTC}{1}
\newcommand{\INSTEE}{2}
\newcommand{\INSTFE}{3}
\newcommand{\INSTD}{4}
\newcommand{\INSTGA}{5}
\newcommand{\INSTI}{6}
\newcommand{\INSTGB}{7}
\newcommand{\INSTFG}{8}
\newcommand{\INSTFH}{9}
\newcommand{\INSTBA}{10}
\newcommand{\INSTEF}{11}
\newcommand{\INSTEG}{12}
\newcommand{\INSTDG}{13}
\newcommand{\INSTCB}{14}
\newcommand{\INSTED}{15}
\newcommand{\INSTEC}{16}
\newcommand{\INSTEI}{17}
\newcommand{\INSTGF}{18}
\newcommand{\INSTBE}{19}
\newcommand{\INSTBF}{20}
\newcommand{\INSTBD}{21}
\newcommand{\INSTEB}{22}
\newcommand{\INSTHA}{23}
\newcommand{\INSTCC}{24}
\newcommand{\INSTCD}{25}
\newcommand{\INSTEJ}{26}
\newcommand{\INSTFC}{27}
\newcommand{\INSTFI}{28}
\newcommand{\INSTJ}{29}
\newcommand{\INSTHB}{30}
\newcommand{\INSTCE}{31}
\newcommand{\INSTDF}{32}
\newcommand{\INSTFJ}{33}
\newcommand{\INSTGJ}{34}
\newcommand{\INSTCF}{35}
\newcommand{\INSTGG}{36}
\newcommand{\INSTBB}{37}
\newcommand{\INSTGC}{38}
\newcommand{\INSTFA}{39}
\newcommand{\INSTE}{40}
\newcommand{\INSTGD}{41}
\newcommand{\INSTBC}{42}
\newcommand{\INSTFB}{43}
\newcommand{\INSTDI}{44}
\newcommand{\INSTEH}{45}
\newcommand{\INSTCH}{46}
\newcommand{\INSTBJ}{47}
\newcommand{\INSTCG}{48}
\newcommand{\INSTGI}{49}
\newcommand{\INSTF}{50}
\newcommand{\INSTB}{51}
\newcommand{\INSTG}{52}
\newcommand{\INSTDJ}{53}
\newcommand{\INSTDH}{54}
\newcommand{\INSTFD}{55}
\newcommand{\INSTGE}{56}
\newcommand{\INSTGH}{57}
\newcommand{\INSTEA}{58}
\newcommand{\INSTH}{59}

\author{
\name{K.\,Abe}{\INSTBJ},
\name{J.\,Adam}{\INSTFJ},
\name{H.\,Aihara}{\INSTCH,\INSTHA},
\name{T.\,Akiri}{\INSTFH},
\name{C.\,Andreopoulos}{\INSTEH},
\name{S.\,Aoki}{\INSTCC},
\name{A.\,Ariga}{\INSTEE},
\name{S.\,Assylbekov}{\INSTFG},
\name{D.\,Autiero}{\INSTJ},
\name{M.\,Barbi}{\INSTE}
% ...
\\
\name{(The T2K Collaboration)}{}
}

\address{
\affil{\INSTC}{{University of Alberta, Centre for Particle Physics, Department of Physics, Edmonton, Alberta, Canada}}
\affil{\INSTEE}{{University of Bern, Albert Einstein Center for Fundamental Physics, Laboratory for High Energy Physics (LHEP), Bern, Switzerland}}
\affil{\INSTFE}{{Boston University, Department of Physics, Boston, Massachusetts, U.S.A.}}
\affil{\INSTD}{{University of British Columbia, Department of Physics and Astronomy, Vancouver, British Columbia, Canada}}
\affil{\INSTGA}{{University of California, Irvine, Department of Physics and Astronomy, Irvine, California, U.S.A.}}
\affil{\INSTI}{{IRFU, CEA Saclay, Gif-sur-Yvette, France}}
\affil{\INSTGB}{{University of Colorado at Boulder, Department of Physics, Boulder, Colorado, U.S.A.}}
\affil{\INSTFG}{{Colorado State University, Department of Physics, Fort Collins, Colorado, U.S.A.}}
\affil{\INSTFH}{{Duke University, Department of Physics, Durham, North Carolina, U.S.A.}}
\affil{\INSTBA}{{Ecole Polytechnique, IN2P3-CNRS, Laboratoire Leprince-Ringuet, Palaiseau, France }}
\affil{\INSTEF}{{ETH Zurich, Institute for Particle Physics, Zurich, Switzerland}}
\affil{\INSTEG}{{University of Geneva, Section de Physique, DPNC, Geneva, Switzerland}}
\affil{\INSTDG}{{H. Niewodniczanski Institute of Nuclear Physics PAN, Cracow, Poland}}
\affil{\INSTCB}{{High Energy Accelerator Research Organization (KEK), Tsukuba, Ibaraki, Japan}}
\affil{\INSTED}{{Institut de Fisica d'Altes Energies (IFAE), Bellaterra (Barcelona), Spain}}
\affil{\INSTEC}{{IFIC (CSIC \& University of Valencia), Valencia, Spain}}
\affil{\INSTEI}{{Imperial College London, Department of Physics, London, United Kingdom}}
\affil{\INSTGF}{{INFN Sezione di Bari and Universit\`a e Politecnico di Bari, Dipartimento Interuniversitario di Fisica, Bari, Italy}}
\affil{\INSTBE}{{INFN Sezione di Napoli and Universit\`a di Napoli, Dipartimento di Fisica, Napoli, Italy}}
\affil{\INSTBF}{{INFN Sezione di Padova and Universit\`a di Padova, Dipartimento di Fisica, Padova, Italy}}
\affil{\INSTBD}{{INFN Sezione di Roma and Universit\`a di Roma ``La Sapienza'', Roma, Italy}}
\affil{\INSTEB}{{Institute for Nuclear Research of the Russian Academy of Sciences, Moscow, Russia}}
\affil{\INSTHA}{{Kavli Institute for the Physics and Mathematics of the Universe (WPI), Todai Institutes for Advanced Study, University of Tokyo, Kashiwa, Chiba, Japan}}
\affil{\INSTCC}{{Kobe University, Kobe, Japan}}
\affil{\INSTCD}{{Kyoto University, Department of Physics, Kyoto, Japan}}
\affil{\INSTEJ}{{Lancaster University, Physics Department, Lancaster, United Kingdom}}
\affil{\INSTFC}{{University of Liverpool, Department of Physics, Liverpool, United Kingdom}}
\affil{\INSTFI}{{Louisiana State University, Department of Physics and Astronomy, Baton Rouge, Louisiana, U.S.A.}}
\affil{\INSTJ}{{Universit\'e de Lyon, Universit\'e Claude Bernard Lyon 1, IPN Lyon (IN2P3), Villeurbanne, France}}
\affil{\INSTHB}{{Michigan State University, Department of Physics and Astronomy,  East Lansing, Michigan, U.S.A.}}
\affil{\INSTCE}{{Miyagi University of Education, Department of Physics, Sendai, Japan}}
\affil{\INSTDF}{{National Centre for Nuclear Research, Warsaw, Poland}}
\affil{\INSTFJ}{{State University of New York at Stony Brook, Department of Physics and Astronomy, Stony Brook, New York, U.S.A.}}
\affil{\INSTGJ}{{Okayama University, Department of Physics, Okayama, Japan}}
\affil{\INSTCF}{{Osaka City University, Department of Physics, Osaka, Japan}}
\affil{\INSTGG}{{Oxford University, Department of Physics, Oxford, United Kingdom}}
\affil{\INSTBB}{{UPMC, Universit\'e Paris Diderot, CNRS/IN2P3, Laboratoire de Physique Nucl\'eaire et de Hautes Energies (LPNHE), Paris, France}}
\affil{\INSTGC}{{University of Pittsburgh, Department of Physics and Astronomy, Pittsburgh, Pennsylvania, U.S.A.}}
\affil{\INSTFA}{{Queen Mary University of London, School of Physics and Astronomy, London, United Kingdom}}
\affil{\INSTE}{{University of Regina, Department of Physics, Regina, Saskatchewan, Canada}}
\affil{\INSTGD}{{University of Rochester, Department of Physics and Astronomy, Rochester, New York, U.S.A.}}
\affil{\INSTBC}{{RWTH Aachen University, III. Physikalisches Institut, Aachen, Germany}}
\affil{\INSTFB}{{University of Sheffield, Department of Physics and Astronomy, Sheffield, United Kingdom}}
\affil{\INSTDI}{{University of Silesia, Institute of Physics, Katowice, Poland}}
\affil{\INSTEH}{{STFC, Rutherford Appleton Laboratory, Harwell Oxford,  and  Daresbury Laboratory, Warrington, United Kingdom}}
\affil{\INSTCH}{{University of Tokyo, Department of Physics, Tokyo, Japan}}
\affil{\INSTBJ}{{University of Tokyo, Institute for Cosmic Ray Research, Kamioka Observatory, Kamioka, Japan}}
\affil{\INSTCG}{{University of Tokyo, Institute for Cosmic Ray Research, Research Center for Cosmic Neutrinos, Kashiwa, Japan}}
\affil{\INSTGI}{{Tokyo Metropolitan University, Department of Physics, Tokyo, Japan}}
\affil{\INSTF}{{University of Toronto, Department of Physics, Toronto, Ontario, Canada}}
\affil{\INSTB}{{TRIUMF, Vancouver, British Columbia, Canada}}
\affil{\INSTG}{{University of Victoria, Department of Physics and Astronomy, Victoria, British Columbia, Canada}}
\affil{\INSTDJ}{{University of Warsaw, Faculty of Physics, Warsaw, Poland}}
\affil{\INSTDH}{{Warsaw University of Technology, Institute of Radioelectronics, Warsaw, Poland}}
\affil{\INSTFD}{{University of Warwick, Department of Physics, Coventry, United Kingdom}}
\affil{\INSTGE}{{University of Washington, Department of Physics, Seattle, Washington, U.S.A.}}
\affil{\INSTGH}{{University of Winnipeg, Department of Physics, Winnipeg, Manitoba, Canada}}
\affil{\INSTEA}{{Wroclaw University, Faculty of Physics and Astronomy, Wroclaw, Poland}}
\affil{\INSTH}{{York University, Department of Physics and Astronomy, Toronto, Ontario, Canada}}
\email{ichikawa@scphys.kyoto-u.ac.jp}
}


\begin{abstract}
The observation of the recent electron neutrino appearance 
in a muon neutrino beam and 
the high-precision measurement of the mixing angle $\theta_{13}$ 
have led to a re-evaluation of the physics potential 
of the T2K long-baseline neutrino oscillation experiment.
Sensitivities are explored for CP violation in neutrinos,
non-maximal $\sin^22\theta_{23}$, the octant of $\theta_{23}$,
and the mass hierarchy,
in addition to the measurements of $\delta_{CP}$, $\sin^2\theta_{23}$,
and $\Delta m^2_{32}$, for various combinations of
$\nu$-mode and \(\bar{\nu}\)-mode data-taking. 

With an exposure of $7.8\times10^{21}$~protons-on-target, T2K can achieve 1-$\sigma$ resolution of 0.050(0.054) on $\sin^2\theta_{23}$ and $0.040(0.045)\times10^{-3}~\rm{eV}^2$ on $\Delta m^2_{32}$ for 100\%(50\%) neutrino beam mode running assuming $\sin^2\theta_{23}=0.5$ and $\Delta m^2_{32} = 2.4\times10^{-3}$ eV$^2$. 
T2K will have sensitivity to the CP-violating phase $\delta_{\rm{CP}}$ 
at 90\% C.L. or better over a significant range.
For example, if $\sin^22\theta_{23}$ is maximal (i.e $\theta_{23}$=$45^\circ$) 
the range is
 $-115^\circ<\delta_{\rm{CP}}<-60^\circ$ for normal hierarchy and $+50^\circ<\delta_{\rm{CP}}<+130^\circ$ 
 for inverted hierarchy.  
When T2K data is combined with data from the NO$\nu$A experiment,
the region of oscillation parameter space 
where there is sensitivity to observe a non-zero $\delta_{CP}$ 
is substantially increased compared to if each experiment is analyzed alone.
\end{abstract}

\maketitle

\linenumbers

\section{Introduction}
\label{introduction}
The experimental confirmation of neutrino oscillations, where neutrinos of a particular flavor 
($\nu_e$,$\nu_{\mu}$,$\nu_{\tau}$) can transmute to another flavor, has profound implications for physics. 
The observation of a zenith-angle-dependent deficit 
in muon neutrinos produced by high-energy proton interactions 
in the atmosphere~\cite{Fukuda:1998mi} confirmed 
the neutrino flavor oscillation hypothesis. 
The ``anomalous'' solar neutrino flux~\cite{Cleveland:1998ApJ} problem
was shown to be due to neutrino oscillation 
by more precise measurements~\cite{Abe:2010hy, Aharmim:2011vm, 
Collaboration:2011nga, Eguchi:2002dm}. 
Atmospheric neutrino measurements have provided further precision 
on the disappearance of muon neutrinos~\cite{Wendell:2010md, Adamson:2013whj} 
and the appearance of tau neutrinos~\cite{Abe:2012jj}. 
Taking advantage of nuclear reactors as intense sources, 
the disappearance of electron antineutrinos has been firmly established 
using both widely distributed multiple sources 
at an average distance of 180~km~\cite{Eguchi:2002dm}
and from specialized detectors placed 
within $\sim2$~km~\cite{An:2012eh, Abe:2012tg, Ahn:2012nd}. 
The development of high-intensity proton accelerators that can produce 
focused neutrino beams with mean energy from a few hundred MeV to tens of GeV 
have enabled measurements of the disappearance of muon-neutrinos 
(and muon antineutrinos)~\cite{Adamson:2013whj, Ahn:2006zza, numurun4} 
and appearance of electron-neutrinos 
(and electron antineutrinos)~\cite{Abe:2011sj, Abe:2013xua, Abe:2013hdq, 
Adamson:2013ue} 
and tau-neutrinos~\cite{Agafonova:2010dc} over distances of hundreds of 
kilometers.

While the early solar and atmospheric oscillation experiments could be described in a two-neutrino framework, recent experiments 
with diverse neutrino sources support a three-flavor oscillation framework.
In this scenario, the three neutrino flavor eigenstates mix 
with three mass eigenstates ($\nu_1$, $\nu_2$, $\nu_3$) 
through the Pontecorvo-Maki-Nakagawa-Sakata \cite{Maki:1962mu} (PMNS) matrix 
in terms of three mixing angles ($\theta_{12}$, $\theta_{23}$, $\theta_{13}$) 
and one complex phase ($\delta_{CP}$). 
The probability of neutrino oscillation depends on these parameters, 
as well as the difference of the squared masses of the mass states 
($\Delta m^2_{21}$, $\Delta m^2_{31}$, $\Delta m^2_{32}$). 
Furthermore, there is an explicit dependence on the energy of the neutrino ($E_\nu$) and the 
distance traveled ($L$) before detection. 
To date, all the experimental results are well-described 
within the neutrino oscillation framework as described 
in Sec.~\ref{sec:oscillation}. 

T2K is a long-baseline neutrino oscillation experiment proposed in 2003~\cite{T2KLOI} 
with three main physics goals that were to be achieved
with data corresponding to $7.8\times 10^{21}$ protons-on-target (POT)
from a 30 GeV proton beam:
\begin{itemize}
\item 
search for \numu{}\goesto{}\nue{} appearance and establish  
that $\theta_{13} \ne 0$ with a sensitivity 
down to $\sstt{}_{13} \sim 0.008 (90\%~\mathrm{C.L.})$;
\item precision measurement of oscillation parameters in \numu{} disappearance 
with $\delta{}(\dms{}_{32}) \sim 10^{-4}$~eV$^2$ 
and $\delta{}(\sstt{}_{23}) \sim 0.01$ ; and 
\item search for sterile components in \numu{} disappearance.
\end{itemize}

The T2K experiment began data taking in 2009~\cite{Abe:2011ks} and 
a major physics goal, the discovery of \numu{}\goesto{}\nue{} appearance, 
has been realized at 7.3 $\sigma$ level of significance with just 8.4\% 
of the total approved POT \cite{Abe:2013hdq}. 
This is the first time an explicit flavor appearance has been observed 
from another neutrino flavor with significance larger than 5$\sigma$.  This
observation opens the door to study CP violation (CPV) in neutrinos as described in Sec.~\ref{sec:oscillation}.
Following this discovery, the primary physics goal 
for the neutrino physics community has become a detailed investigation 
of the three-flavor paradigm which requires 
determination of the CP-violating phase $\delta_{CP}$, 
resolution of the mass hierarchy (MH),
precise measurement of $\theta_{23}$ to determine
how close $\theta_{23}$ is to $45^\circ$,
and determination of the $\theta_{23}$ octant, {\it i.e.}, whether the 
mixing angle $\theta_{23}$ is less than or greater than $45^\circ$. 
T2K, along with the \nova \cite{novaTDR} experiment 
that recently began operation, 
will lead in the determination of these parameters 
for at least a decade.

This paper provides a comprehensive update of the anticipated sensitivity of the
T2K experiment to the oscillation parameters as given in the original T2K
proposal \cite{T2KLOI},
and includes an investigation of the enhancements from
performing combined fits including the projected \nova sensitivity.
It starts with a brief overview of the neutrino oscillation framework in Sec.\
\ref{sec:oscillation}, and a description of the T2K experiment in Sec.\ \ref{t2k}.  Updated T2K
sensitivities are given in Sec.\ \ref{sec:t2ksensitivity}, while sensitivities when results from
T2K are combined with those from the \nova experiment are given in Sec.\ \ref{pfst2knova}.  Finally,
results of a study of the optimization of the \(\nu\) and \(\bar{\nu}\) running time for both
T2K and \nova are given in Sec.\ \ref{sec:runratio}.


\section{Neutrino Mixing and Oscillation Framework}
\label{sec:oscillation}
Three-generation neutrino mixing can be described by a unitary
matrix, often referred to as the
PMNS matrix. The weak flavor
eigenstates, \nue{}, \numu{}, and \nutau{} are related to the mass
eigenstates, $\nu_1$, $\nu_2$, and $\nu_3$, by the unitary mixing
matrix $U$:
\begin{equation}
  \left(
    \begin{array}{c}
      \nue \\ \numu{}\\ \nutau
    \end{array}
  \right)
  =
  \left[
    \begin{array}{ccc}
      U_{e1} & U_{e2} & U_{e3} \\
      U_{\mu1} & U_{\mu2} & U_{\mu3} \\
      U_{\tau1} & U_{\tau2} & U_{\tau3} \\
    \end{array}
  \right]
  \left(
    \begin{array}{c}
      \nu_1 \\ \nu_2 \\ \nu_3
    \end{array}
  \right)
\end{equation}
where the matrix is commonly parameterized as
\begin{equation}
  U_{PMNS}
  =
  \left[
    \begin{array}{ccc}
      1 & 0 & 0 \\
      0 & C_{23} & S_{23} \\
      0 & -S_{23} & C_{23} \\
    \end{array}
  \right]
  \left[
    \begin{array}{ccc}
      C_{13} & 0 & S_{13}e^{-i\delta_{CP}} \\
      0 & 1 & 0 \\
      - S_{13}e^{+i\delta_{CP}} & 0 & C_{13} \\
    \end{array}
  \right]
  \left[
    \begin{array}{ccc}
      C_{12} & S_{12} & 0 \\
      - S_{12} & C_{12} & 0 \\
      0 & 0 & 1 \\
    \end{array}
  \right]
\end{equation}
with $C_{ij}$ ($S_{ij}$) representing $\cos{\theta_{ij}}$
($\sin{\theta_{ij}}$), where $\theta_{ij}$ is the mixing angle
between the generations $i$ and $j$.  There is one
irreducible phase, $\delta_{CP}$, allowed in a unitary 3$\times$3
mixing matrix.\footnote{If the neutrino is a Majorana particle, two additional 
phases are allowed that have no consequences for neutrino oscillations.} 
After neutrinos propagate through vacuum, the probability that they will 
interact via one of the three flavors will depend on the values of these
mixing angles.
As neutrinos propagate through matter, coherent forward scattering 
of electron-neutrinos causes a change in the effective neutrino mass 
that leads to a modification of the oscillation probability. 
This is the so-called {\it matter effect}.
Interference between multiple
terms in the transition probability can lead to CP violation in
neutrino mixing if the phase $\delta_{CP}$ is non-zero.

For T2K, the neutrino oscillation modes of interest are
the $\nu_\mu \rightarrow \nu_e$ appearance mode and the
$\nu_\mu$ disappearance mode. 
The $\nu_\mu \rightarrow \nu_e$ appearance 
oscillation probability 
(to first order approximation in the matter effect\cite{Arafune:1997hd}) 
is given by
\begin{eqnarray}
\label{eq:numu2nue}
P(\nu_\mu \rightarrow \nu_e) = & 4C_{13}^2S_{13}^2S_{23}^2\sin^2\Phi_{31}
(1+\frac{2a}{\Delta m_{31}^2}(1-2S_{13}^2))\nonumber\\
 & +8C_{13}^2S_{12}S_{13}S_{23}(C_{12}C_{23}\cos\delta_{CP}-S_{12}S_{13}S_{23})
\cos\Phi_{32}\sin\Phi_{31}\sin\Phi_{21} \nonumber\\
 & -8C_{13}^2C_{12}C_{23}S_{12}S_{13}S_{23}\sin\delta_{CP}\sin\Phi_{32}
\sin\Phi_{31}\sin\Phi_{21} \\
 & +4S_{12}^2C_{13}^2(C_{12}^2C_{23}^2+S_{12}^2S_{23}^2S_{13}^2-2C_{12}C_{23}
S_{12}S_{23}S_{13}\cos\delta_{CP})\sin^2\Phi_{21} \nonumber\\
 & -8C_{13}^2S_{13}^2S_{23}^2(1-2S_{13}^2)\frac{aL}{4E_\nu}\cos\Phi_{32}
\sin\Phi_{31}\nonumber\nonumber,
\end{eqnarray}
where $\Phi_{ji}=\Delta m^2_{ji}L/4E_\nu$.
The terms that include
$a\equiv 2\sqrt{2}G_Fn_eE_\nu
=7.56\times 10^{-5}$[eV$^2$]$(\frac{\rho}{[g/cm^3]})(\frac{E_\nu}{[GeV]})$
are a consequence of the matter effect,
where $n_e$ and $\rho$ are the electron and matter densities, respectively. 
The equivalent expression for antineutrino appearance, $\bar{\nu}_\mu \rightarrow \bar{\nu}_e$, 
is obtained by reversing the signs of terms proportional to 
$\sin\delta_{CP}$ and $a$.
The first and fourth terms of Eq.\ref{eq:numu2nue} come from oscillations
induced
by $\theta_{13}$ and $\theta_{12}$, respectively, in the presence
of non-zero $\theta_{23}$.
The second and third terms come from interference caused by these oscillations.
At the T2K peak energy of $\sim 0.6$~GeV and baseline length of $L=$295~km,
$\cos\Phi_{32}$ is nearly zero and the second and fifth terms vanish.
The fourth term, to which solar neutrino disappearance is attributed,
is negligibly small. Hence, the dominant contribution for $\nu_e$ appearance
in the T2K experiment comes from the first and third terms.
The contribution from the matter effect is about 10\% of
the first term without the matter effect.
Since the third term contains $\sin\delta_{CP}$,
it is called the `CP-violating' term.
It is as large as 27\% of the first term without the matter effect
when $\sin\delta_{CP}=1$ and $\sin^22\theta_{23}=1$,
meaning that the CP-violating term makes a
non-negligible contribution to the total \(\nu_e\) appearance probability.
The measurement of $\theta_{13}$ from the reactor experiments is independent of
the CP phase, and future measurements from Daya Bay~\cite{An:2012eh}, 
Double Chooz~\cite{Abe:2012tg} and RENO~\cite{Ahn:2012nd}
will reduce the \(\theta_{13}\) uncertainty such that 
the significance of the CP-violating term will be enhanced for T2K.
It is also important to recognize that since the sign of the CP-violating term 
is opposite for neutrino and  antineutrino oscillations, 
data taken by T2K with an antineutrino beam for comparison to neutrino data 
may allow us to study CP violation effects directly.

The $\nu_\mu$ disappearance oscillation 
probability is given by
\begin{eqnarray}
\label{eq:numu2numu}
1 - P(\nu_\mu \rightarrow \nu_\mu) = (C_{13}^4\sin^22\theta_{23}+S_{23}^2
\sin^22\theta_{13})\sin^2\Phi_{32}
\end{eqnarray}
(where other matter effect and $\Delta m^2_{21}$ terms can be neglected). 
The $\nu_\mu$ disappearance measurement is
sensitive to \(\sin^22\theta_{23}\) and \(\Delta m^2_{32}\).
Currently, the measured value of $\sin^22\theta_{23}$ is consistent
with full mixing, but 
more data are required to know if that is the case. 
If the mixing is not maximal, 
the $\nu_e$ appearance data, together
with the $\nu_\mu$ disappearance data, have the potential to resolve 
the $\theta_{23}$ octant degeneracy because the first term of
Eq.\ref{eq:numu2nue} is proportional to $\sin^2\theta_{23}$.

The \nova experiment is similar to T2K in the basic goals 
to measure $\nu_\mu$ disappearance and $\nu_e$ appearance 
in an off-axis muon neutrino beam. 
The most important difference between the two experiments 
is the distance from the neutrino source to the far detector, 
810 km for \nova and 295 km for T2K, 
with a correspondingly higher peak neutrino beam energy for \nova 
to maximize the appearance probability.
\nova is projected to have similar sensitivity compared to T2K for 
$\theta_{23}$, $\theta_{13}$, and $\delta_{CP}$,
but better sensitivity to the sign of $\Delta m_{32}^2$ since, 
as can be seen in $a$ in Eq.~\ref{eq:numu2nue}, 
the size of the matter effect is proportional to the distance 
$L$. 
The combination of results from the two experiments at different baselines will 
further improve the sensitivity to the sign of $\Delta m_{32}^2$ and to $\delta_{\rm CP}$.
  
In this paper we present the updated T2K sensitivity 
to neutrino oscillation parameters using a large value of $\sin^22\theta_{13}$
similar to that measured by the reactor experiments, together with the sensitivity 
when projected T2K and $\rm NO \nu A$ results are combined.

The latest measured values of the neutrino mixing parameters 
($\theta_{12}$, $\theta_{23}$, $\theta_{13}$, $|\Delta m_{32}^2|$, $\Delta m_{21}^2$, $\delta_{CP}$) 
are listed in Table~\ref{ta:osc:para}~\cite{Beringer:1900zz}.
The CP-violating phase, $\delta_{CP}$, is not yet well constrained, nor is the sign of $\Delta m_{32}^2\equiv m_3^2-m_2^2$ known.
The sign of $\Delta m_{32}^2$ is related 
to the ordering of the three mass eigenstates; 
the positive sign is referred to as the normal MH (NH) 
and the negative sign as the inverted MH (IH).
Of the mixing angles, the angle $\theta_{23}$ is measured
with the least precision; the value of $\sin^22 \theta_{23}$ 
in Table~\ref{ta:osc:para}
corresponds to $0.4 < \sin^2(\theta_{23}) < 0.6$.
Many theoretical models, e.g.\ some based on flavor symmetries
and some on random draws on parameter spaces, sometimes try to explain
the origin of the PMNS matrix together with
the Cabibbo-Kobayashi-Maskawa matrix, which describes mixing
in the quark sector.
Precise determination of how close
this mixing angle is to $45^\circ$ would be an important piece of understanding
the origin of flavor mixing of both quarks and leptons.
\begin{table}[h]
\caption{Neutrino oscillation parameters from~\cite{Beringer:1900zz}.} 
\label{ta:osc:para}
\begin{center}
\begin{tabular}{c|c}
\hline \hline
Parameter & Value \\
\hline
$\sin^22 \theta_{12}$ & $0.857 \pm 0.024$ \\
$\sin^22 \theta_{23}$ & $> 0.95$ \\
$\sin^22 \theta_{13}$ & $0.095 \pm 0.010$ \\
$\Delta m_{21}^2$ & $(7.5 \pm 0.20) \times 10^{-5}$~$\rm eV^2$ \\
$|\Delta m_{32}^2|$ & $(2.32^{+0.12}_{-0.08}) \times 10^{-3}$~$\rm eV^2$ \\
$\delta_{CP}$ & unknown \\
\hline \hline
\end{tabular}
\end{center}
\end{table}%

\section{T2K Experiment}
\label{t2k}
The T2K experiment~\cite{Abe:2011ks} uses a 30-GeV proton
beam accelerated by the J-PARC accelerator facility. This
is composed of (1) the muon neutrino beamline, 
(2) the near detector complex, which is located 280 m downstream 
of the neutrino production target, monitors the beam, 
and constrains the neutrino flux parameterization and cross sections, 
and (3) the far detector, Super-Kamiokande (Super-K), which detects neutrinos 
at a baseline distance of 295~km from the target. The neutrino beam 
is directed 2.5$^\circ$ away from Super-K, producing a narrow-band 
$\nu_{\mu}$ beam~\cite{PhysRevD.87.012001} 
 at the far detector.  The off-axis angle is chosen such that the 
energy peaks at  $E_\nu$=$\Delta m^2_{32}L/2\pi$ $\approx 0.6$~GeV, 
which corresponds to the first oscillation minimum of the $\nu_\mu$ 
survival probability at Super-K. 
This enhances the sensitivity to  $\theta_{13}$ and $\theta_{23}$ 
and reduces backgrounds from higher-energy neutrino 
interactions at Super-K.

The J-PARC main ring accelerator provides a fast-extracted high-intensity 
proton beam to a graphite target located in the first of three consecutive
electro-magnetic horns.
Pions and kaons produced in the target are focused by the 
horns and decay in flight to muons and \(\nu_\mu\)'s in the helium-filled 
96-m-long decay tunnel.
This is followed by a beam dump and
a set of muon monitors, which are used to monitor the direction and stability 
of the neutrino beam.

The near detector complex contains an on-axis Interactive Neutrino 
Grid detector (INGRID)~\cite{Abe2012} and an off-axis magnetized detector, 
ND280.  
INGRID 
measures the neutrino interaction event rate at
various positions from 0$^\circ$ to $\sim1^\circ$ around the beam axis, and
provides monitoring of the intensity, 
direction, profile, and stability of the neutrino beam.
The ND280 off-axis detector measures neutrino beam properties
and neutrino interactions at approximately the same off-axis angle
as Super-K.
It is enclosed in a 0.2-T magnet that contains 
a subdetector optimized to measure $\pi^0$s (P$\O$D)~\cite{Assylbekov201248},
three time projection chambers (TPC1,2,3)~\cite{Abgrall:2010hi} 
alternating with two one-ton fine-grained detectors 
(FGD1,2)~\cite{Amaudruz:2012pe}, 
and an electromagnetic calorimeter (ECal) that surrounds the 
TPC, FGD, and P$\O$D detectors. A side muon range detector 
(SMRD)~\cite{Aoki:2012mf} built into slots in the magnet
return-yoke steel detects muons that exit or 
stop in the magnet steel.
A schematic diagram of the detector layout has been published 
elsewhere~\cite{Abe:2011ks}.

The Super-K water Cherenkov far detector~\cite{Ashie:2005ik} has a fiducial 
mass of 22.5 kt contained within a cylindrical inner detector (ID) 
instrumented with 11,129 inward facing 20-inch phototubes. Surrounding 
the ID is a 2-meter wide outer detector (OD) with 1,885 outward-facing 
8-inch phototubes. 
A Global Positioning System receiver with $<$150 ns precision 
synchronizes the timing between reconstructed Super-K events and the J-PARC  
beam spill.   
 
T2K employs various analysis methods to estimate oscillation parameters from the data, but
in general it is done
by comparing the observed and
predicted $\nu_e$ and $\nu_\mu$ interaction rates and 
energy spectra at the far detector. 
The rate and spectrum
depend on the oscillation parameters, the incident neutrino flux,
neutrino interaction cross sections,
and the detector response.
The initial estimate of the neutrino flux is determined from detailed
simulations incorporating
proton beam measurements, INGRID measurements,
and pion and kaon production measurements from 
the NA61/SHINE~\cite{Abgrall:2011ae, Abgrall:2011ts}
experiment. The ND280 detector measurement of  $\nu_\mu$
charged current (CC) events is used to constrain the initial flux estimates
and
parameters of the neutrino interaction models that affect the
predicted rate and
spectrum of neutrino interactions at both ND280 and Super-K.
At Super-K,  $\nu_e$ and $\nu_\mu$ charged current 
quasi-elastic (CCQE) events, for which the neutrino energy can be 
reconstructed using simple kinematics, are selected.
Efficiencies and backgrounds are determined through detailed 
simulations tuned to control samples which account for 
final state interactions (FSI) inside the nucleus
and secondary hadronic interactions (SI) in the detector material.
These combined results are used in a fit 
to determine the oscillation parameters.

As of May 2013, T2K has accumulated $6.57 \times 10^{20}$~POT, which corresponds to 
about 8.4\% of the total approved data. 
Results from this dataset on the measurement of $\theta_{23}$ 
and $|\Delta m_{32}^2|$ by $\nu_\mu$ disappearance \cite{numurun4},
and of $\theta_{13}$ and $\delta_{CP}$ by $\nu_e$ appearance 
have been published \cite{Abe:2013hdq}.
It is reported in \cite{Abe:2013hdq} that combining the T2K result with
the world average value of $\theta_{13}$ from reactor experiments
leads to some values of $\delta_{CP}$ being disfavored at 90\% CL.

\section{T2K Projected Sensitivities to Neutrino Oscillation Parameters}
\label{sec:t2ksensitivity}
To demonstrate the T2K physics potential,
we have performed sensitivity studies using combined fits 
to the reconstructed energy spectra of $\nu_e$($\bar{\nu_e}$)
and $\nu_\mu$($\bar{\nu_\mu}$) events observed at Super-K
with both \(\nu\)-mode beam, and \(\bar{\nu}\)-mode beam
in the three-flavor mixing model.
Results shown here generally use the systematic errors established for the 2012 oscillation 
analyses \cite{Abe:2012gx,Abe:2013xua} as described below, although, in addition, we have studied cases 
with projected systematic errors
as described in Sec.~\ref{sec:pfssys}.

Since the sensitivity depends on the true values of the oscillation parameters,
a set of oscillation parameters ($\theta$) is chosen as a test point 
for each study
and is used to generate simulated `observed' reconstructed energy spectra.  
Then, a hypothesis test for the set of parameters of interest ($H_0$)
is applied 
using
\begin{equation}
\Delta\chi^2 = \chi^2(H_0)-\chi^2_{min}.
\label{eq:chi2test}
\end{equation}
The value of $\chi^2(H_0)$ is calculated as $-2\ln\mathcal{L}(\theta|H_0)$, 
where $\mathcal{L}(\theta|H_0)$ is the likelihood to observe the
spectrum generated at $\theta$ when the `true' oscillation parameters are given
by $H_0$.
The minimum value of $\chi^2$ 
in the oscillation parameter space is given by $\chi^2_{min}$.
The oscillation parameter set which gives $\chi^2_{min}$ is equivalent 
to $\theta$, since spectra are generated without statistical fluctuations
in this analysis.
When we test only one or two of the five varied oscillation parameters 
(\(\sin^22\theta_{13}\), 
\(\delta_{CP}\), \(\sin^2\theta_{23}\), \(\Delta m^2_{32}\), and the MH), the
tested parameters are fixed at a set of test points, and the remaining oscillation parameters are fit to give 
a minimized $\chi^2(H_0)$.

In most cases, this $\Delta\chi^2$ closely resembles 
a $\chi^2$ distribution for $n$ degrees of freedom, where
$n$ corresponds to the number of tested oscillation parameters.
Then, critical $\chi^2$ values for Gaussian distributed variables
($\Delta\chi^2_{critical}$)
can be used for determining confidence level (C.L.) regions~\cite{PDG2013}.
Each simulated spectrum is generated at the MC sample statistical mean, 
and therefore the results of this test represent the median sensitivity.
Thus the results of these studies indicate 
that half of experiments are expected to be able to reject
$H_0$ at the reported C.L.
This is accurate if two conditions are met: 
(1) the probability density function (pdf) for $\Delta\chi^2$ follows 
a true $\chi^{2}$ distribution, and 
(2) the $\Delta\chi^2$ value calculated with the MC sample statistical mean
spectra ($\bar{\Delta\chi^2}$) is equivalent to the median of
the $\Delta\chi^2$ pdf.
Then, $\bar{\Delta\chi^2}$ can be used to construct
median sensitivity C.L. contours.
Studies using ensembles of toy MC experiments where statistical
fluctuations expected at a given POT and
systematic fluctuations are included have shown
that calculating C.L.s
  by applying a $\Delta\chi^2_{critical}$ value to $\Delta\chi^2$ 
  gives fairly consistent C.L.s, and that $\bar{\Delta\chi^2}$
  is in good agreement with the median $\Delta\chi^2$ value of each ensemble of
  toy MC experiments, except in the case of a mass hierarchy determination.
Therefore, in this paper we show C.L.s constructed by applying the
  $\Delta\chi^2_{critical}$ value to $\bar{\Delta\chi^2}$
  as our median sensitivity.
  The exception of the MH case will be discussed in detail
in Sec.~\ref{pfst2knova}.

\subsection{Expected observables and summary of current systematic errors}
Our sensitivity studies are based on the signal efficiency, background,  and 
systematic errors established for the T2K 2012 oscillation analyses\cite{Abe:2012gx,Abe:2013xua};
however, we note that errors are lower
in more recent published analyses.  
Since official T2K systematic errors are used, these errors have been
reliably estimated based on data analysis, unlike previous sensitivity studies which use 
errors based only on simulation and estimations \cite{T2KLOI}.
Systematic errors therefore include both normalization and shape errors, 
and are implemented as a covariance matrix for these studies, where full
correlation between \(\nu\)- and \(\bar{\nu}\)-modes is generally assumed.

For the \(\nu_e\) sample, 
interaction candidate events fully contained in the fiducial volume with a
single electron-like Cherenkov ring are selected.  
The visible energy is required to
exceed 100 MeV/c, events with a delayed electron signal are rejected, and 
events with an invariant mass near that of the \(\pi^0\) are rejected, where the invariant mass is reconstructed assuming the existence of a second ring.
Finally, events are required to have a reconstructed neutrino energy below 1250 MeV.
The efficiency of the event selection for the CC \(\nu_e\) signal
is 62\% and the fraction of CCQE events in the signal is 80\%.
For the \(\nu_\mu\) sample, again events must be fully contained in the fiducial
volume, but they must now have a single muon-like Cherenkov ring with a momentum
exceeding 200 MeV/c.  There must be either zero or one delayed electron. 
The efficiency and purity of \(\nu_\mu\) CCQE events
are estimated to be 72\% and 61\%, respectively.

\begin{figure}[tbp]
\centering \begin{subfigure}[b]{7cm}
\includegraphics[width=7cm]{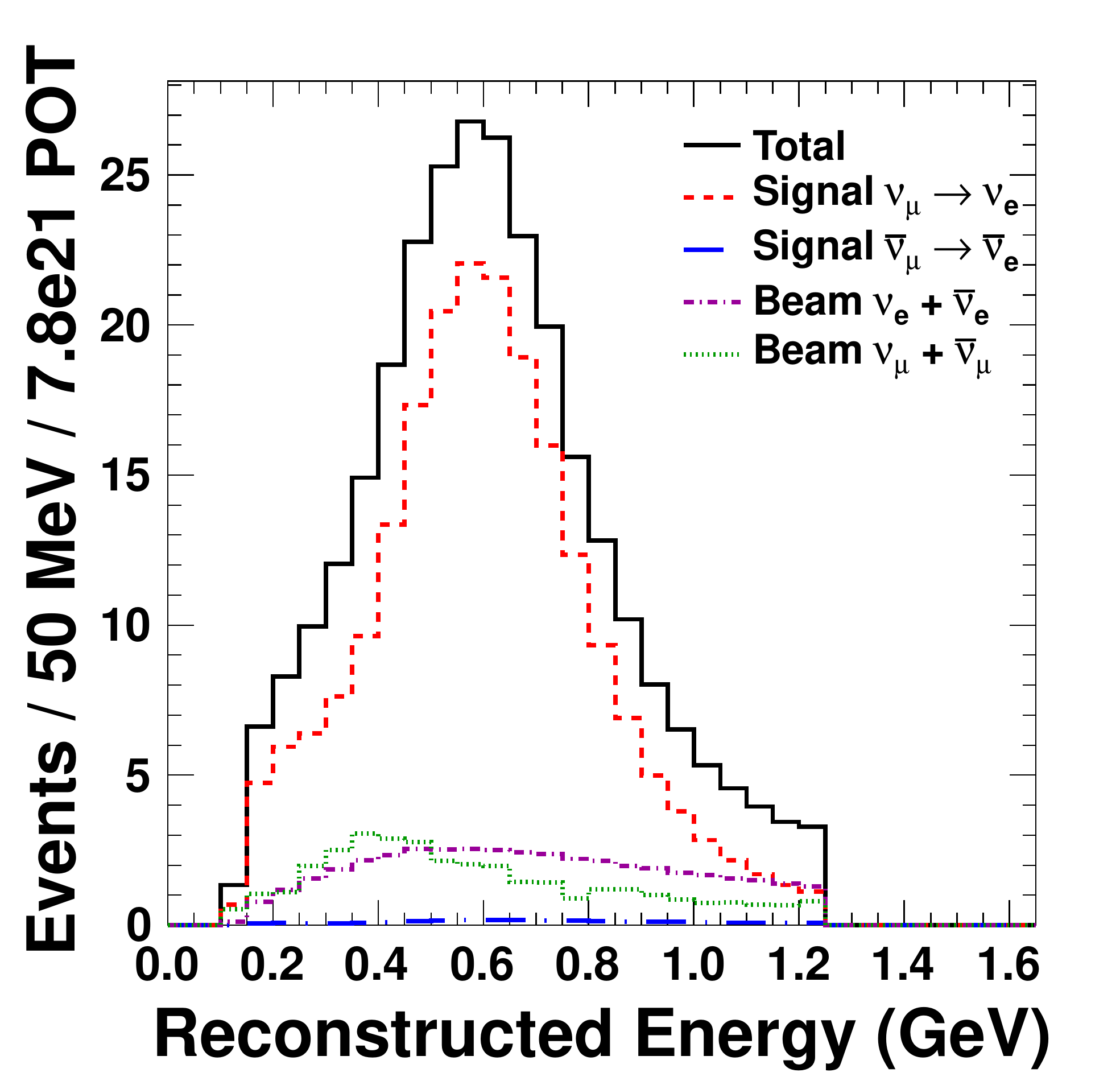}
\caption{\(\nu_e\)
appearance reconstructed energy spectrum, 100\% \(\nu\)-mode running.} 
\label{fig:appRecEspect}
\end{subfigure} \quad
\begin{subfigure}[b]{7cm}
\includegraphics[width=7cm]{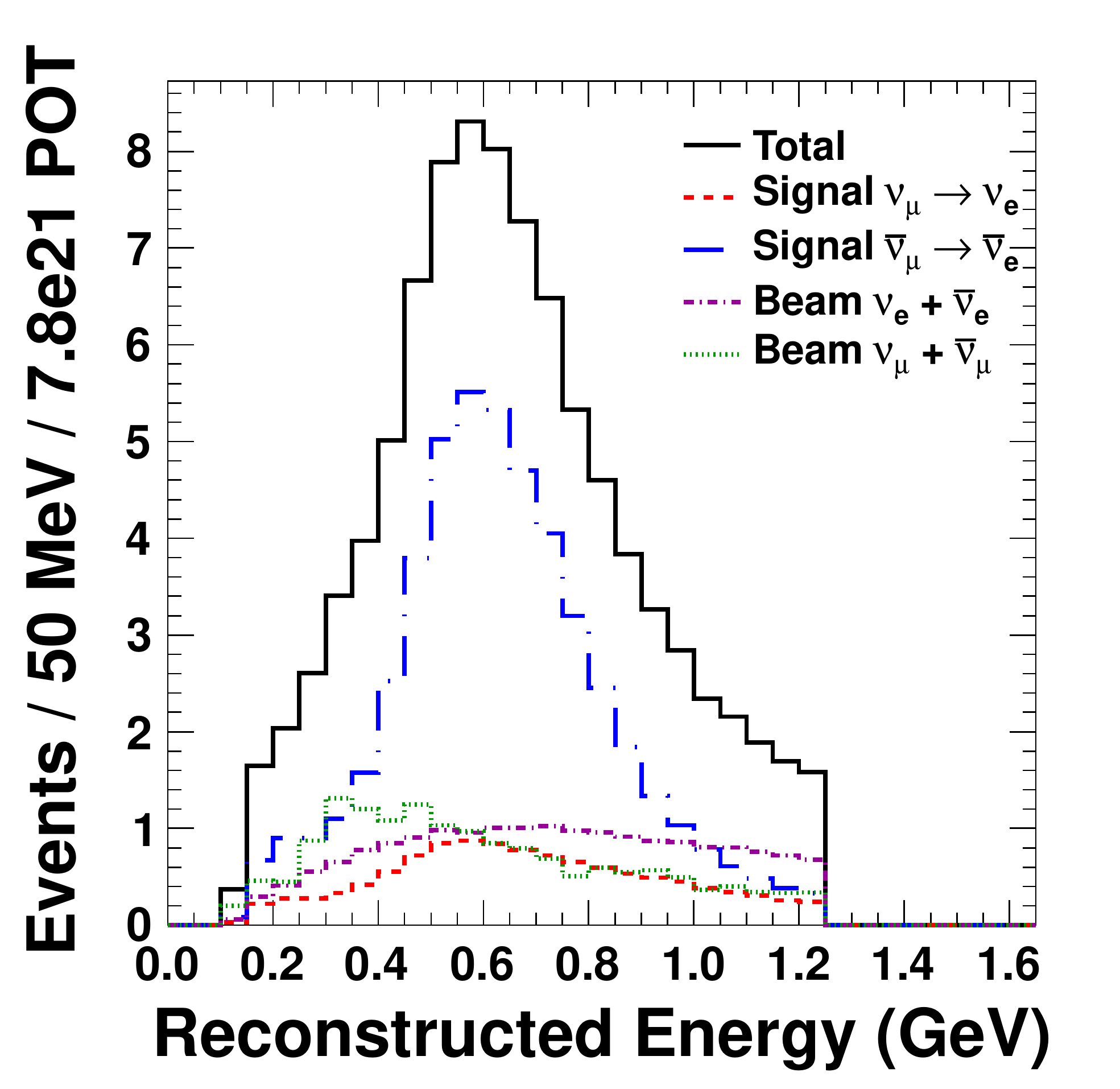}
\caption[]{\(\bar{\nu}_e\)
appearance reconstructed energy spectrum, 100\% \(\bar{\nu}\)-mode running.}
\label{fig:appAntiRecEspect}
\end{subfigure} \quad
\begin{subfigure}[b]{7cm}\includegraphics[width=7cm]{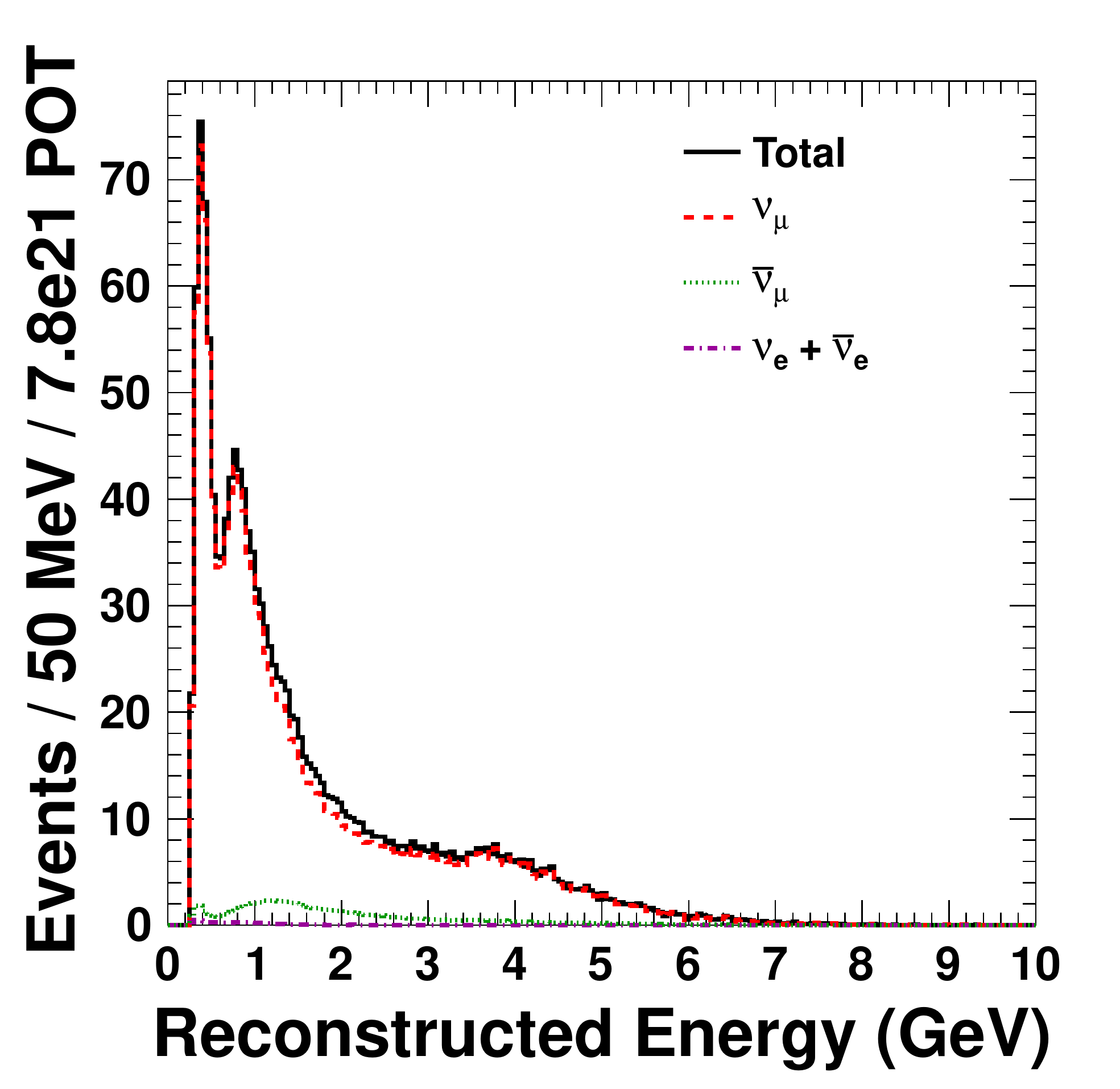}\caption[]{\(\nu_\mu\)
disappearance reconstructed energy spectrum, 100\% \(\nu\)-mode running.}
\end{subfigure} \quad
\begin{subfigure}[b]{7cm}\includegraphics[width=7cm]{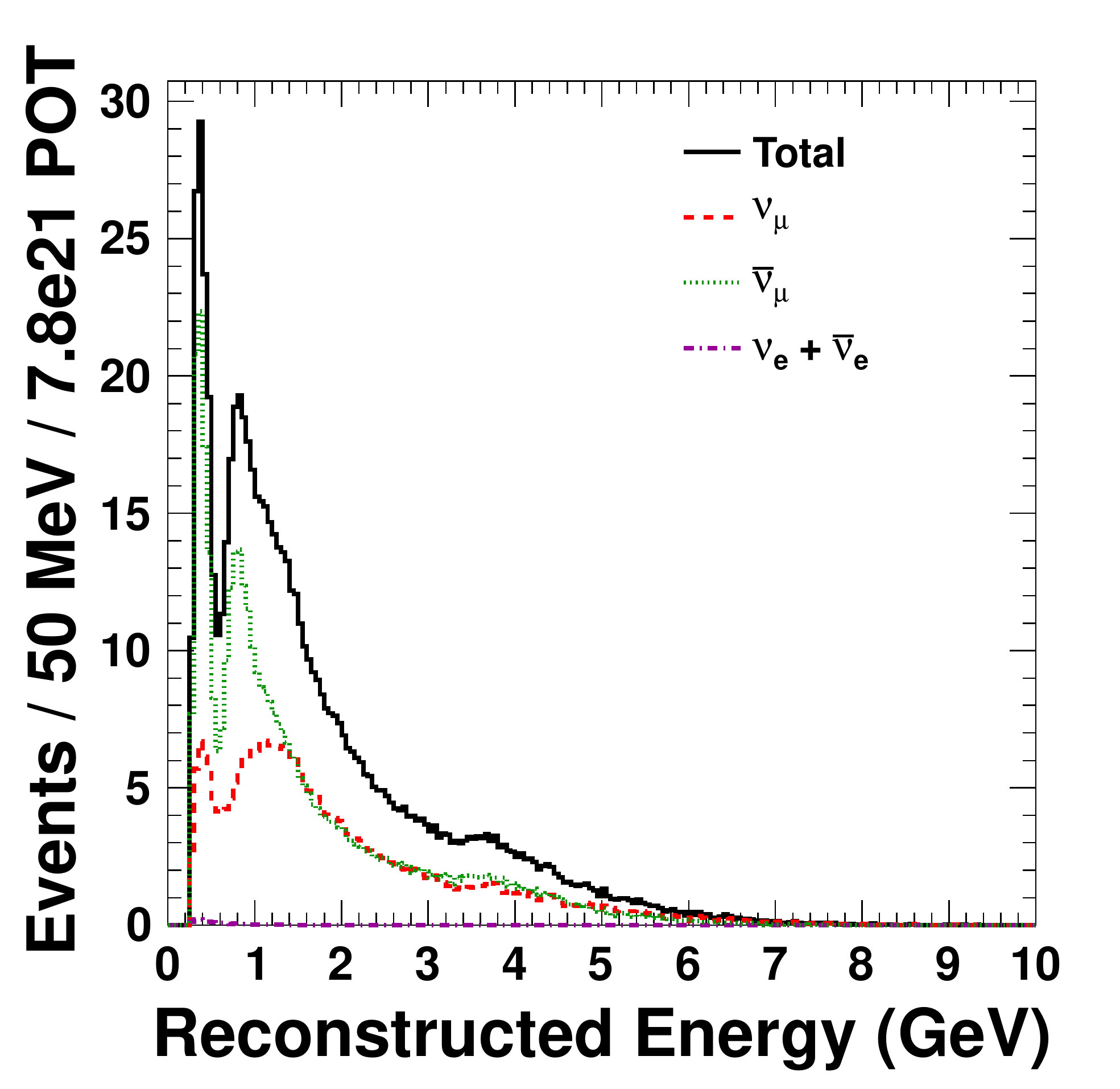}\caption[]{\(\bar{\nu}_\mu\)
disappearance reconstructed energy spectrum, 100\% \(\bar{\nu}\)-mode running.}
\end{subfigure} 
\caption[Reconstructed Energy Spectra]{Appearance and disappearance 
reconstructed energy spectra in Super-K for \(\nu_e\), \(\nu_\mu\), \(\bar{\nu}_e\), 
and \(\bar{\nu}_\mu\) at \(7.8\times10^{21}\)~POT 
for the nominal oscillation parameters as given in Table \ref{tab:truepars} 
}
\label{fig:RecEspect}
\end{figure}

Fits are performed by calculating
\(\Delta\chi^2\) using a binned likelihood method for the appearance and
disappearance reconstructed energy spectra in Super-K.  
Reconstructed appearance and disappearance energy spectra generated 
for the approved full T2K statistics, \(7.8\times10^{21}\)~POT, 
assuming a data-taking condition of either 100\% \(\nu\)-mode
or 100\% \(\bar{\nu}\)-mode
are given in Fig.\ \ref{fig:RecEspect}.  
These spectra are generated assuming the nominal oscillation parameters given 
in Table \ref{tab:truepars}.

Although errors on the shape of the reconstructed energy spectra are used for 
the analysis described in Sec.\ \ref{sec:t2ksensitivity}, the 
total error on the number of events at Super-K 
is given in Table~\ref{tab:nsker}.
This includes uncertainties on the flux prediction,
uncertainties on \(\nu\) interactions both constrained by the near detector and measured by
external experiments, Super-K detector errors, and final state interaction
uncertainties, all of which can cause fluctuations in the shape of the final reconstructed
energy spectra.

\begin{table}[htbp]
\caption[Systematic errors on the Super-K event numbers]
{The systematic errors in percentage on the predicted number of events 
at Super-K (assuming the oscillation parameters given 
in Table \ref{tab:truepars} are the true values of the oscillation 
parameters) as used in the 2012 oscillation analyses.}
\begin{center}
\begin{tabular}{l|c c} \hline
 & Appearance  & Disappearance \\ \hline \hline
Flux and cross section constrained by the near detector
         &   5.0 \% & 4.2 \%  \\ 
Cross section not constrained by the near detector    &  7.4 \% & 6.2 \%  \\ 
Super-K detector and FSI    &  3.9 \% & 11.0 \%  \\  \hline
Total &  9.7 \% & 13.3 \% \\ \hline
\end{tabular}
\end{center}
\label{tab:nsker}
\end{table}

\begin{figure}[htbp]
\centering 
\begin{subfigure}[b]{8cm}\includegraphics[width=8cm]{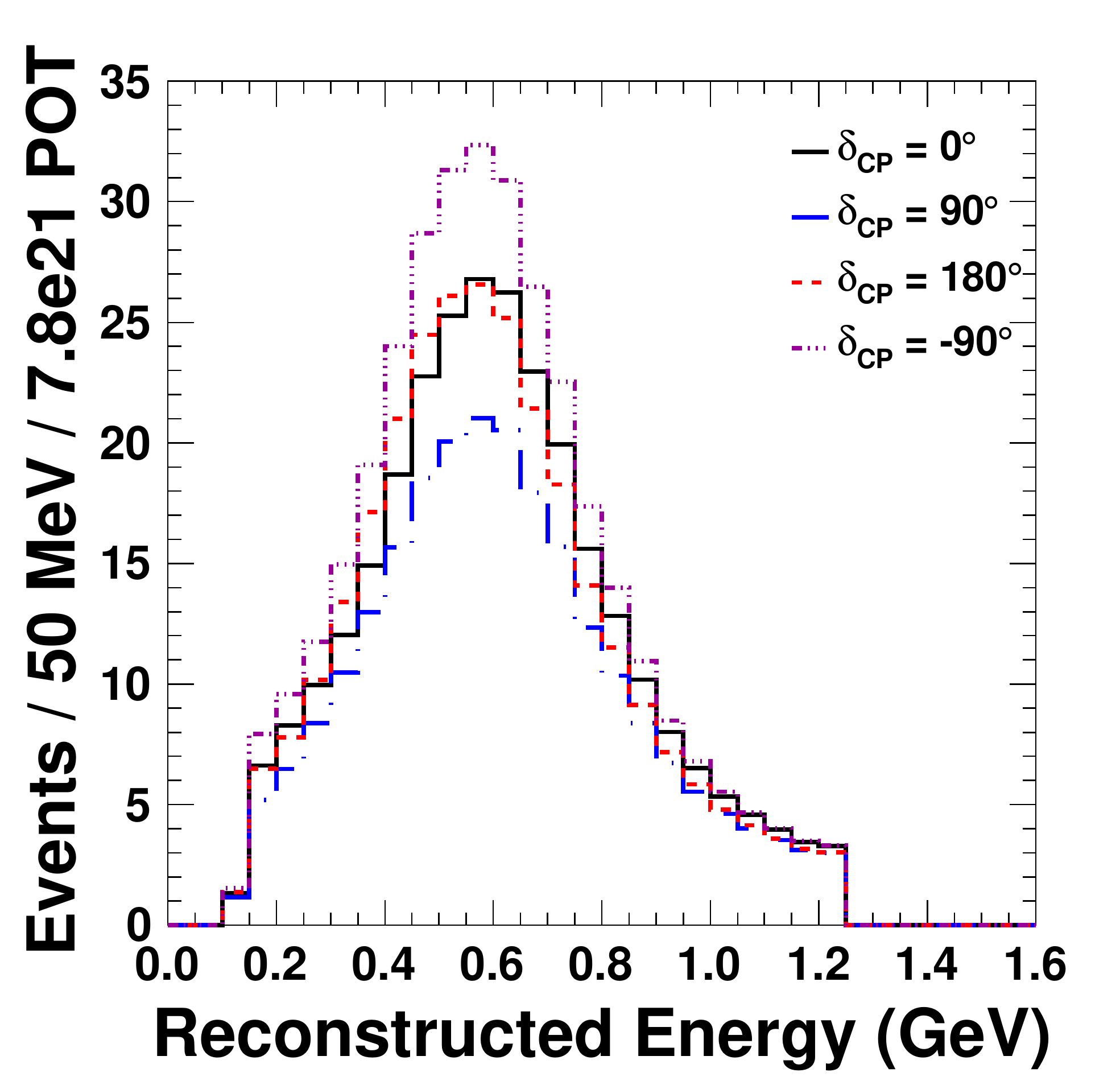}\caption[]{
$\nu$-mode running}
\end{subfigure} \quad
\begin{subfigure}[b]{8cm}\includegraphics[width=8cm]{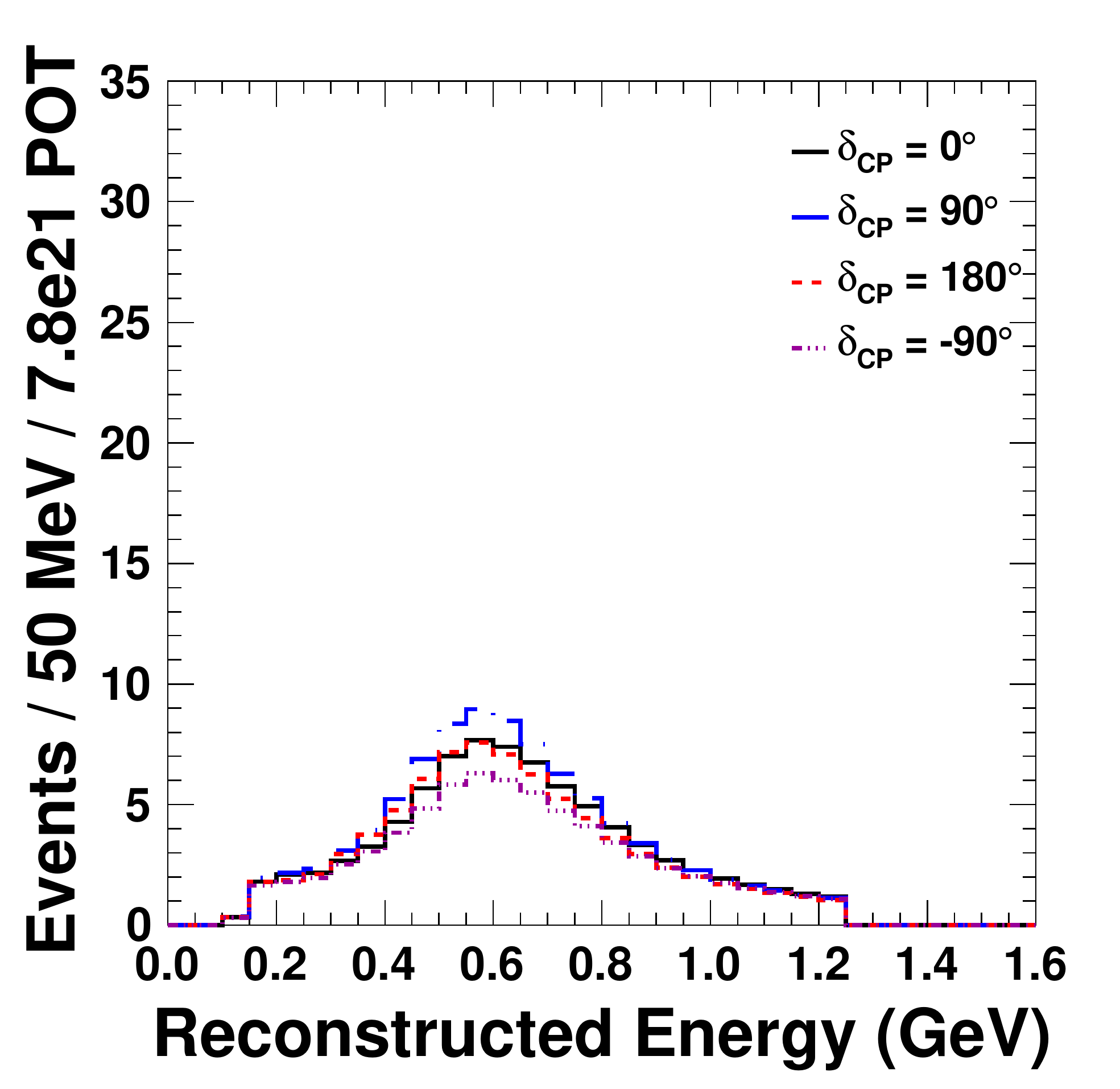}\caption[]{
$\bar{\nu}$-mode running}
\end{subfigure} 
\caption[]{
$\nu_e$ appearance reconstructed energy spectra in Super-K 
for \(7.8\times10^{21}\)~POT in either \(\nu\)-mode or \(\bar{\nu}\)-mode at various values of assumed true \(\delta_{CP}\) with \(\sin^2\theta_{23}=0.5\).
} 
\label{fig:appRecEspect_CPvals}
\end{figure}

When performing fits, the oscillation parameters \(\delta_{CP}\),
\(\sin^22\theta_{13}\), \(\sin^2\theta_{23}\), and \(\Delta m^2_{32}\) are
considered unknown unless otherwise stated, while \(\sin^22\theta_{12}\) and \(\Delta m^2_{21}\) are
assumed fixed to the values given in this table.
Tables \ref{tab:nevapp} and \ref{tab:nevdis} give the number 
of events expected  with the T2K full statistics.
Fig.~\ref{fig:appRecEspect_CPvals} shows the dependence
of the $\nu_e$ appearance reconstructed energy spectrum
on $\delta_{CP}$. 
Some of the sensitivities are enhanced by constraining the error on $\sin^22\theta_{13}$
based on the projected precision of reactor measurements. 
For this study, the
uncertainty (referred to as the ultimate reactor error) on
$\sin^22\theta_{13}$ is chosen to be 0.005, which corresponds to the 2012
systematic error only of the Daya Bay experiment\cite{DB13}
\footnote{The statistical error is 0.010 for \cite{DB13}}. 
\begin{table}[htbp]
\caption[Nominal Oscillation Parameter Values]{Nominal values of the oscillation parameters. When the reactor constraint is used, 
we assume 0.005 as the expected uncertainty of the reactor measurement.}
\begin{center}
\begin{tabular}{  l || c | c | c | c | c | c | c } \hline
Parameter & \(\sin^22\theta_{13}\) & \(\delta_{CP}\) & \(\sin^2\theta_{23}\) &
\(\Delta m^2_{32}\) &
Hierarchy & \(\sin^22\theta_{12}\) & \(\Delta m^2_{21}\)\\ \hline
Nominal & 0.1 & 0 & 0.5 & \(2.4\times10^{-3}\) & normal & 0.8704 &
\(7.6\times10^{-5}\) \\
Value & &  &  &  eV\(^2\) & & &  eV\(^2\) \\ \hline
\end{tabular}
\end{center}  
\label{tab:truepars} \end{table}
\begin{table}[htbp] 
\caption[Number of $\nu$ Appearance Events]{Expected numbers of 
$\nu_e$ or $\bar{\nu}_e$ appearance events at $7.8\times10^{21}$~POT.  The
number of events is broken down into those coming from: appearance
signal or intrinsic beam background events that undergo charged current (CC)
interactions in Super-K, or beam background events that undergo neutral current (NC) interactions.}
\begin{center}
\begin{tabular}{  c | c | c | c | c | c | c | c  } \hline 
& & & Signal & Signal & Beam CC & Beam CC & \\
& \(\delta_{CP}\) & Total & \(\nu_{\mu} \rightarrow \nu_e\) & \(\bar{\nu}_{\mu} \rightarrow
\bar{\nu}_e\) & \(\nu_e + \bar{\nu}_e \) & \(\nu_{\mu} + \bar{\nu}_{\mu} \) &
NC\\ \hline \hline
100\% $\nu$-mode  & 0\(\degree\) & 291.5 & 211.9 & 2.4 & \multirow{2}{*}{41.3} & \multirow{2}{*}{1.4} & \multirow{2}{*}{34.5} \\ 
100\% $\nu$-mode  & -90\(\degree\) & 341.8 & 262.9 & 1.7 & & & \\ \hline 
100\% $\bar{\nu}$-mode & 0\(\degree\) & 94.9 & 11.2 &  48.8 & \multirow{2}{*}{17.2} & \multirow{2}{*}{0.4} & \multirow{2}{*}{17.3} \\ 
100\% $\bar{\nu}$-mode & -90\(\degree\) & 82.9 & 13.1 & 34.9 &  &  &  \\ \hline 
\end{tabular}
\end{center} 
\label{tab:nevapp} \end{table}
%
\begin{table}[htbp] 
\caption[Number of $\nu$ Disappearance Events]{Expected numbers 
of \(\nu_\mu\) or $\bar{\nu}_\mu$ disappearance events 
for \(7.8\times10^{21}\)~POT.  The first two columns show the
number of \(\nu_\mu\) and \(\bar{\nu}_\mu\) events, broken down into those that undergo charged-current quasi-elastic (CCQE)
scattering at Super-K, and those that undergo other types of CC scattering (CC
non-QE).  The third column shows CC \(\nu_e\) and \(\bar{\nu}\) events, both
from intrinsic beam backgrounds and oscillations, while
the fourth column shows NC events.}
\begin{center}
\begin{tabular}{  c | c | c | c | c | c  } 
\hline
& & CCQE & CC non-QE & CC \(\nu_e + \bar{\nu}_e\) &  \\
& Total & \(\nu_{\mu} (\bar{\nu}_{\mu}\)) & \(\nu_{\mu} (\bar{\nu}_{\mu}\)) & CC \(\nu_\mu (\bar{\nu}_{\mu}) \rightarrow \nu_e (\bar{\nu}_e)\) & NC \\
\hline \hline
100\% running in $\nu$-mode& 1,493 & 782(48) &  544 (40) & 4 & 75 \\ 
\hline
100\% running in $\bar{\nu}$-mode & 715 & 130(263) & 151(138) & 0.5 & 33\\
\hline
\end{tabular}
\end{center} 
\label{tab:nevdis} 
\end{table}

\subsection{Expected 90\% C.L.\ regions}
In this section we show expected 90\% C.L.\ intervals
for the T2K full statistics of \(7.8\times10^{21}\)~POT.  
Contours showing both the T2K
sensitivity for \(\delta_{CP}\) vs.\ \(\sin^22\theta_{13}\) and for \(\Delta
m^2_{32}\) vs.\ \(\sin^2\theta_{23}\) are provided, 
where the assumed true value of the 
oscillation parameters is indicated by a black cross.
The oscillation parameters \(\delta_{CP}\),
\(\sin^22\theta_{13}\), \(\sin^2\theta_{23}\), and \(\Delta m^2_{32}\) are
considered unknown, as stated above.
Both the NH and IH are considered, and
\(\Delta\chi^2\) values are calculated from the minimum \(\chi^2\) value for
both MH assumptions.  The blue curves are generated assuming the correct MH 
and the red curves are generated 
assuming the incorrect MH, such that if an experiment or combination of
experiments from the global neutrino
community were to determine the MH
the red contour would be eliminated.  
A contour consisting of the outermost edge of all contours in each plot can be
considered as the T2K sensitivity assuming an unknown MH.
For the sake of brevity, 
only results assuming true NH are shown; similar conclusions can be
drawn from plots assuming true IH. 

\begin{figure}[htbp]
\centering 
\begin{subfigure}[t]{0.62\textwidth}
\centering
\includegraphics[width=0.62\textwidth]
{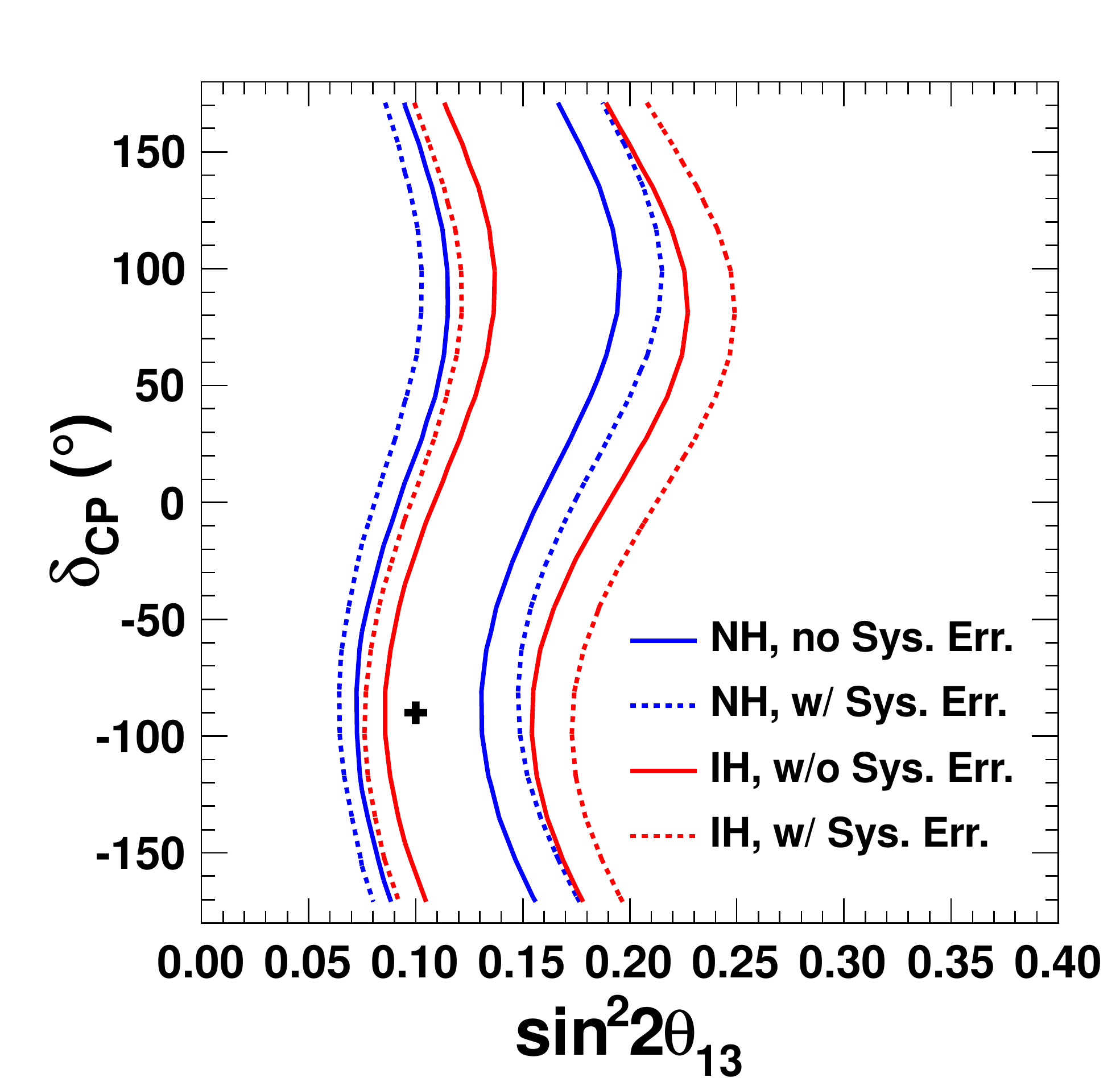}
\caption{50\% \(\nu\)-mode only.} 
\end{subfigure} 
\begin{subfigure}[t]{0.62\textwidth}
\centering
\includegraphics[width=0.62\textwidth]
{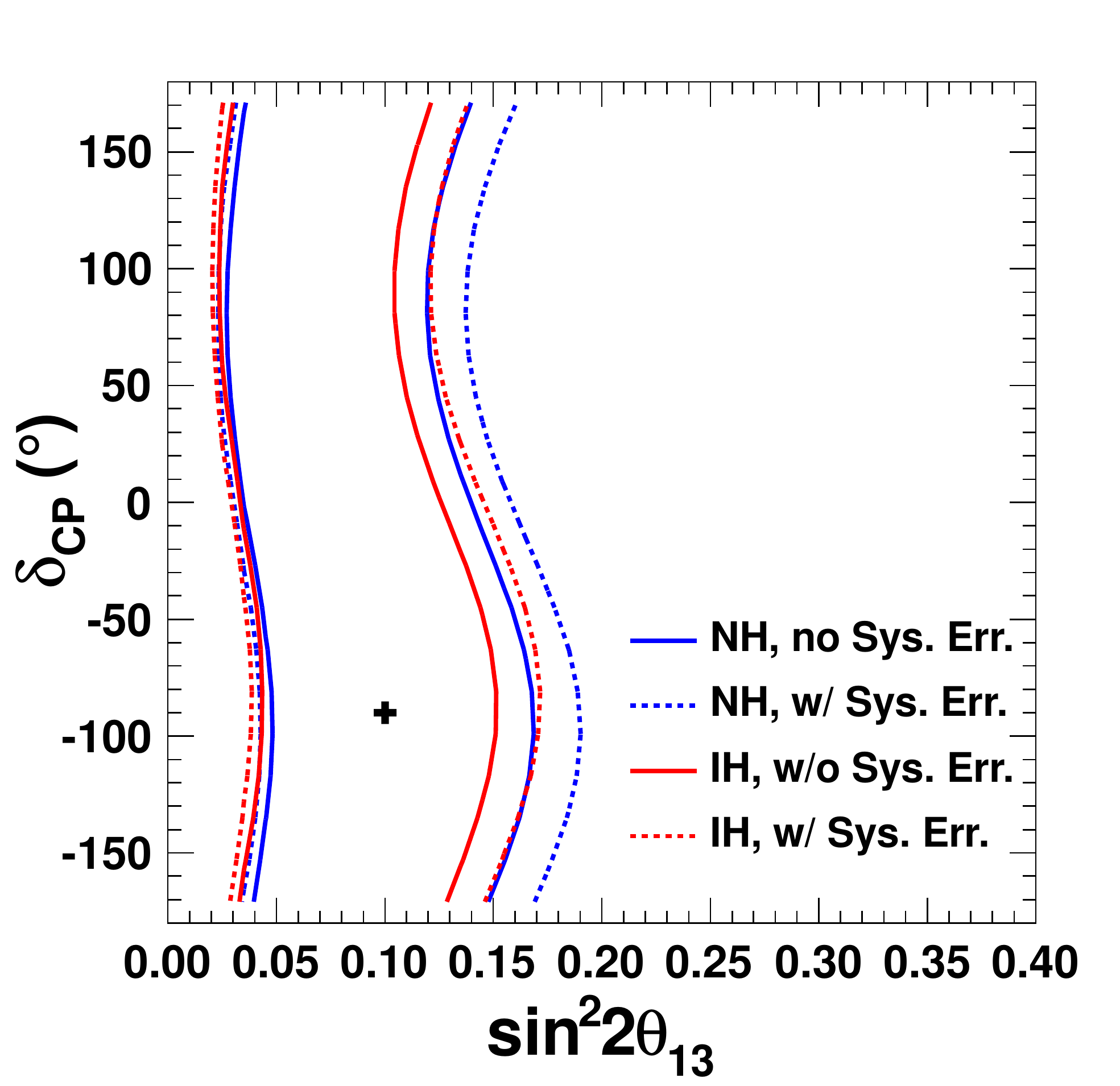}
\caption{50\% \(\bar{\nu}\)-mode only.} 
\end{subfigure} 
\begin{subfigure}[t]{0.62\textwidth}
\centering
\includegraphics[width=0.62\textwidth]
{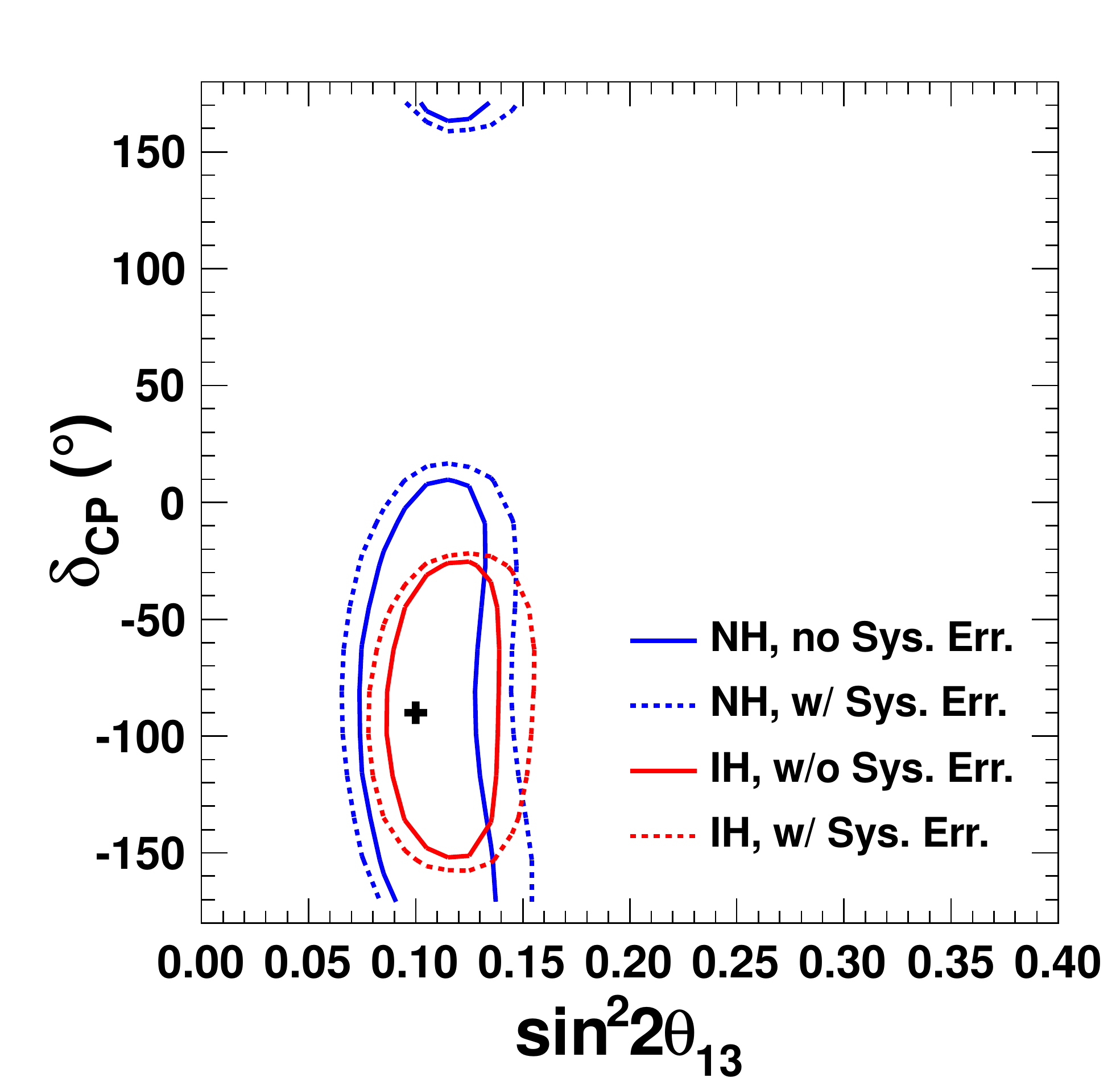}
\caption{50\% \(\nu\)-, 50\% \(\bar{\nu}\)-mode.} 
\end{subfigure} 
\caption[Comparison of $\nu$ and $\bar{\nu}$ 90\% C.L. Regions]
{Expected \(\delta_{CP}\) vs.\ \(\sin^22\theta_{13}\) 90\% C.L.\ 
intervals, where (a) and (b) are each given for 50\% 
of the full T2K POT, and (c) demonstrates the sensitivity of the total T2K
POT with 50\% \(\nu\)-mode plus 50\% \(\bar{\nu}\)-mode running.  
Contours are plotted for the case of true \(\delta_{CP} = -90\degree\) and NH.  
The blue curves are fit  assuming the correct MH(NH)}, while the red are fit 
assuming the incorrect MH(IH), and contours are plotted from the minimum \(\chi^2\)
value for both MH assumptions.
The solid contours are with statistical error only, while the dashed contours 
include the systematic errors used in the 2012 oscillation analysis
assuming full correlation between \(\nu\)- and \(\bar{\nu}\)-mode running errors.
\label{fig:app2dregions_showeffects_text}
\end{figure}

Figure~\ref{fig:app2dregions_showeffects_text} gives an example of the difference in the shape of the T2K
sensitive region for \(\nu\)- vs.\ \(\bar{\nu}\)-mode at true \(\delta_{CP} =
-90\degree\) (and the other oscillation parameters as given in Table \ref{tab:truepars}) by comparing the \(\nu\)-mode  
--~Fig.\ \ref{fig:app2dregions_showeffects_text} (a)~-- and 
\(\bar{\nu}\)-mode --~Fig.\ \ref{fig:app2dregions_showeffects_text} (b)~-- C.L.\ contours without a reactor constraint 
at 50\% of the full T2K POT.  
These two contours are then combined in 
Fig.\ \ref{fig:app2dregions_showeffects_text} (c), which shows 
the 90\% C.L.\ region for 50\% \(\nu\)- plus 50\% \(\bar{\nu}\)-mode running 
to achieve the full T2K POT.  This demonstrates that $\delta_{CP}$ 
can be constrained by combining \(\nu\)-mode and \(\bar{\nu}\)-mode data.

\begin{figure}[htbp]
\centering 
\begin{subfigure}[t]{7cm}
\includegraphics[width=7cm]{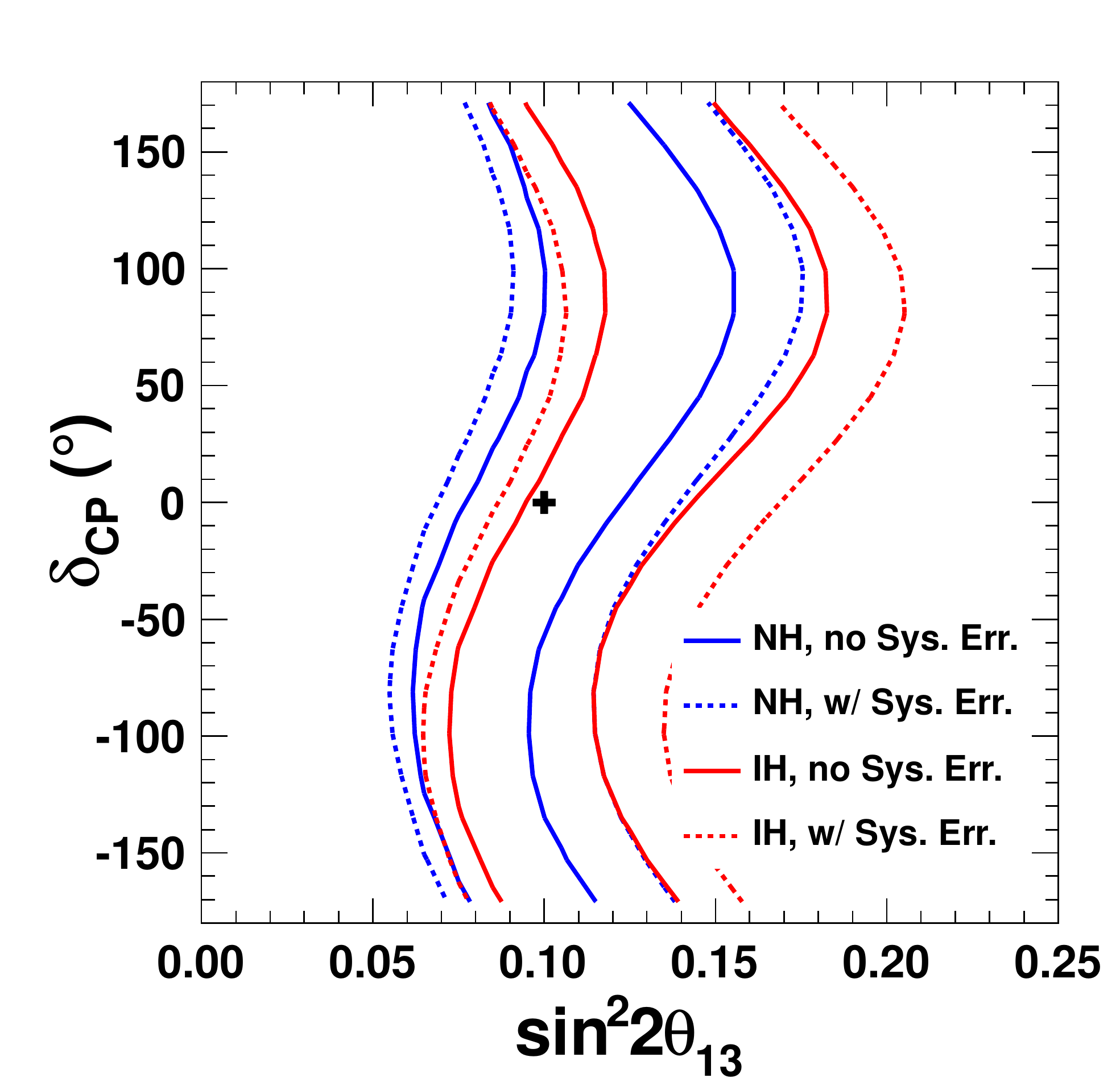}\caption{100\% \(\nu\)-mode.} 
\end{subfigure}
\begin{subfigure}[t]{7cm}
\includegraphics[width=7cm]{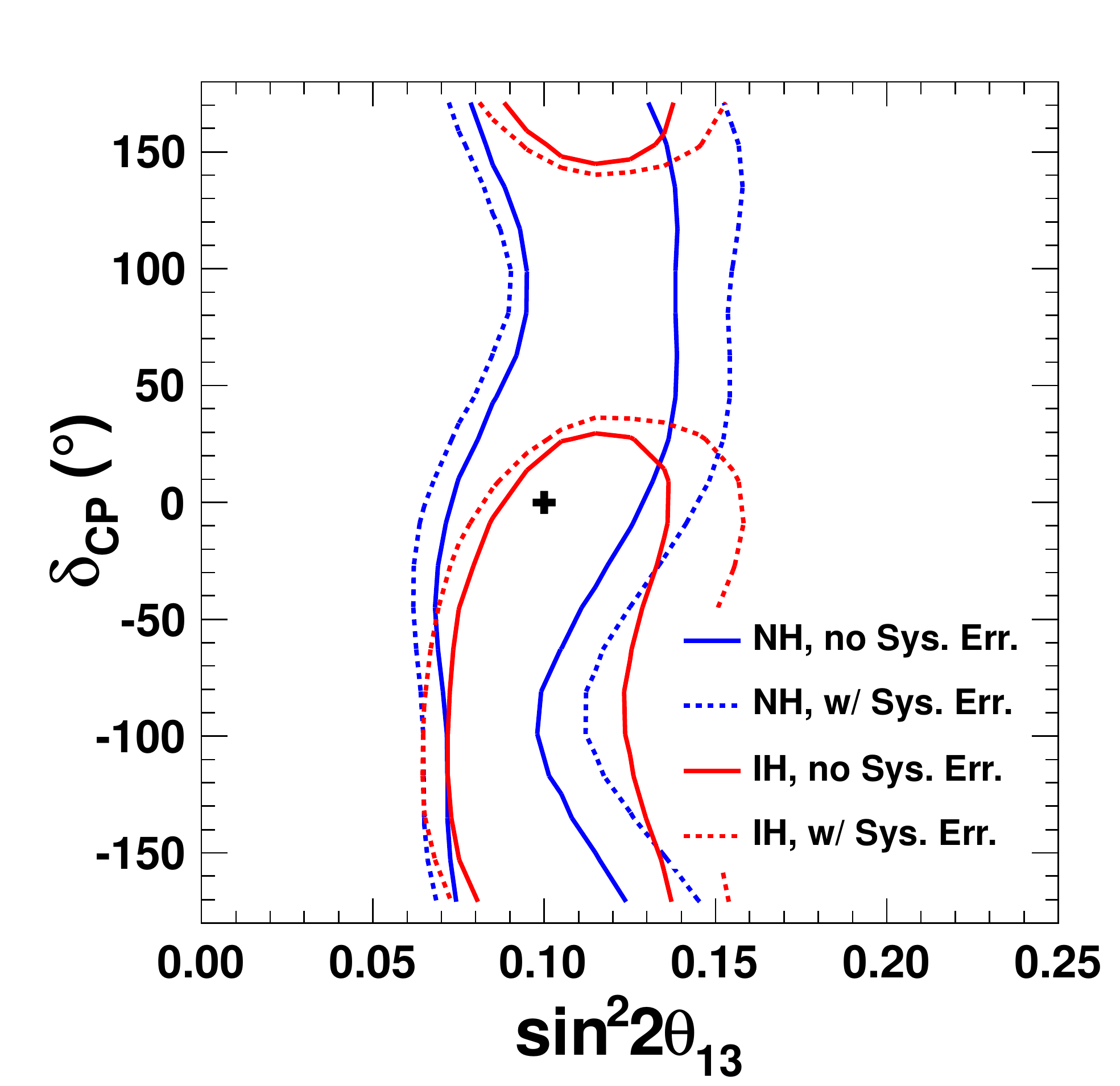}\caption{50\% \(\nu\)-, 50\% \(\bar{\nu}\)-mode.} 
\end{subfigure} 
\begin{subfigure}[t]{7cm}
\includegraphics[width=7cm]{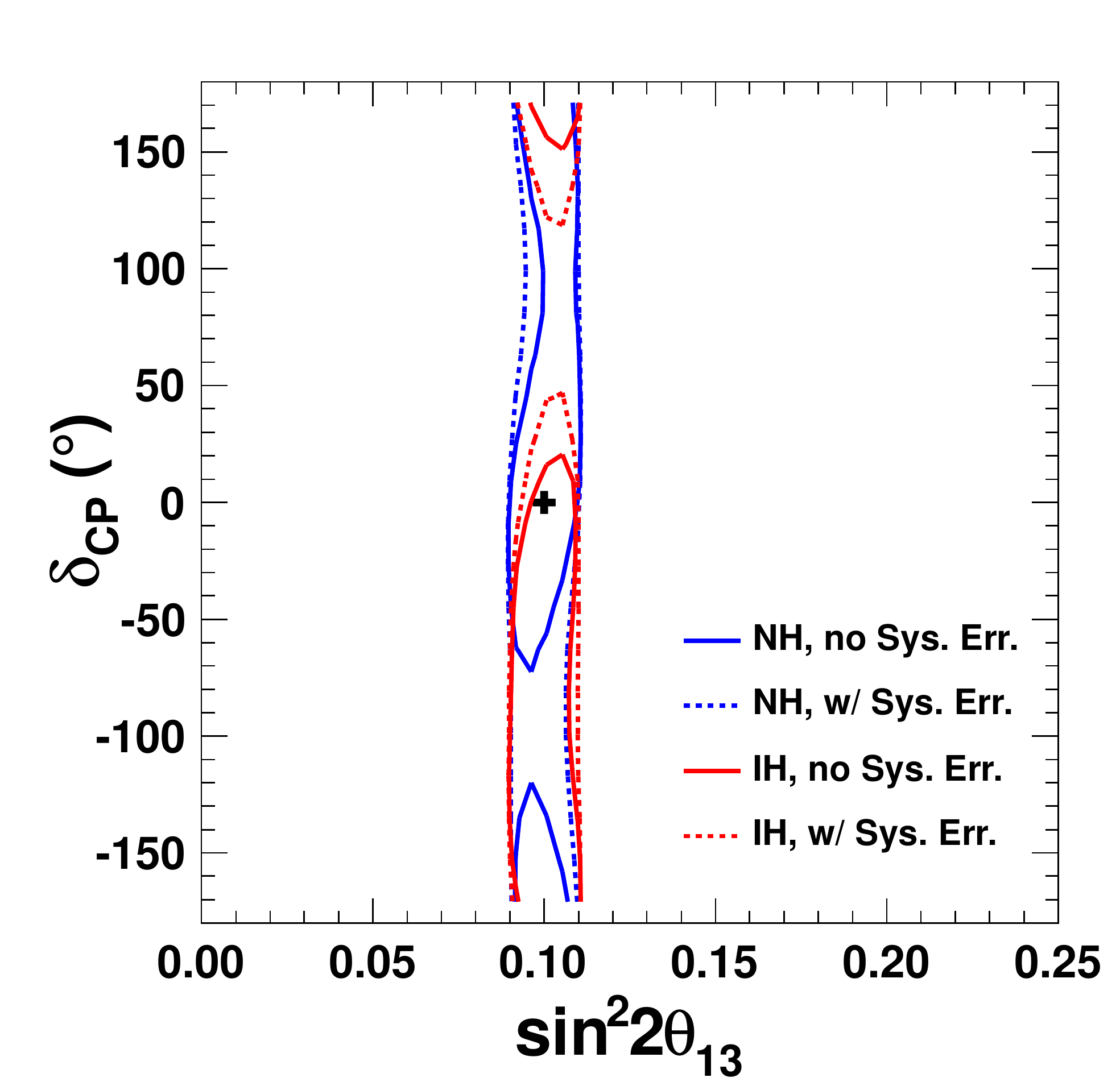}\caption{100\% \(\nu\)-mode, with ultimate reactor constraint.} 
\end{subfigure}
\begin{subfigure}[t]{7cm}
\includegraphics[width=7cm]{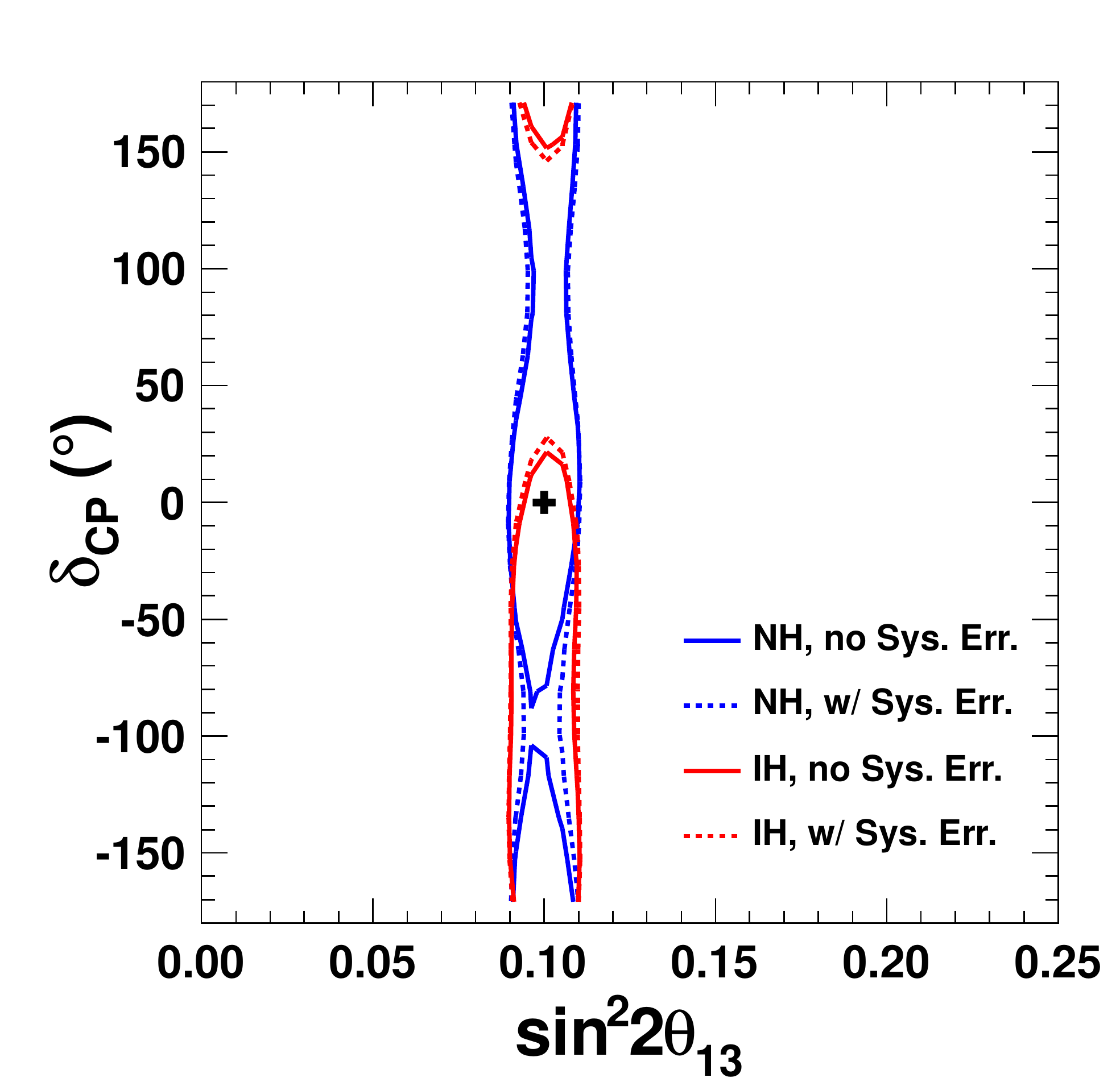}
\caption{50\% \(\nu\)-, 50\% \(\bar{\nu}\)-mode, 
with ultimate reactor constraint.} 
\end{subfigure}
\caption[Appearance 90\% C.L. Regions with Nominal True Parameters]
{\(\delta_{CP}\) vs.\ \(\sin^22\theta_{13}\) 90\% C.L.\ intervals 
for \(7.8\times10^{21}\)~POT.  
Contours are plotted for the case of true $\delta_{CP} = 0\degree$ and NH.
The blue curves are fit 
assuming the correct MH(NH), while the red are fit assuming the incorrect MH(IH), and
contours are plotted from the minimum \(\chi^2\) value for both MH assumptions.  
The solid contours are with statistical error only, while the dashed contours 
include the 2012 systematic errors fully correlated 
between \(\nu\)- and \(\bar{\nu}\)-mode.
\label{fig:app2dregionsCP0}} \end{figure}

\begin{figure}[htbp]
\centering 
\begin{subfigure}[t]{7cm}
\includegraphics[width=7cm]{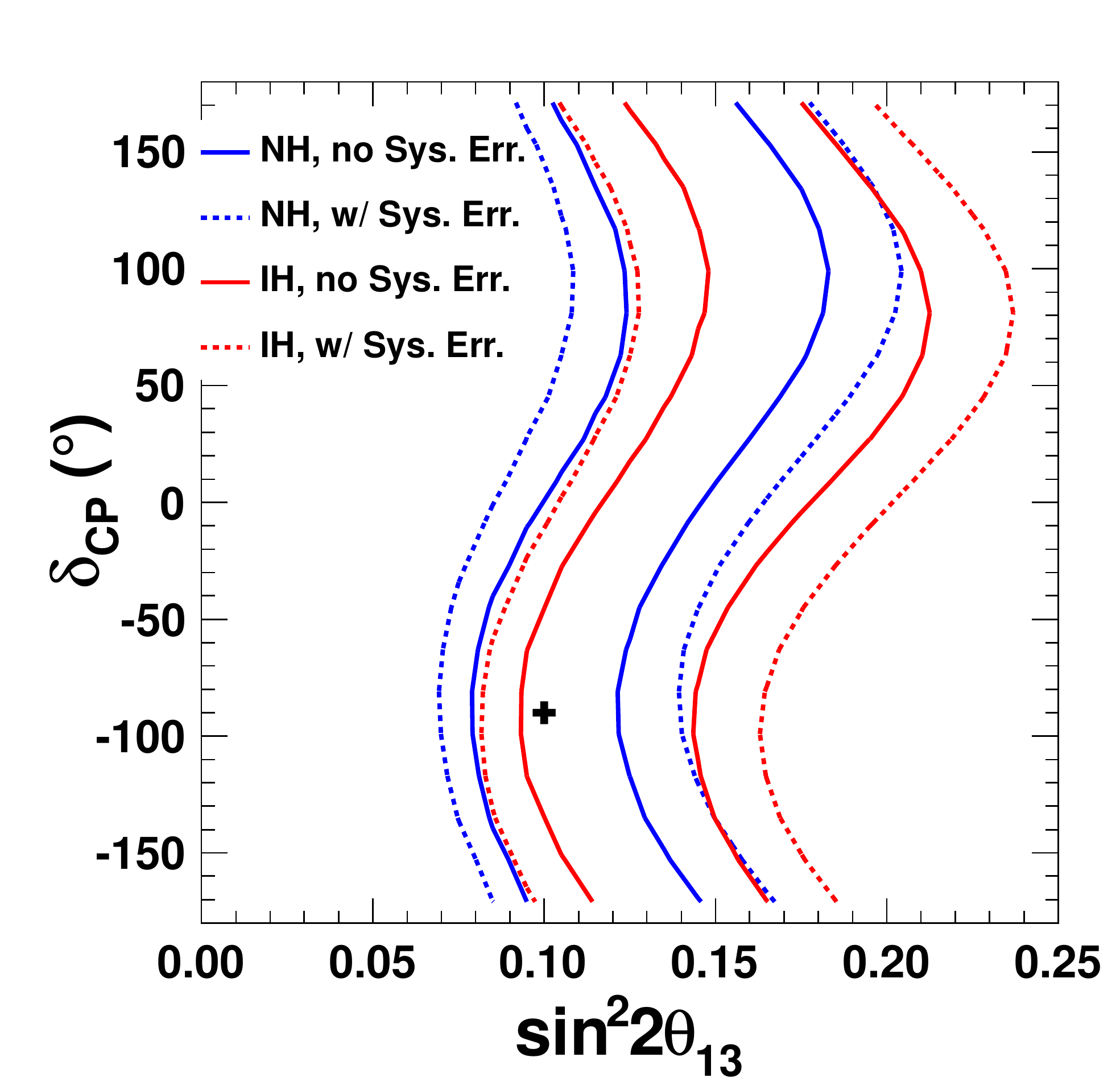}\caption{100\% \(\nu\)-mode.} 
\end{subfigure}
\begin{subfigure}[t]{7cm}
\includegraphics[width=7cm]{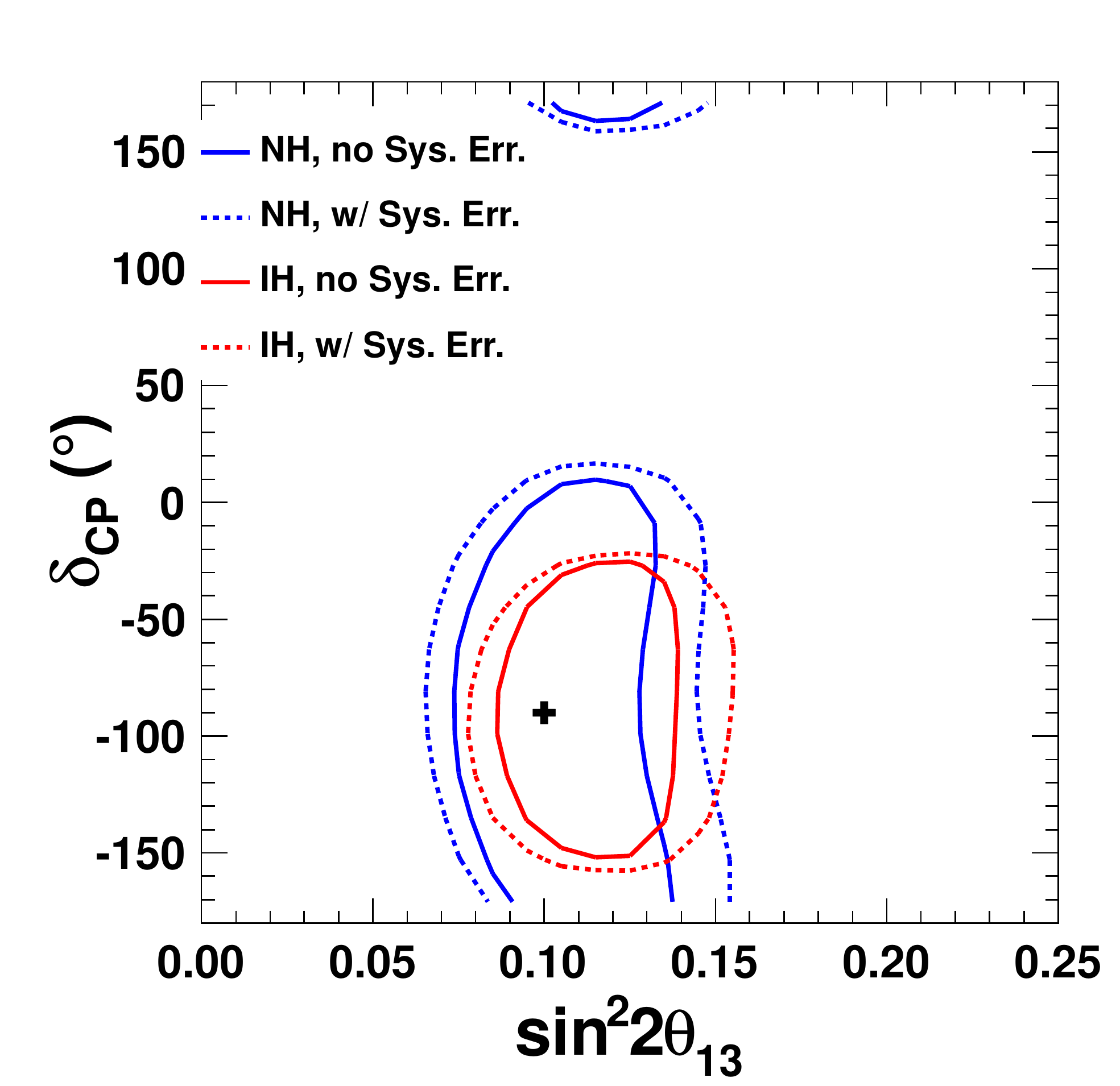}\caption{50\% \(\nu\)-, 50\% \(\bar{\nu}\)-mode.}
\end{subfigure} 
\begin{subfigure}[t]{7cm}
\includegraphics[width=7cm]{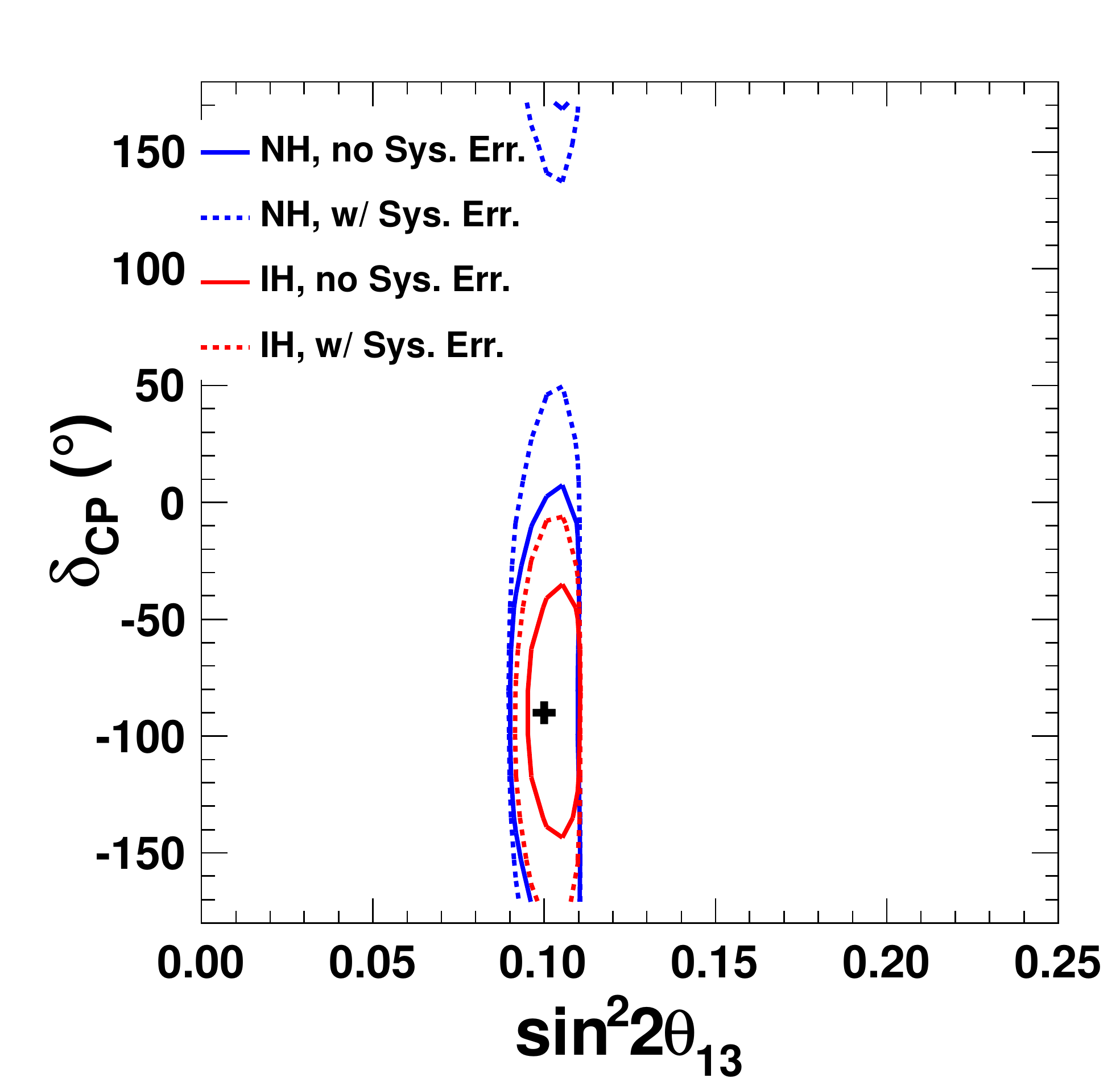}\caption{100\% \(\nu\)-mode, with ultimate reactor constraint.} 
\end{subfigure}
\begin{subfigure}[t]{7cm}
\includegraphics[width=7cm]{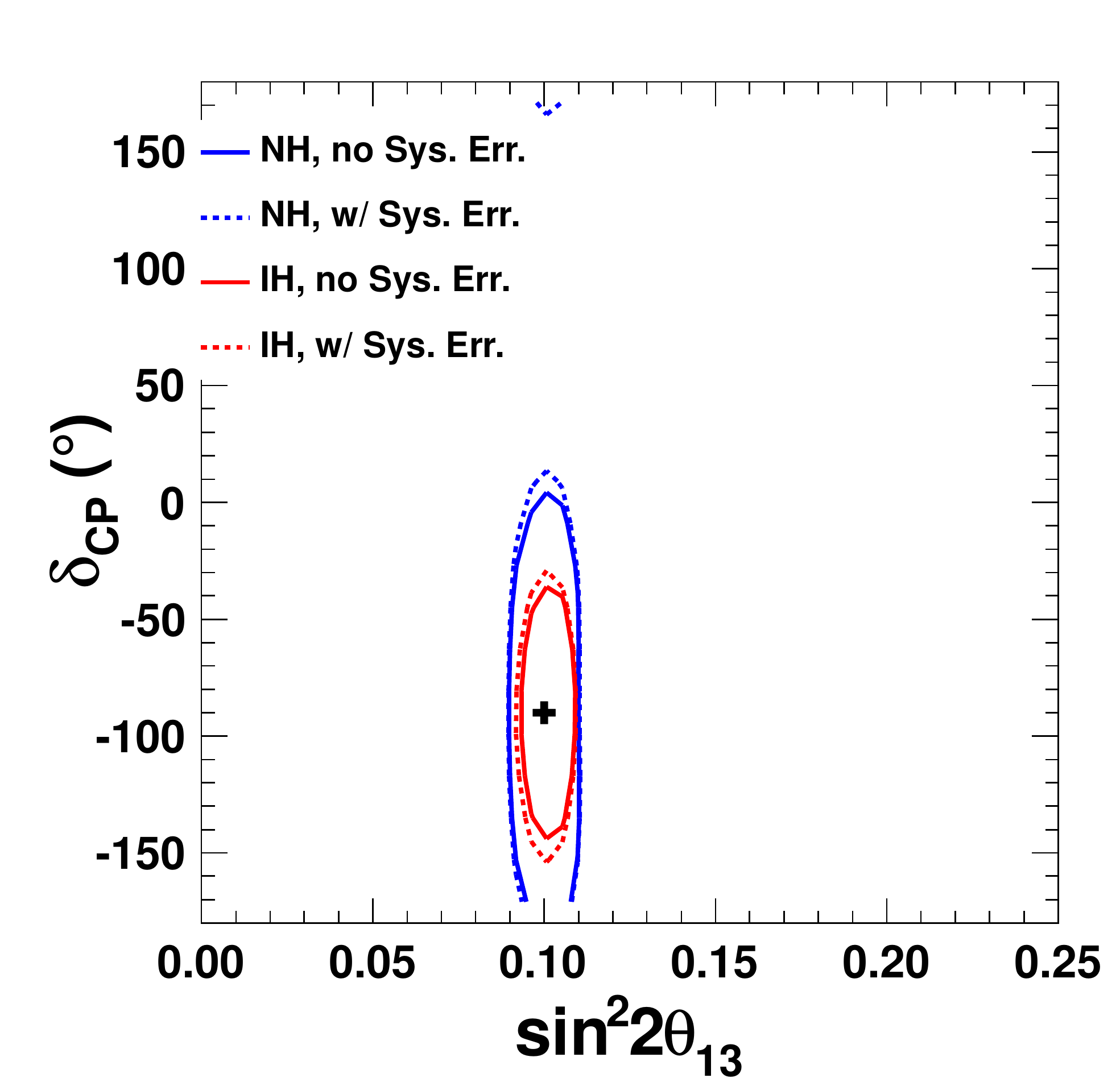}
\caption{50\% \(\nu\)-, 50\% \(\bar{\nu}\)-mode, with ultimate reactor constraint.} 
\end{subfigure}
\caption[Appearance 90\% C.L. Regions with Nominal True Parameters]{\(\delta_{CP}\) vs.\ \(\sin^22\theta_{13}\) 90\% C.L.\ intervals for \(7.8\times10^{21}\)~POT.  
Contours are plotted for the case of true \(\delta_{CP} = -90\degree\) and NH.  
The blue curves are fit 
assuming the correct MH(NH), while the red are fit assuming the incorrect MH(IH), and
contours are plotted from the minimum \(\chi^2\) value for both MH assumptions.  
The solid contours are with statistical error only, while the dashed contours 
include the 2012 systematic errors fully correlated between \(\nu\)- and \(\bar{\nu}\)-mode.
\label{fig:app2dregionsCP-90}} 
\end{figure}

Figures~\ref{fig:app2dregionsCP0} and \ref{fig:app2dregionsCP-90}
show example 90\% C.L.\ 
regions for $\delta_{CP}$ vs.\ $\sin^2{2\theta_{13}}$
at the full T2K statistics, both for T2K alone and including an extra constraint 
on the T2K predicted data fit based on the
ultimate reactor error $\delta(\sin^22\theta_{13}) = 0.005$ as discussed above,
for true $\delta_{CP}$ of $0\degree$ and $-90\degree$, respectively.
In the case of $\delta_{CP}=-90\degree$, we start to have sensitivity to 
resolve $\delta_{CP}$ without degeneracies.

\begin{figure}[htbp]
\centering 
\begin{subfigure}[t]{7cm} 
\includegraphics[width=7cm]{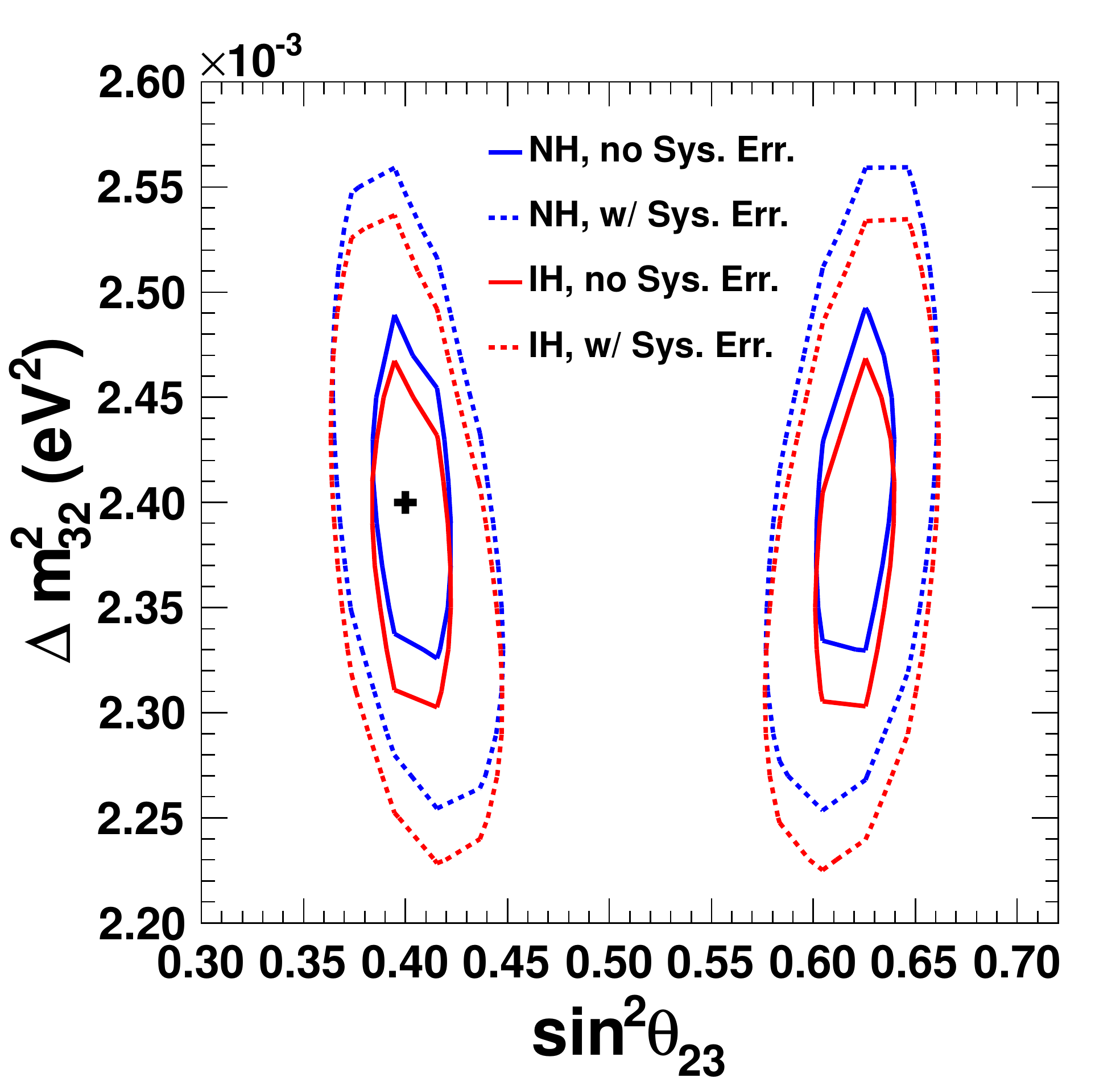}
\caption{100\% \(\nu\)-mode.} 
\end{subfigure} \quad
\begin{subfigure}[t]{7cm}
\includegraphics[width=7cm]{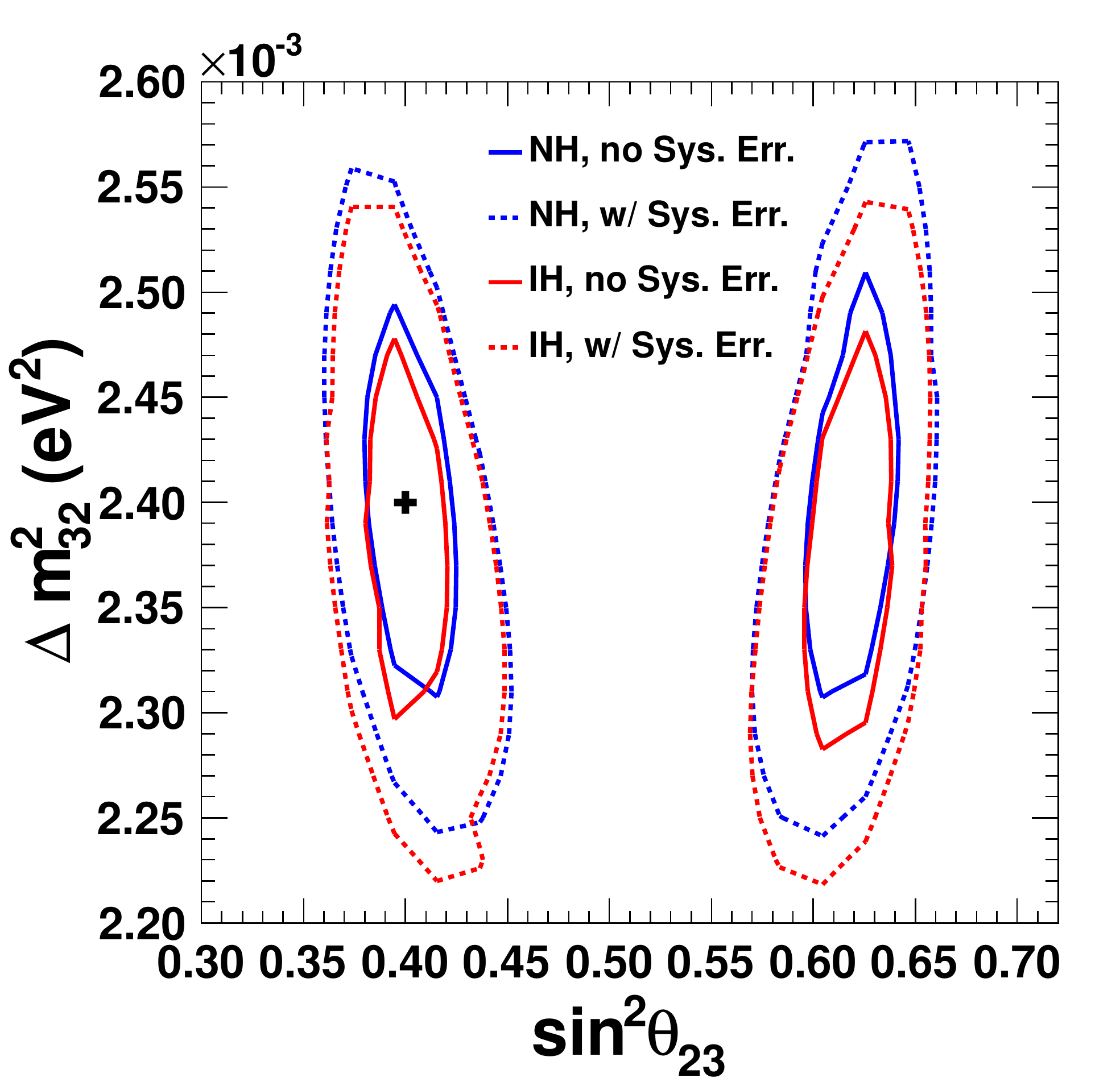}
\caption{50\% \(\nu\)-, 50\% \(\bar{\nu}\)-mode.} 
\end{subfigure} 
\begin{subfigure}[t]{7cm}
\includegraphics[width=7cm]{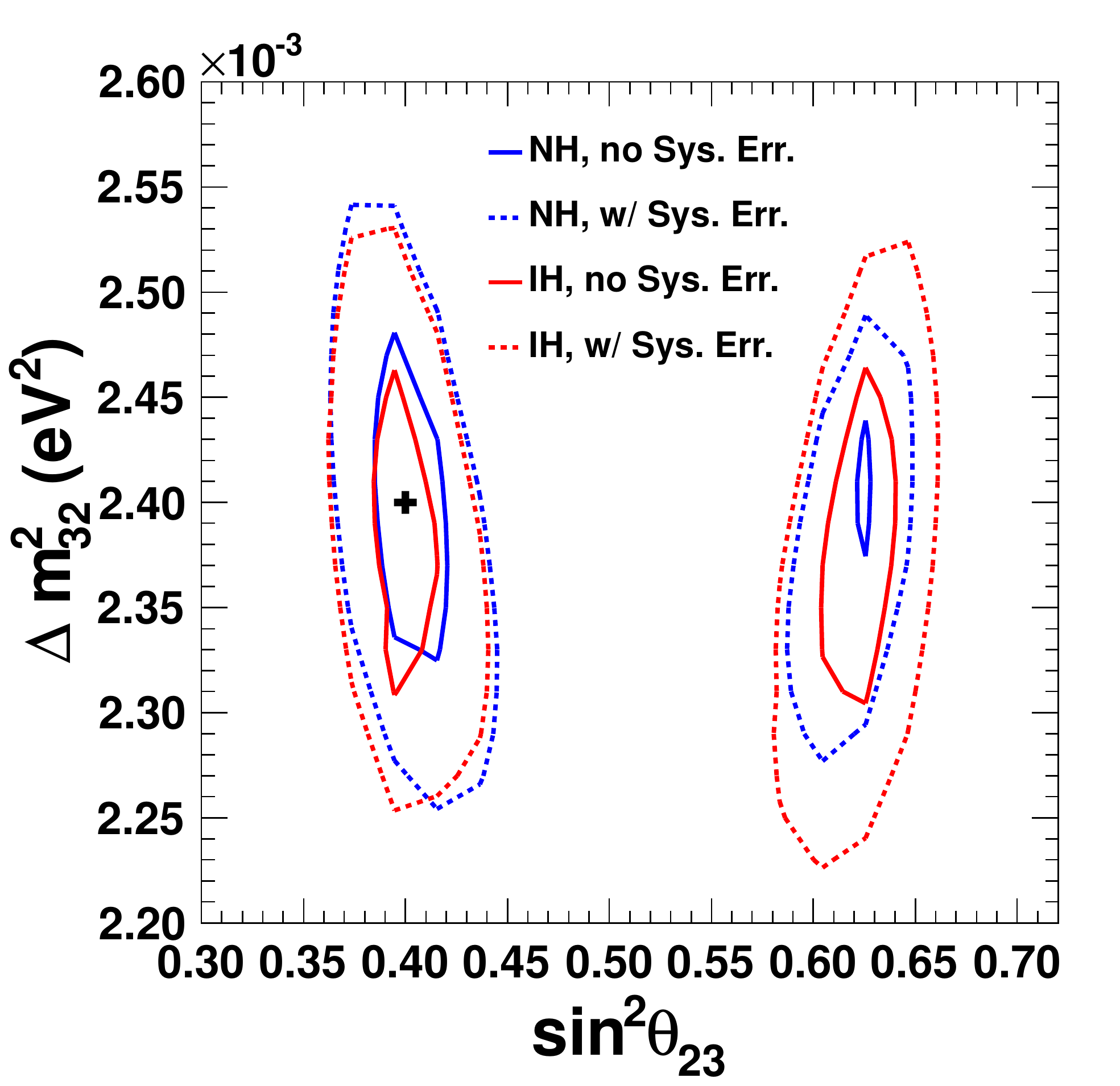}
\caption{100\% \(\nu\)-mode, with ultimate reactor error.} 
\end{subfigure} \quad
\begin{subfigure}[t]{7cm}
\includegraphics[width=7cm]{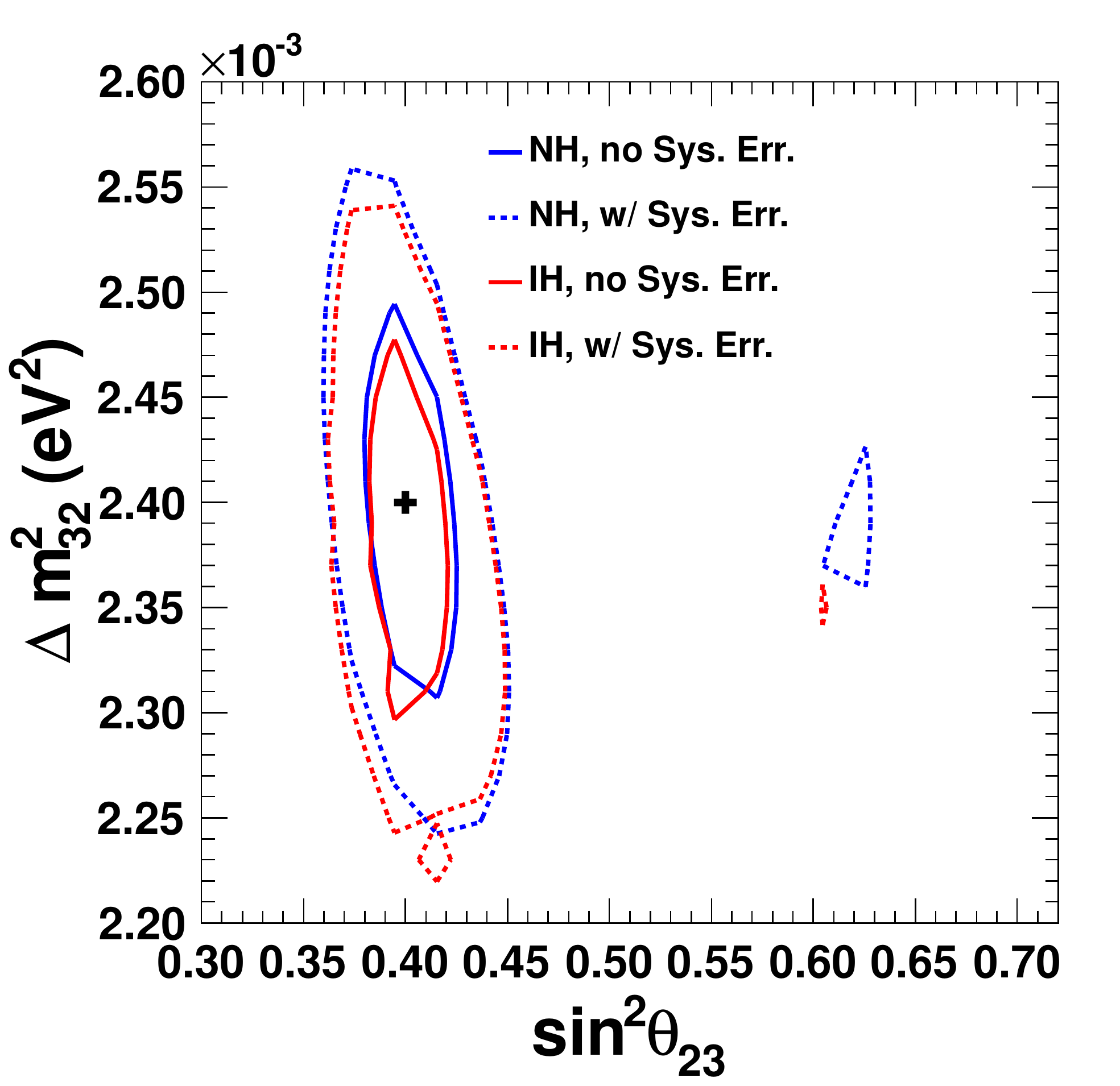}
\caption{50\% \(\nu\)-, 50\% \(\bar{\nu}\)-mode, 
with ultimate reactor error.} 
\end{subfigure} 
\caption[Disappearance 90\% C.L. Regions with $\sin^2\theta_{23}=0.4$]{\(\Delta m^2_{32}\) vs.\ \(\sin^2\theta_{23}\) 90\% C.L.\ intervals for \(7.8\times10^{21}\)~POT.  Contours are plotted for the case of true \(\delta_{CP} = 0\degree\), 
\(\sin^2\theta_{23}=0.4\), $\Delta m^2_{32} = 2.4\times10^{-3}~\mbox{eV}^2$ and NH. 
The blue curves are fit
assuming the correct MH(NH), while the red are fit assuming the incorrect MH(IH), and
contours are plotted from the minimum \(\chi^2\)
value for both MH assumptions.
The solid contours are with statistical error only, while the dashed contours
include the 2012 systematic errors fully correlated between \(\nu\)- and
\(\bar{\nu}\)-mode.
\label{fig:dis2dregionspt39S2Th23_text}} \end{figure}

Figure~\ref{fig:dis2dregionspt39S2Th23_text} shows example 90\% C.L.\ 
regions for $\Delta m_{32}^2$ vs.\ $\sin^2{\theta_{23}}$
at the full T2K statistics 
for $\sin^2{\theta_{23}}=0.4$. 
The $\theta_{23}$ octant can be resolved in this case by combining both $\nu$-mode and $\bar{\nu}$-mode
data and also including a reactor constraint on $\theta_{13}$, where this 
combination of inputs 
is required to resolve degeneracies between the oscillation parameters 
\(\sin^2\theta_{23}\), \(\sin^22\theta_{13}\), and \(\delta_{CP}\),
demonstrating the importance of the reactor constraint in this case.
 
\subsection{Sensitivities for CP-violating term, non-maximal $\theta_{23}$,
and $\theta_{23}$ octant}
\begin{figure}[htbp]
\centering 
\begin{subfigure}[t]{7cm}
\includegraphics[width=7cm]{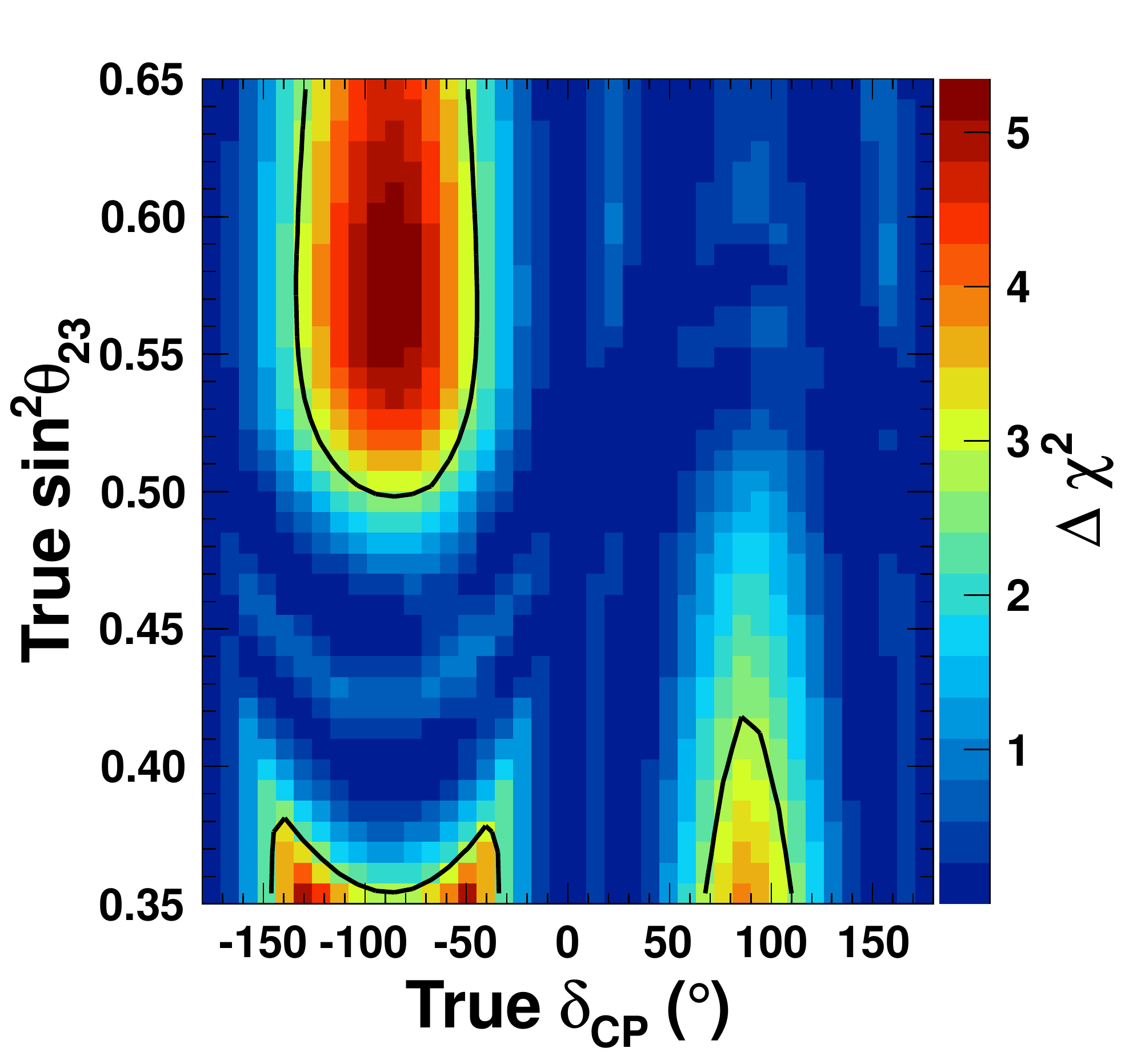}
\caption{Normal mass hierarchy. \\100\% \(\nu\)-mode.} 
\end{subfigure} \quad
\begin{subfigure}[t]{7cm}
\includegraphics[width=7cm]{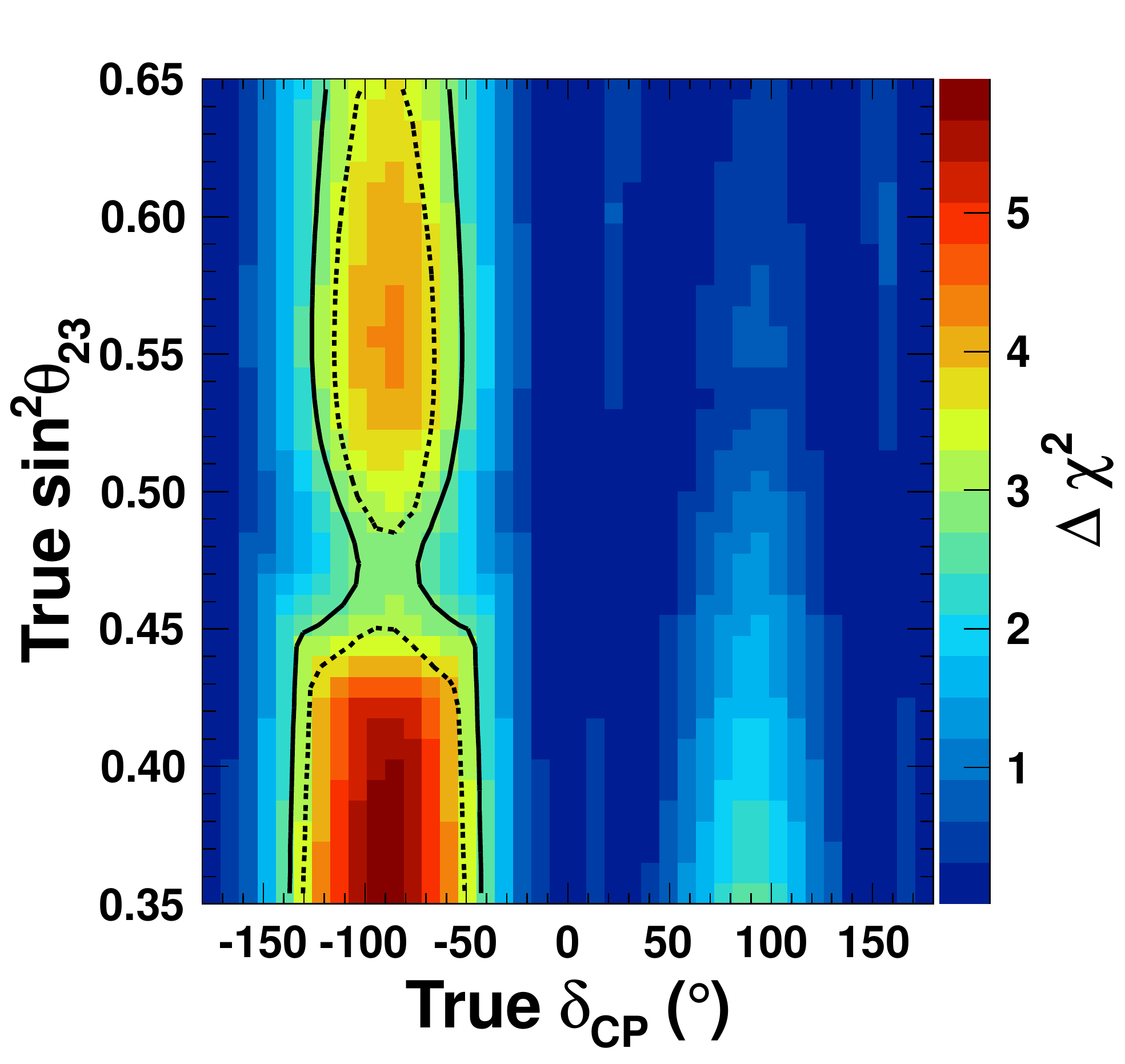}
\caption{Normal mass hierarchy. \\50\% \(\nu\)-, 50\% \(\bar{\nu}\)-mode.}
\end{subfigure} 
\begin{subfigure}[t]{7cm}
\includegraphics[width=7cm]{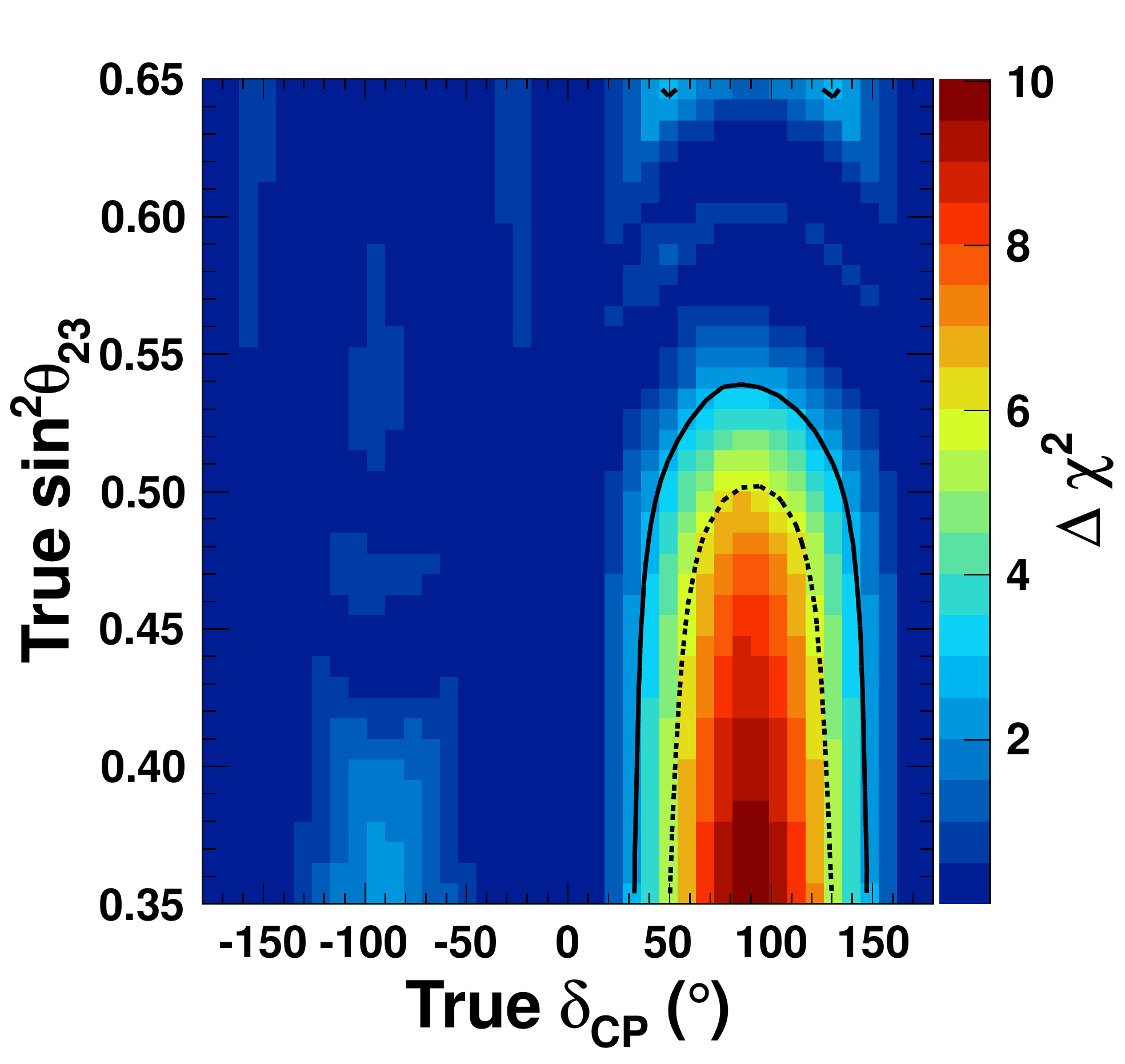}
\caption{Inverted mass hierarchy. \\100\% \(\nu\)-mode.} 
\end{subfigure} \quad
\begin{subfigure}[t]{7cm}
\includegraphics[width=7cm]{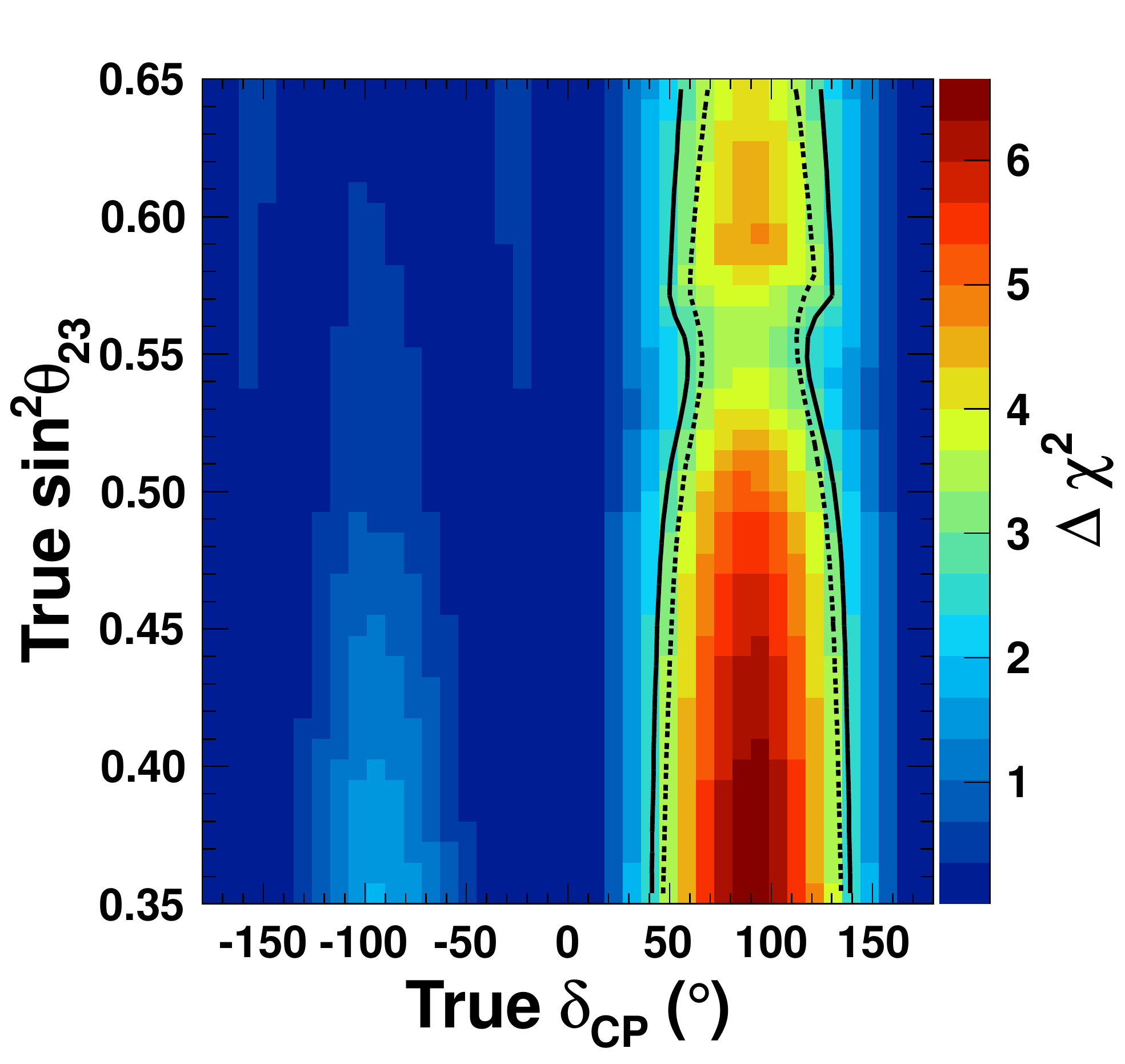}
\caption{Inverted mass hierarchy. \\50\% \(\nu\)-, 50\% \(\bar{\nu}\)-mode.}
\end{subfigure} 
\caption[$\sin\delta_{CP}$ Resolved Contours]{
The expected $\Delta\chi^2$ for the $\sin\delta_{CP} = 0$ hypothesis, in the
\(\delta_{CP}-\sin^2\theta_{23}\) plane.
The $\Delta\chi^2$ map shown in color is calculated assuming 
no systematic errors.
The solid contours show the 90\% C.L.\ sensitivity with statistical error only, 
while the dashed contours include the 2012 T2K systematic error.
The dashed contour does not appear in (a) because T2K does not have
90\% C.L.\ sensitivity in this case.
\label{fig:cpresolvedcontour}} \end{figure}

\begin{figure}[htbp]
\centering 
\begin{subfigure}[t]{7cm}
\includegraphics[width=7cm]
{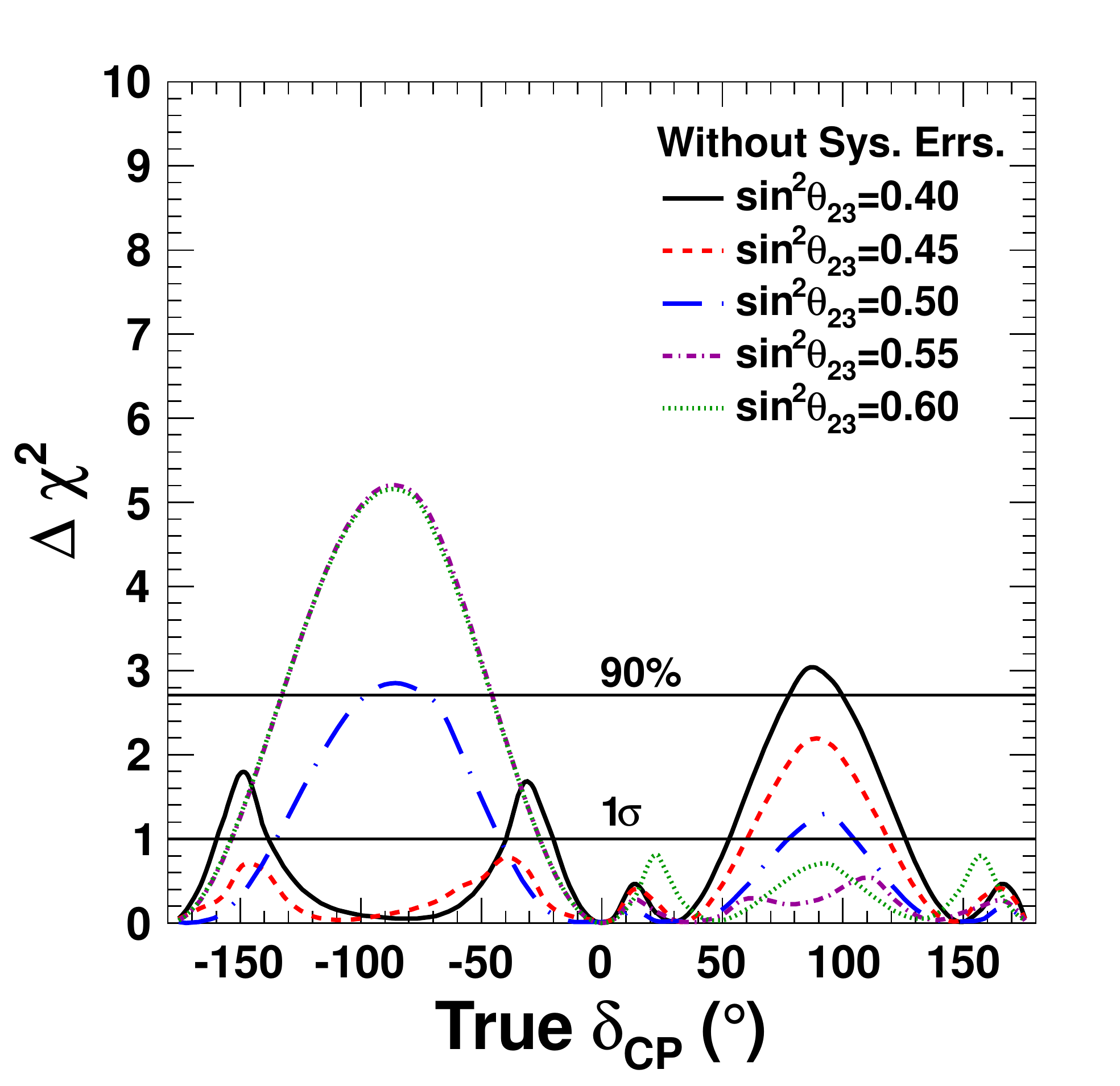}
\caption{100\% \(\nu\)-mode, \\statistical error only.} 
\end{subfigure} \quad
\begin{subfigure}[t]{7cm}
\includegraphics[width=7cm]
{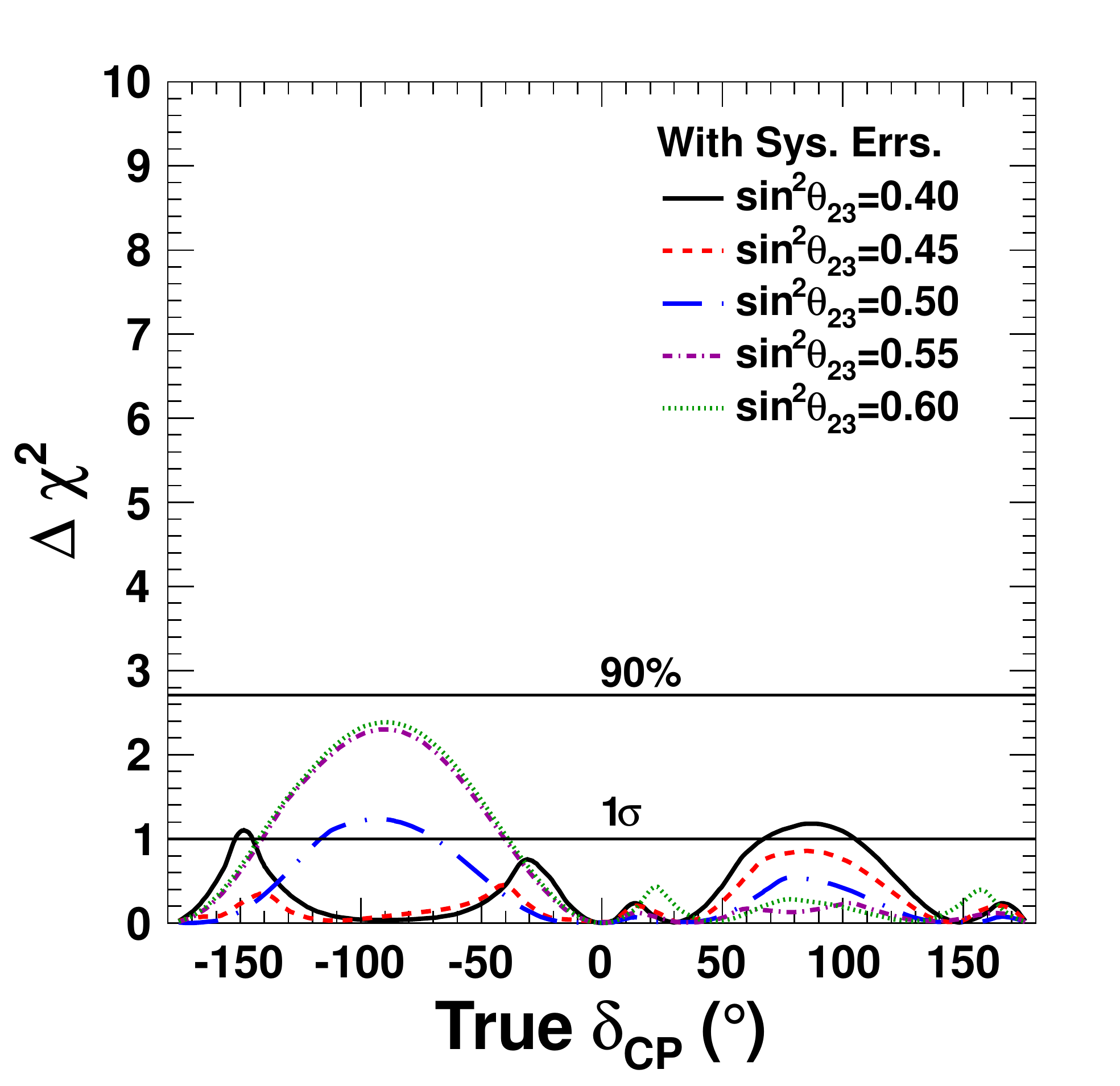}
\caption{100\% \(\nu\)-mode, \\
with the 2012 systematic errors.}
\end{subfigure}
\begin{subfigure}[t]{7cm}
\includegraphics[width=7cm]
{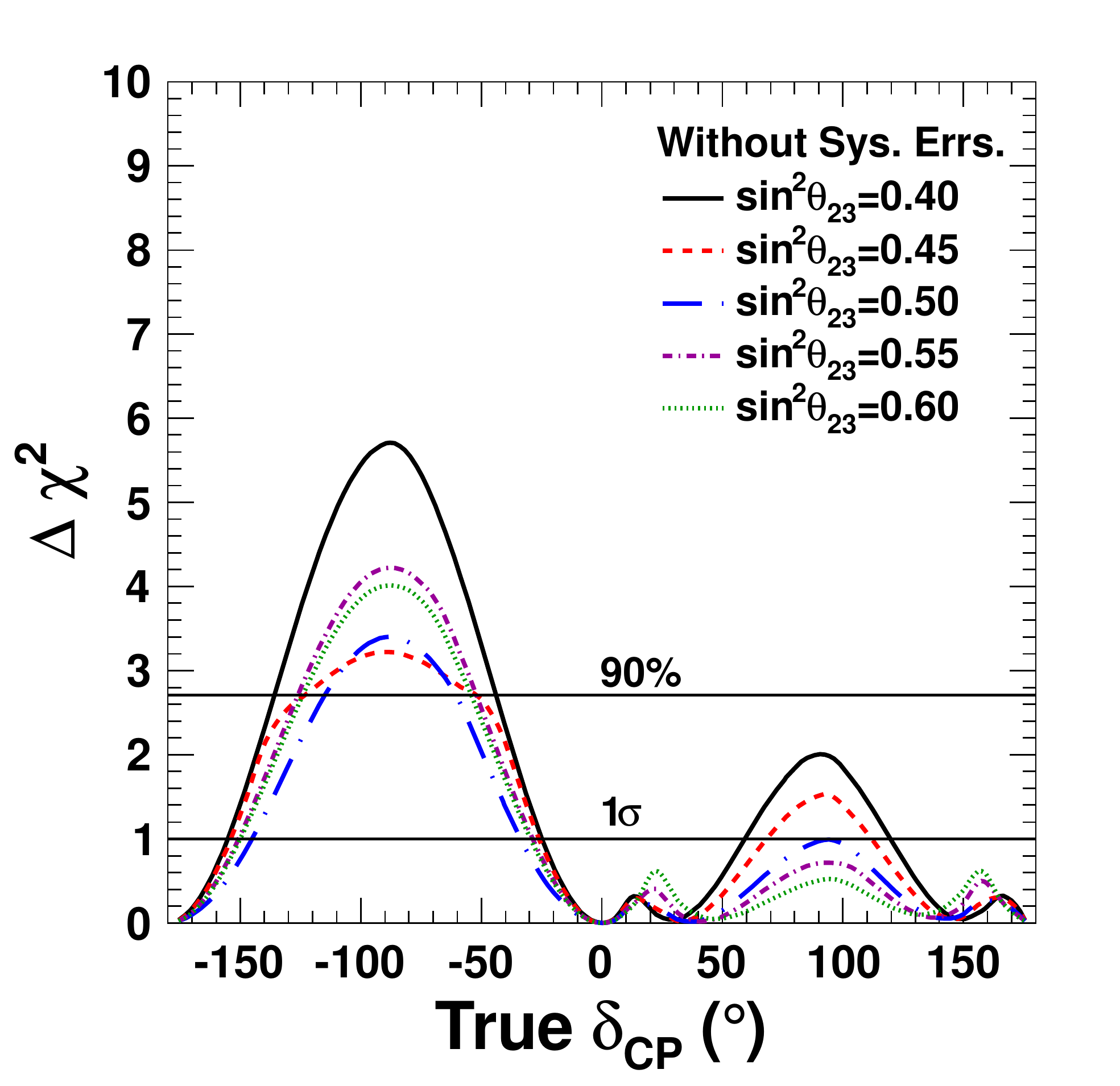}
\caption{50\% \(\nu\), 50\% $\bar{\nu}$-mode,\\ 
statistical error only.} 
\end{subfigure} \quad
\begin{subfigure}[t]{7cm}
\includegraphics[width=7cm]
{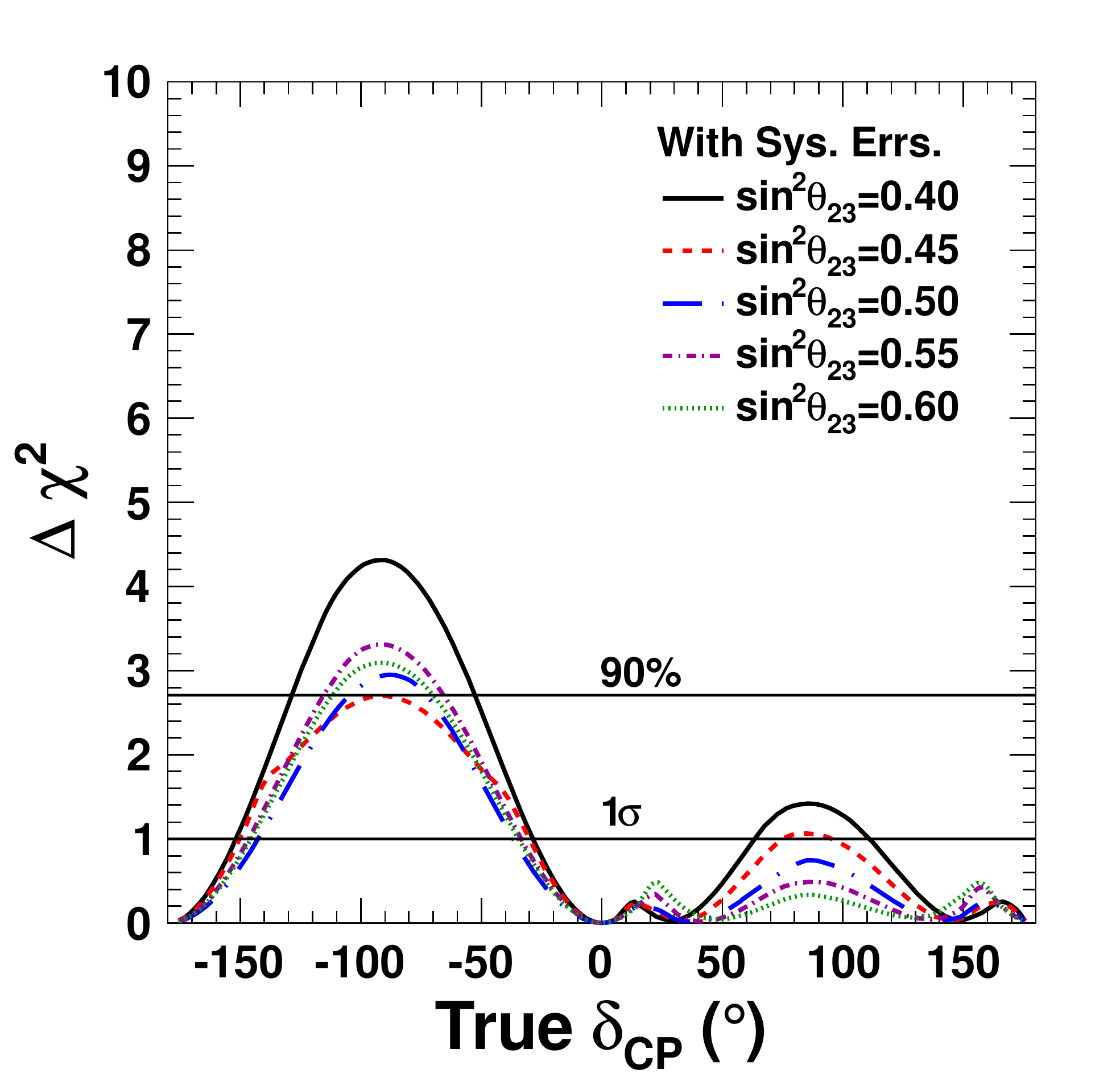}
\caption{50\% \(\nu\)-, 50\% $\bar{\nu}$-mode, \\
with the 2012 systematic errors.} 
\end{subfigure}
\caption[$\sin\delta_{CP}$ sensitivity]{
The expected $\Delta\chi^2$ for the $\sin\delta_{CP} = 0$ hypothesis, plotted
as a function of \(\delta_{CP}\) for various values of 
\(\sin^2\theta_{23}\) (given in the legend) in the case of normal mass
hierarchy.
\label{fig:cpresolvedvscpnh}} \end{figure}

\begin{figure}[htbp]
\centering 
\begin{subfigure}[t]{7cm}
\includegraphics[width=7cm]
{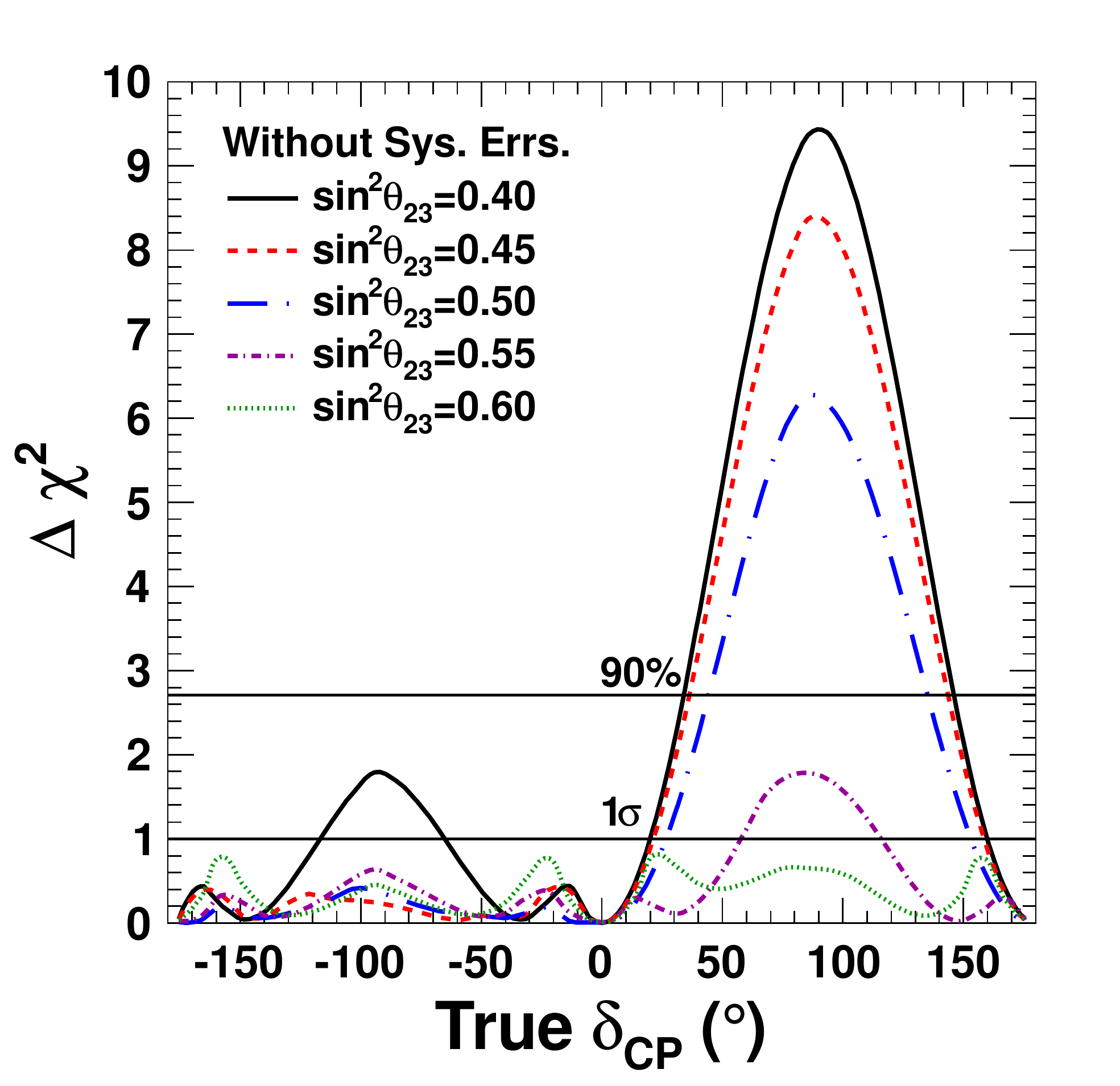}
\caption{100\% \(\nu\)-mode, \\statistical error only.} 
\end{subfigure} \quad
\begin{subfigure}[t]{7cm}
\includegraphics[width=7cm]
{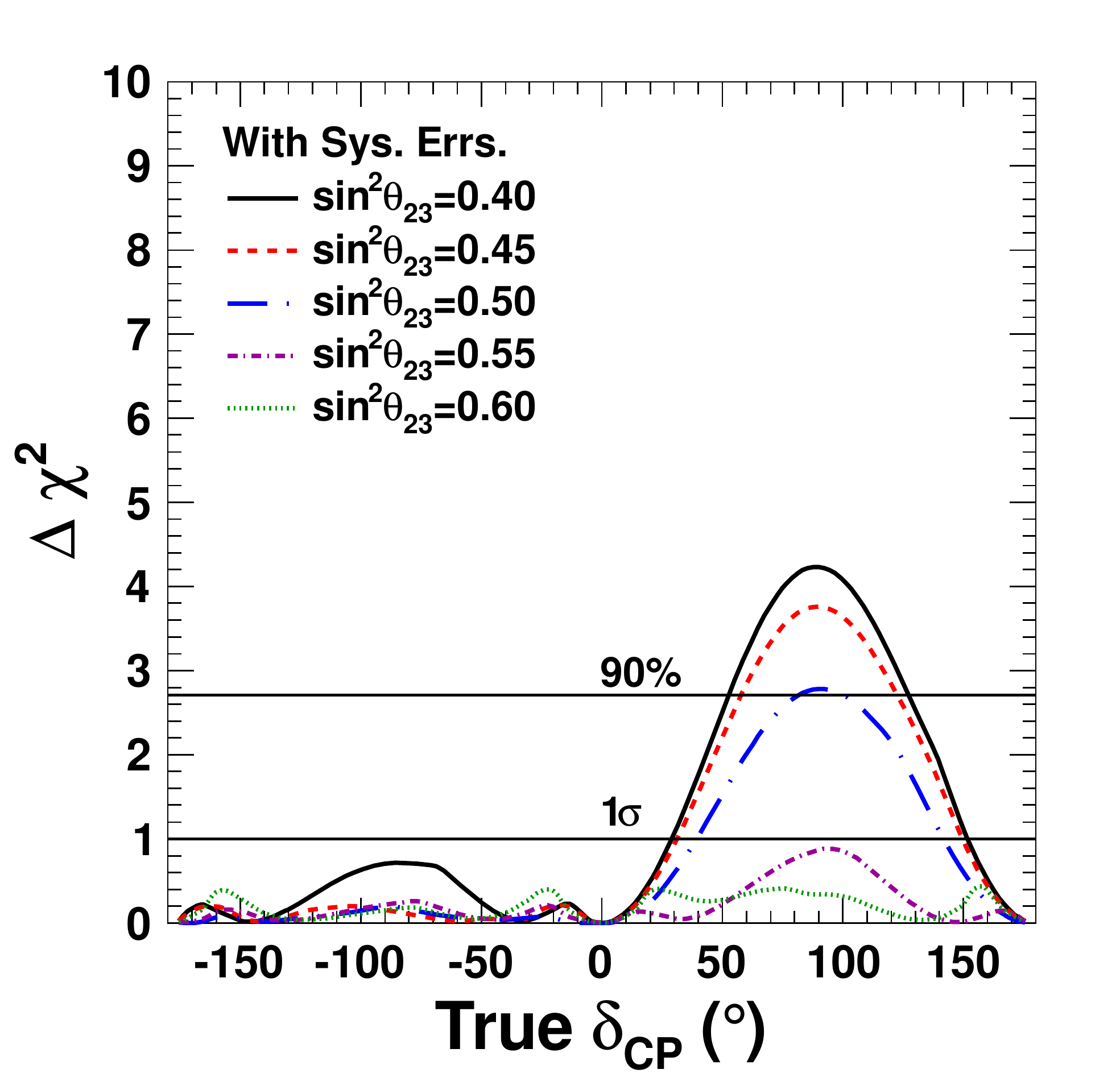}
\caption{100\% \(\nu\)-mode, \\with the 2012 systematic errors.}
\end{subfigure}
\begin{subfigure}[t]{7cm}
\includegraphics[width=7cm]
{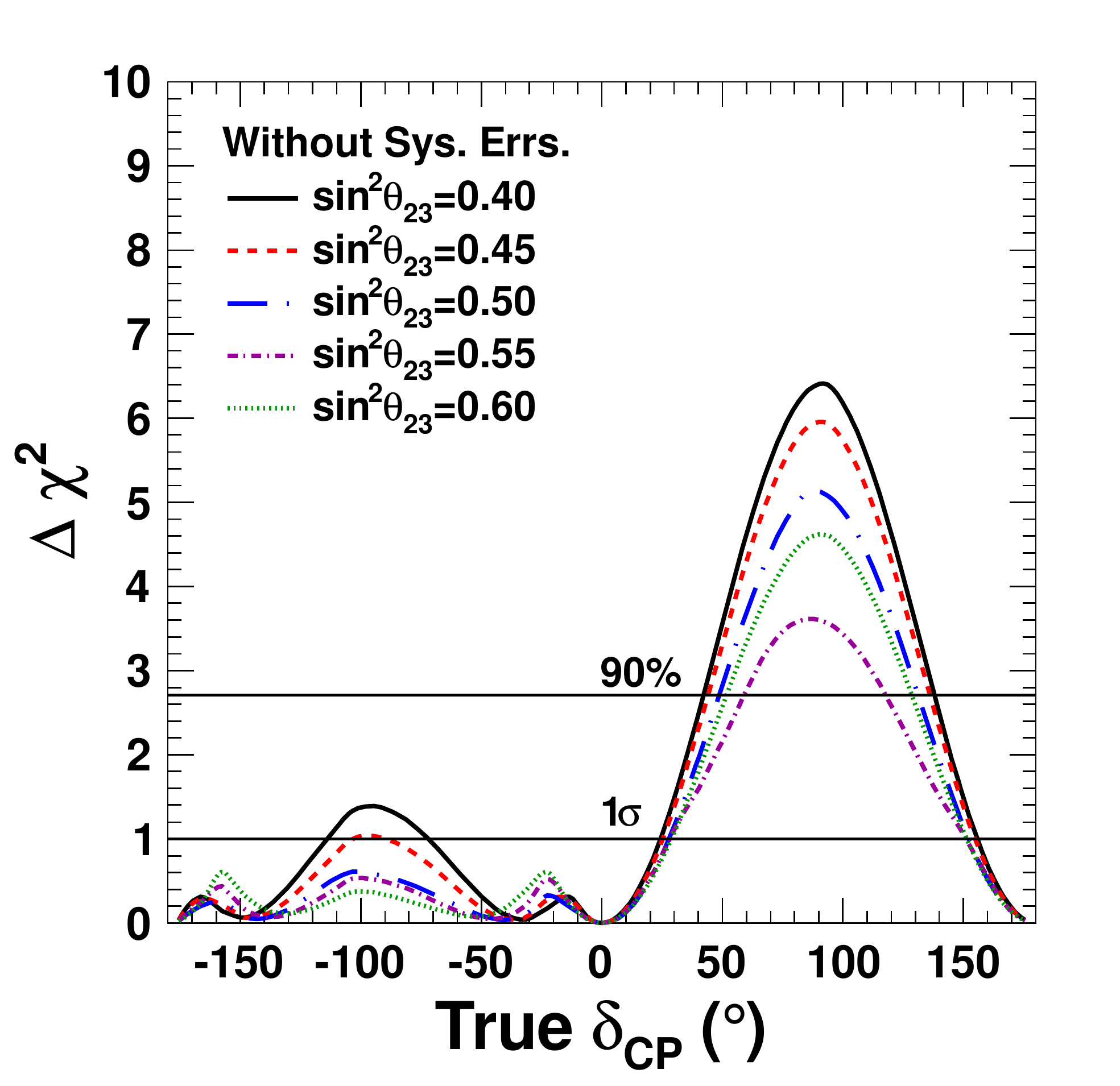}
\caption{50\% \(\nu\), 50\% $\bar{\nu}$-mode, \\statistical error only.} 
\end{subfigure} \quad
\begin{subfigure}[t]{7cm}
\includegraphics[width=7cm]
{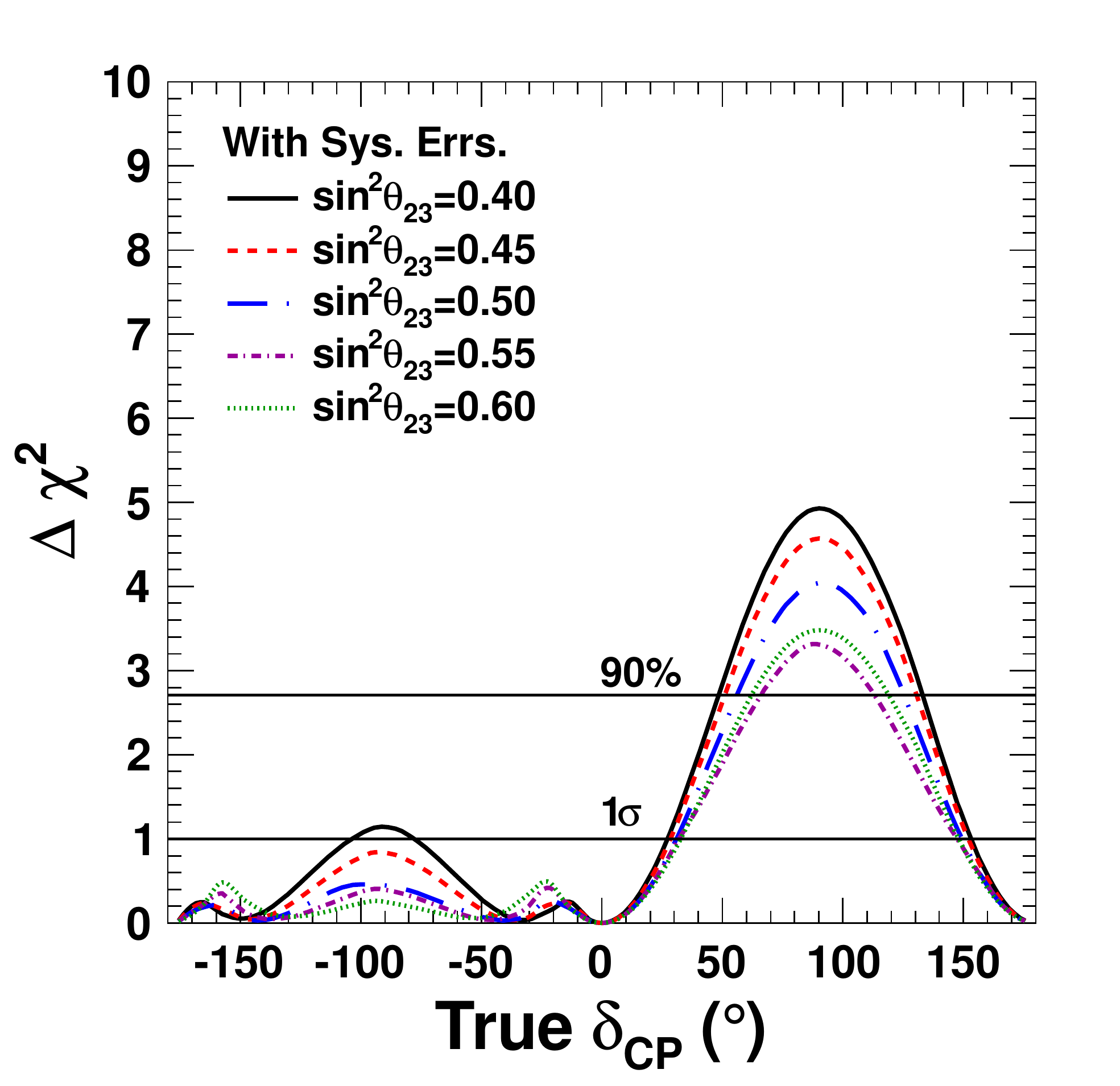}
\caption{50\% \(\nu\)-, 50\% $\bar{\nu}$-mode,\\ 
with the 2012 systematic errors.} 
\end{subfigure}
\caption[$\sin\delta_{CP}$ sensitivity]{
The expected $\Delta\chi^2$ for the $\sin\delta_{CP} = 0$ hypothesis, plotted
as a function of \(\delta_{CP}\) for various values of 
\(\sin^2\theta_{23}\) (given in the legend) in the case of inverted mass
hierarchy.
\label{fig:cpresolvedvscpih}} \end{figure}

The sensitivities for CP violation, non-maximal $\theta_{23}$, and the octant of 
$\theta_{23}$ (i.e., whether the mixing
angle \(\theta_{23}\) is less than or greater than 45\(\degree\)) 
depend on the true oscillation parameter values.
Fig.~\ref{fig:cpresolvedcontour} shows the expected 
$\Delta\chi^2$ for
the $\sin\delta_{CP}=0$ hypothesis, for various true values of \(\delta_{CP}\) and 
\(\sin^2\theta_{23}\).
To see the dependence more clearly, $\Delta\chi^2$ 
is plotted as a function of $\delta_{CP}$ for various values of 
$\sin^2\theta_{23}$ in Fig.~\ref{fig:cpresolvedvscpnh} (normal MH case)
and Fig.~\ref{fig:cpresolvedvscpih} (inverted MH case).  For 
favorable sets of the oscillation parameters and mass hierarchy,
T2K will have greater than 90\% C.L.\ sensitivity to non-zero
$\sin\delta_{\rm{CP}}$.

\begin{figure}[htbp]
\centering 
\begin{subfigure}[t]{7cm}
\includegraphics[width=7cm]{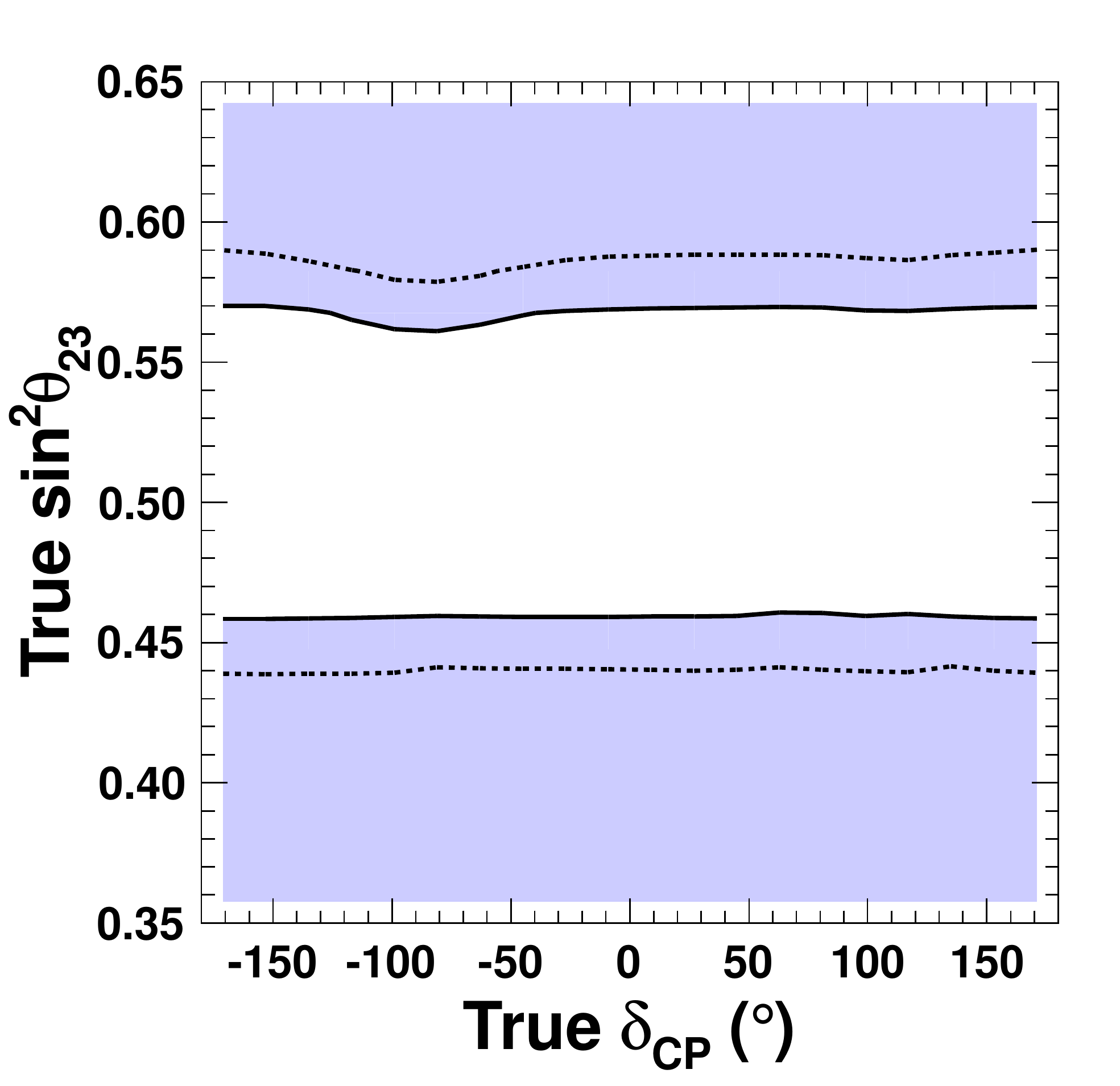}
\caption{Normal mass hierarchy. \\100\% \(\nu\)-mode} 
\end{subfigure} \quad
\begin{subfigure}[t]{7cm}
\includegraphics[width=7cm]{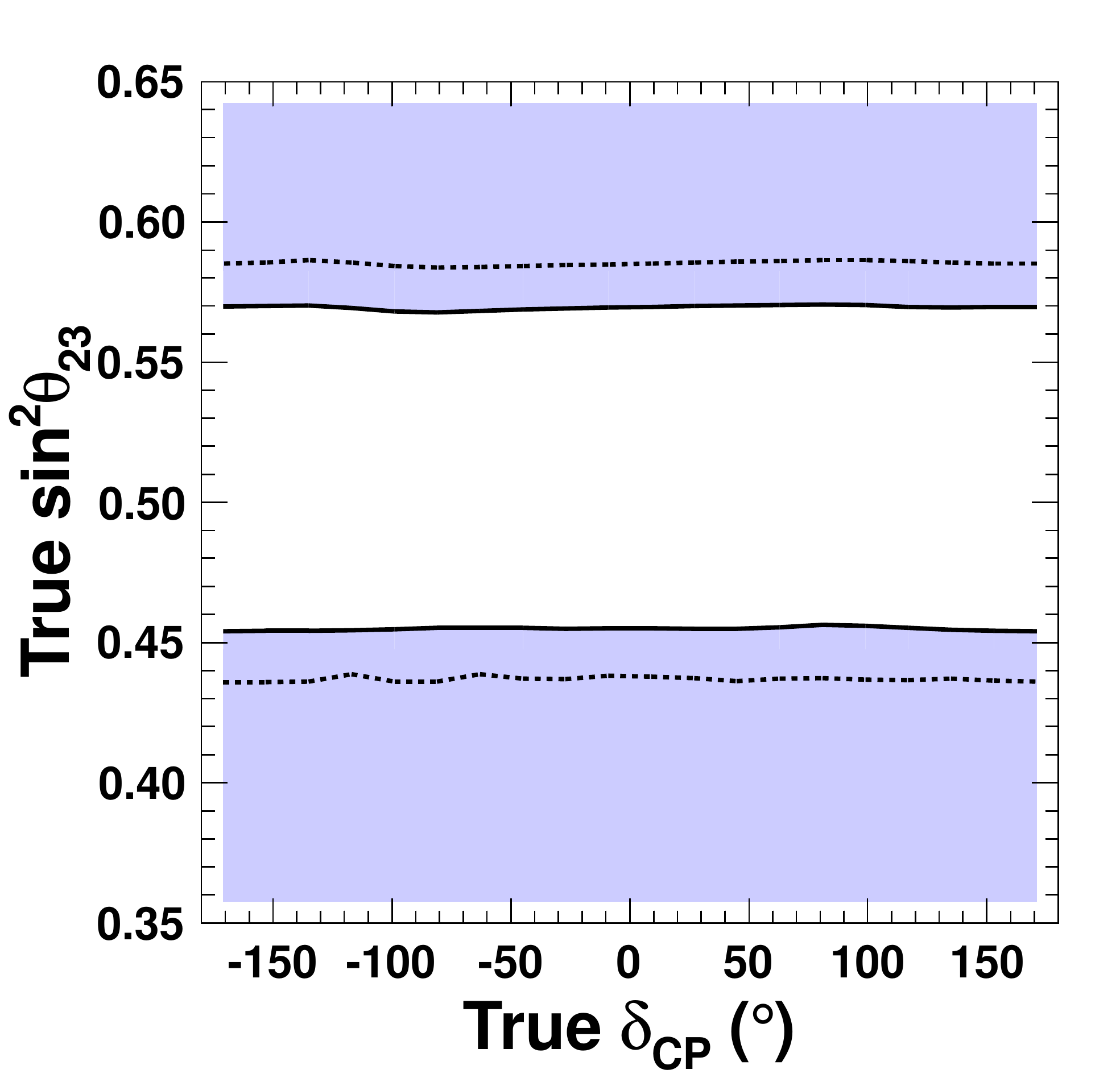}
\caption{Normal mass hierarchy. \\50\% \(\nu\)-, 50\% \(\bar{\nu}\)-mode.} 
\end{subfigure} 
\begin{subfigure}[t]{7cm}
\includegraphics[width=7cm]{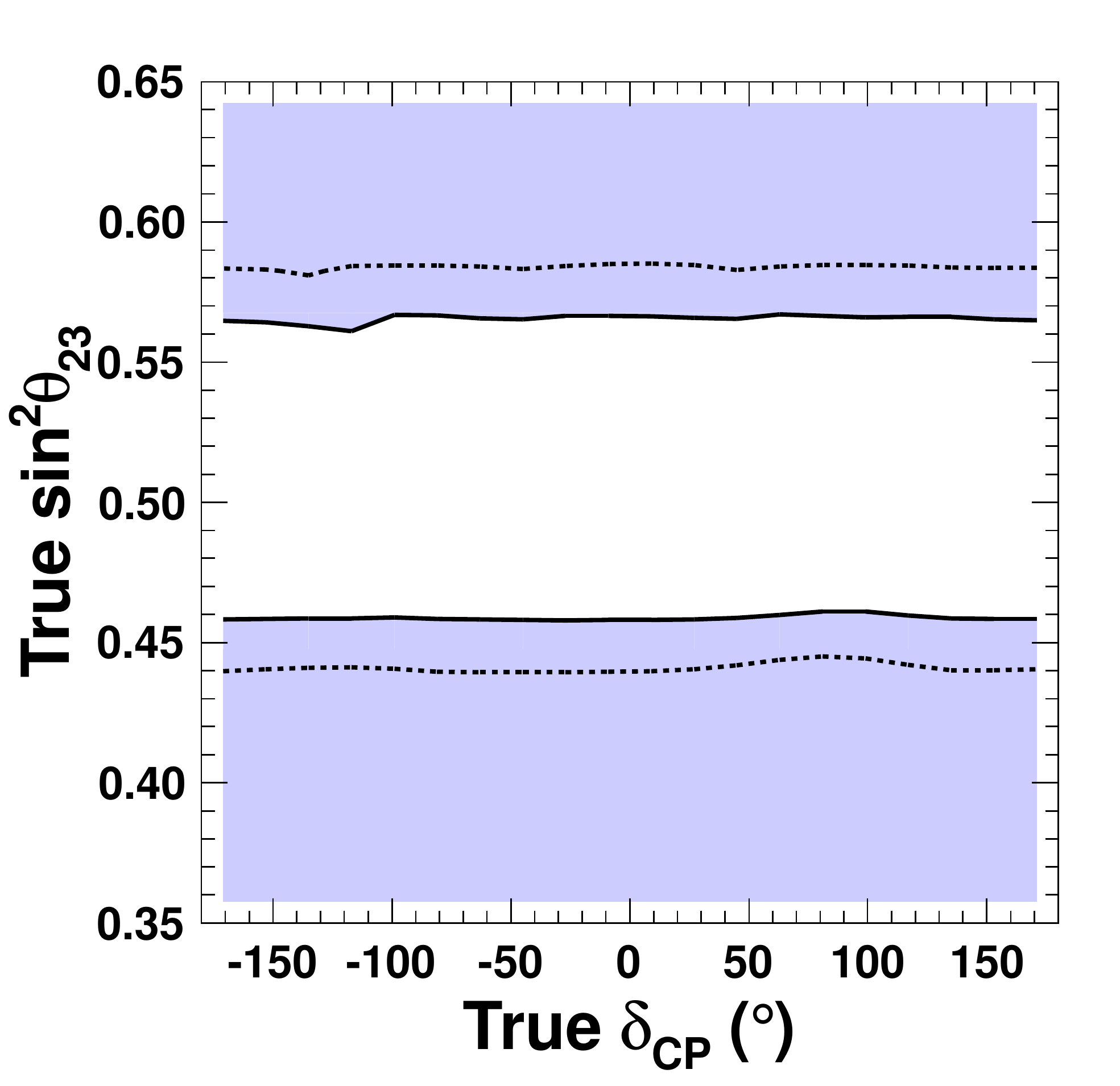}
\caption{Inverted mass hierarchy. \\100\% \(\nu\)-mode} 
\end{subfigure} \quad
\begin{subfigure}[t]{7cm}
\includegraphics[width=7cm]{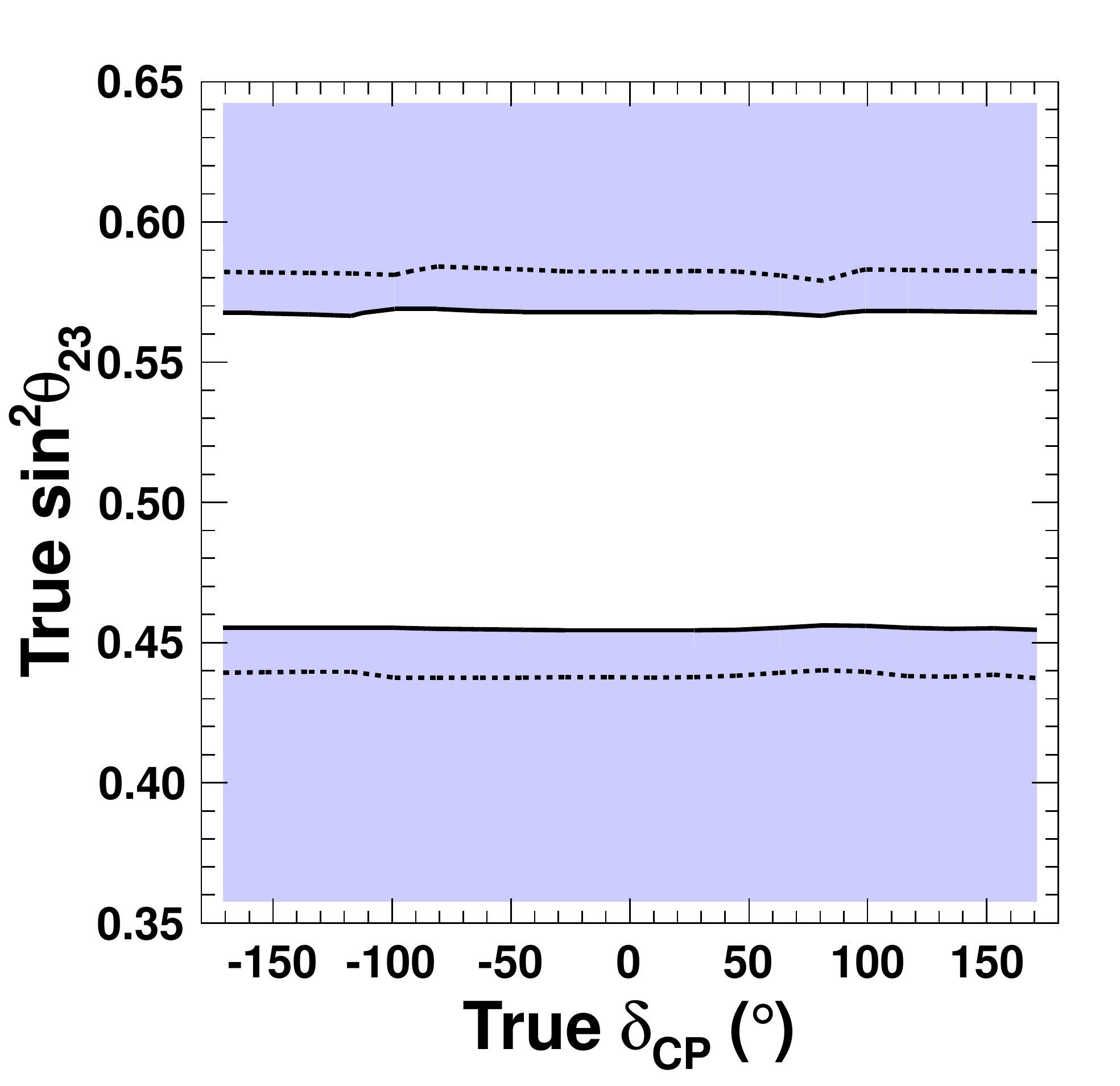}
\caption{Inverted mass hierarchy. \\50\% \(\nu\)-, 50\% \(\bar{\nu}\)-mode.} 
\end{subfigure} 
\caption[$\theta_{23}\neq\pi/4$ Resolved Contours]
{The region, shown as a shaded area, where T2K has more than a 
90 \% C.L. sensitivity to reject maximal mixing.
The shaded region is calculated assuming no systematic errors (the 
solid contours show the 90\% C.L.\ sensitivity with statistical error 
only), and the dashed contours show the sensitivity including the 2012 systematic errors.
\label{fig:nonmaxresolvedcontour}} \end{figure}

\begin{figure}[htbp]
\centering \begin{subfigure}[t]{7cm}
\includegraphics[width=7cm]{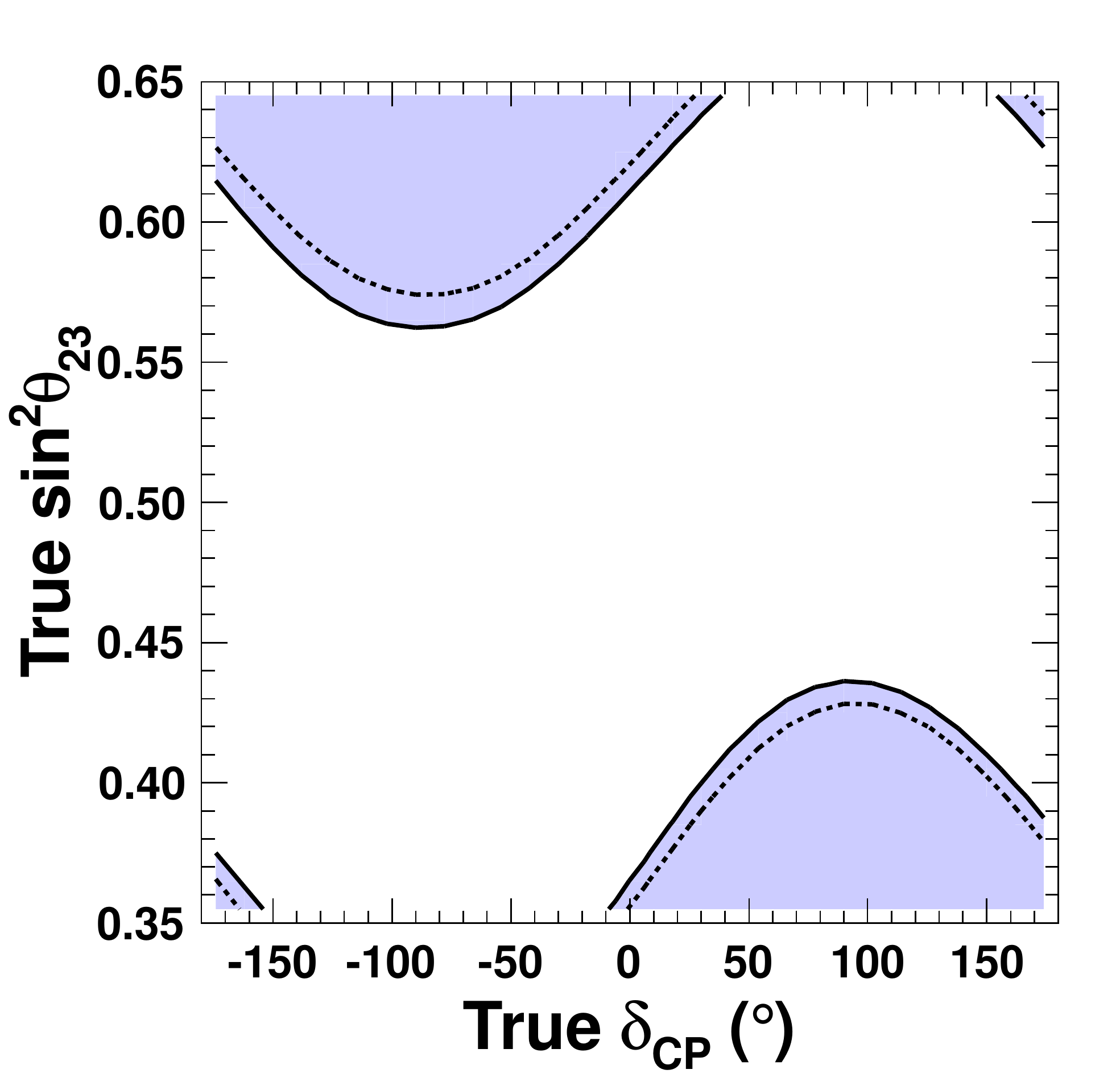}
\caption{Normal mass hierarchy. \\100\% \(\nu\)-mode.} 
\end{subfigure} \quad
\begin{subfigure}[t]{7cm}
\includegraphics[width=7cm]{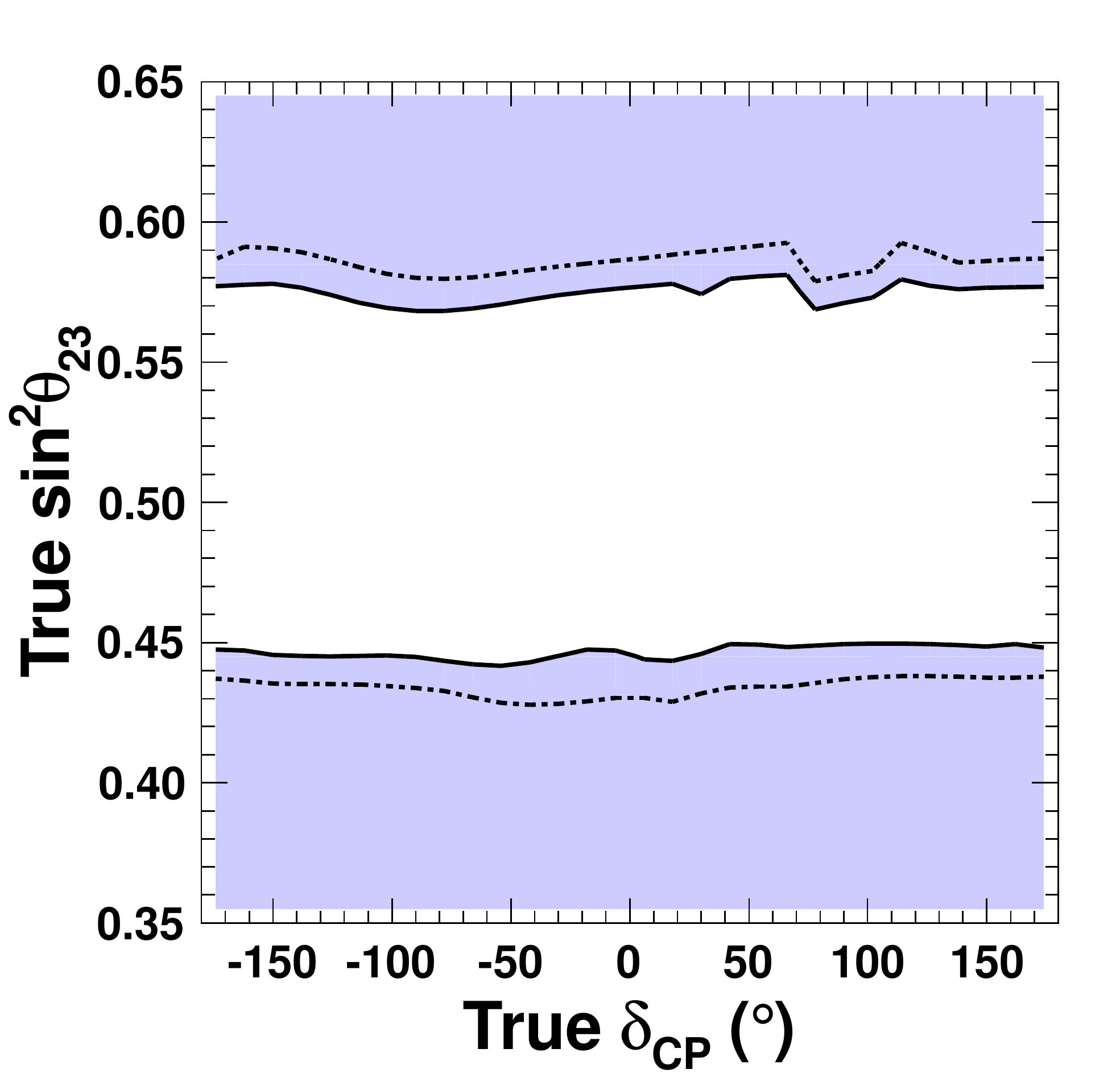}
\caption{Normal mass hierarchy. \\
50\% \(\nu\)-, 50\% \(\bar{\nu}\)-mode.} 
\end{subfigure} 
\begin{subfigure}[t]{7cm}
\includegraphics[width=7cm]{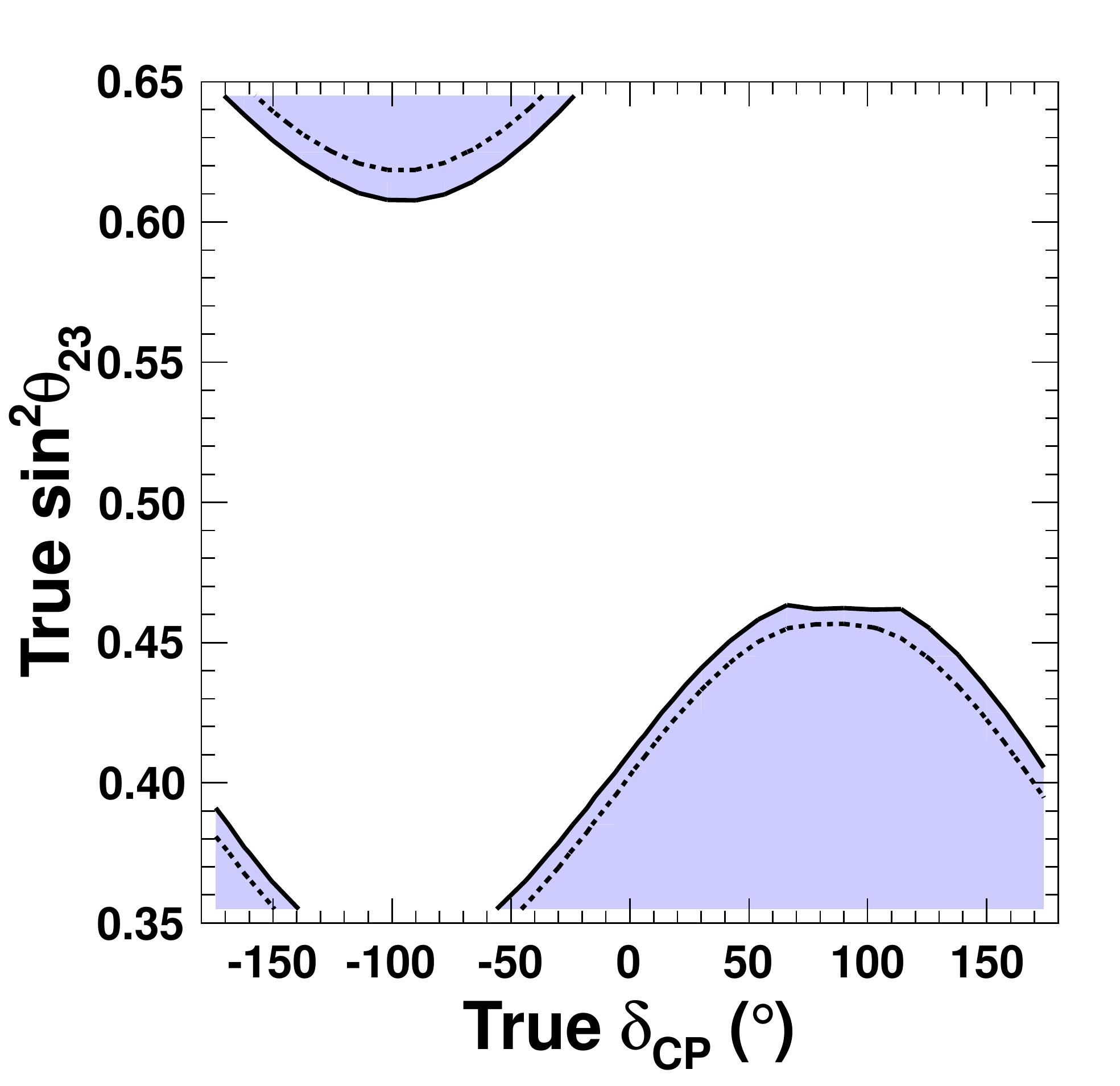}
\caption{Inverted mass hierarchy. \\100\% \(\nu\)-mode.}
\end{subfigure} \quad
\begin{subfigure}[t]{7cm}
\includegraphics[width=7cm]{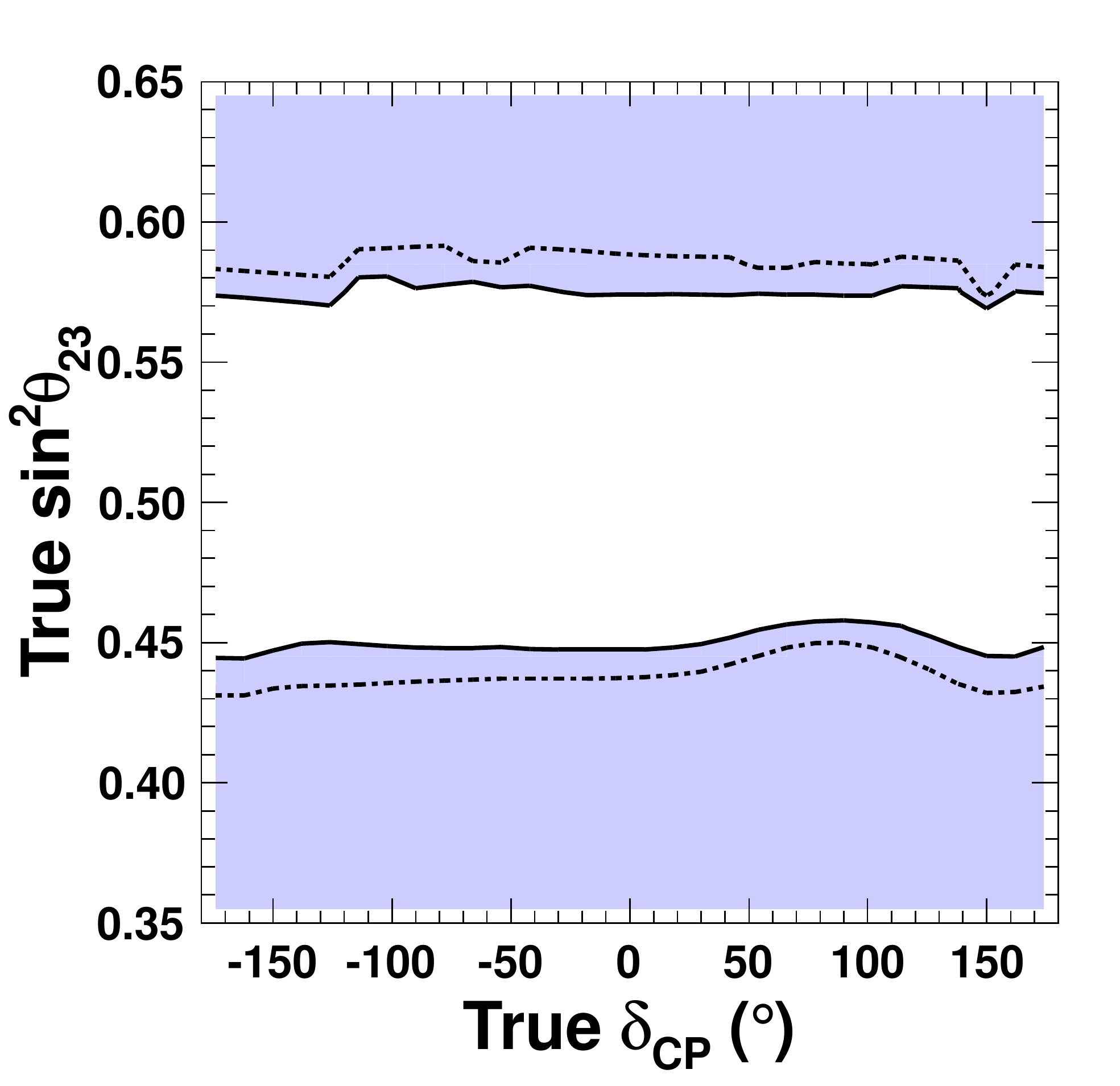}
\caption{Inverted mass hierarchy. \\
50\% \(\nu\)-, 50\% \(\bar{\nu}\)-mode.} 
\end{subfigure} 
\caption[Octant Resolved Contours]
{The region, shown as a shaded area,  where T2K has more than 
a 90\% C.L. sensitivity 
to reject one of the octants of $\theta_{23}$.
The shaded region is calculated assuming no systematic errors (the
solid contours show the 90\% C.L.\ sensitivity with statistical error 
only), and the dashed contours show the sensitivity including the 2012 T2K
systematic errors.
\label{fig:octantresolvedcontour}} \end{figure}

Figures~\ref{fig:nonmaxresolvedcontour} and \ref{fig:octantresolvedcontour} 
show the $\sin^2\theta_{23}$ vs.\ $\delta_{CP}$ regions where T2K has more 
than a 90\% C.L. 
sensitivity to reject maximal mixing 
or reject one octant of $\theta_{23}$.
In each of these figures, the oscillation parameters \(\delta_{CP}\),
\(\sin^22\theta_{13}\), \(\sin^2\theta_{23}\), \(\Delta m^2_{32}\), and 
the MH are considered unknown 
and a constraint based on the ultimate reactor error is used.
Note that the T2K sensitivity to reject maximal mixing is roughly independent of
\(\nu-\bar{\nu}\) running ratio, while the sensitivity to reject one
octant is better when \(\nu\)- and \(\bar{\nu}\)-modes are combined.
Again, the combination of \(\nu\)- and \(\bar{\nu}\)-modes, as well as the tight
constraint on \(\theta_{13}\) from the reactor measurement, are all required to
resolve the correct values for the parameters \(\sin^2\theta_{23}\),
\(\sin^22\theta_{13}\), and \(\delta_{CP}\) from many possible solutions.
Resolving the values of these three oscillation parameters is required in order to
also resolve the \(\theta_{23}\) octant.

These figures show that by running with a significant amount of
$\bar{\nu}$-mode, T2K has sensitivity to the CP-violating term
and octant of $\theta_{23}$ for a wider region of oscillation 
parameters ($\delta_{CP}, \theta_{23}$) and for both mass hierarchies,
particularly when systematic errors are taken into account.
The optimal running ratio is discussed in more detail
in Sec.~\ref{sec:runratio}.
%

\subsection{Precision or sensitivity vs. POT}
\label{sec:pfsvspot}
\begin{figure}[htbp]
\centering 
\begin{subfigure}[t]{7cm}
\includegraphics[width=7cm]
{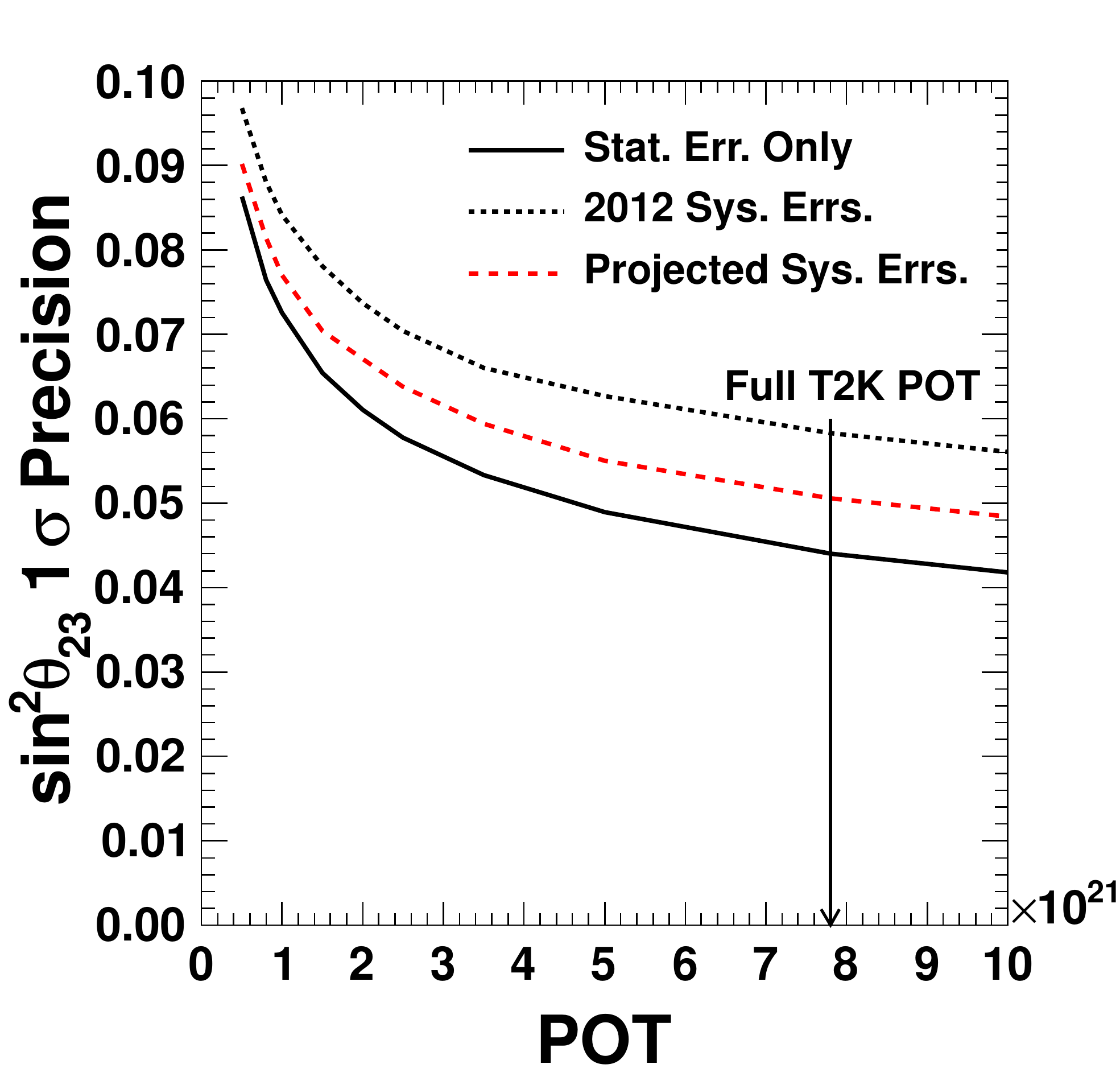}
\caption{100\% \(\nu\)-mode.} 
\end{subfigure} \quad
\begin{subfigure}[t]{7cm}
\includegraphics[width=7cm]
{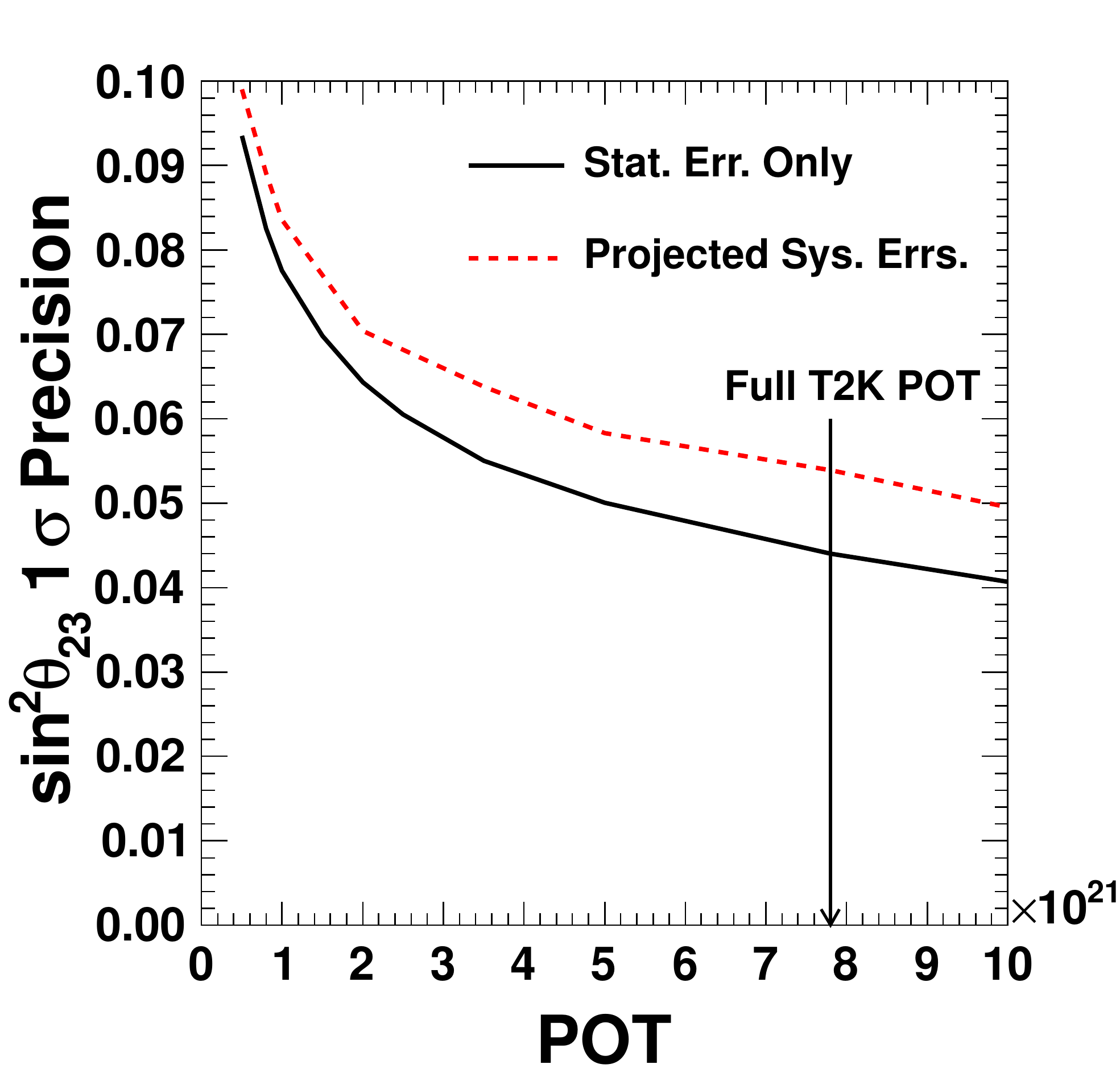}
\caption{50\% \(\nu\), 50\% \(\bar{\nu}\)-mode.} 
\end{subfigure} 
\begin{subfigure}[t]{7cm}
\includegraphics[width=7cm]
{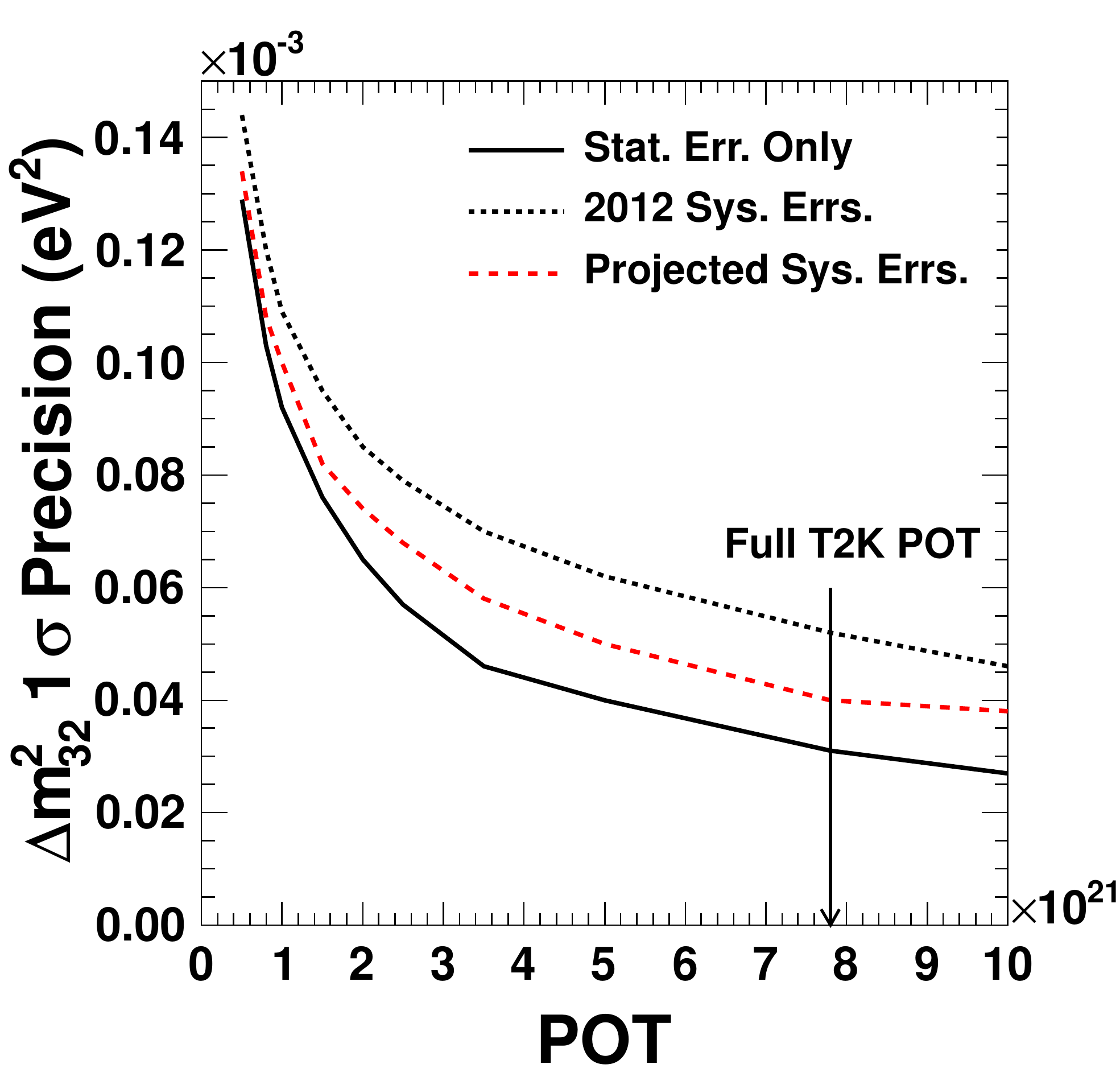}
\caption{100\% \(\nu\)-mode.} 
\end{subfigure} \quad
\begin{subfigure}[t]{7cm}
\includegraphics[width=7cm]
{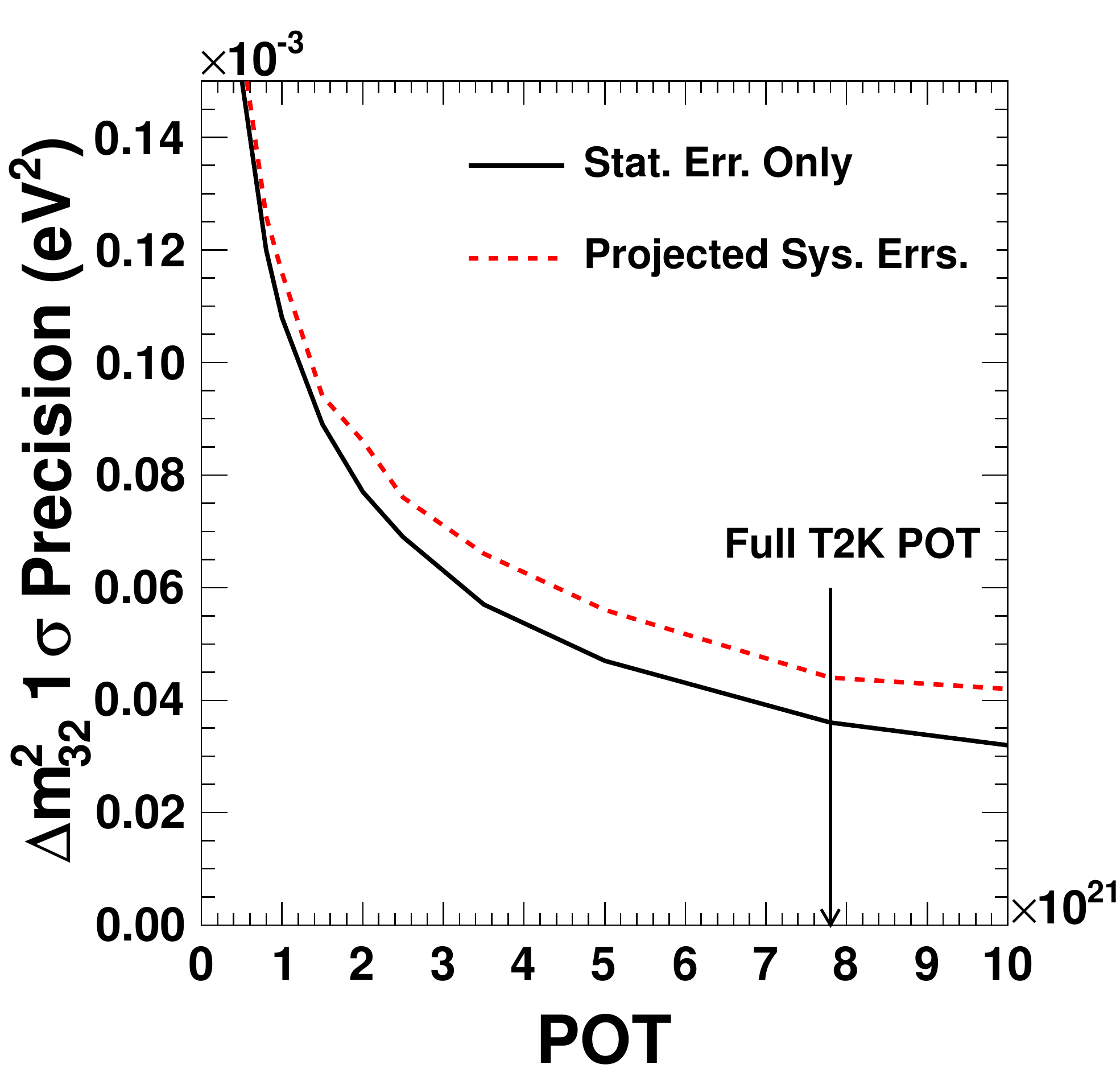}
\caption{50\% \(\nu\), 50\% \(\bar{\nu}\)-mode.} 
\end{subfigure} 
\caption[Uncertainty of $\sin^2\theta_{23}$ and $\Delta m^2_{32}$ vs.\ POT]
{The uncertainty on \(\sin^2\theta_{23}\) and $\Delta m^2_{32}$ plotted as 
a function of T2K POT.  
Plots assume the true oscillation parameters given in Table \ref{tab:truepars}.
The solid curves include statistical errors only, 
while the dashed curves assume the 2012 systematic errors (black) 
or the projected systematic errors (red).  
A constraint 
based on the ultimate reactor precision is included.
\label{fig:dispvsPOT}} \end{figure}

The T2K uncertainty (i.e.\ precision) vs.\ POT for 
\(\sin^2\theta_{23}\) and \(\Delta m^2_{32}\)
is given in Fig.~\ref{fig:dispvsPOT}
for the 100\% $\nu$-mode running case and the 50\% plus 50\% $\nu-\bar{\nu}$-mode running 
case.  
The precision includes either statistical errors only, 
statistical errors combined with the 2012 systematic errors, or statistical errors combined with 
conservatively-projected systematic errors for the full POT.  
See Sec.~\ref{sec:pfssys} for details about the projected systematic errors 
used.

Generally, the effect of the systematic errors is reduced by running 
with combined $\nu$-mode and $\bar{\nu}$-mode.
When running 50\% in $\nu$-mode and 50\% in $\bar{\nu}$-mode, 
the statistical 1$\sigma$ uncertainty of $\sin^2\theta_{23}$ and 
$\Delta m_{32}^2$ is 0.045 and $0.04\times10^{-3}$~eV$^2$,
respectively, at the T2K full statistics.

It should be noted that the sensitivity to \(\sin^2\theta_{23}\) shown here for the
current exposure (\(6.57\times10^{20}\)~POT) is significantly worse than the most 
recent T2K result \cite{numurun4}, and in fact the recent result is quite close to 
the final sensitivity (at \(7.8\times10^{21}\)~POT) shown.   This apparent
discrepancy comes from three factors.  About half of the difference between the
expected sensitivity and observed result is due to an apparent statistical fluctuation, where fewer T2K
\(\nu_\mu\) events have been observed than expected.  Of the remaining difference, half comes
from the use of a Feldman-Cousins statistical analysis for the T2K official oscillation result
which this sensitivity study does not use. The rest comes from the
location of the best fit point: the expected error depends on the true
value of \(\sin^2\theta_{23}\) because a local minimum in each octant on each side of the 
point of maximal disappearance, \(\sin^2\theta_{23} \simeq 0.503\) for \(\sin^22\theta_{13} = 0.1\), 
increases the full width of the \(\Delta\chi^2\)
curve such that the farther the true point is from maximal disappearance, 
the larger the error on \(\sin^2\theta_{23}\) becomes (where the studies here assume a true value of 
\(\sin^2\theta_{23}\) slightly lower than the point of maximal disappearance --
\(\sin^2\theta_{23}=0.5\)).
Therefore, if results from future running continue to favor maximal
disappearance we expect modest improvements in our current constraints,
eventually approaching a value close to, and possibly slightly better than, 
the predicted final sensitivity shown here.

\begin{figure}[htbp]
\centering 
\begin{subfigure}[t]{7cm}
\includegraphics[width=7cm]{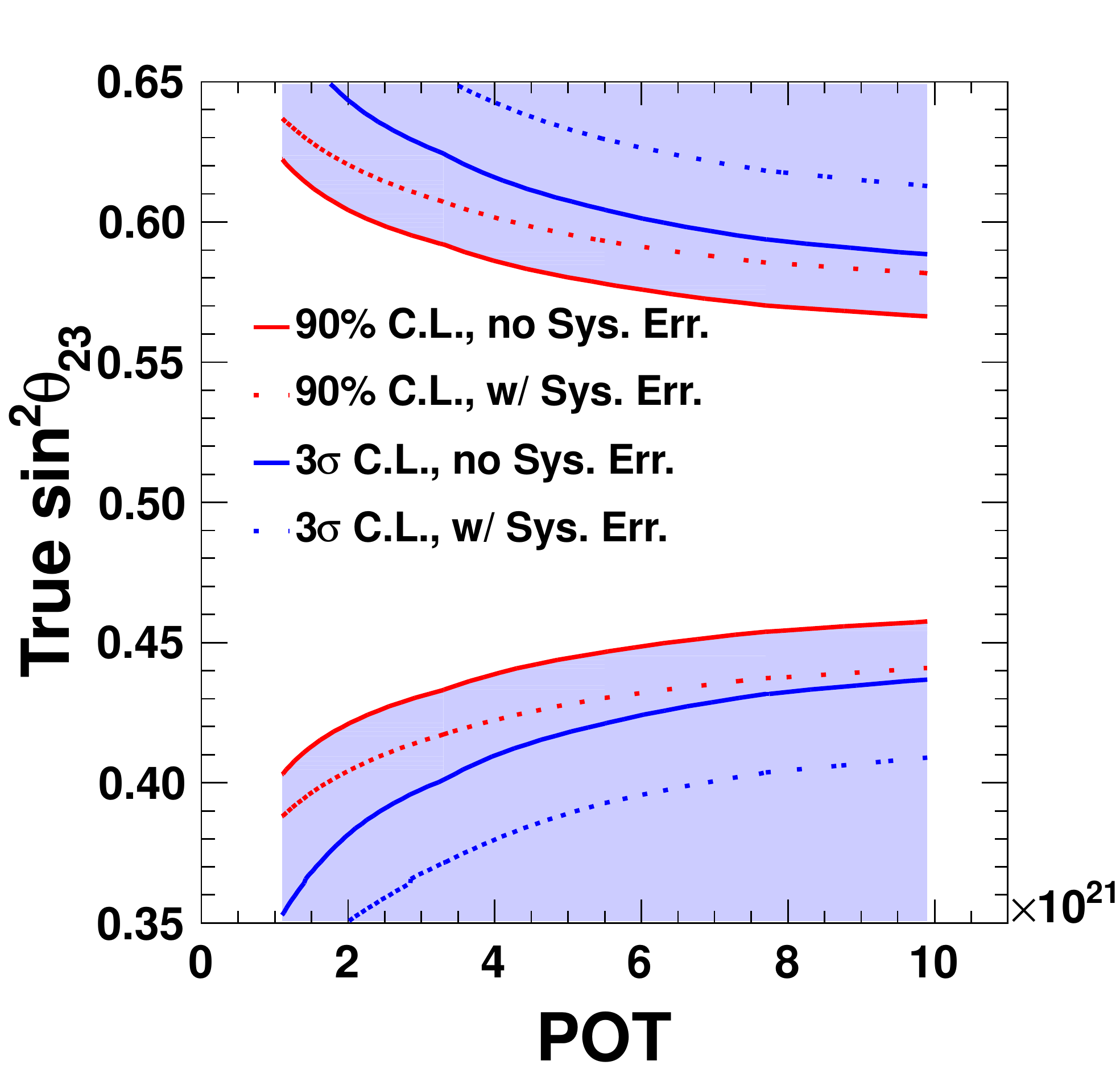}
\caption{$\theta_{23}\neq\pi/4$}
\end{subfigure} \quad
\begin{subfigure}[t]{7cm}
\includegraphics[width=7cm]{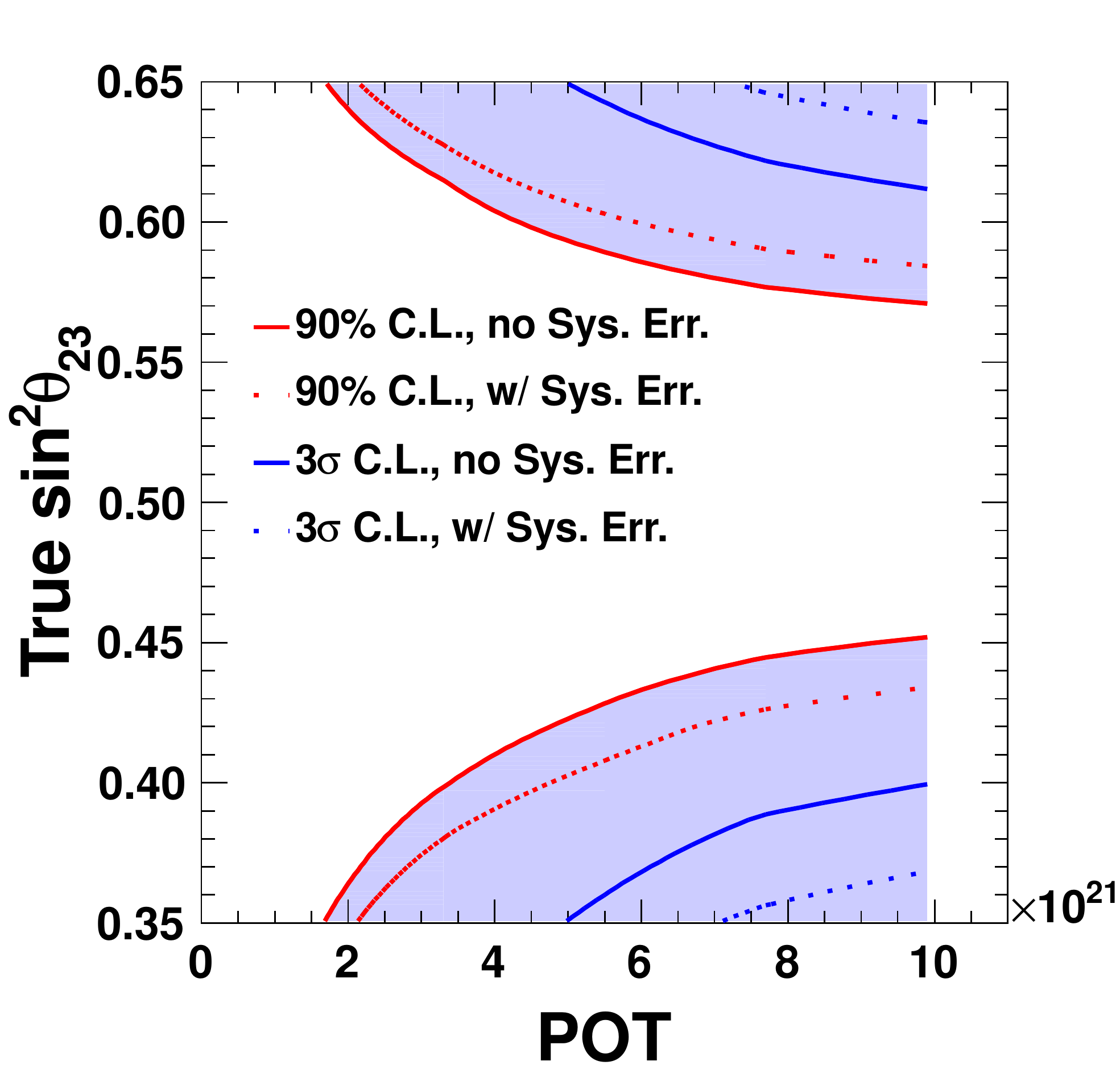}
\caption{$\theta_{23}$ Octant}
\end{subfigure}
\caption[Non-maximal or Octant of $\theta_{23}$ Resolved vs.\ POT]{
The region where maximal mixing or one $\theta_{23}$ octant 
can be rejected at the stated confidence levels (given by the shaded region), 
as a function of POT in the case of 
50\% \(\nu\)-, 50\% \(\bar{\nu}\)-mode.
These plots are made under the condition that
the true mass hierarchy is normal and $\delta_{CP}=0$.
The dashed contours
include the 2012 systematic errors fully correlated between \(\nu\) and
\(\bar{\nu}\).
A constraint based on the ultimate reactor precision is included.
\label{fig:th23resolved}} \end{figure}

Figure~\ref{fig:th23resolved} shows the $\sin^2\theta_{23}$ region where
maximal mixing or one of the $\theta_{23}$ octants can be rejected, 
as a function of POT in the case of 50\% \(\nu\)- plus 50\% \(\bar{\nu}\)-mode running.
Although these plots are made under the condition that
the true mass hierarchy is normal and $\delta_{CP}=0\degree$, 
dependence on these conditions is moderate in the case of 
50\% \(\nu\)- plus 50\% \(\bar{\nu}\)-mode running.

\begin{figure}[htbp]
\centering 
\begin{subfigure}[t]{7cm}
\includegraphics[width=7cm]
{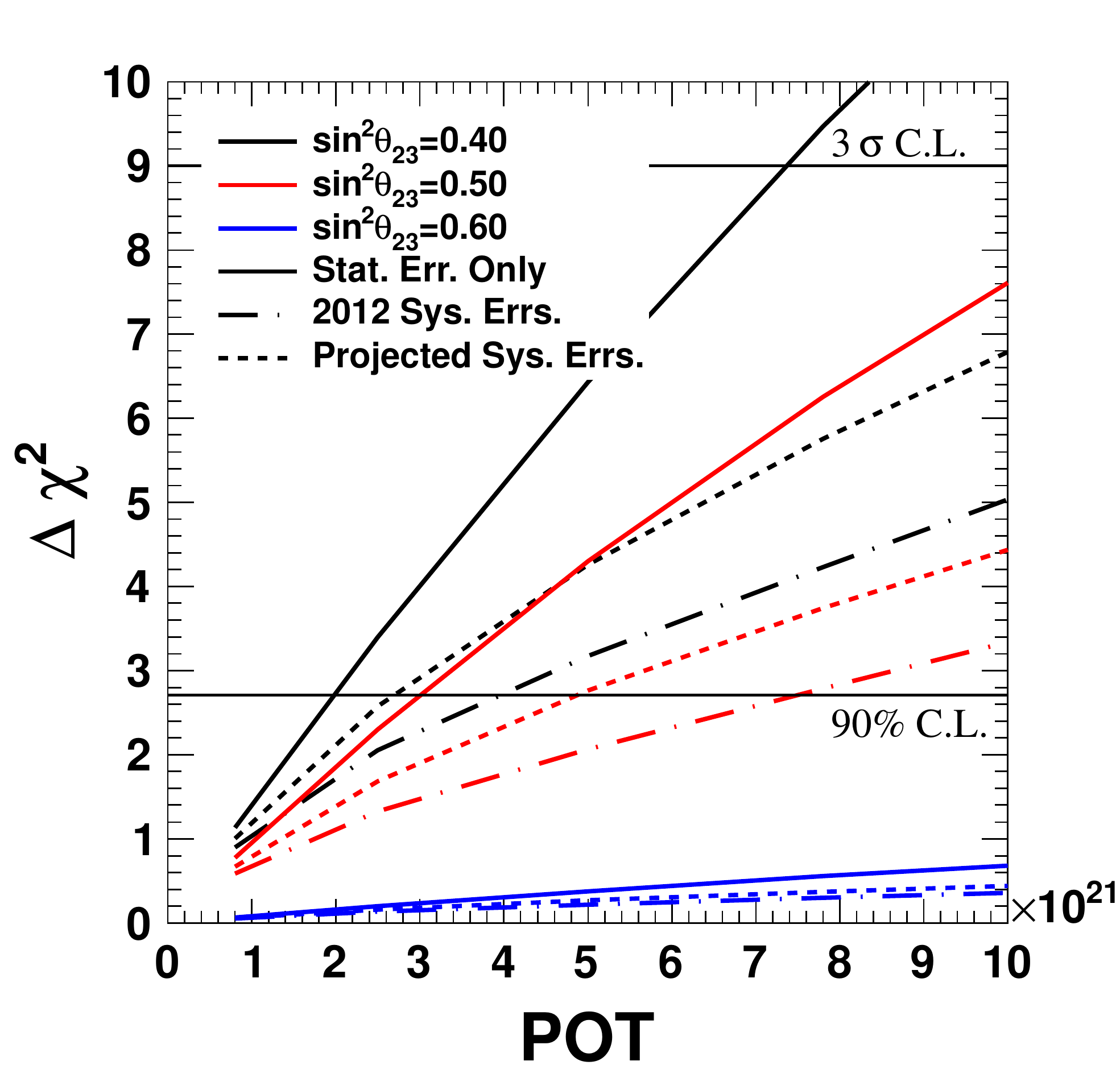}
\caption{100\% \(\nu\)-mode, \(\delta_{CP}=90\degree\), IH.} 
\end{subfigure} \quad
\begin{subfigure}[t]{7cm}
\includegraphics[width=7cm]
{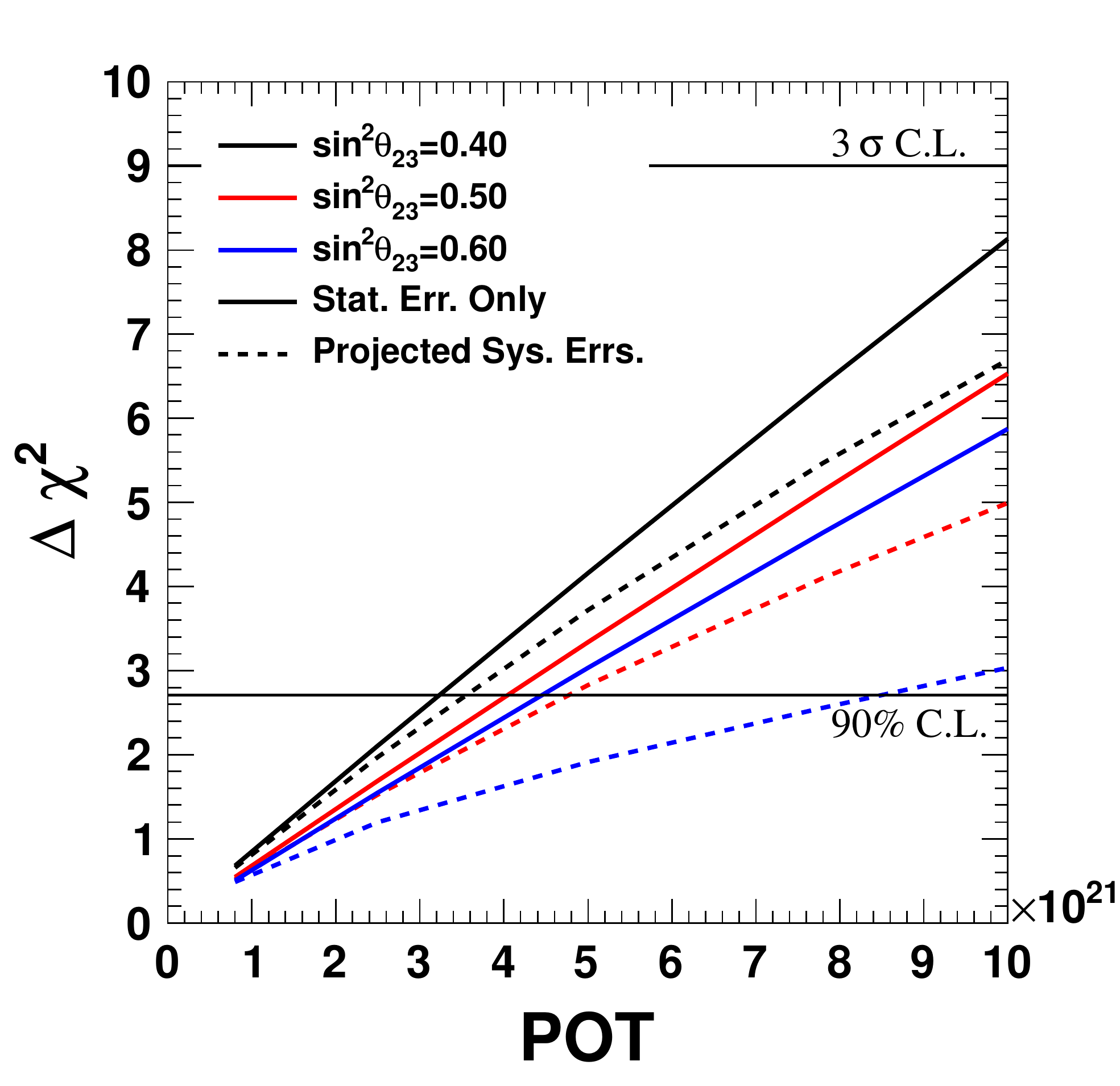}
\caption{50\% \(\nu\)-, 50\% $\bar{\nu}$ running, \(\delta_{CP}=90\degree\), IH.} 
\end{subfigure} \quad
\begin{subfigure}[t]{7cm}
\includegraphics[width=7cm]
{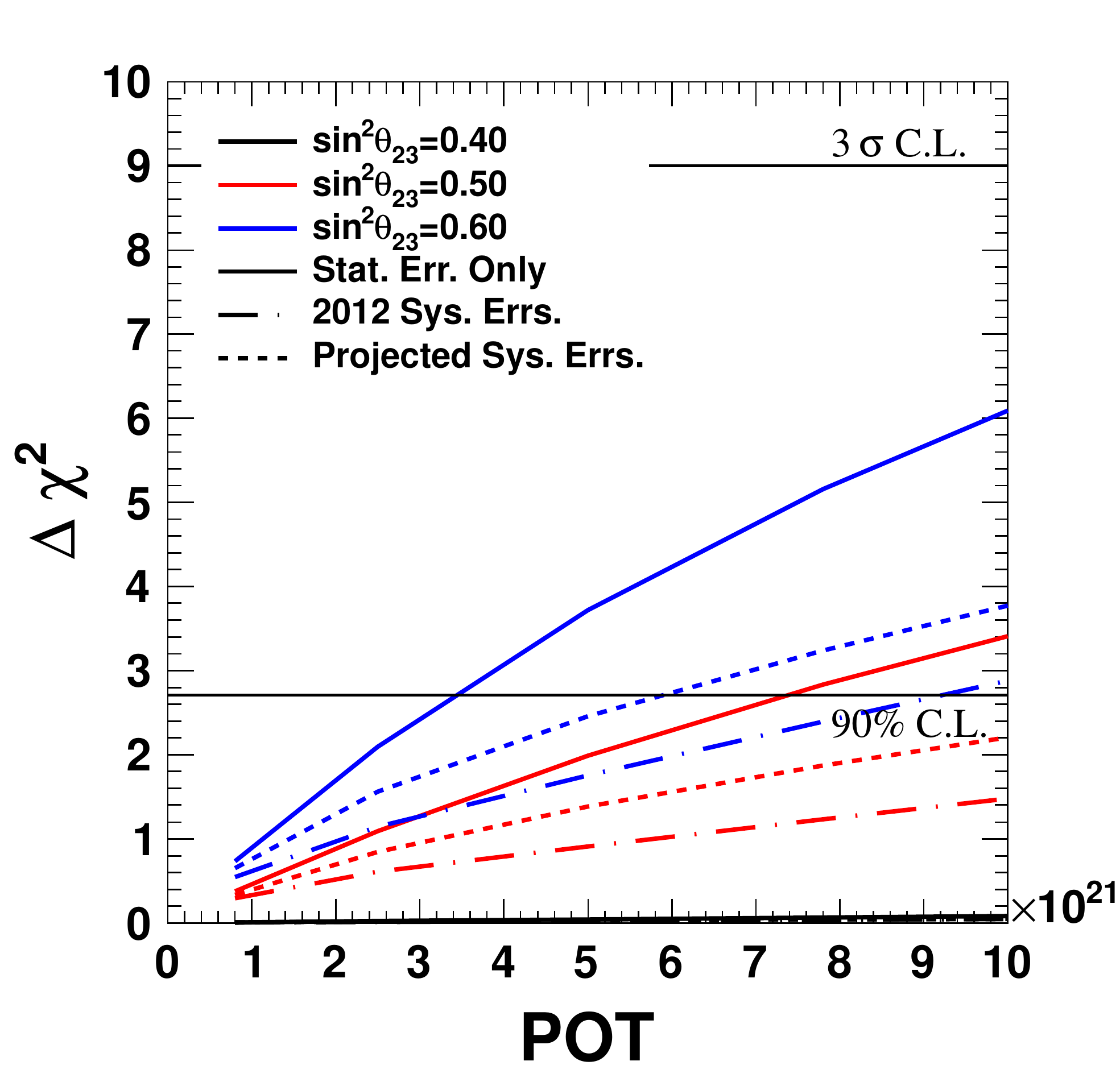}
\caption{100\% \(\nu\)-mode, \(\delta_{CP}=-90\degree\), NH.} 
\end{subfigure} \quad
\begin{subfigure}[t]{7cm}
\includegraphics[width=7cm]
{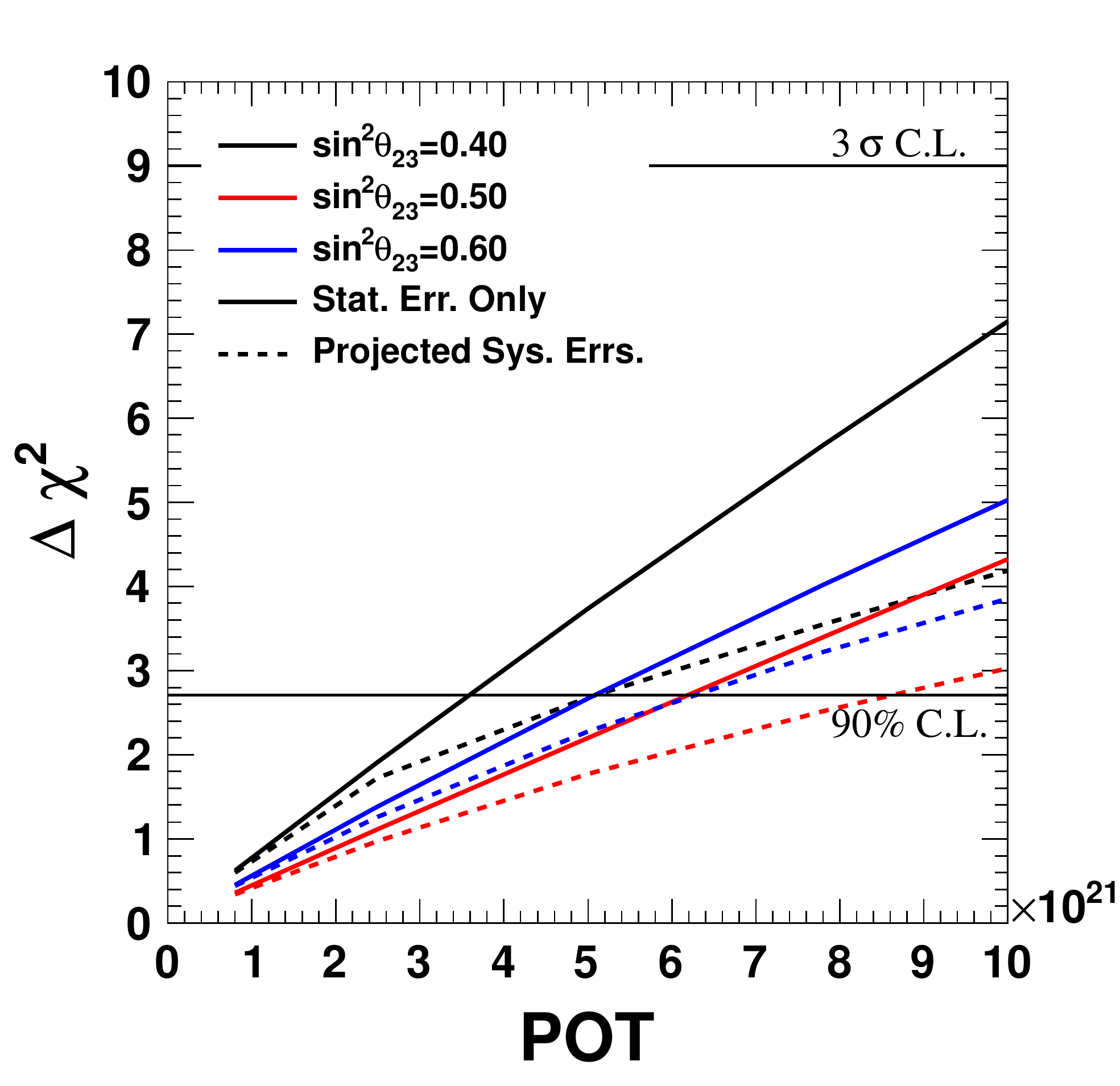}
\caption{50\% \(\nu\)-, 50\% $\bar{\nu}$ running, \(\delta_{CP}=-90\degree\), NH.} 
\end{subfigure} \quad
\caption
[$\sin\delta_{CP}$ Resolved vs.\ POT for True $\delta_{CP} = 90\degree$ and IH]
{The expected $\Delta\chi^2$ for the \(\sin\delta_{CP}=0\) hypothesis, plotted
as a function of POT.  Plots assume true \(\sin^22\theta_{13}=0.1\), 
various true values of \(\sin^2\theta_{23}\) (as given in the plot legends), and
\(\delta_{CP}\) and the MH as given in the figure captions.
The solid curves include statistical errors only, 
while the dash-dotted (dashed) curves assume the 2012 systematic errors  
(the projected systematic errors).  
Note that the sensitivity heavily depends on the assumed conditions, and that
the conditions applied for these figures 
correspond to the cases where the sensitivity for \(\sin\delta_{CP}\neq0\)
is maximal.
\label{fig:CPresolvedvsPOT}} \end{figure}

The sensitivity to reject the null hypothesis $\sin\delta_{CP}=0$
depends 
on the true oscillation parameters and is expected to be greatest for 
the case $\delta_{CP}=+90\degree$ and inverted MH. 
Figure~\ref{fig:CPresolvedvsPOT} shows how the expected 
$\Delta\chi^2$ evolves as a function of POT in this case, as well as for
$\delta_{CP}=-90\degree$ and normal MH, another case in which the sensitivity is
high.
These plots indicate the earliest case for T2K to observe CP violation.
If the systematic error size is negligibly small, T2K may reach
a higher sensitivity at an earlier stage by running in 100\% $\nu$-mode, since
higher statistics are expected in this case.
However, with projected systematic errors, 100\% $\nu$-mode and 50\% $\nu$-mode + 
50\% $\bar{\nu}$-mode running give essentially equivalent sensitivities.

\subsection{Effect of reduction of the systematic error size}
\label{sec:pfssys}
An extensive study of the effect of the systematic error 
size was performed.
Although the actual effect depends on the details of the errors, here we 
summarize the results of the study.
As given in Table~\ref{tab:nsker}, the systematic error on the predicted number
of events in Super-K in the 2012
oscillation analysis is 9.7\% for the $\nu_e$ appearance sample and 13\% for the
$\nu_\mu$ disappearance sample.

In Sec.\ \ref{sec:pfsvspot} we showed the T2K sensitivity with projected
systematic errors which are estimated based on a conservative expectation of
T2K systematic error reduction. In this case the systematic error on the predicted
number of events in Super-K
is about 7\% for the $\nu_{\mu}$ and $\nu_e$ samples and about
14\%  for the $\bar{\nu}_{\mu}$ and $\bar{\nu}_e$ samples.
These errors were calculated by reducing the 2012 oscillation analysis errors by removing 
certain interaction model and cross section uncertainties from both the $\nu_e$- and $\nu_{\mu}$-mode errors, and by additionally
scaling all $\nu_{\mu}$-mode errors down by a factor of two.  Errors for the $\bar{\nu}_\mu$- 
and $\bar{\nu}_e$-modes were estimated to be twice those of the $\nu_{\mu}$- and $\nu_e$-modes, respectively.
These reduced \(\nu\)-mode errors are in fact very close
to the errors used for the oscillation results reported by T2K in 2014, where
the T2K oscillation analysis errors have similarly been reduced by 
improvements in understanding the relevant interactions and cross sections.

For the measurement of $\delta_{CP}$, studies have shown that it is desirable to reduce 
this to 5$\sim$8\% for the $\nu_e$ sample and $\sim$10\% for the $\bar{\nu}_e$ sample
to maximize the T2K sensitivity with full statistics.
The measurement of $\delta_{CP}$ is nearly independent of the size of the error on the $\nu_{\mu}$ 
and $\bar{\nu}_\mu$ samples as long as we can achieve uncertainty
on $\bar{\nu}_\mu$ similar to the current uncertainty on $\nu_{\mu}$.
For the measurement of $\theta_{23}$ and $\Delta m^2_{32}$, 
the systematic error sizes are significant compared
to the statistical error, and the result
would benefit from
systematic error reduction even for uncertainties as small as 5\%.

These error reductions may also be achievable with the implementation of further T2K and 
external cross section and hadron production measurements, which continue to 
be made with improved precision.

\section{T2K and \nova Combined Sensitivities}
\label{pfst2knova}
The ability of T2K to measure the value of $\delta_{CP}$ (or
determine if CPV exists in the lepton sector) is greatly 
enhanced by the determination of the MH. 
This enhancement results from 
the nearly degenerate $\nu_{e}$ appearance event rate predictions at Super-K
in the normal hierarchy with positive values of $\delta_{CP}$ compared 
to the inverted hierarchy with negative values of $\delta_{CP}$.
Determination of the MH thus breaks the 
degeneracy, enhancing the $\delta_{CP}$ resolution for $\sim$50\% of
$\delta_{CP}$ values. 
T2K does not have sufficient sensitivity to determine the 
mass hierarchy by itself. The
\nova experiment~\cite{novaTDR}, which started operating in 2014,
has a longer baseline ($810$ km) and higher peak neutrino energy 
($\sim2$ GeV) than T2K. Accordingly,  the impact of
the matter effect on the predicted far detector event spectra is larger 
in \nova $\sim 30\%$) than in T2K ($\sim 10\%$),
leading to a greater sensitivity to the mass hierarchy.
Because of the complementary nature of these two experiments,
better constraints on the oscillation parameters,
$\delta_{CP}$, $\sin^2\theta_{23}$ and the MH can be obtained 
by comparing 
the $\nu_\mu \rightarrow \nu_e$ oscillation probability of 
the two experiments.
To evaluate the benefit of combining the two experiments, 
we have developed a code based on GLoBES~\cite{globes1,globes2}. 
The studies using projected T2K and \nova data samples 
show the full physics reach for the two 
experiments, individually and combined, along with studies aimed at
optimization of the $\nu$-mode to $\bar{\nu}$-mode running ratios of the two experiments.

\begin{figure}[htbp]
  \centering
  \begin{subfigure}[b]{0.49\textwidth}
    \includegraphics[trim=0.1cm 0.1cm 0.1cm 0.1cm, clip=true,width=\textwidth]{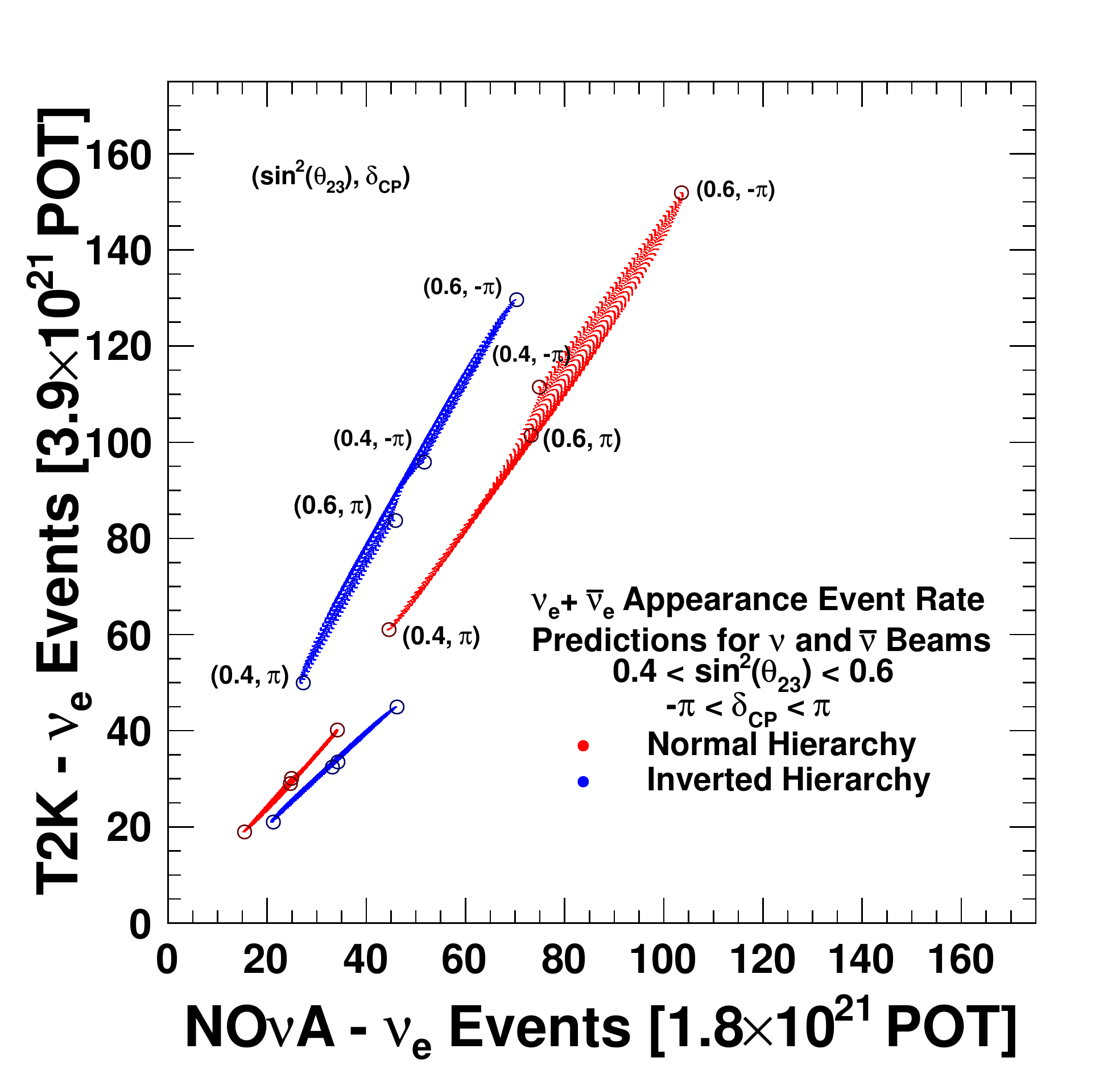}
    \label{fig:t2k_nova_event1}
  \end{subfigure}
  \begin{subfigure}[b]{0.49\textwidth}
    \includegraphics[trim=0.1cm 0.1cm 0.1cm 0.1cm, clip=true,width=\textwidth]{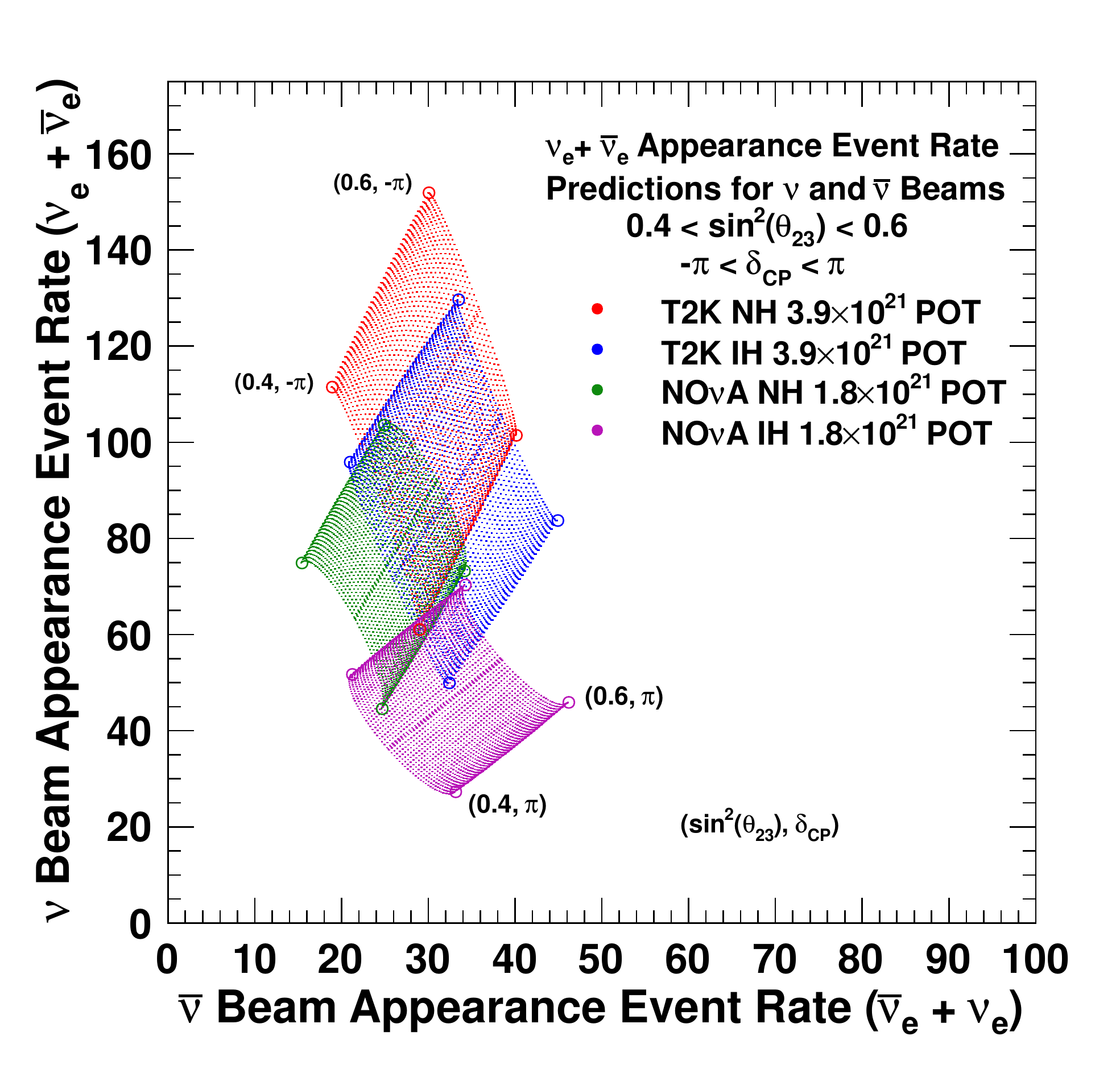}
    \label{fig:t2k_nova_event2}
  \end{subfigure}
  \caption[T2K vs. \nova and $\nu$ beam vs. $\bar{\nu}$ beam event rates]
{Relation between the expected number of $\nu_e + \bar{\nu}_e$ signal events produced by neutrino-mode running and antineutrino mode running in T2K and \nova, for various values of $\delta_{CP}$, $\sin^2\theta_{23}$ and mass hierarchy.
In the plot of predicted T2K rate versus the predicted \nova rate (left) the blue (IH) and red (NH) upper bands are for neutrino-mode running while the red (NH) and blue (IH) bottom bands are for the antineutrino mode running. The predicted number of $\nu_e + \bar{\nu}_e$ events produced in neutrino-mode running versus events produced in antineutrino mode running (right) are shown for T2K in red (NH) and blue (IH), and for \nova in green (NH) and magenta (IH). Representative points at the edges of the $\delta_{CP}$ and $\sin^2\theta_{23}$ ranges are highlighted. Systematic and statistical uncertainties are not included.
}
\label{fig:t2k_nova_event}
\end{figure}

Figure \ref{fig:t2k_nova_event} shows the relation between 
the expected number of events of T2K and \nova for various values of 
$\delta_{CP}, \sin^2\theta_{23}$ and mass hierarchies. The NH and IH
predictions occupy distinct regions in the plot suggesting how a 
combined analysis T2K-\nova fit leads to increased sensitivity.
However, this plot does not include 
the (statistical + systematic) uncertainties
on measurements of these event rates. 
This would result in regions of overlap where the MH can not be determined, 
and the sensitivity to $\delta_{CP}$ is degraded.
In order to evaluate the effect of combining the 
results from T2K and \nova quantitatively, we  
have conducted a T2K-\nova combined sensitivity study.
The GLoBES~\cite{globes1,globes2} software package was used to
fit oscillation parameters based on the reconstructed neutrino energy spectra of the two experiments.
The fits were conducted by minimizing $\Delta\chi^{2}$
which is calculated from spectra generated with different
sets of oscillation parameters, and includes penalty terms
for deviations of the signal and background normalizations from nominal.
The best-fit $\Delta\chi^{2}$ calculated by GLoBES, was the metric chosen to characterize sensitivity, 
as it is related to the probability that a given data set can result from two different hypotheses.

GLoBES combines flux, cross section, energy resolution/bias and efficiency information for an experiment to estimate
energy spectra of neutrino interaction samples used for analyses. 
Then GLoBES uses a full three-flavor oscillation probability formulation 
to fit analysis spectra generated assuming different oscillation parameters to each other 
(varying oscillation parameter values and parameters accounting for systematic uncertainties within their uncertainties). 
The oscillation parameters, unless otherwise stated,
are those shown in Table~\ref{tab:truepars}. 
The GLoBES three-flavor analysis package works very similarly to fitter used for the studies presented in Section 4. 
Several validation studies were done to ensure that the two methods produced the same results when given the same inputs.

The T2K, \nova, and combined sensitivities 
were generated using a modified version of GLoBES
that allowed for use of inputs generated from Monte Carlo simulations of 
T2K neutrino interactions in the Super-Kamiokande detector.
The inputs describing the \nova experiment were developed in conjunction 
with \nova collaborators, and validated against official \nova sensitivity 
plots~\cite{novaplotsandfigures,novaneutrino2012,novaheavyquarks2012}.
We assume the same run plan as presented in \nova's TDR:  
$1.8\times 10^{21}$ POT for $\nu$ and $1.8\times 10^{21}$ POT for 
$\bar{\nu}$ modes, corresponding to 3 years of running in each mode.

The GLoBES inputs defining the analysis sample acceptances for the signal,
the NC background, the $\nu_\mu$ CC background, and the $\nu_e$ CC background
were tuned to match this official event rate prediction from \nova. 
For example, Table \ref{tbl:nova_rates} summarizes
the expected number of $\nu_e$ appearance events for \nova
\cite{novaheavyquarks2012} when
$\sin^22\theta_{13}=0.95$ is assumed and
the solar oscillation terms or matter effects in the oscillation probability
are neglected.

\begin{table}[htbp]
    \caption[\nova expected number of events]
{Expected number of $\nu_e$ appearance signal and background events for \nova
at $1.8\times 10^{21}$ POT for each of $\nu$ 
and $\bar{\nu}$ modes\cite{novaheavyquarks2012}.
The oscillation probabilities used to calculate
the predicted number of events assumed $\sin^22\theta_{13}=0.095$ and 
do not include the solar oscillation terms or matter effects.}
    \begin{center}
    \begin{tabular}{lccccc}
    \hline
    Beam            & Signal & NC Bkg & $\nu_\mu$ CC & $\nu_e$ CC & Total Bkg \\ 
\hline
    $\nu$-mode      & 72.6   & 20.8   & 5.2          & 8.4        & 34.5      \\
    $\bar{\nu}$-mode & 33.8   & 10.6   & 0.7          & 5.0        & 16.3   \\
\hline
    \end{tabular}
    \end{center}
    \label{tbl:nova_rates}
\end{table}


Since \nova has only recently began taking data, 
detailed evaluation of systematic uncertainties is not yet published.
Therefore, the combined sensitivity studies used 
a simplified systematics treatment for both T2K and \nova: 
a 5\% normalization uncertainty on signal events and 
a 10\% normalization uncertainty on background events 
for both appearance and disappearance spectra. 
Uncertainties that impact the spectral shape are not considered. 
This is a reasonable choice since both experiments use a narrow band beam 
and much of the oscillation sensitivity comes from the measured event rates. 
The uncertainties are assumed to be uncorrelated  for $\nu_e$ appearance, 
$\bar{\nu}_e$ appearance, $\nu_\mu$ disappearance, 
and $\bar{\nu}_\mu$ disappearance. 
This simple systematics implementation, referred to in the rest of the paper as ``normalization systematics'', is the same as the one adopted 
in the \nova TDR and is also a reasonable representation of 
the projected uncertainties at T2K.
The sensitivities shown here are obtained assuming $\sin^22\theta_{13}=0.1$ 
with the projected reactor constraint of 5\%.

When determining the MH,
$\Delta\chi^2$ is not distributed according to a $\chi^{2}$ distribution
because the MH is a discrete, rather than a continuous, variable.
Toy MC studies, where many pseudo-experiments are generated 
with statistical and systematic fluctuations,
were used to evaluate the validity 
of applying a $\Delta\chi^2$ test statistic, as given in Eq.\ \ref{eq:chi2test}, for the MH determination.

\begin{figure}[htbp]
  \centering

  \begin{subfigure}[b]{0.35\textwidth}
    \includegraphics[trim=0.3cm 0cm 0.5cm 0cm, clip=true,width=\textwidth]{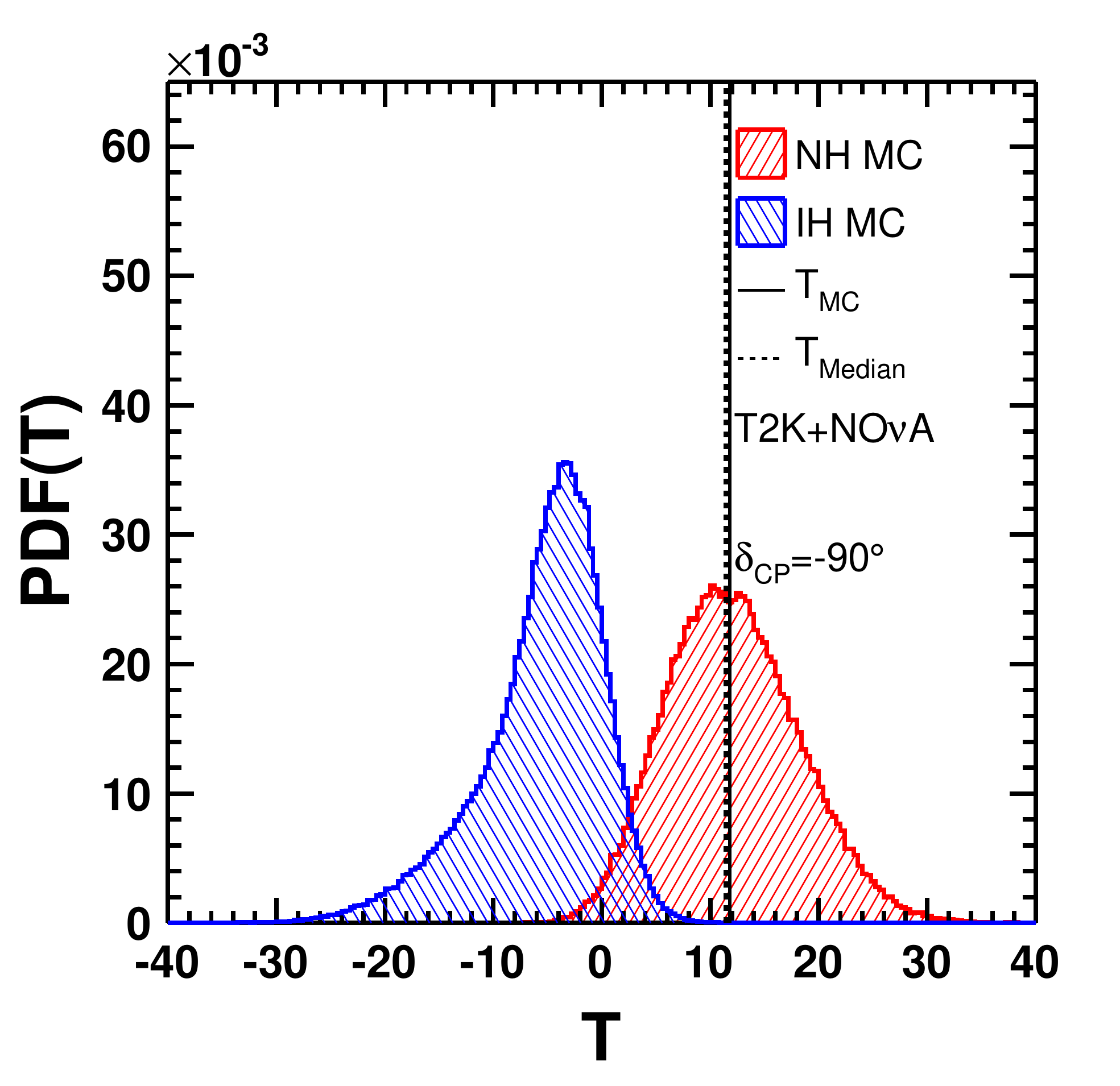}
    \caption{$\delta_{CP} = -90^\circ$, NH}
  \end{subfigure}
  \begin{subfigure}[b]{0.35\textwidth}
    \includegraphics[trim=0.3cm 0cm 0.5cm 0cm, clip=true,width=\textwidth]{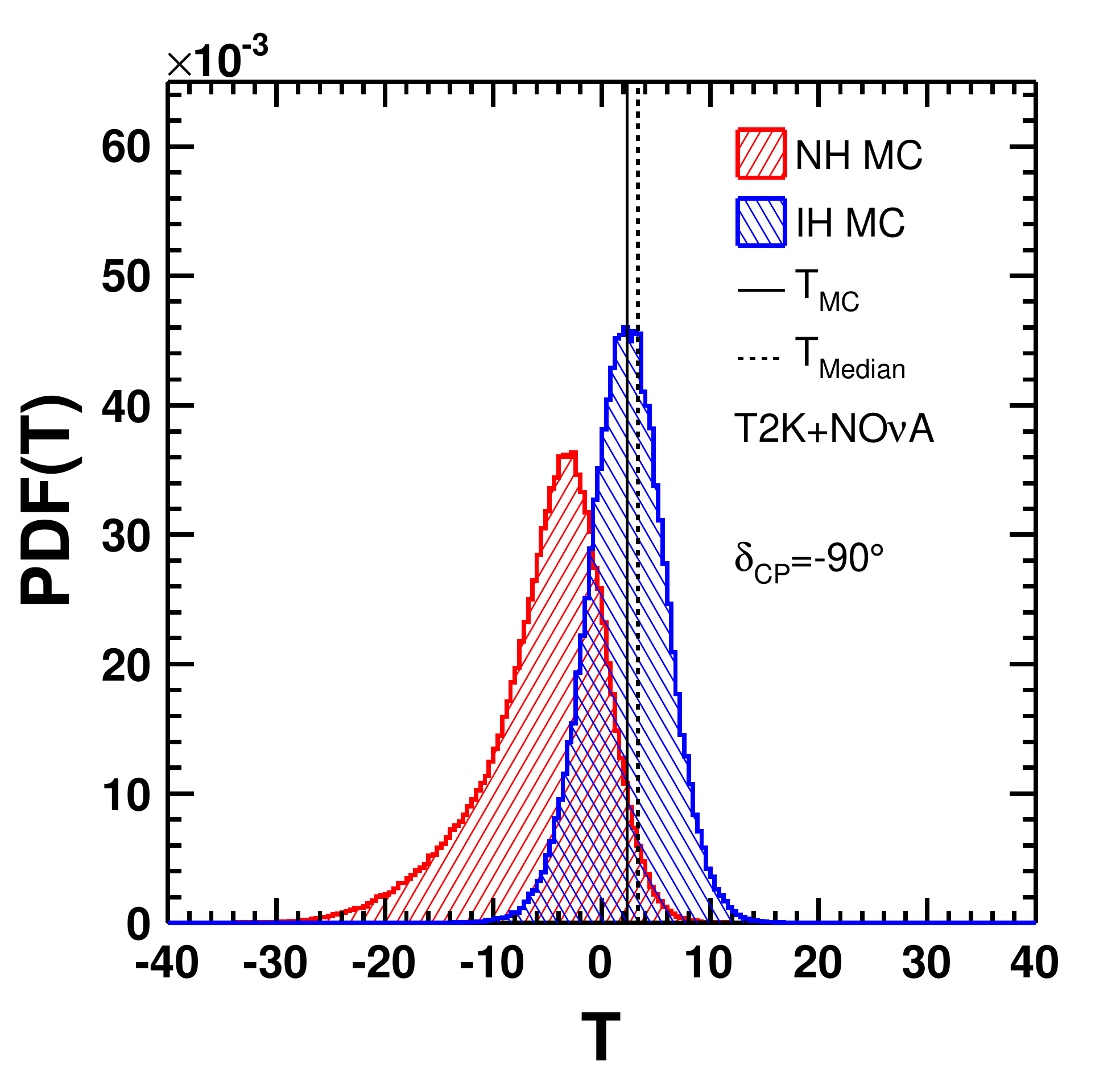}
    \caption{$\delta_{CP} = -90^\circ$, IH}
  \end{subfigure}
  \begin{subfigure}[b]{0.35\textwidth}
    \includegraphics[trim=0.3cm 0cm 0.5cm 0cm, clip=true,width=\textwidth]{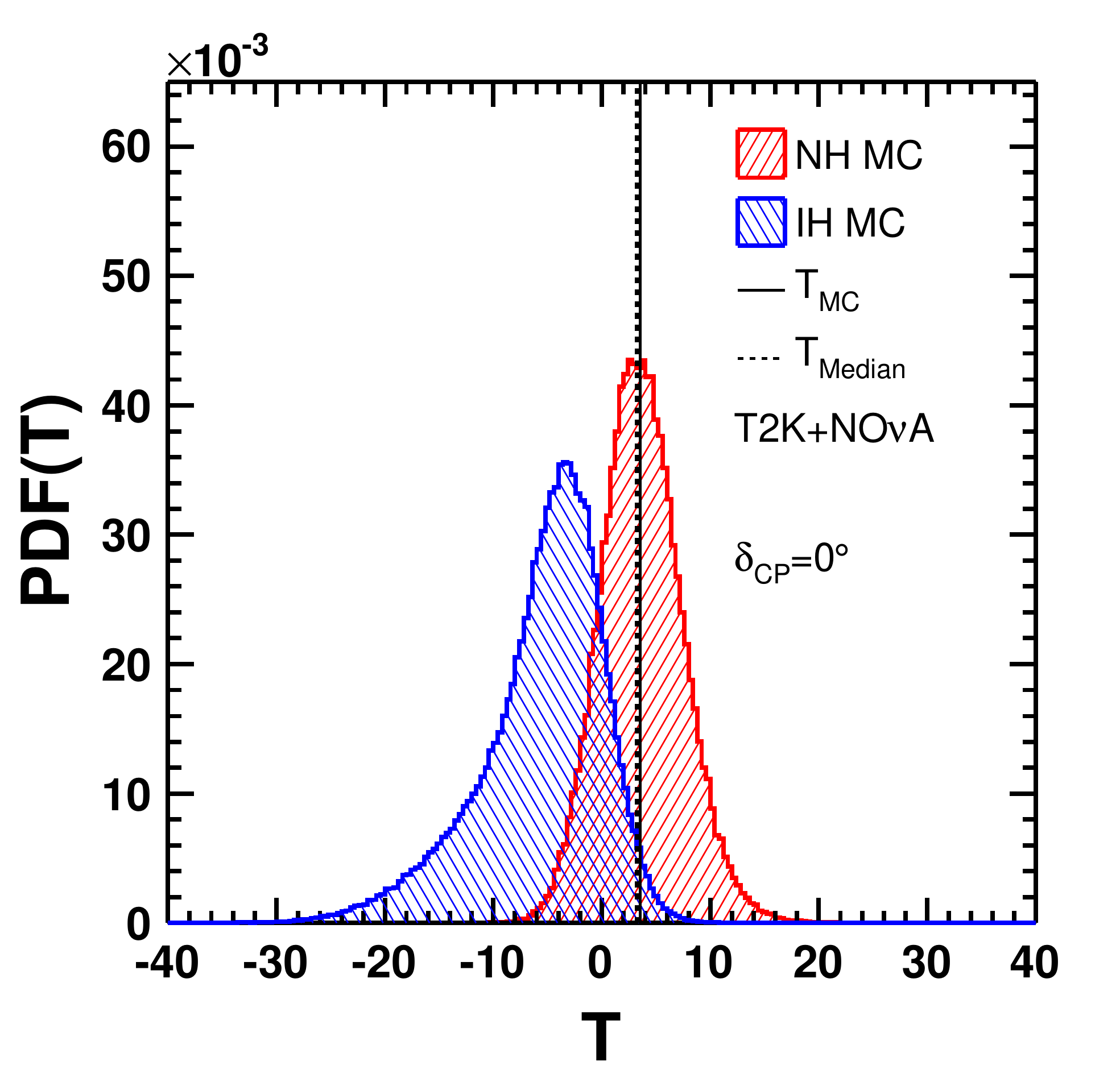}
    \caption{$\delta_{CP} = 0^\circ$, NH}
  \end{subfigure}
  \begin{subfigure}[b]{0.35\textwidth}
    \includegraphics[trim=0.3cm 0cm 0.5cm 0cm, clip=true,width=\textwidth]{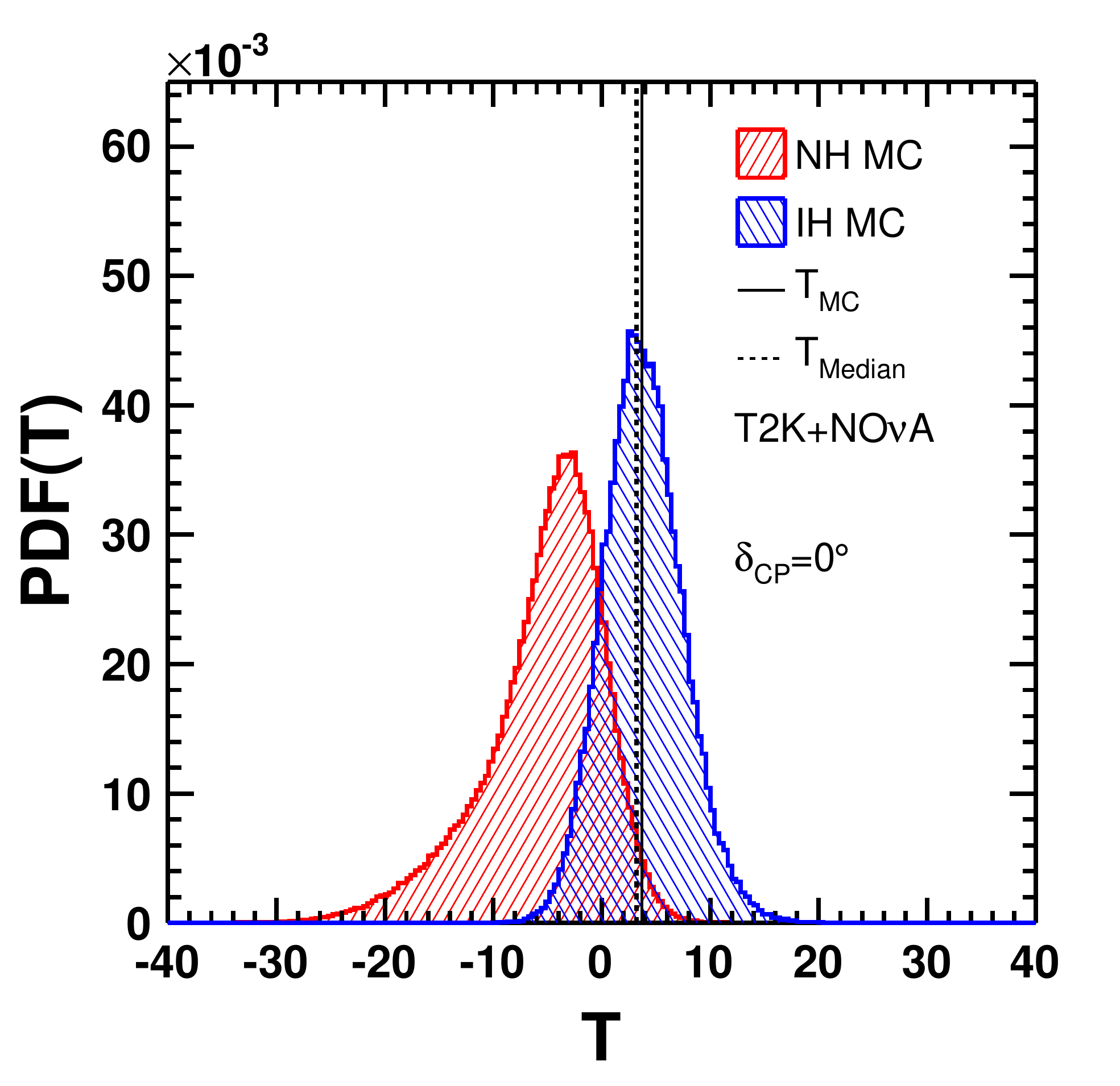}
    \caption{$\delta_{CP} = 0^\circ$, IH}
  \end{subfigure}
  \begin{subfigure}[b]{0.35\textwidth}
    \includegraphics[trim=0.3cm 0cm 0.5cm 0cm, clip=true,width=\textwidth]{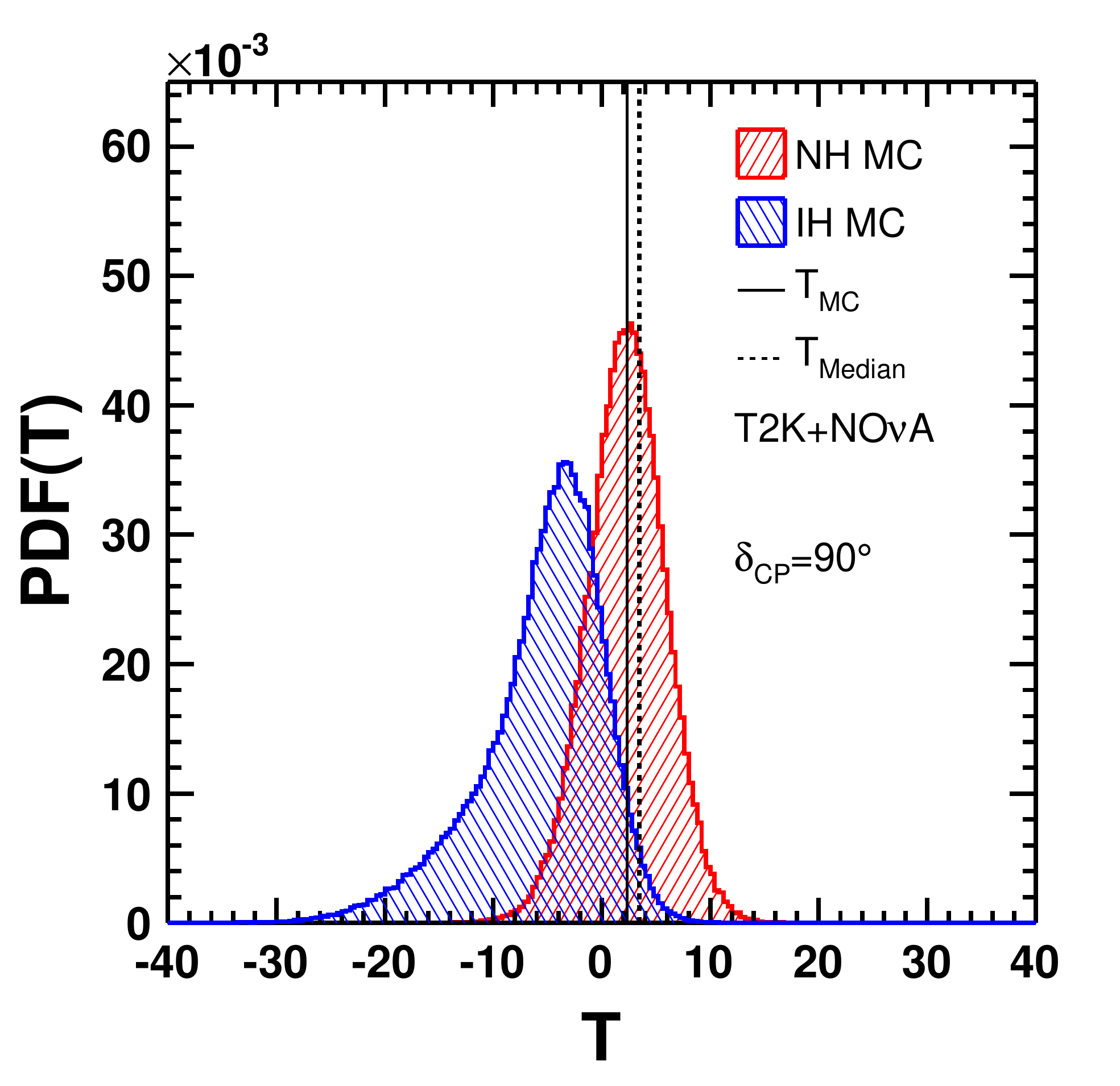}
    \caption{$\delta_{CP} = 90^\circ$, NH}
  \end{subfigure}
  \begin{subfigure}[b]{0.35\textwidth}
    \includegraphics[trim=0.3cm 0cm 0.5cm 0cm, clip=true,width=\textwidth]{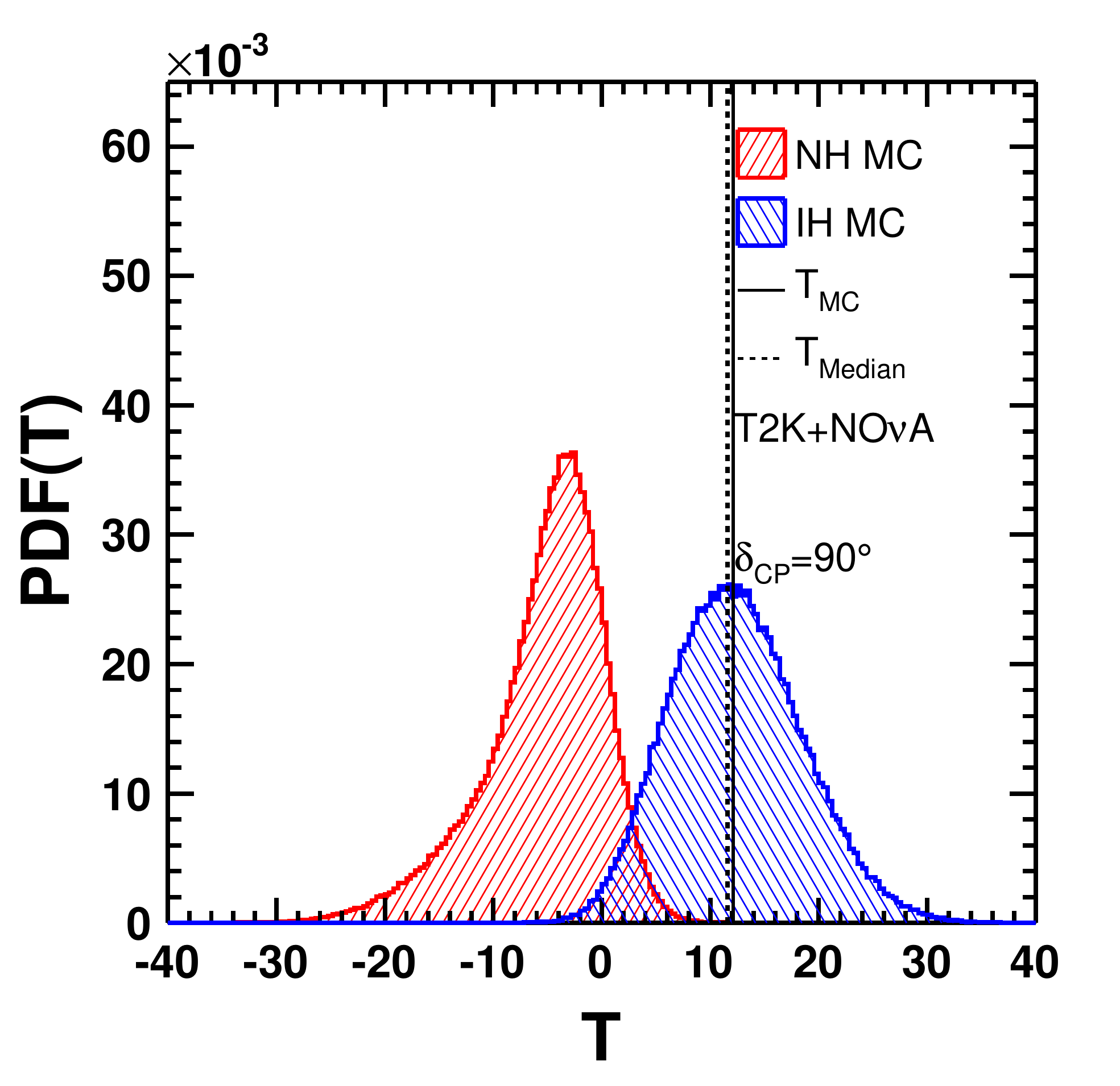}
    \caption{$\delta_{CP} = 90^\circ$, IH}
  \end{subfigure}
\caption{
Distributions of the test statistic, $T$ for toy MC experiments
with the null hypothesis $H_0=$ I(N)H are shown in the left (right) column. 
Toy MC experiments are generated with the nominal oscillation parameters
except for the MH and $\delta_{CP}$;
those generated with NH are indicated in red and those with IH in blue.
The value of $\delta_{CP}$ is fixed to the value indicated in the sub-captions
when $H_0 = (\mathrm{correct\ MH})$, but thrown 
when $H_0 = (\mathrm{incorrect\ MH})$, where the correct MH is also given in the
sub-captions.
Solid lines indicate the value of the MH determination sensitivity metric 
used in this paper (calculated using the spectra at the MC
sample statistical mean), 
and dashed lines indicate the $T$-value
for the median of the toy MC distribution.
}
\label{fig:TestMH}
\end{figure}

The left column of Fig.~\ref{fig:TestMH} shows distributions 
for a test static for $H_0=$ IH:
\begin{equation}
T = \chi^2_{IH} - \chi^2_{NH},
\end{equation}
where $\chi^2_{IH}$ and $\chi^2_{NH}$ are 
the minimum $\chi^2$ values obtained by fitting the oscillation parameters
while fixing the MH to the inverted or normal mass hierarchy, respectively.
This $T$ is plotted here instead of $\Delta\chi^2$
for easier interpretation.
In the figure, the blue (red) distributions are 
for the case where test or `observed' spectra were generated 
for the inverted (normal) mass hierarchy 
with statistical and systematic fluctuations.
Except for $\delta_{CP}$, the test oscillation parameters were fixed to the nominal values 
given in Table~\ref{tab:truepars}.
The value of $\delta_{CP}$ was fixed to that given in each caption
for the NH, while it was thrown over all values of \dcp for the IH.
This is done in order to calculate the p-value for $H_0=$ IH with unknown \dcp 
when the test point is in the NH~\cite{PhysRevD.86.113011}.
The right column of Figure~\ref{fig:TestMH} is the same, but with 
the opposite MH hypothesis test ($H_0=$ NH):
\begin{eqnarray}
T = \chi^2_{NH} - \chi^2_{IH}
\end{eqnarray}
with a test point in the IH.
The $T$-value calculated using the spectrum generated 
from the MC sample statistical mean ($T_{MC}$),
which is generally used in this paper, is compared 
with the median $T$-value for the ensemble of toy MC experiments ($T_{median}$)
in Table~\ref{tab:ToyMC} for different oscillation parameter sets.
The p-values calculated for $T_{MC}$, assuming that $\Delta\chi^2$ 
follows a true $\chi^{2}$ distribution, compared with the p-values
calculated as the fraction of the $T$ distribution 
for $H_0=(\mathrm{correct\ MH})$ above $T_{median}$ are also given. 

\begin{table}
  \centering
    \caption{
Values of $T_{MC}$ and $T_{Median}$ and their associated p-values. 
The $T$ values correspond to the vertical lines shown in Fig.\ref{fig:TestMH}. 
The p-values are computed either with a $\chi^2$ distribution 
for one degree of freedom from the spectra at the toy MC statistical mean
or using an ensemble of toy MC experiments. 
}
\begin{tabular}{l|cccccc}
 & \multicolumn{2}{c}{by MC mean spectra} & \multicolumn{2}{c}{by toy MC experiments}\\
~   & $T_{MC}$  & p-value($\chi^2$) & $T_{Median}$ & p-value(toy MC) \\
\hline
NH, $\delta_{CP}=-90^\circ$ & 11.4 & 0.00073 & 11.8 & 0.000065 \\
NH, $\delta_{CP}=0^\circ$   & 3.22 & 0.073   & 3.57 & 0.019 \\
NH, $\delta_{CP}=+90^\circ$ & 3.47 & 0.063   & 2.34  & 0.040\\
IH, $\delta_{CP}=-90^\circ$ & 3.33 & 0.068 & 2.30 & 0.042\\
IH, $\delta_{CP}=0^\circ$   & 3.19 & 0.074 & 3.79 & 0.015\\
IH, $\delta_{CP}=+90^\circ$ & 11.6 & 0.00067 & 12.5 & 0.000031 \\
\end{tabular}
\label{tab:ToyMC}
\end{table}

\begin{figure}[htbp]
  \centering

 \begin{subfigure}[b]{0.90\textwidth}
   \hfill
   \includegraphics[trim=7cm 6.5cm 1.0cm 0cm, clip=true,width=1.4in]{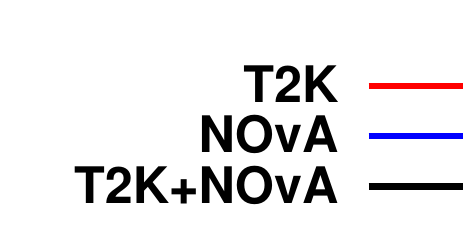}
   \vspace{-2.5\baselineskip}
 \end{subfigure}
 \begin{subfigure}[b]{0.49\textwidth}
    \includegraphics[trim=0.7cm 0.5cm 1.3cm 0.5cm, clip=true,width=\textwidth]{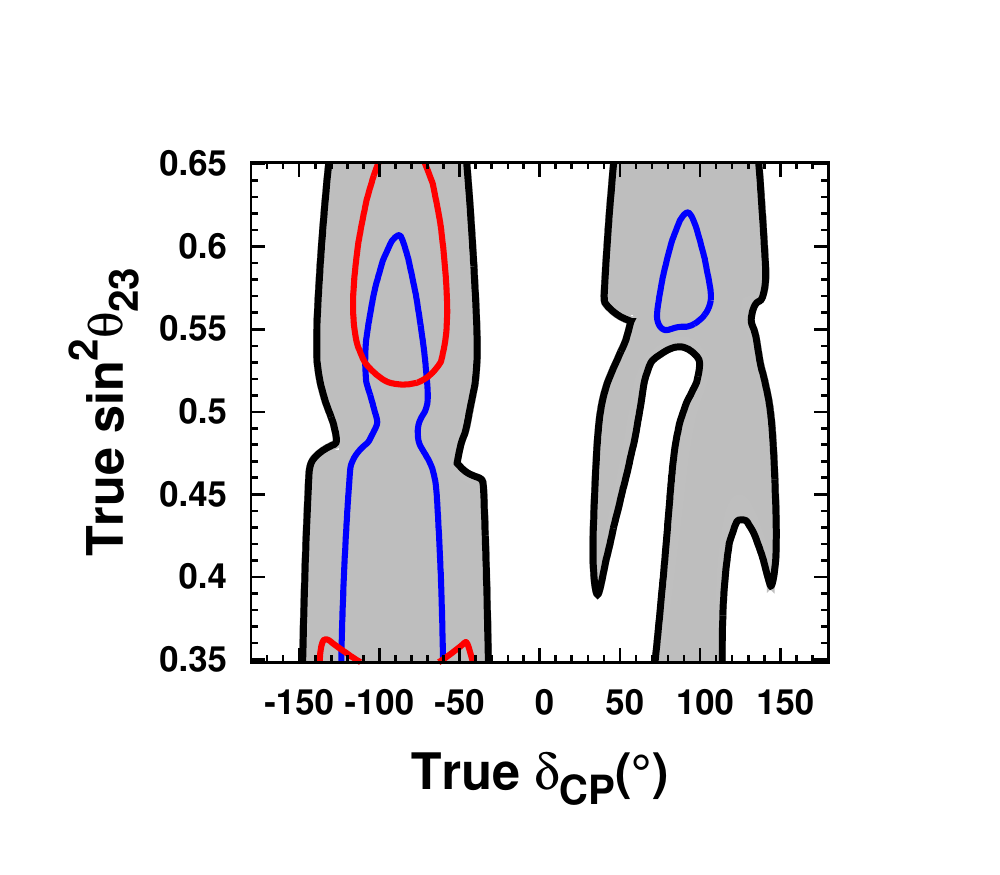}
    \caption{1:0 T2K, 1:1 \nova \nn, NH}
  \end{subfigure}
  \begin{subfigure}[b]{0.49\textwidth}
    \includegraphics[trim=0.7cm 0.5cm 1.3cm 0.5cm, clip=true,width=\textwidth]{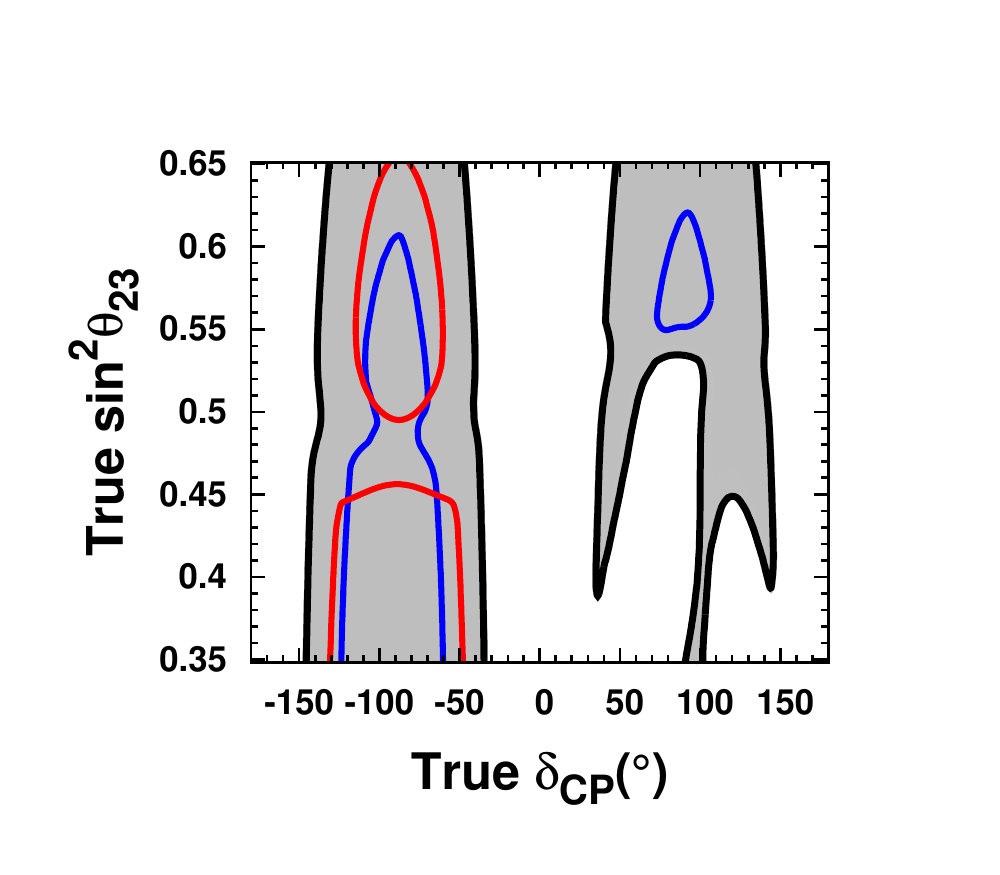}
    \caption{1:1 T2K, 1:1 \nova \nn, NH}
  \end{subfigure}
  \begin{subfigure}[b]{0.49\textwidth}
    \includegraphics[trim=0.7cm 0.5cm 1.3cm 0.5cm, clip=true,width=\textwidth]{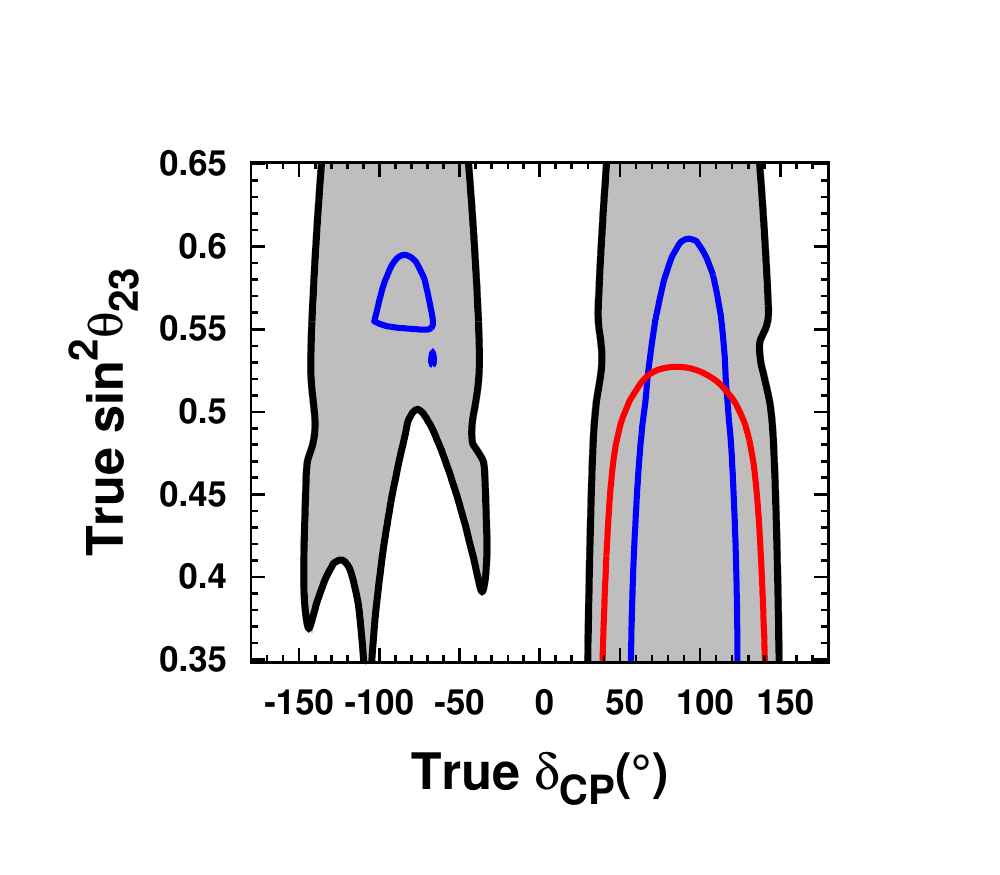}
    \caption{1:0 T2K, 1:1 \nova \nn, IH}
  \end{subfigure}
  \begin{subfigure}[b]{0.49\textwidth}
    \includegraphics[trim=0.7cm 0.5cm 1.3cm 0.5cm, clip=true,width=\textwidth]{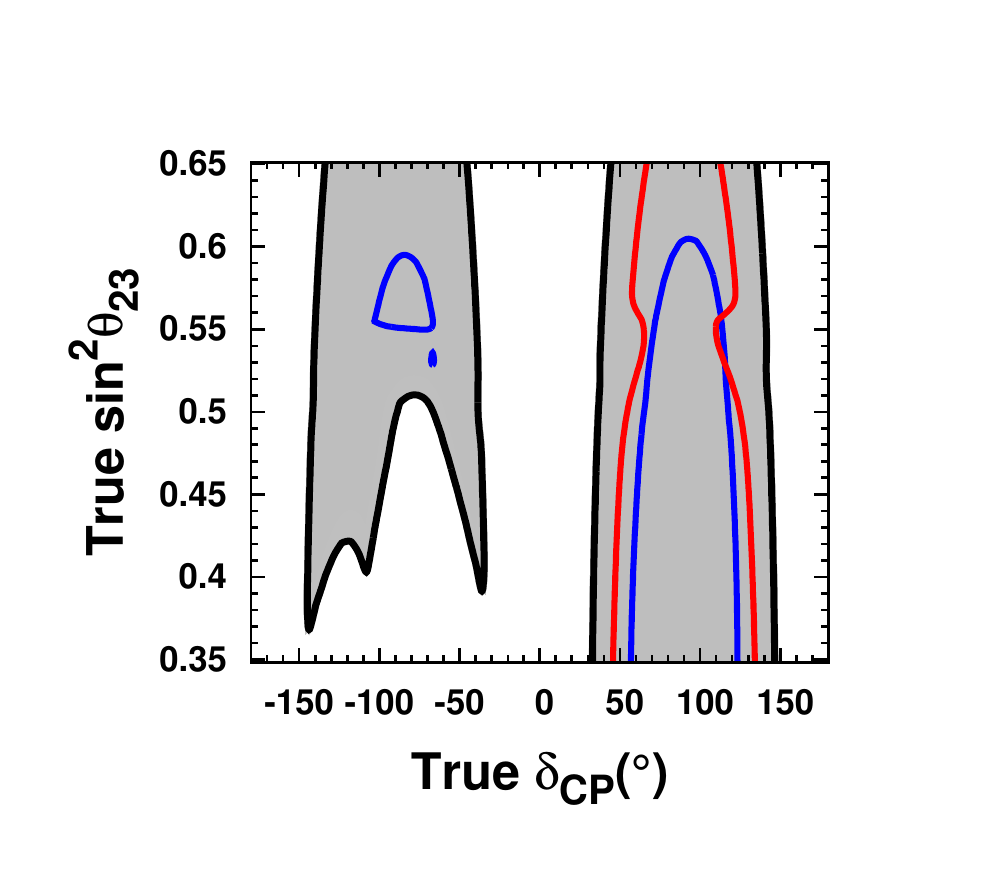}
    \caption{1:1 T2K, 1:1 \nova \nn, IH}
  \end{subfigure}
\caption[T2K and \nova 90\% sensitivity region for \sdnz in \qtt vs \dcp]
{Regions where T2K~(red), \nova~(blue), and T2K+\nova~(black) 
is predicted to rule out $\sin\delta_{CP}=0$ at 90\% C.L. 
Points within the gray regions are where $\sin\delta_{CP}=0$ 
is predicted to be rejected at 90\% C.L. 
for T2K+\nova, assuming simple normalization systematics as described in the text.}
  \label{fig:dcpvst23_cpv_t2knova}
\end{figure}

\begin{figure}[htbp]
  \centering
 \begin{subfigure}[b]{0.90\textwidth}
   \hfill
   \includegraphics[trim=7cm 6.5cm 1.0cm 0cm, clip=true,width=1.4in]{ocm_key}
   \vspace{-2.5\baselineskip}
 \end{subfigure}
  \begin{subfigure}[b]{0.49\textwidth}
    \includegraphics[trim=0.7cm 0.5cm 1.3cm 0.5cm, clip=true,width=\textwidth]
{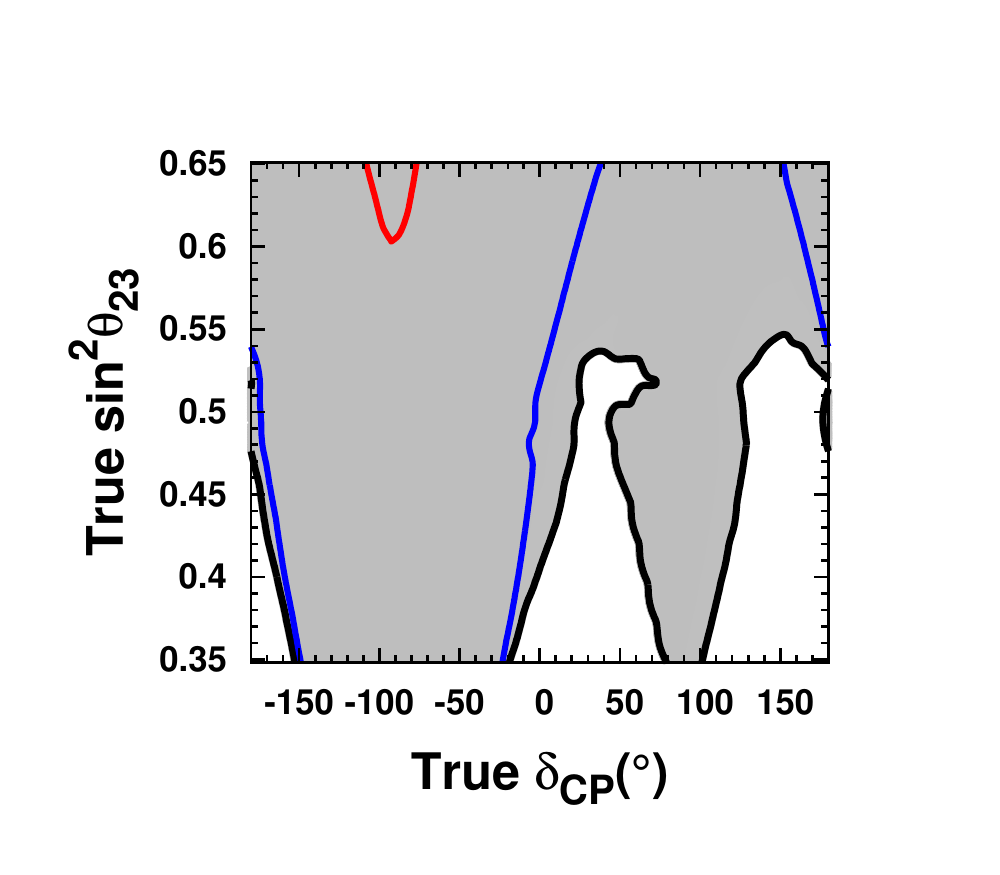}
    \caption{1:0 T2K, 1:1 \nova \nn, NH}
  \end{subfigure}
  \begin{subfigure}[b]{0.49\textwidth}
    \includegraphics[trim=0.7cm 0.5cm 1.3cm 0.5cm, clip=true,width=\textwidth]
{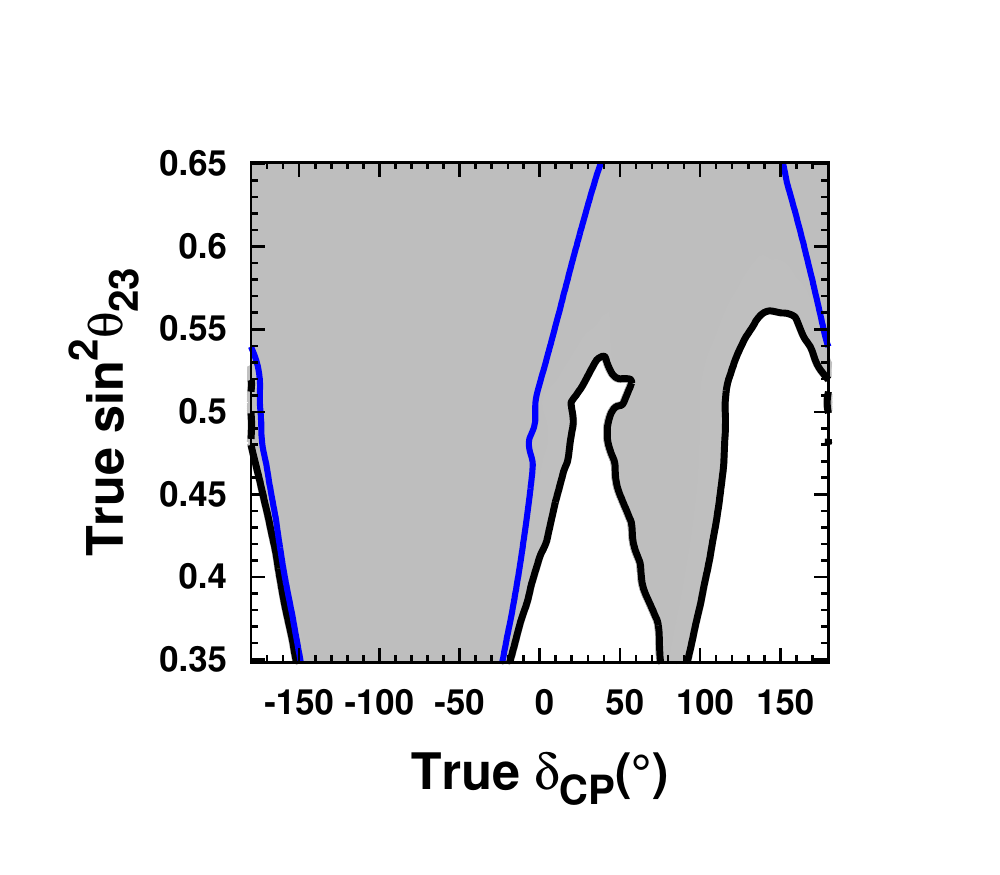}
    \caption{1:1 T2K, 1:1 \nova \nn, NH}
  \end{subfigure}
  \begin{subfigure}[b]{0.49\textwidth}
    \includegraphics[trim=0.7cm 0.5cm 1.3cm 0.5cm, clip=true,width=\textwidth]
{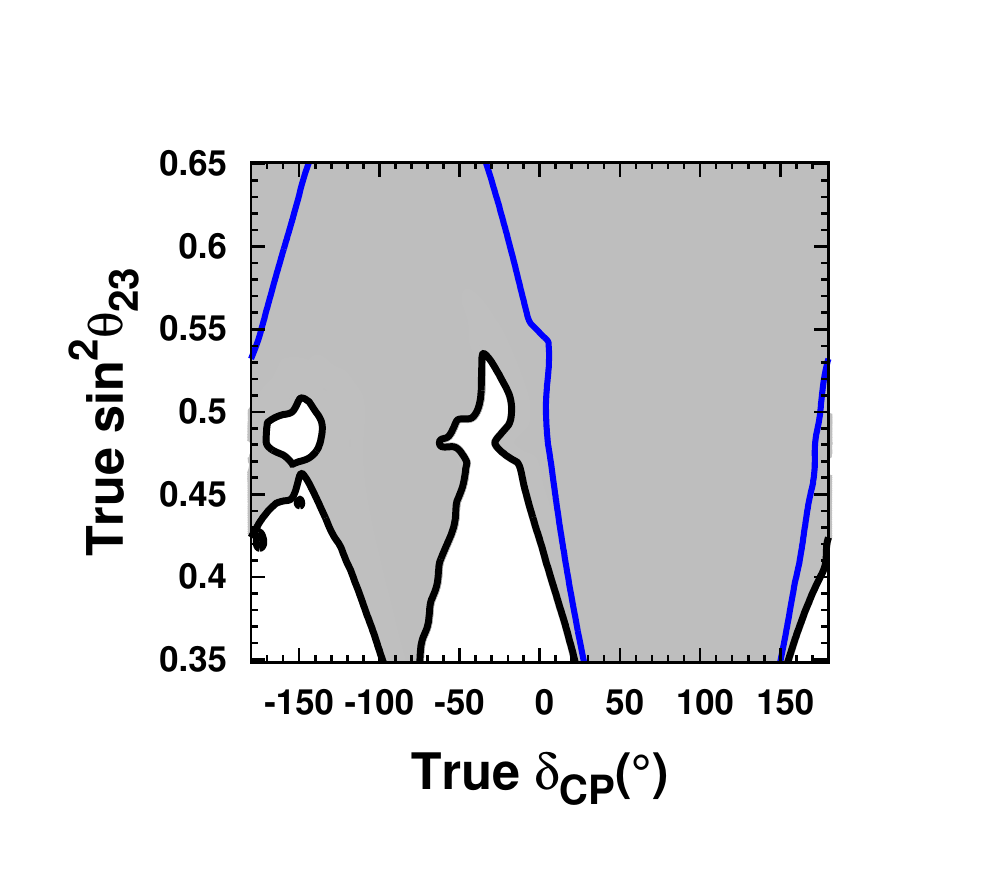}
    \caption{1:0 T2K, 1:1 \nova \nn, IH}
  \end{subfigure}
  \begin{subfigure}[b]{0.49\textwidth}
    \includegraphics[trim=0.7cm 0.5cm 1.3cm 0.5cm, clip=true,width=\textwidth]
{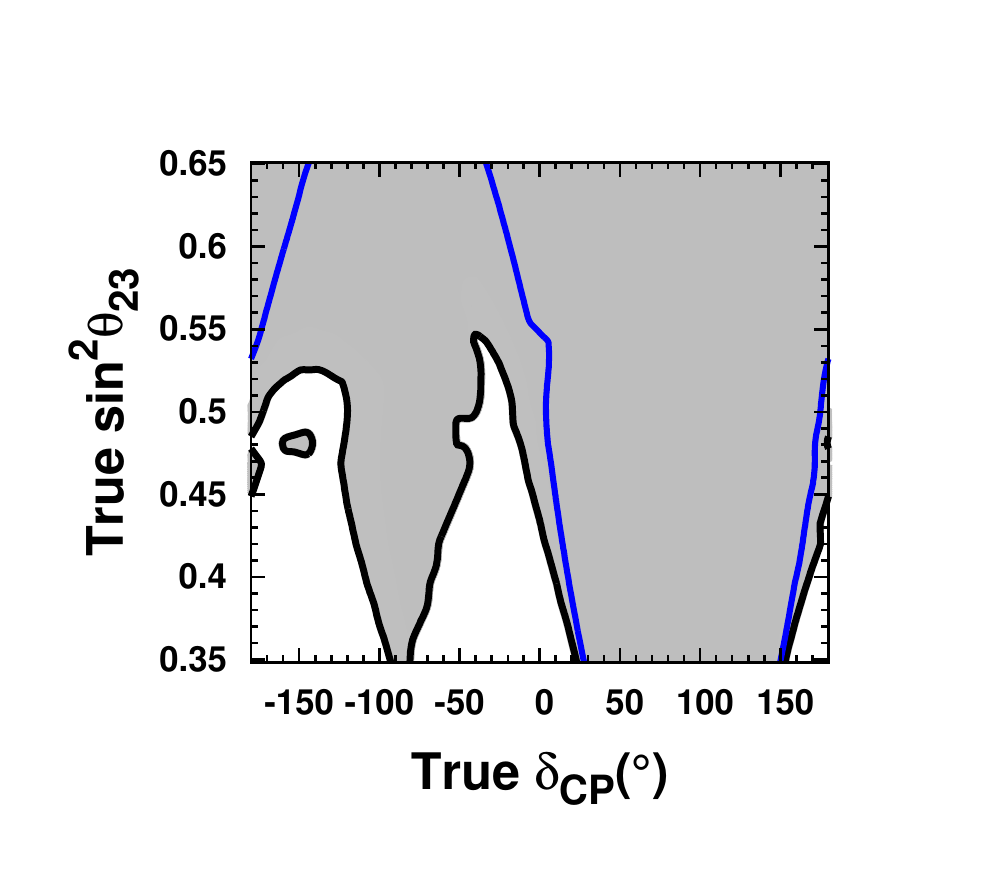}
    \caption{1:1 T2K, 1:1 \nova \nn, IH}
  \end{subfigure}

  \caption[T2K and \nova 90\% CL mass hierarchy sensitivity regions 
in \qtt vs \dcp.]
{Regions for T2K~(red), \nova~(blue), and T2K+\nova~(black)
where the incorrect Mass Hierarchy is predicted to be rejected at 
90\% C.L. Points within the gray regions are where the incorrect
mass hierarchy is predicted to be rejected at 90\% C.L. for T2K+\nova, assuming
simple normalization systematics as described in the text.}
  \label{fig:dcpvst23_mh_t2knova}
\end{figure}

\begin{figure}[htbp]
  \centering
 \begin{subfigure}[b]{0.90\textwidth}
   \hfill
   \includegraphics[trim=7cm 6.5cm 1.0cm 0cm, clip=true,width=1.4in]{ocm_key}
   \vspace{-2.5\baselineskip}
 \end{subfigure}
  \begin{subfigure}[b]{0.49\textwidth}
    \includegraphics[trim=0.7cm 0.5cm 1.3cm 0.5cm, clip=true,width=\textwidth]{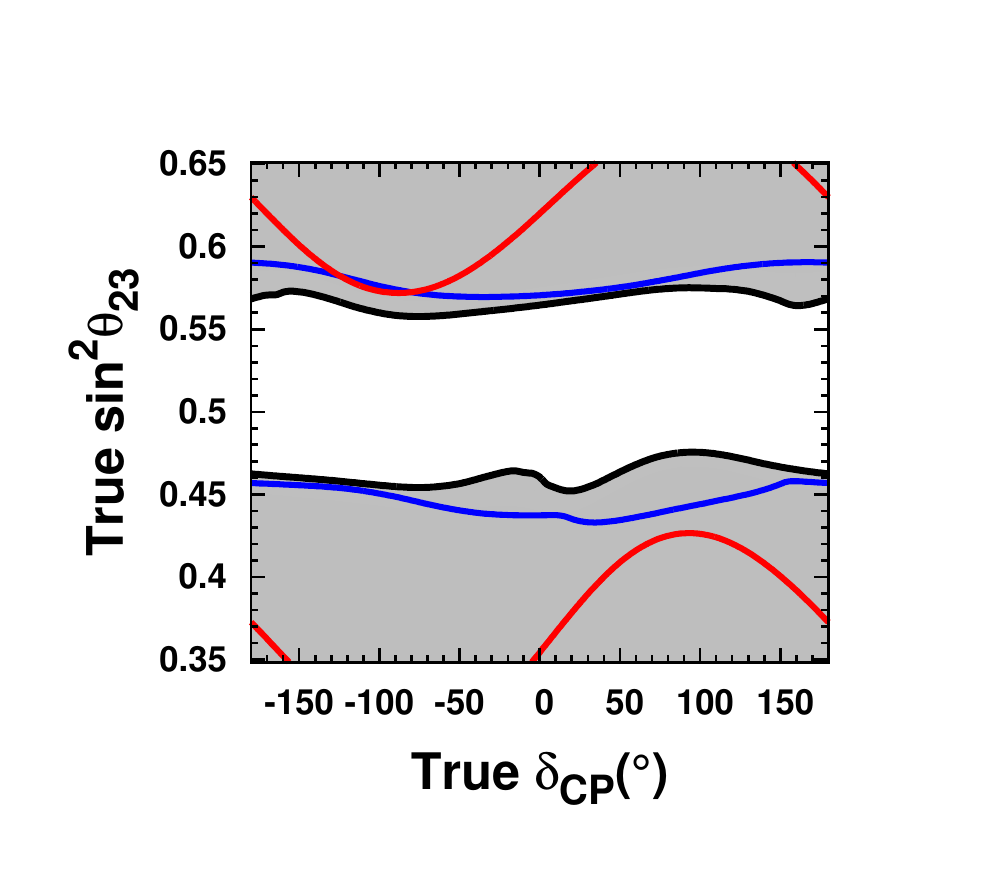}
    \caption{1:0 T2K, 1:1 \nova \nn, NH}
  \end{subfigure}
  \begin{subfigure}[b]{0.49\textwidth}
    \includegraphics[trim=0.7cm 0.5cm 1.3cm 0.5cm, clip=true,width=\textwidth]{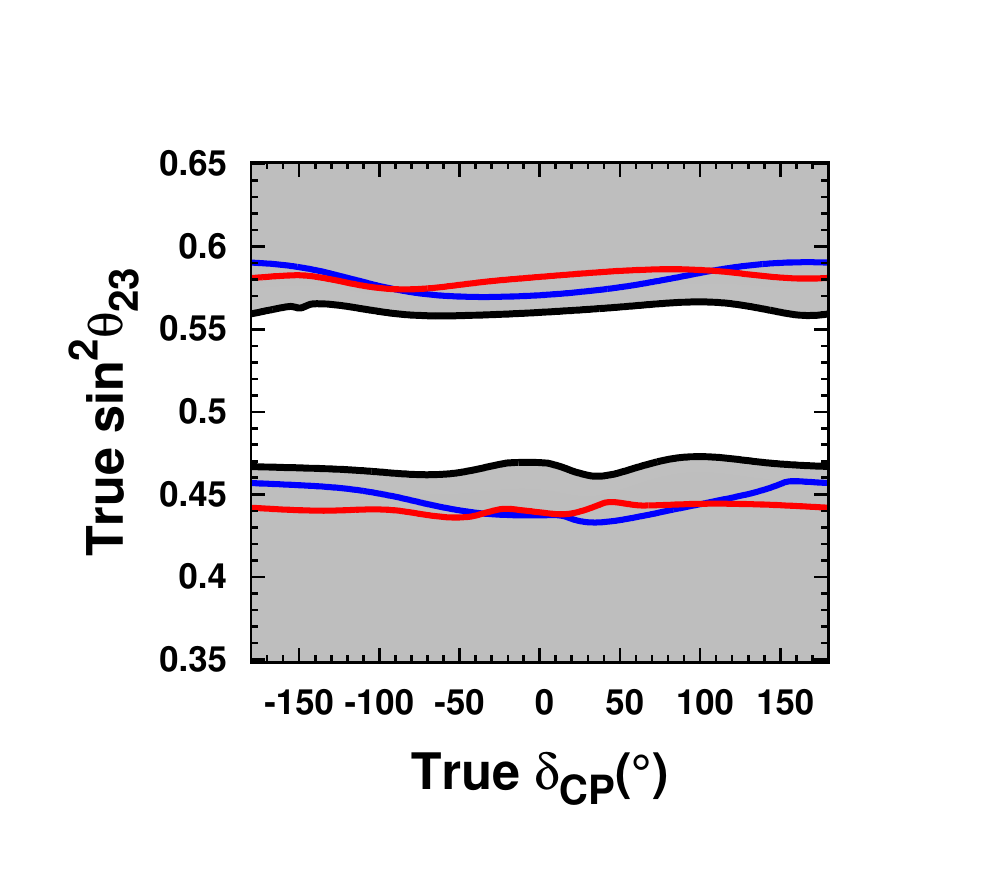}
    \caption{1:1 T2K, 1:1 \nova \nn, NH}
  \end{subfigure}
  \begin{subfigure}[b]{0.49\textwidth}
    \includegraphics[trim=0.7cm 0.5cm 1.3cm 0.5cm, clip=true,width=\textwidth]{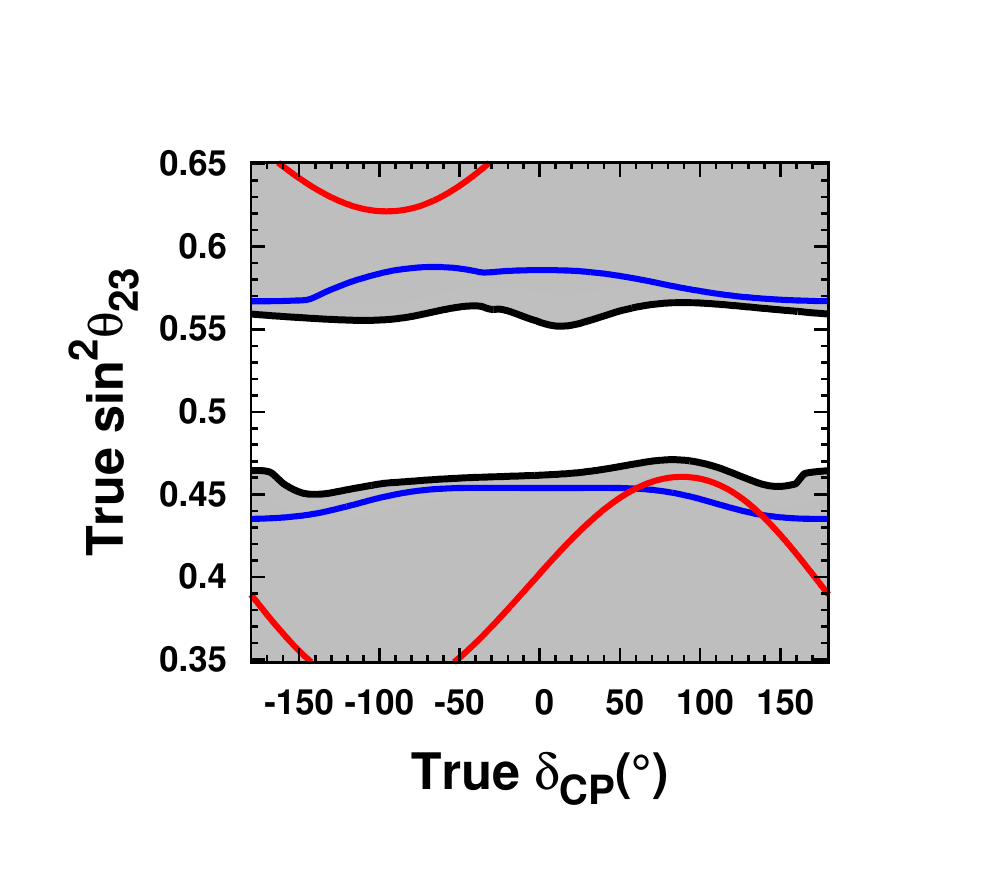}
    \caption{1:0 T2K, 1:1 \nova \nn, IH}
  \end{subfigure}
  \begin{subfigure}[b]{0.49\textwidth}
    \includegraphics[trim=0.7cm 0.5cm 1.3cm 0.5cm, clip=true,width=\textwidth]{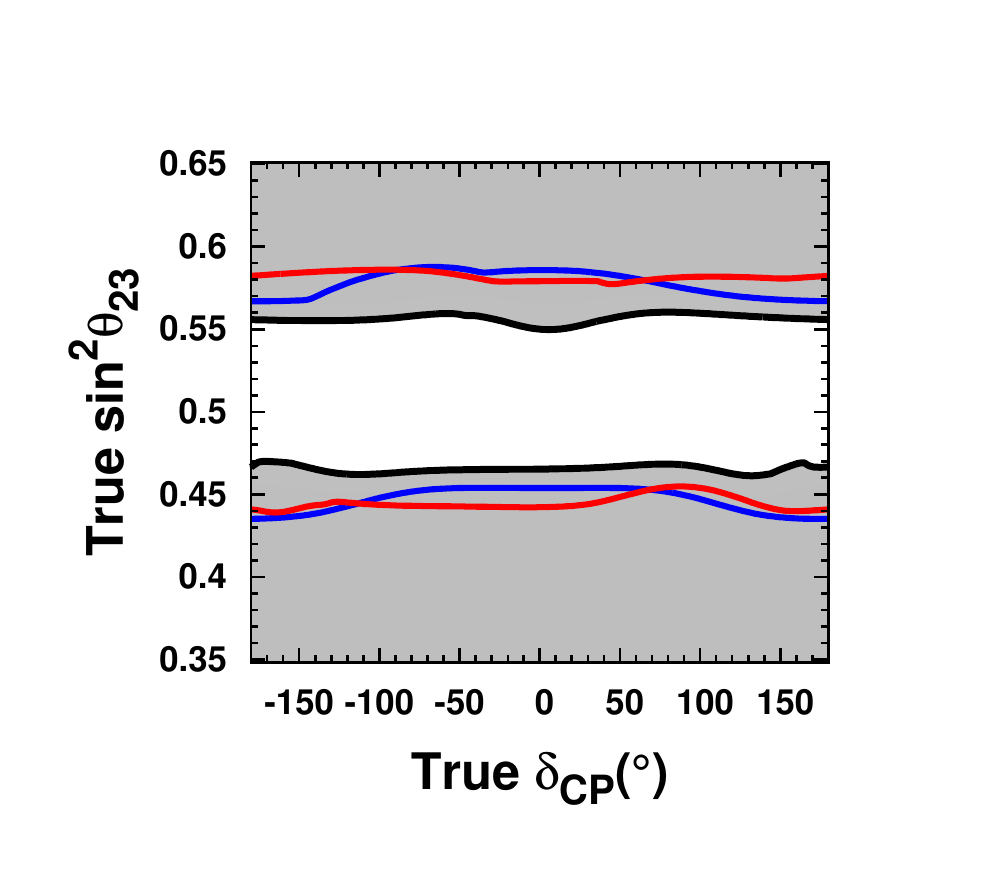}
    \caption{1:1 T2K, 1:1 \nova \nn, IH}
  \end{subfigure}

  \caption[T2K and \nova 90\% CL 
$\theta_{23}$ octant sensitivity regions in \qtt vs \dcp.]
{Regions for T2K~(red), \nova~(blue), and T2K+\nova~(black) 
where the incorrect octant is predicted to be rejected at 90\% C.L. 
Points inside the gray regions are where the incorrect octant 
is predicted to be rejected at 90\% C.L. for T2K+\nova
assuming simple normalization systematics as described in the text.}
  \label{fig:dcpvst23_oct_t2knova}
\end{figure}

Figures~\ref{fig:dcpvst23_cpv_t2knova} through~\ref{fig:dcpvst23_oct_t2knova} show
plots of expected C.L. contours for T2K, \nova and a T2K-\nova 
combined fits as functions of $\sin^2\theta_{23}$ vs. $\delta_{CP}$.
Regions where $\sin\delta_{CP}=0$, one MH and one $\theta_{23}$ octant
are expected to be ruled out at the 90\% C.L are shown.
Significantly wider regions are covered by combining the results
from T2K and \nova.

\begin{figure}[htbp]
  \centering
  \begin{subfigure}[b]{0.49\textwidth}
    \includegraphics[trim=0cm 0cm 0cm 0cm, clip=true,width=\textwidth]{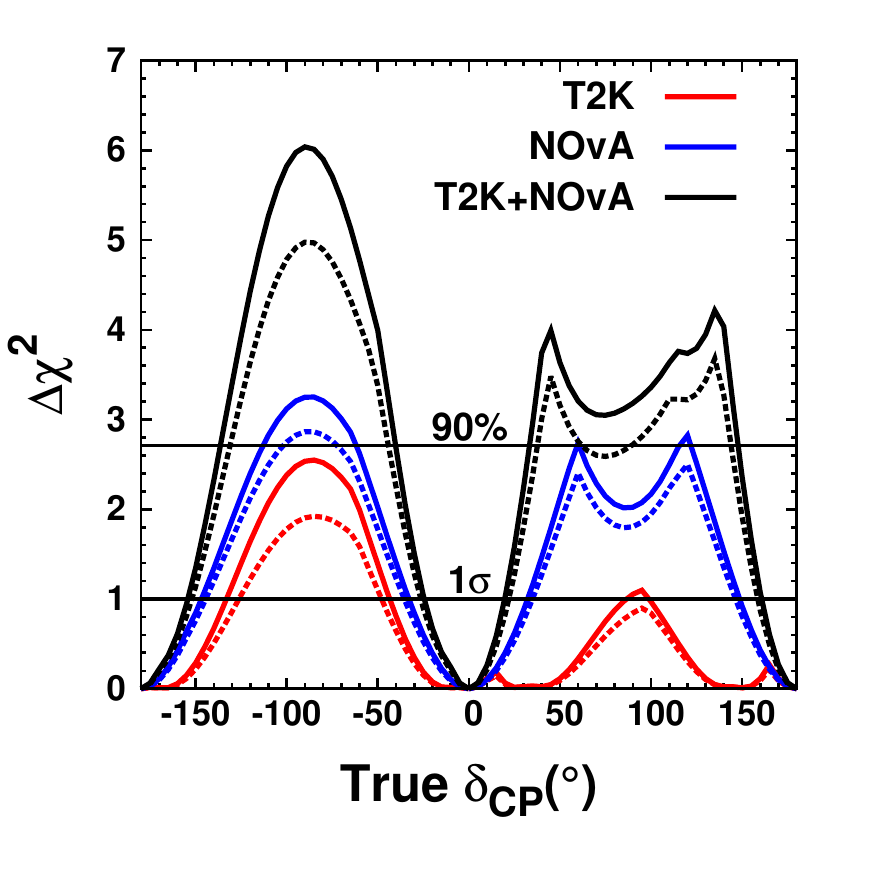}
    \caption{1:0 T2K, 1:1 \nova \nn, NH}
  \end{subfigure}
  \begin{subfigure}[b]{0.49\textwidth}
    \includegraphics[trim=0cm 0cm 0cm 0cm, clip=true,width=\textwidth]{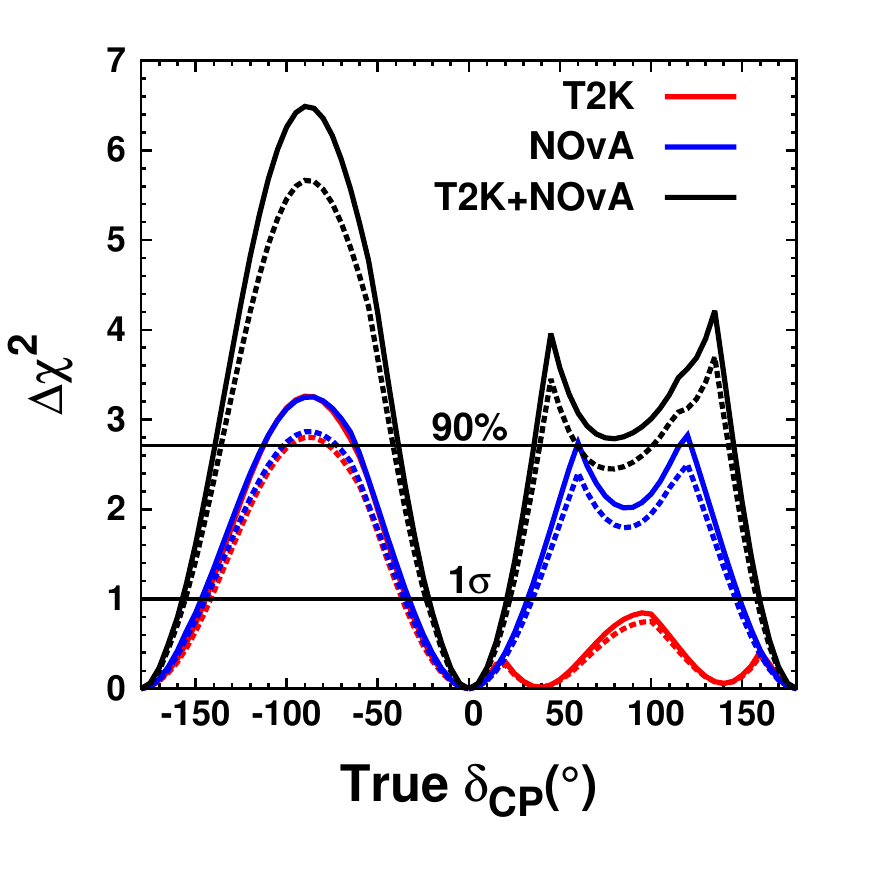}
    \caption{1:1 T2K, 1:1 \nova \nn, NH}
  \end{subfigure}
  \\
  \begin{subfigure}[b]{0.49\textwidth}
    \includegraphics[trim=0cm 0cm 0cm 0cm, clip=true,width=\textwidth]{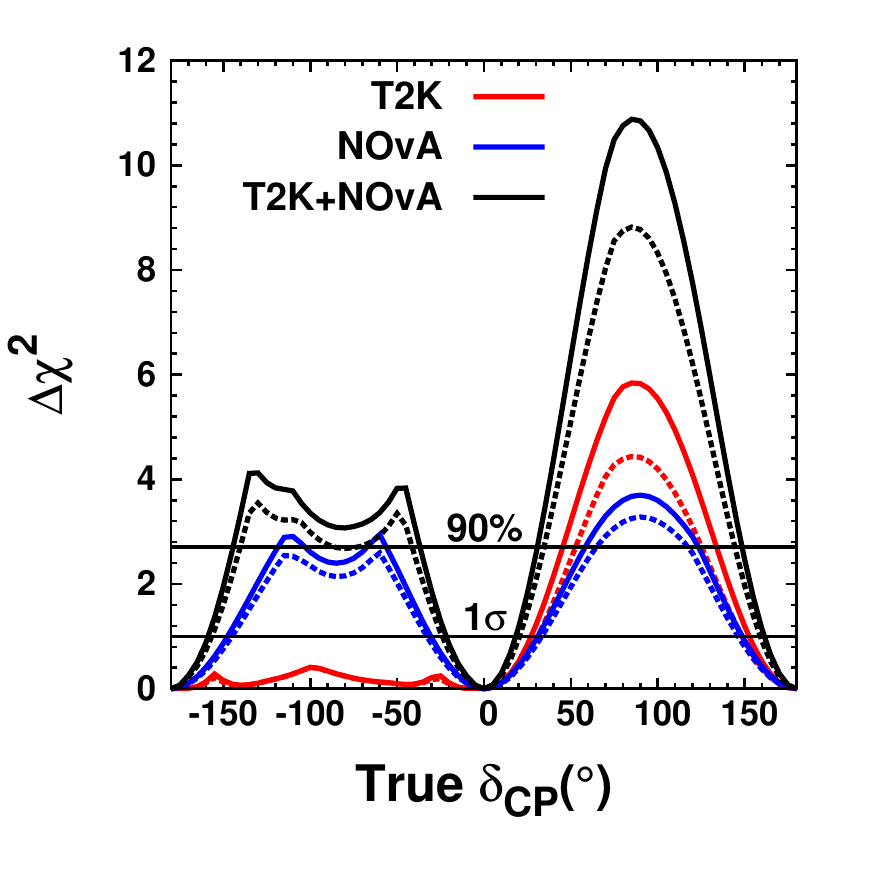}
    \caption{1:0 T2K, 1:1 \nova \nn, IH}
  \end{subfigure}
  \begin{subfigure}[b]{0.49\textwidth}
    \includegraphics[trim=0cm 0cm 0cm 0cm, clip=true,width=\textwidth]{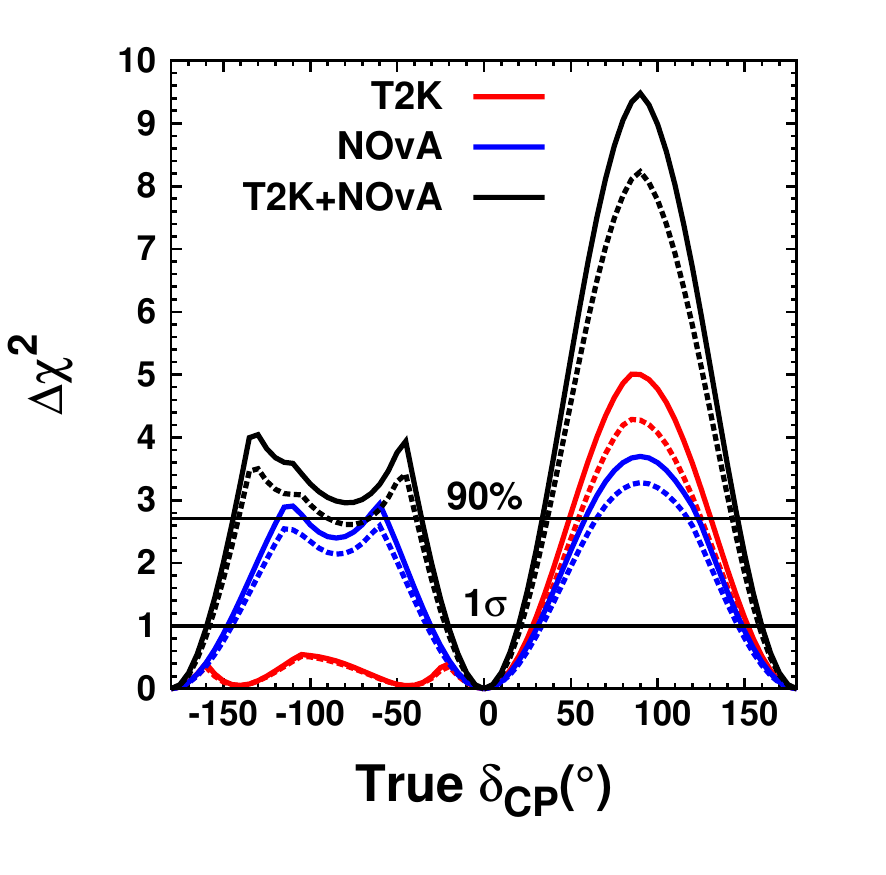}
    \caption{1:1 T2K, 1:1 \nova \nn, IH}
  \end{subfigure}
  \caption[T2K and \nova $\sin\delta_{CP}$ sensitivity]
{The predicted $\Delta\chi^2$ for rejecting $\sin\delta_{CP}=0$ hypothesis, 
as a function of \(\delta_{CP}\) for T2K~(red), \nova~(blue), 
and T2K+\nova~(black). Dashed (solid) curves indicate studies where normalization systematics are (not)
considered. The `true' value of \qtt is assumed to be 0.5, and the `true' MH is assumed to be the NH (top)
or the IH (bottom). The `test' MH is unconstrained.}
%
  \label{fig:sigdcpcpv_t2k_nova_t2knova}
\end{figure}

\begin{figure}[htbp]
  \centering
  \begin{subfigure}[b]{0.49\textwidth}
    \includegraphics[trim=0cm 0cm 0cm 0cm, clip=true,width=\textwidth]{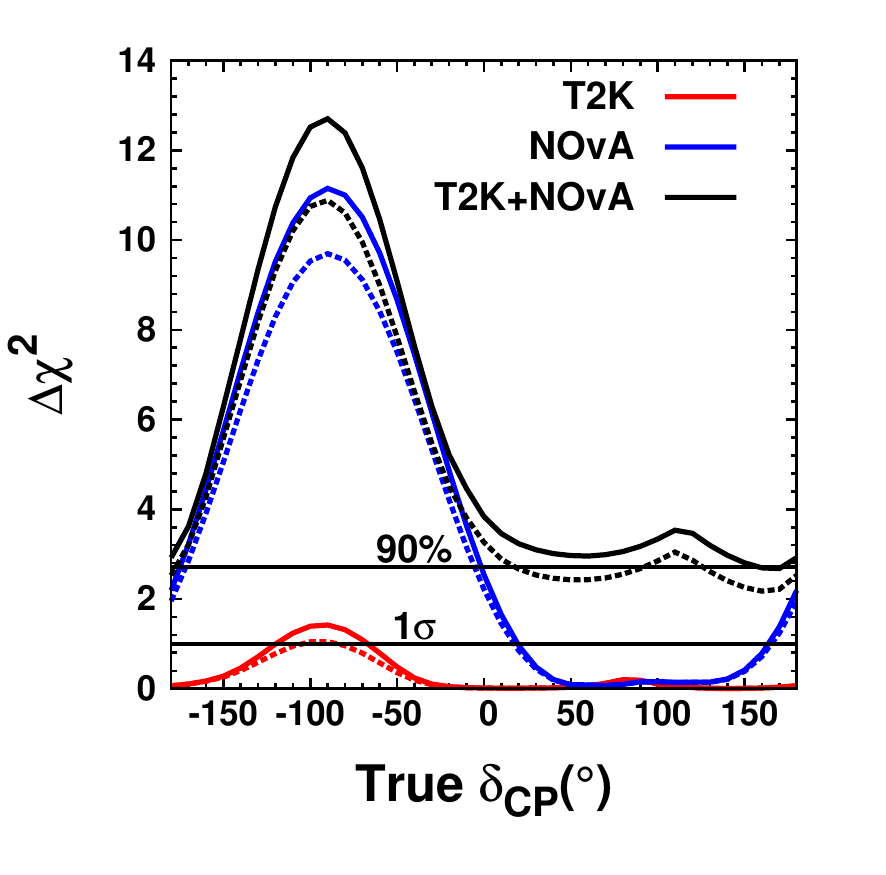}
    \caption{1:0 T2K, 1:1 \nova \nn, NH}
  \end{subfigure}
  \begin{subfigure}[b]{0.49\textwidth}
    \includegraphics[trim=0cm 0cm 0cm 0cm, clip=true,width=\textwidth]{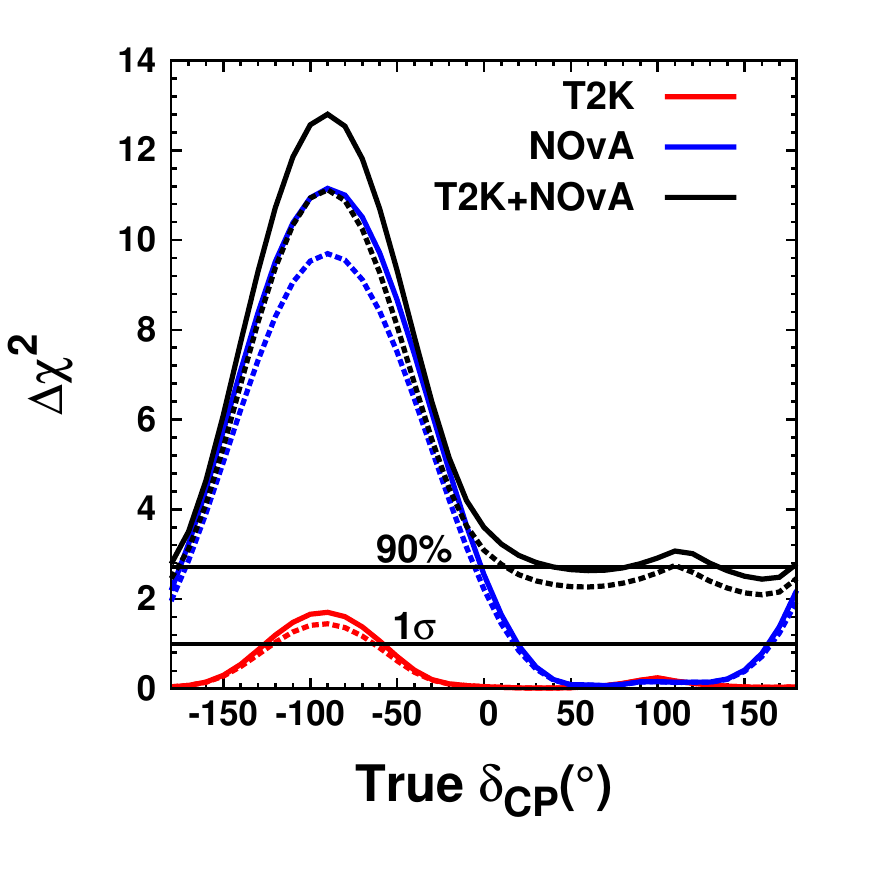}
    \caption{1:1 T2K, 1:1 \nova \nn, NH}
  \end{subfigure}
  \\
  \begin{subfigure}[b]{0.49\textwidth}
    \includegraphics[trim=0cm 0cm 0cm 0cm, clip=true,width=\textwidth]{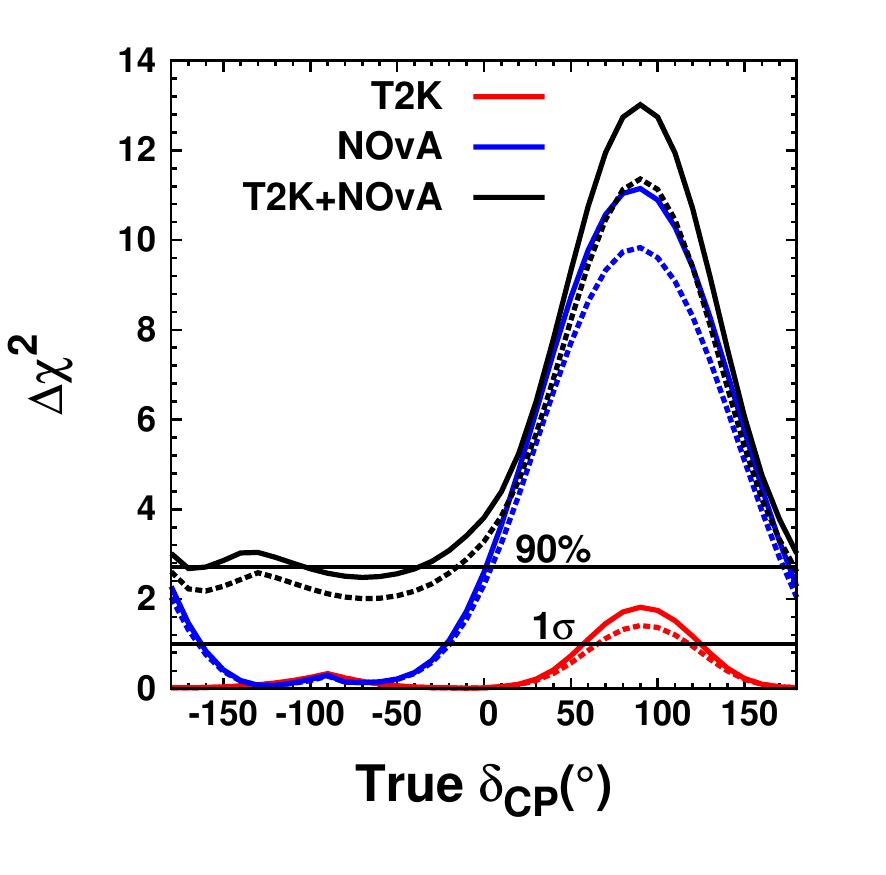}
    \caption{1:0 T2K, 1:1 \nova \nn, IH}
  \end{subfigure}
  \begin{subfigure}[b]{0.49\textwidth}
    \includegraphics[trim=0cm 0cm 0cm 0cm, clip=true,width=\textwidth]{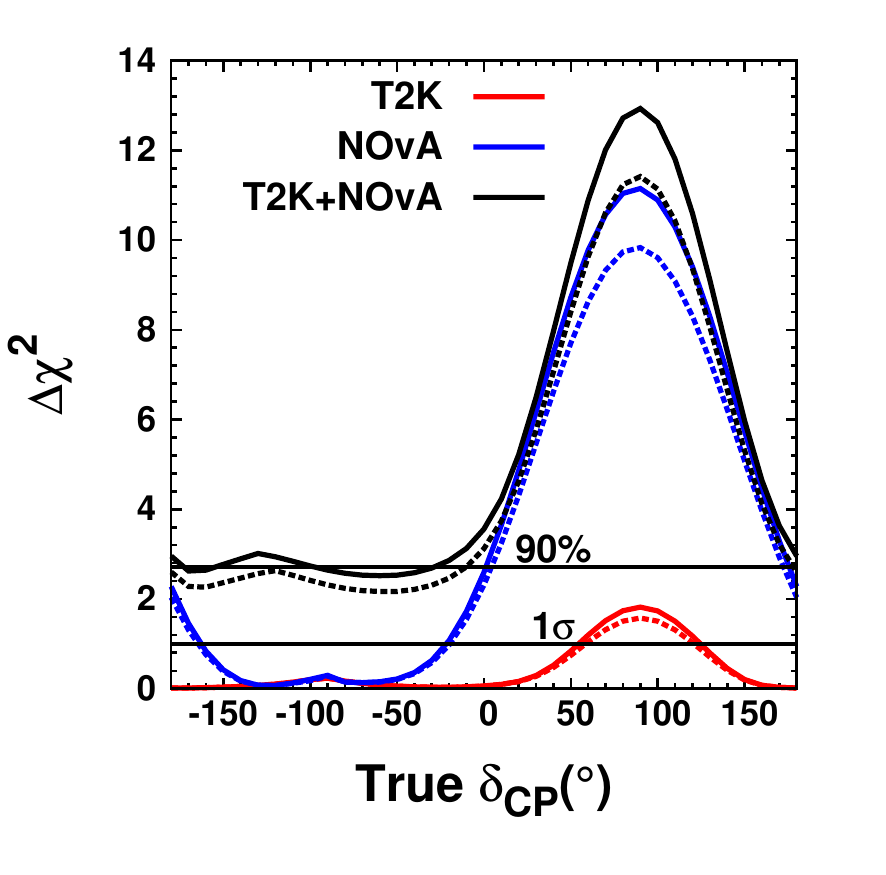}
    \caption{1:1 T2K, 1:1 \nova \nn, IH}
  \end{subfigure}
  \caption[T2K and MH sensitivity]
{The predicted $\Delta\chi^2$ for rejecting the incorrect MH hypothesis,
as a function of \(\delta_{CP}\) for T2K~(red), \nova~(blue),
and T2K+\nova~(black). Dashed (solid) curves indicate studies where normalization systematics are (not)
considered. The `true' value of \qtt is assumed to be 0.5, and the `true' MH is assumed to be the NH (top)
or the IH (bottom). The `test' MH is unconstrained.
}
  \label{fig:sigdcpmh_t2k_nova_t2knova}
\end{figure}

In Figures~\ref{fig:sigdcpcpv_t2k_nova_t2knova} and~\ref{fig:sigdcpmh_t2k_nova_t2knova} 
the \dc for $\sin\delta_{CP}=0$ and for each MH is plotted as a function of 
`true' \dcp in case of \sqtt.
The `true' value of \sqtt was chosen to present a simplified view of 
the sensitivities for maximal mixing.
The T2K's \dc is smaller at $\delta_{CP}=+90^\circ(-90^\circ)$
compared to that at the opposite sign of $\delta_{CP}=-90^\circ(+90^\circ)$
for NH(IH) case while those are similar for \nova.
This comes from the large degeneracy between the CP-violating term
and the matter effect for T2K.
In case of \nova, the matter effect is large enough that
the degenerate parameters space is much smaller as can be seen in Fig.~\ref{fig:t2k_nova_event}.
The complex structure for positive (negative) values of \dcp 
with a true NH (IH) is also due to the fact that 
\dc calculation profiles over MH, and the expected number of $\nu_{e}$ appearance
events is nearly degenerate in these regions.
T2K would perform better than or comparable to \nova, if the MH was assumed to be known. However, there is no experiment, besides \nova, that expects to determine the MH on the relevant time scale, thus the case of a known MH is not presented. 
These figures demonstrate the sensitivity of the two experiments,
as well as the benefit of combined analysis of the two data sets on the
ability to determine MH and CPV.

\section{Neutrino Mode and Antineutrino Mode Running Time Optimization}
\label{sec:runratio}
As previously shown in Sec.~\ref{sec:t2ksensitivity},
a significant fraction of $\bar{\nu}$-mode running improves
the sensitivity to CP violation, especially when systematic uncertainties are taken into account.
In this section studies of the \nn running ratios are shown for T2K, \nova, and combined fits of T2K+\nova simulated data
using the tools developed in Sec.~\ref{pfst2knova}.
A set of metrics are defined  
that characterize the ability of each experiment 
or a combined fit of both experiments to constrain \dcp, reject 
$\delta_{CP}=0$, or determine the MH. 
The following metrics are used in these studies:

\begin{itemize}
  \item {$\delta_{CP}$ half-width}: 
        The \os half-width is defined as half of the \os Confidence Interval 
(C.I.) about the true value of \dcp.
        In some cases there are degenerate \os C.I. regions in \dcp that are disconnected from the central value.
In this case half of the width of the degenerate region is added to this metric. This is a measure of
the precision that can be acheived in measurment of \dcp.

  \item {Median \dc for $\delta_{CP}=0$}: 
This metric defines the \dc value for which 50\% of true \dcp values 
can be distinguished from $\delta_{CP}=[0,\pi]$. This is a measure of
sensitivity to CPV.
  \item {Lowest \dc for mass hierarchy determination}: 
This metric defines the \dc value at which the mass hierarchies can be 
distinguished for 100\% of true \dcp values.
\end{itemize}

\begin{figure}[htbp]
  \centering
  \includegraphics[trim=0.08cm 0.1cm 0.1cm 1.5cm, clip=true,width=0.55\textwidth]{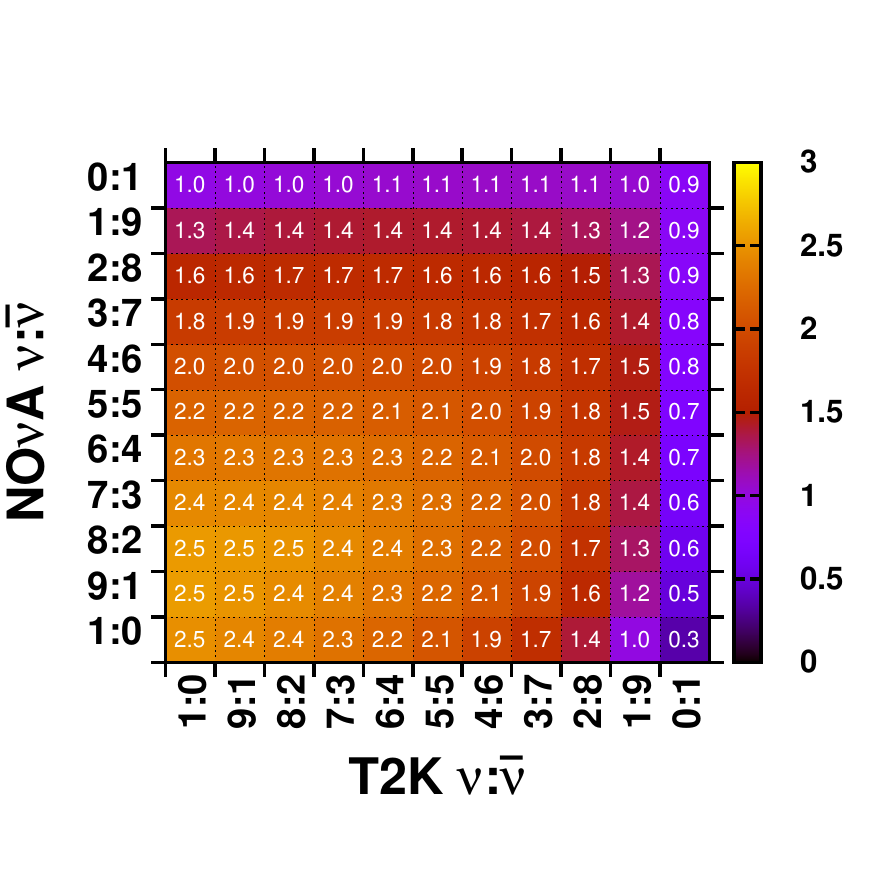}
  \caption[T2K+\nova MH determination at various \nn running ratios.]
{Lowest \dc for a combined T2K+\nova fit to determine 
the mass hierarchy for various \nn running ratios. 
True values are assumed to be: MH=NH, \sqtt.
           Normalization systematics are assumed. }
  \label{fig:runratio_mhfrac_systs_corr_minmh}
\end{figure}

Each metric is calculated for a T2K+\nova combined analysis 
for various \nn run ratios.
Figure \ref{fig:runratio_mhfrac_systs_corr_minmh} gives the lowest \dc values 
for mass hierarchy determination for \nn variations
in a combined T2K+\nova fit.
They are computed from the results of studies like the one shown in
Fig.~\ref{fig:sigdcpmh_t2k_nova_t2knova} and conservatively summarize 
the content of the plot in one data point.
For example, the lowest \dc value for mass hierarchy determination
at 1:0 (100\% $\nu$ running) T2K, 5:5 (50\% $\nu$ / 50\% $\bar{\nu}$ running) \nova running is the lowest \dc from 
Fig.~\ref{fig:sigdcpmh_t2k_nova_t2knova}(a) ($\Delta \chi^2=2.19$).

\begin{figure}[htbp]
  \centering
  \includegraphics[trim=0.08cm 0.1cm 0.1cm 1.5cm, clip=true,width=0.55\textwidth]{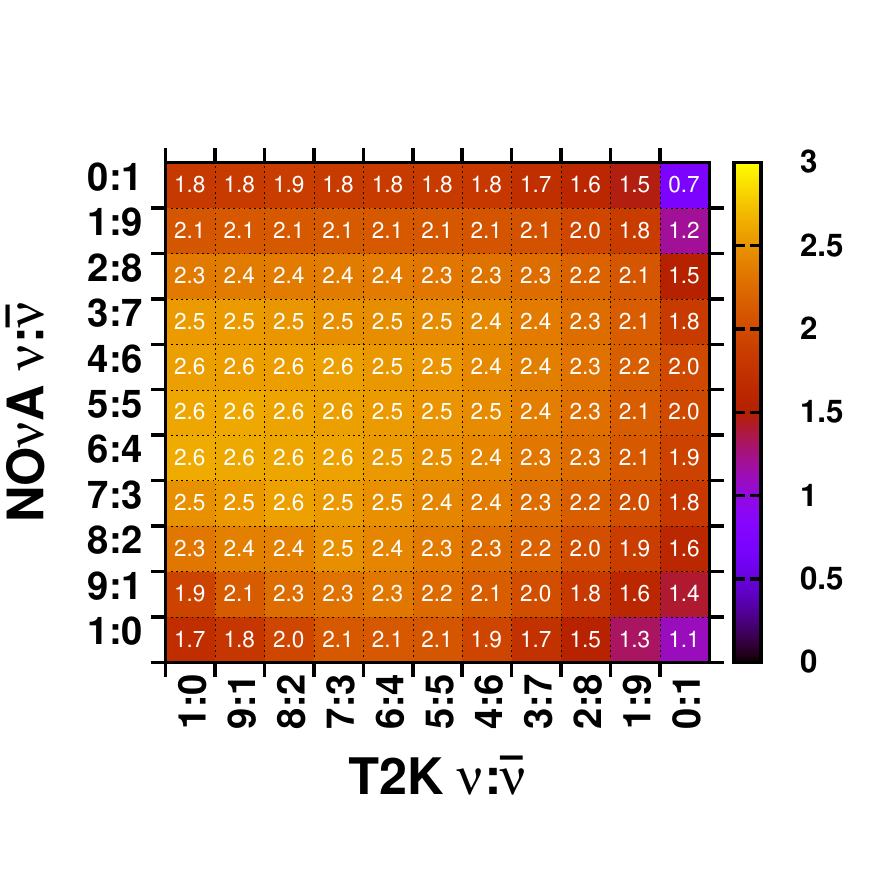}
  \caption[T2K+\nova \sdnz sensitivity for various \nn running ratios]
{Median \dc for $\sin\delta_{CP}=0$ for a combined T2K+\nova fit. 
           True values are assumed to be: MH=NH, \sqtt.
           Normalization systematics are assumed.
}
  \label{fig:runratio_cpvfrac_systs_corr_minmh}
\end{figure}

Similarly, Fig.~\ref{fig:runratio_cpvfrac_systs_corr_minmh} gives 
the median \dc values for $\sin\delta_{CP}=0$ for \nn variations
in a combined T2K+\nova fit. 
These values are computed from studies like the ones presented 
in Fig.~\ref{fig:sigdcpcpv_t2k_nova_t2knova}.
The $\sin\delta_{CP}=0$ median \dc value at 1:0 T2K, 5:5 \nova running is the median \dc from Fig.~\ref{fig:sigdcpcpv_t2k_nova_t2knova}(a) ($\Delta \chi^2=2.6$).

\begin{figure}[htbp]
  \centering
  \begin{subfigure}[b]{0.49\textwidth}
    \includegraphics[trim=0.1cm 0.1cm 0.1cm 0.1cm, clip=true,width=\textwidth]{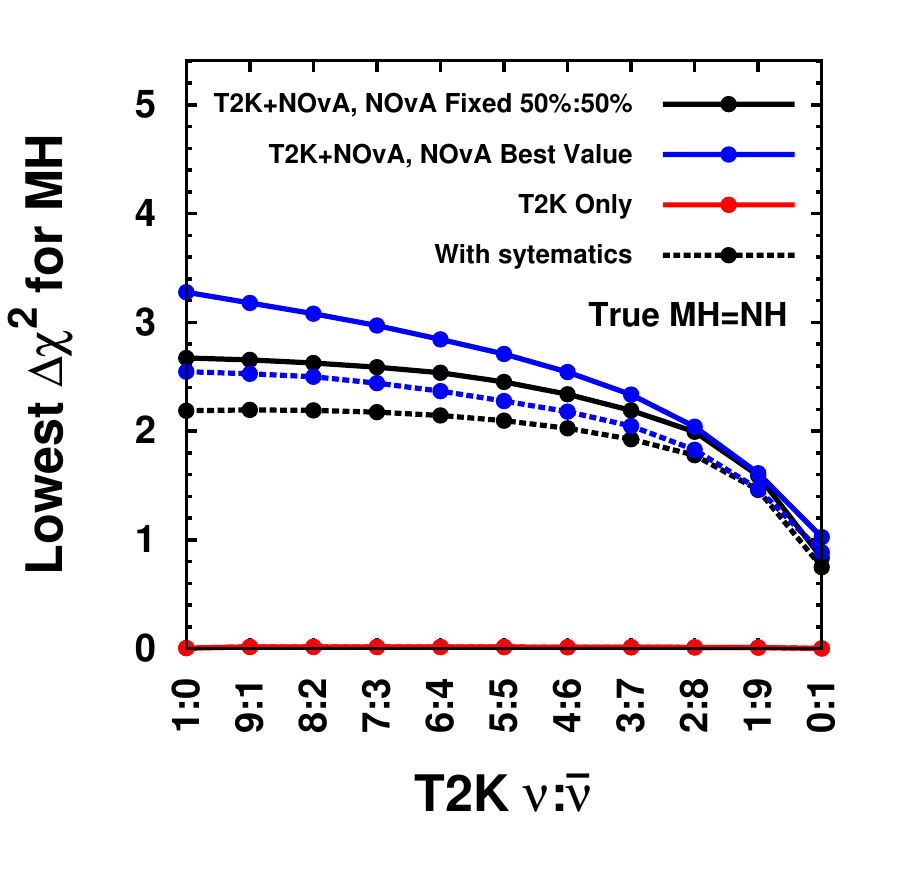}
    \label{fig:runratio_MH_summary}
  \end{subfigure}
  \begin{subfigure}[b]{0.49\textwidth}
    \includegraphics[trim=0.1cm 0.1cm 0.1cm 0.1cm, clip=true,width=\textwidth]{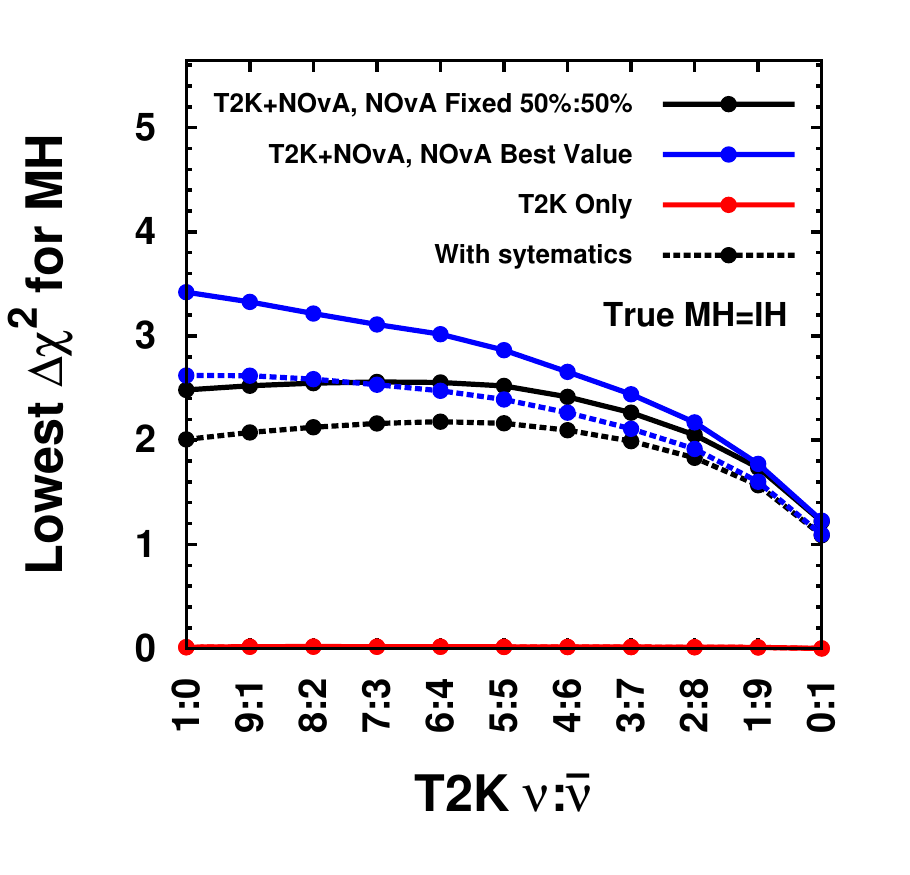}
    \label{fig:runratio_MH_ih_summary}
  \end{subfigure}
  \caption[T2K and T2K+\nova MH determination at various \nn running ratios.]
{Lowest $\Delta \chi^2$ for mass hierarchy determination
in a combined, T2K+\nova, fit as a function of T2K \nn running ratio 
for true MH=NH (left) and IH (right).
           Curves are given for the \dc value at nominal 5:5 \nova running~(black), best case T2K+\nova running~(blue), and T2K only running~(red).
           Dashed~(solid) curves indicate studies performed~(without) assuming normalization systematics. }
  \label{fig:runratio_MH_sum}
\end{figure}

\begin{figure}[htbp]
  \centering
  \begin{subfigure}[b]{0.49\textwidth}
    \includegraphics[trim=0.1cm 0.1cm 0.1cm 0.1cm, clip=true,width=\textwidth]{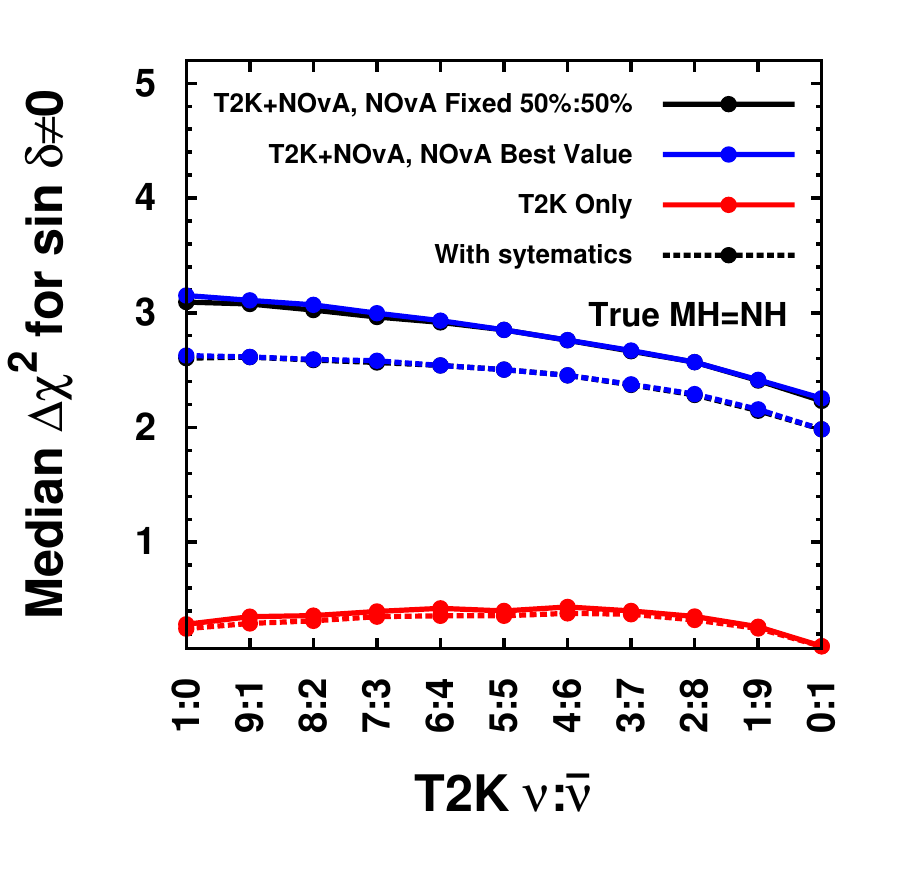}
    \label{fig:runratio_cpvfrac_summary}
  \end{subfigure}
  \begin{subfigure}[b]{0.49\textwidth}
    \includegraphics[trim=0.1cm 0.1cm 0.1cm 0.1cm, clip=true,width=\textwidth]
{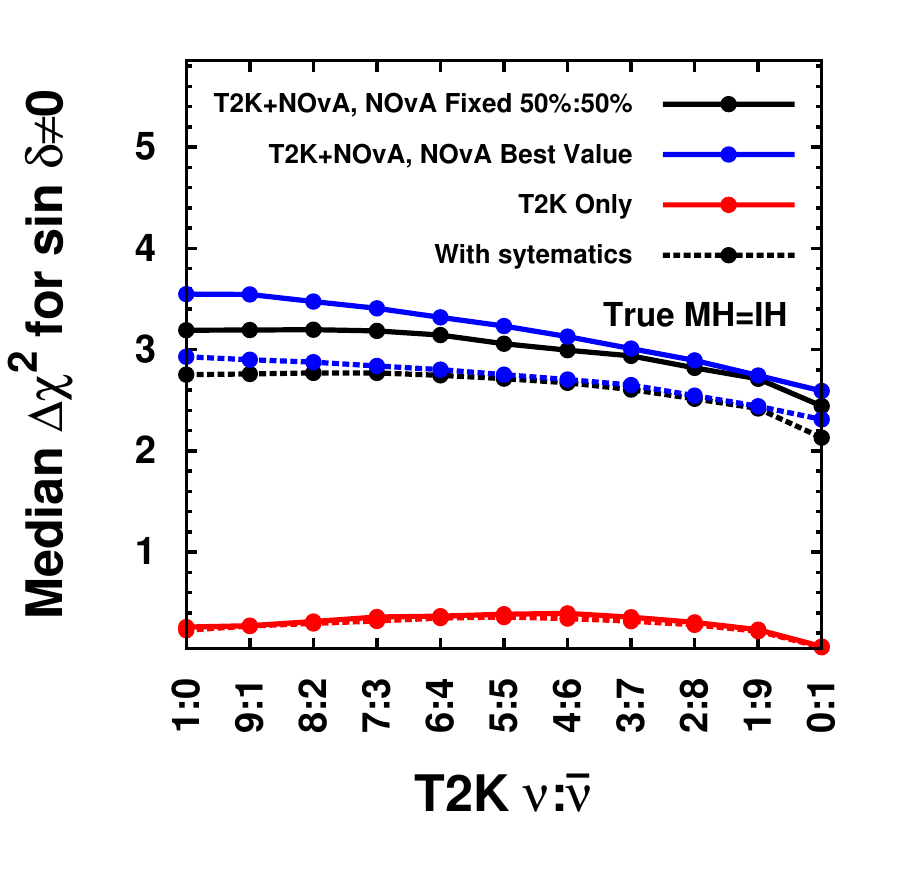}
    \label{fig:runratio_cpvfrac_ih_summary}
  \end{subfigure}
  \caption[T2K and T2K+\nova \sdnz sensitivity at various \nn running ratios.]
{Median $\Delta \chi^2$ for $\sin\delta_{CP}=0$ 
in a combined, T2K+\nova, fit as a function of T2K \nn running ratio for true MH=NH (left) and IH (right).
           Curves are given for the \dc value at nominal 5:5 \nova running (black), best case T2K+\nova running (blue), and T2K only running (red).
           Dashed~(solid) curves indicate studies performed~(without) assuming normalization systematics. }
  \label{fig:runratio_cpvfrac_sum}
\end{figure}

Figure \ref{fig:runratio_MH_sum} summarizes the data in Fig.~\ref{fig:runratio_mhfrac_systs_corr_minmh} and compares it with the
metric calculated for T2K only running. 
The black curve gives the lowest \dc for MH determination in a combined, 
T2K+\nova, fit as a function of T2K \nn running ratio 
with the \nova running fixed at 1:1.
As shown previously, the T2K data set alone has almost no sensitivity to the MH
determination.
The curves for 5:5 \nova running with systematics (black dashed) shows 
an optimal T2K running ratio of around 6:4 for a combined fit.
However, the metric is very flat with respect to the T2K \nn run ratio for $\nu$ running greater than 50\%.
Figure \ref{fig:runratio_cpvfrac_sum} shows the summary for 
median \dc for $\sin\delta_{CP}=0$.
T2K run ratios between 1:0 and 5:5 produce relatively similar values of median \dc for the combined fit.
This is also true for combined T2K+\nova running independent of the \nova run plan optimization.
There is a slight preference for all neutrino running in T2K 
in the combined fit.

\begin{figure}[htbp]
  \centering
  \begin{subfigure}[b]{0.49\textwidth}
    \includegraphics[trim=0.1cm 0.1cm 0.1cm 0.1cm, clip=true,width=\textwidth]{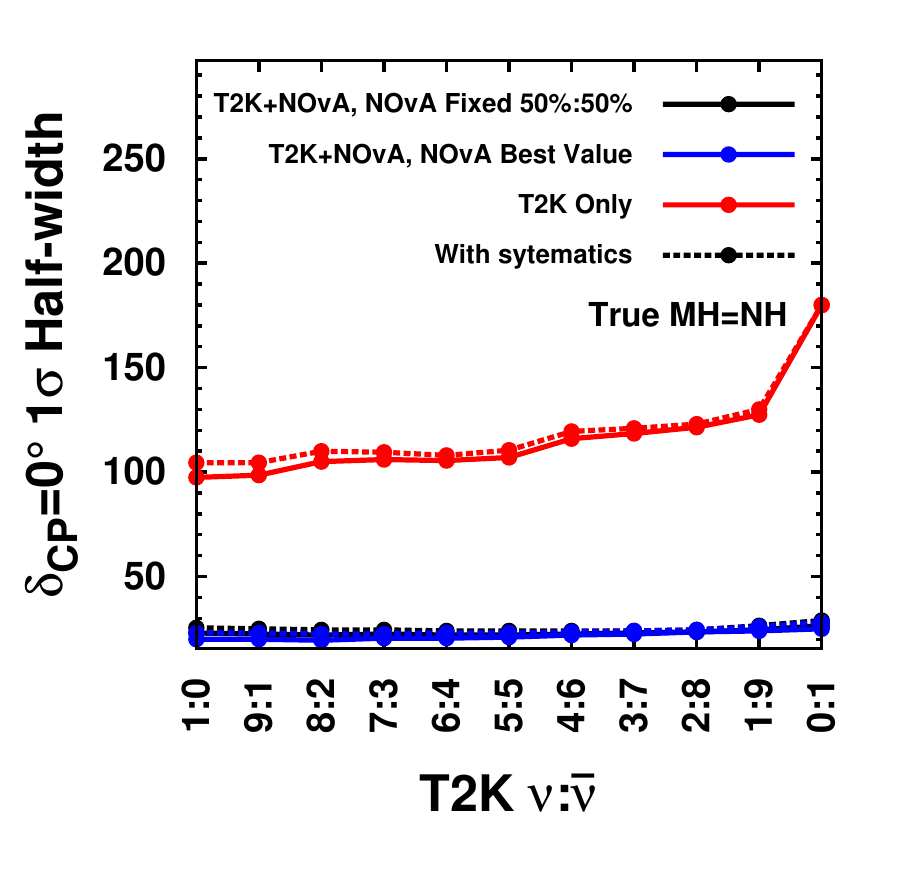}
  \caption{$\delta=0\degree$, NH}
  \end{subfigure}
  \begin{subfigure}[b]{0.49\textwidth}
    \includegraphics[trim=0.1cm 0.1cm 0.1cm 0.1cm, clip=true,width=\textwidth]
{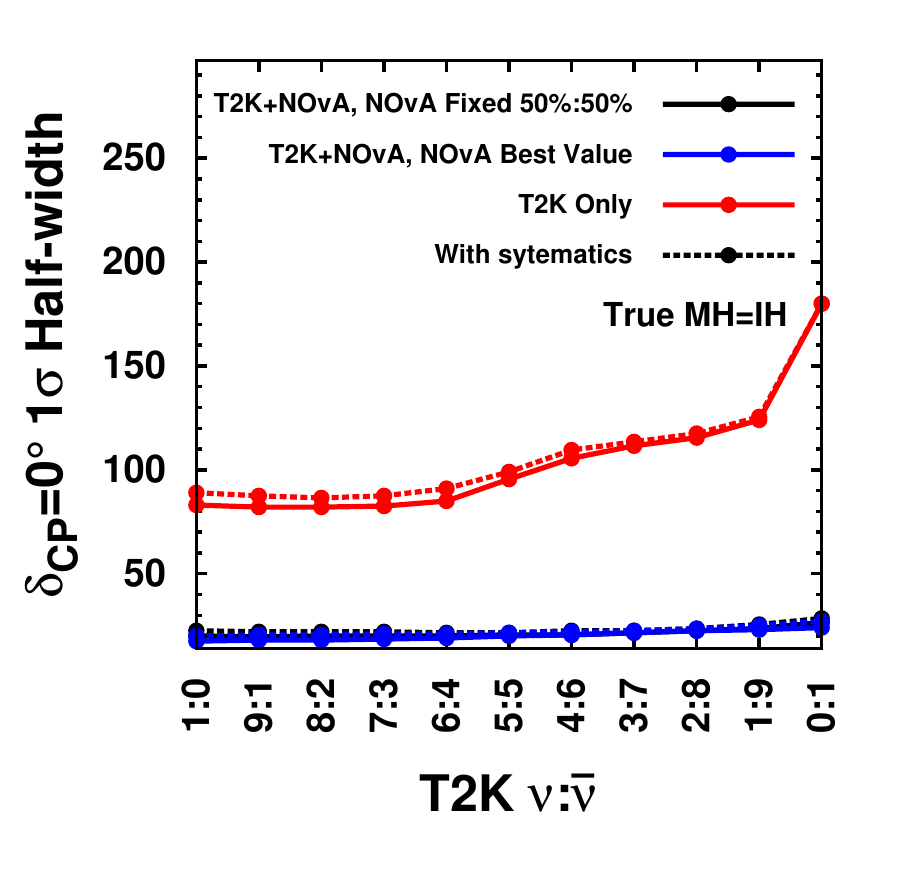}
  \caption{$\delta=0\degree$, IH}
  \end{subfigure}
  \begin{subfigure}[b]{0.49\textwidth}
    \includegraphics[trim=0.1cm 0.1cm 0.1cm 0.1cm, clip=true,width=\textwidth]{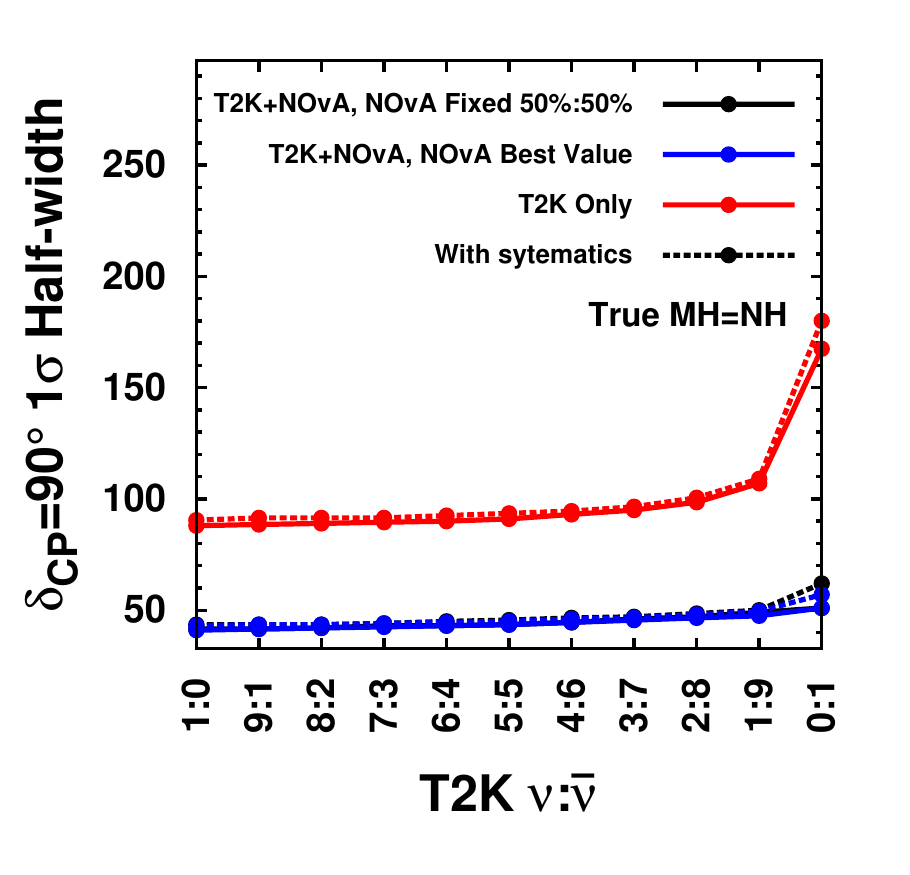}
  \caption{$\delta=90\degree$, NH}
  \end{subfigure}
  \begin{subfigure}[b]{0.49\textwidth}
    \includegraphics[trim=0.1cm 0.1cm 0.1cm 0.1cm, clip=true,width=\textwidth]
{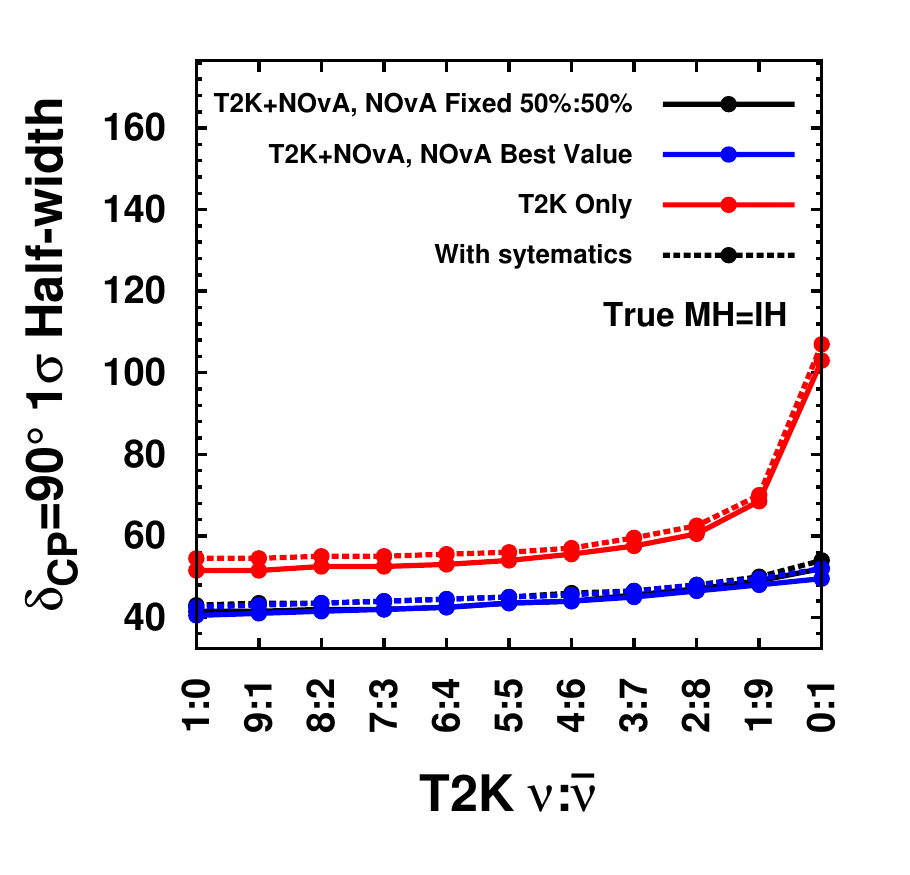}
  \caption{$\delta=90\degree$, IH}
  \end{subfigure}
  \caption[T2K and T2K+\nova 1$\sigma$ width at various \nn running ratios.]
{$\delta_{CP}$ resolution in a combined, T2K+\nova, fit 
as a function of T2K \nn running ratio.
           Curves are given for the resolution value, in degress, at nominal 5:5 \nova running (black), best case T2K+\nova running (blue), and T2K only running (red).
           Dashed~(solid) curves indicate studies performed~(without) assuming normalization systematics. }
  \label{fig:runratio_dcpwd_sum}
\end{figure}

\begin{figure}[htbp]
  \centering
  \begin{subfigure}[b]{0.49\textwidth}
    \includegraphics[trim=0.1cm 0.1cm 0.1cm 0.1cm, clip=true,width=\textwidth]{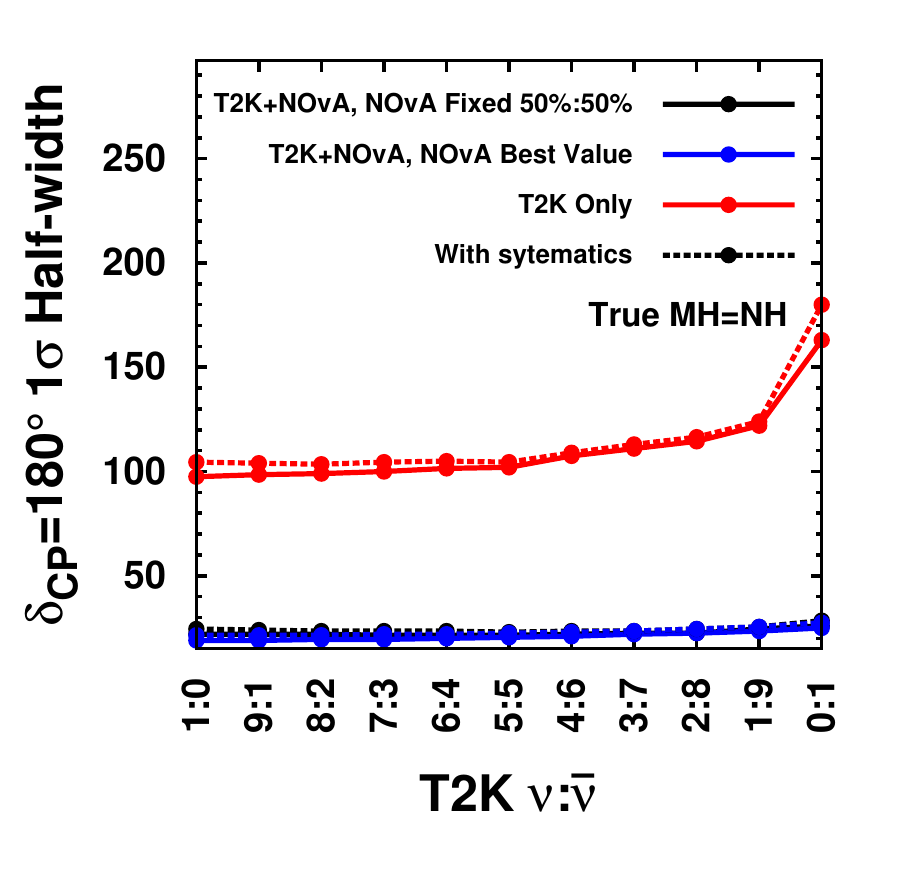}
  \caption{$\delta=180\degree$, NH}
  \end{subfigure}
  \begin{subfigure}[b]{0.49\textwidth}
    \includegraphics[trim=0.1cm 0.1cm 0.1cm 0.1cm, clip=true,width=\textwidth]
{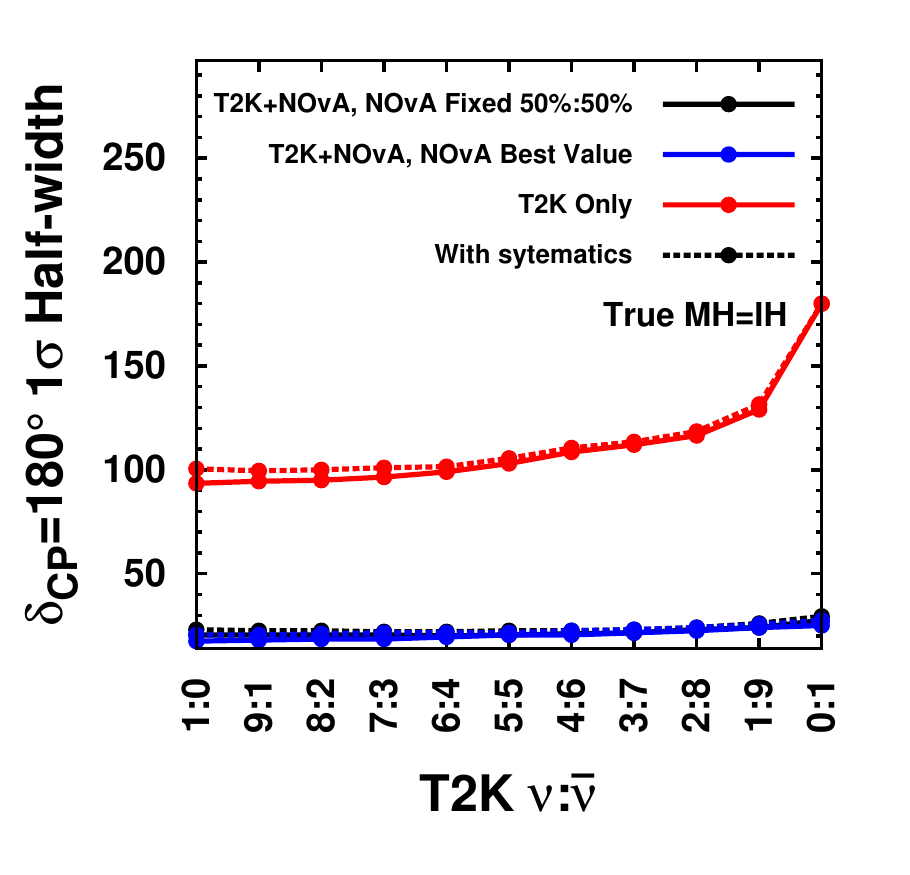}
  \caption{$\delta=180\degree$, IH}
  \end{subfigure}
  \begin{subfigure}[b]{0.49\textwidth}
    \includegraphics[trim=0.1cm 0.1cm 0.1cm 0.1cm, clip=true,width=\textwidth]{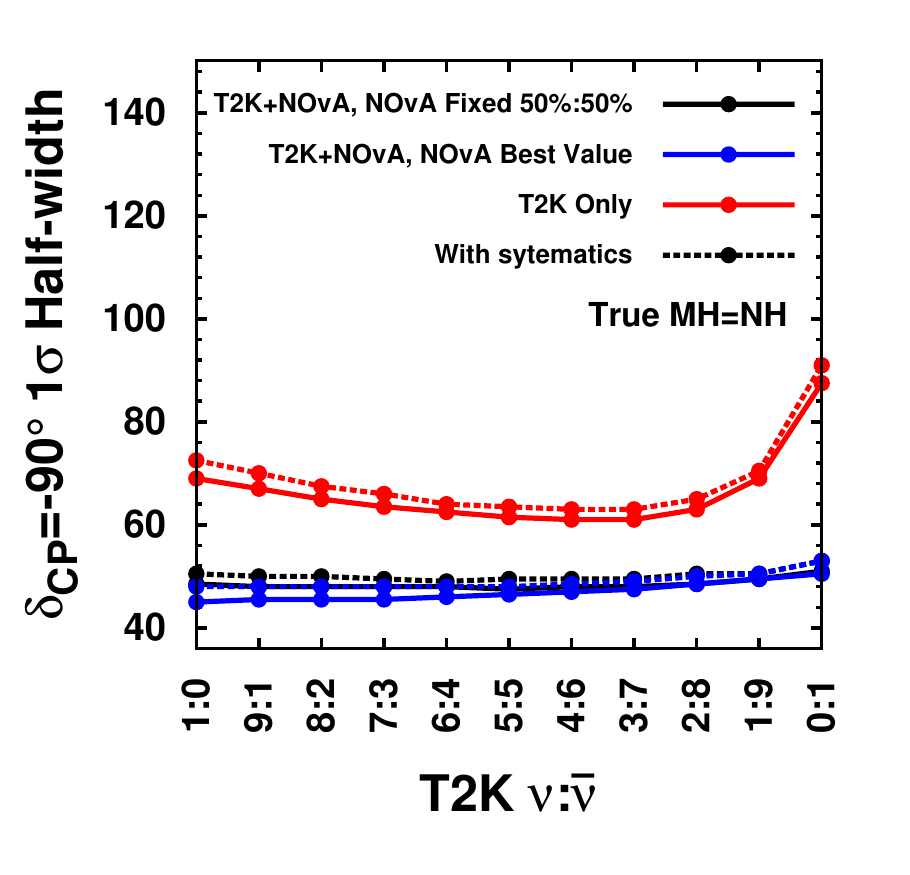}
  \caption{$\delta=-90\degree$, NH}
  \end{subfigure}
  \begin{subfigure}[b]{0.49\textwidth}
    \includegraphics[trim=0.1cm 0.1cm 0.1cm 0.1cm, clip=true,width=\textwidth]
{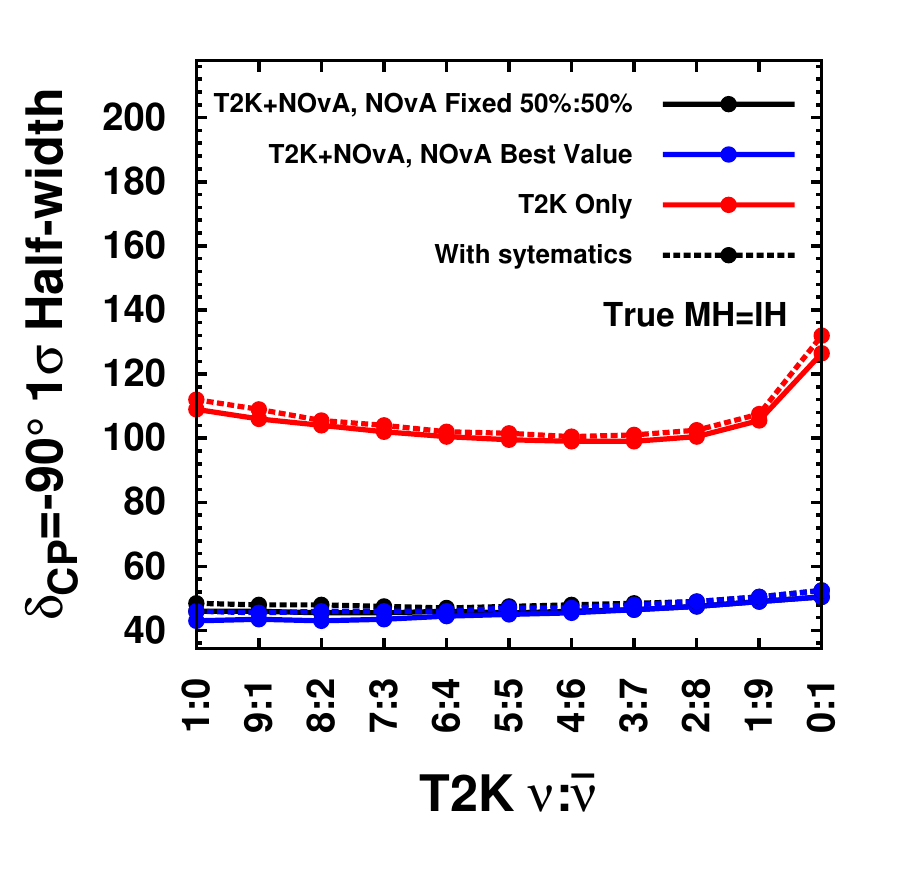}
  \caption{$\delta=-90\degree$, IH}
  \end{subfigure}
  \caption[T2K and T2K+\nova 1$\sigma$ width at various \nn running ratios.]
{
Same as Fig.~\ref{fig:runratio_dcpwd_sum}, but for different
$\delta_{CP}$ values.
}
  \label{fig:runratio_dcpwd_sum_2}
\end{figure}

Figure \ref{fig:runratio_dcpwd_sum} and \ref{fig:runratio_dcpwd_sum_2}
summarize the $\delta_{CP}\ 1\sigma$ width at various values of $\delta_{CP}$.
Again, relatively similar values of $\delta_{CP}\ 1\sigma$ width
are expected for the T2K run ratios between 1:0 and 1:9.

All of the metrics demonstrate a relatively flat response between approximately 7:3 and 3:7
for T2K and for T2K+\nova(5:5) with systematics, with a worse response outside that range.
These results are consistent with several other studies not shown in this paper (e.g. 
the measures of the precision on $\sin^2\theta_{13}$ in $\nu$-mode and in $\bar{\nu}$-mode).
The results are also robust with respect to reasonable variations in $\sin^2\theta_{23}$, 
$\delta_{CP}$ and the MH. Thus, the results suggest that T2K run with a $\nu$-mode to 
$\bar{\nu}$-mode at ratio of 1:1 with an allowed variation of $\pm20\%$ of 
the total exposure.
The variation can be used to optimize the experiment to any one analysis without
significant degradation of the sensitivity to any other analysis. A more detailed
optimization of the $\nu$:$\bar{\nu}$ run ratio will require tighter constraints on
oscillation parameters from future analyses, a more detailed treatment of systematic
uncertainties from both T2K and \nova, and a clear prioritization of analysis goals 
from the T2K and \nova collaborations.


\section {Summary}
In this paper we have presented studies of the T2K experiment sensitivity 
to oscillation parameters by performing a three-flavor analysis 
combining appearance and disappearance, for both \(\nu\)-mode, 
and \(\bar{\nu}\)-mode assuming the expected full statistics 
of $7.8\times10^{21}$ POT. 
The T2K precision study includes either statistical errors only, 
systematic errors established for the 2012 oscillation analyses, 
or conservatively projected systematic errors, 
and takes into consideration signal efficiency and background. 
We have derived the sensitivity to the oscillation parameters 
$\sin^22\theta_{13}$, $\delta_{CP}$, $\sin^22\theta_{23}$, and $\Delta m^2_{32}$ 
for a range of the true parameter values 
and using constraints from other experiments.
For example, with equal exposure of $\nu$-mode and $\bar{\nu}$-mode
and using signal efficiency from the 2012 analysis we project a dataset of approximately 100 $\nu_e$  and 25 $\bar{\nu_e}$ appearance events and 390 (270)  $\nu_\mu$  and 130 (70) $\bar{\nu_\mu}$ CCQE (CC non-QE) events.   
From these data, with the projected systematic uncertainties we would achieve 
a 1-$\sigma$ resolution of 0.050(0.054) on $\sin^2\theta_{23}$ 
and $0.040(0.045)\times10^{-3} \rm{eV}^2$ on $\Delta m^2_{32}$ 
for 100\%(50\%) neutrino beam mode running. 
T2K will also have sensitivity to the CP-violating phase $\delta_{\rm{CP}}$ 
at 90\% C.L. or higher over a significant range.
For example, if $\sin^2\theta_{23}$ is maximal (i.e $\theta_{23}$=$45^\circ$) 
the range is $-115^\circ<\delta_{\rm{CP}}<-60^\circ$ for normal hierarchy 
and $+50^\circ<\delta_{\rm{CP}}<+130^\circ$ for inverted hierarchy.  

Since the ability of T2K to measure the value of $\delta_{\rm{CP}}$ 
is greatly enhanced by the knowledge of the mass hierarchy 
we have also incorporated the expected data from the NO$\nu$A experiment 
into our projections using the GLoBES tools.  
With the same normalization uncertainties of 5\% on the signal and 10\% 
on the background for both experiments we find, for example, 
that the predicted $\Delta\chi^2$ for rejecting 
the $\delta_{\rm{CP}}=0$ hypothesis 
for $\delta_{\rm{CP}}=+90^\circ$, IH and $\sin^2\theta_{23}=0.5$ 
from the combined experiment fit is 8.2 compared to 4.3 and 3.2 
for T2K and NO$\nu$A alone, respectively. 
The region of oscillation parameter space 
where there is sensitivity to observe a non-zero $\delta_{CP}$ 
is substantially increased compared to if each experiment is analyzed alone.

From the investigation of dividing the running time between ${\nu}$- and $\bar{\nu}$-modes
we found that an even split gives the best sensitivity for a wider region of the oscillation parameter space for both T2K data alone, and for T2K data in combination with NO$\nu$A, though the dependence on the ratio is not strong.
 
It is anticipated that the results of these studies will help to guide the optimization of the future run plan for T2K.

\bigskip
\bigskip
\centerline{\bf Acknowledgment}
We thank the J-PARC staff for superb accelerator performance
and the CERN NA61 collaboration for providing valuable particle production data.
We acknowledge the support of MEXT, Japan;
NSERC, NRC and CFI, Canada;
CEA and CNRS/IN2P3, France;
DFG, Germany;
INFN, Italy;
National Science Centre (NCN), Poland;
RAS, RFBR and MES, Russia;
MICINN and CPAN, Spain;
SNSF and SER, Switzerland;
STFC, UK; and
DOE, USA.
We also thank CERN for the UA1/NOMAD magnet,
DESY for the HERA-B magnet mover system,
NII for SINET4,
the WestGrid and SciNet consortia in Compute Canada,
GridPP, UK.
In addition participation of individual researchers
and institutions has been further supported by funds from:
ERC (FP7), EU;
JSPS, Japan;
Royal Society, UK;
DOE Early Career program, USA. 
\bibliographystyle{ptephy} 
\bibliography{1409-036-3C-AtsukoIchikawa}  

\begin{thebibliography}{10}

\bibitem{Fukuda:1998mi}
Y.~Fukuda et~al., Phys. Rev. Lett., {\bf 81}, 1562--1567 (1998).

\bibitem{Cleveland:1998ApJ}
B.T. Cleveland, Timothy Daily, Jr. Davis, Raymond, James~R. Distel, Kenneth
  Lande, et~al., Astrophys.J., {\bf 496}, 505--526 (1998).

\bibitem{Abe:2010hy}
K.~Abe et~al., Phys. Rev. D, {\bf 83}, 052010 (2011).

\bibitem{Aharmim:2011vm}
B.~Aharmim et~al., Phys. Rev. C, {\bf 88}, 025501 (2013).

\bibitem{Collaboration:2011nga}
G.~Bellini et~al., Phys. Rev. Lett., {\bf 108}, 051302 (2012).

\bibitem{Eguchi:2002dm}
K.~Eguchi et~al., Phys.Rev.Lett., {\bf 90}, 021802 (2003).

\bibitem{Wendell:2010md}
R.~Wendell et~al., Phys.Rev., {\bf D81}, 092004 (2010).

\bibitem{Adamson:2013whj}
P.~Adamson et~al., Phys. Rev. Lett., {\bf 110}, 251801 (2013).

\bibitem{Abe:2012jj}
K.~Abe et~al., Phys. Rev. Lett., {\bf 110}, 181802 (2013).

\bibitem{An:2012eh}
F.~P. An et~al., Phys. Rev. Lett., {\bf 108}, 171803 (2012).

\bibitem{Abe:2012tg}
Y.~Abe et~al., Phys. Rev. D, {\bf 86}, 052008 (2012).

\bibitem{Ahn:2012nd}
J.~K. Ahn and other, Phys. Rev. Lett., {\bf 108}, 191802 (2012).

\bibitem{Ahn:2006zza}
M.~H. Ahn et~al., Phys. Rev. D, {\bf 74}, 072003 (2006).

\bibitem{numurun4}
K.~Abe et~al., Phys.Rev.Lett., {\bf 112}, 181801 (2014).

\bibitem{Abe:2011sj}
K.~Abe et~al., Phys. Rev. Lett., {\bf 107}, 041801 (2011).

\bibitem{Abe:2013xua}
K.~Abe et~al., Phys. Rev. D, {\bf 88}, 032002 (2013).

\bibitem{Abe:2013hdq}
K.~Abe et~al., Phys.Rev.Lett., {\bf 112}, 061802 (2014).

\bibitem{Adamson:2013ue}
P.~Adamson et~al., Phys. Rev. Lett., {\bf 110}, 171801 (2013).

\bibitem{Agafonova:2010dc}
N.~Agafonova et~al., Physics Letters B, {\bf 691}(3), 138 -- 145 (2010).

\bibitem{Maki:1962mu}
Ziro Maki, Masami Nakagawa, and Shoichi Sakata, Prog.Theor.Phys., {\bf 28},
  870--880 (1962).

\bibitem{T2KLOI}
Letter of intent: Neutrino oscillation experiment at {JHF} (2003),
\newblock \url{http://neutrino.kek.jp/jhfnu/loi/loi_JHFcor.pdf}.

\bibitem{Abe:2011ks}
K.~Abe et~al., Nucl. Instrum. Methods Phys. Res., Sect. A, {\bf 659}, 106--135,
  See Figure 16 for a schematic diagram of the ND280 detector. (2011).

\bibitem{novaTDR}
No$\nu$a technical design report (),
\newblock \url{http://www-nova.fnal.gov/nova_cd2_review/tdr_oct_23/tdr.htm}.

\bibitem{Arafune:1997hd}
J.~Arafune, M.~Koike, and J~Sato, Phys.Rev., {\bf D56}, 3093--3099 (1997).

\bibitem{Beringer:1900zz}
J.~Beringer et~al., Phys. Rev. D, {\bf 86}, 010001 (2012).

\bibitem{PhysRevD.87.012001}
K.~Abe et~al., Phys. Rev. D, {\bf 87}, 012001 (2013).

\bibitem{Abe2012}
K.~Abe et~al., Nucl. Instrum. Methods Phys. Res., Sect. A, {\bf 694}, 211--223
  (2012).

\bibitem{Assylbekov201248}
S.~Assylbekov et~al., Nucl.Instrum.Meth., {\bf A686}(0), 48 -- 63 (2012).

\bibitem{Abgrall:2010hi}
N.~Abgrall et~al., Nucl. Instrum. Methods Phys. Res., Sect. A, {\bf 637},
  25--46 (2011).

\bibitem{Amaudruz:2012pe}
P.A. Amaudruz et~al., Nucl. Instrum. Methods Phys. Res., Sect. A, {\bf 696},
  1--31 (2012).

\bibitem{Aoki:2012mf}
S.~Aoki et~al., Nucl.Instrum.Meth., {\bf A698}, 135--146 (2013).

\bibitem{Ashie:2005ik}
Y.~Ashie et~al., Phys. Rev. D, {\bf 71}, 112005 (2005).

\bibitem{Abgrall:2011ae}
N.~Abgrall et~al., Phys. Rev. C, {\bf 84}, 034604 (2011).

\bibitem{Abgrall:2011ts}
N.~Abgrall et~al., Phys. Rev. C, {\bf 85}, 035210 (2012).

\bibitem{Abe:2012gx}
K.~Abe et~al., Phys. Rev. D, {\bf 85}, 031103 (2012).

\bibitem{PDG2013}
J.~Beringer et~al. (Particle Data~Group), Phys. Rev., {\bf D86}, 010001, 2013
  partial update for the 2014 edition (2012).

\bibitem{DB13}
F.~P. An et~al., Chinese Physics C, {\bf 37}(1), 011001 (2013).

\bibitem{globes1}
P.~Huber, M.~Lindner, and W.~Winter, Computer Physics Communications, {\bf
  167}(3), 195 -- 202 (2005).

\bibitem{globes2}
P.~Huber, J.~Kopp, M.~Lindner, M.~Rolinec, and W.~Winter, Computer Physics
  Communications, {\bf 177}(5), 432 -- 438 (2007).

\bibitem{novaplotsandfigures}
\url{http://www-nova.fnal.gov/plots_and_figures/plot_and_figures.html} ().

\bibitem{novaneutrino2012}
\url{http://nova-docdb.fnal.gov/cgi-bin/ShowDocument?docid=7546} ().

\bibitem{novaheavyquarks2012}
\url{http://nova-docdb.fnal.gov/cgi-bin/ShowDocument?docid=7552} ().

\bibitem{PhysRevD.86.113011}
X.~Qian, A.~Tan, W.~Wang, J.~J. Ling, R.~D. McKeown, and C.~Zhang, Phys. Rev.
  D, {\bf 86}, 113011 (Dec 2012).

\end{thebibliography}
\end{document}